\documentclass[12pt,a4paper,twoside,normalheadings]{scrbook}

\newcommand{\captionfonts}{\small}
\makeatletter  % Allow the use of @ in command names
\long\def\@makecaption#1#2{%
  \vskip\abovecaptionskip
  \sbox\@tempboxa{{\captionfonts #1: #2}}%
  \ifdim \wd\@tempboxa >\hsize
    {\captionfonts #1: #2\par}
  \else
    \hbox to\hsize{\hfil\box\@tempboxa\hfil}%
  \fi
  \vskip\belowcaptionskip}
\makeatother   % Cancel the effect of \makeatletter

\makeatletter                                  %% MAKRO
\newcommand{\thefigurename}{Fig.}              %% Standard: Figure
\def\fnum@figure{\textit{\textbf{              %% Neudefinition fnum@figure
                \thefigurename\ \thefigure}}}
\newcommand{\thetablename}{Tab.}               %% Standard: Table
\def\fnum@table{\textit{\textbf{               %% Neudefinition fnum@table
                \thetablename\ \thetable}}}

\newlength{\offset}
\newcommand{\LL}[1]{\smash{\lower 1.5ex \hbox{#1}}}

\setcapindent{0mm}

\usepackage[pdftex]{graphics}
\usepackage[english]{babel}
\usepackage{german}
\usepackage{booktabs}
\usepackage{here}
\usepackage{textcomp}
\usepackage{amssymb}
\usepackage{amsmath}
\usepackage{graphicx}
\usepackage{picins}
\usepackage{setspace}
\usepackage{color}
\usepackage{wrapfig}
\usepackage{rotating}
\definecolor{DarkBlue}{rgb}{0,0,0.5}
\usepackage[colorlinks=false, linkcolor=DarkBlue, citecolor=DarkBlue, urlcolor=DarkBlue,
pdfauthor=Henning\ Gast\ ,
pdftitle=PEBS
pdfstartview=FitV,
pdfpagemode=None,
pdfkeywords=PEBS\ Dark\ Matter]
{hyperref}

\begin{document}

%~~~~~~~~~~~~~~~~~~~~~~~~~~~~~~~~
\pagestyle{empty}
\begin{titlepage} 
\vspace*{0.5cm}

{\centering \sf \Huge \bfseries
Towards a precise\\
measurement of the\\
cosmic-ray\\
positron fraction\\
}

\vspace{4.0cm}

{\centering \normalsize

% F"ur die ersten 3 Exemplare:
 {Der Fakult\"at f\"ur Mathematik, Informatik und Naturwissenschaften der\\
% Rheinisch-Westf\"alischen Technischen Hochschule Aachen\\
RWTH Aachen University\\
 vorgelegte Dissertation zur Erlangung des akademischen Grades eines\\
 Doktors der Naturwissenschaften \\ 
 }
}

%}

\vspace{1.5cm}
 
{\centering \normalsize von \\ }

\vspace{1.5cm}

{\centering \normalsize
Diplom--Physiker\\
Henning Gast\\
\vspace{1cm}
aus D\"usseldorf \\
\vspace{1cm}
} 
 
\vspace{2cm}

%\begin{tabbing} 
%\hspace{2.5cm} \= Berichter:  \qquad \= Universit\"atsprofessor Dr.~Stefan Schael \\
%               \>                    \> Universit\"atsprofessor Dr.~Who \\
%\\ 
%               \> Tag der m\"undlichen Pr\"ufung: \qquad wann\\
%\end{tabbing}

\end{titlepage}

\cleardoublepage
% the following are from the koma-script manual
% to kill the f&%$#@ page number on the 1st toc page
\begin{center}
{\bf Abstract}
\end{center}
\begin{small}
This thesis deals with detector concepts aiming at a precise
measurement of the cosmic-ray positron fraction extending to an as yet
unreached range of energy. The indirect search for dark matter is the
main motivation for this endeavour.\\
\par
\noindent The evolution and large-scale structure of the Universe is described
by the Hot Big Bang model which is well supported by observational
evidence. In this model, roughly a quarter of the energy density of
the Universe must be made up of the elusive dark matter. Evidence for
its existence comes from, for example, the observation of galactic rotation curves,
dynamics of clusters of galaxies, and the pattern of anisotropies in the cosmic
microwave background. While the nature of dark matter is as yet
unknown, the most popular candidate for its constituents is the neutralino included
in supersymmetric extensions to the standard model of particle
physics. Neutralinos annihilating in the Galactic halo are considered as
a potential primary source of cosmic-ray positrons. For the
calculation of the expected secondary background of positrons, a
common cosmic-ray propagation model has been adopted and its uncertainties
have been assessed. The cosmic-ray positron fraction data available so far
indicate an excess over the expectation for purely secondary
production, a trend recently confirmed and intensified by measurements
of the PAMELA satellite detector. The AMS-02 detector will be ready
for installation on the International Space
Station in 2010 and is designed to perform precision spectroscopy of
many different cosmic-ray species including positrons.\\
\par
\noindent Here, a design concept for a new detector, called
Positron Electron Balloon Spectrometer (PEBS), is presented. Intended
for a measurement of the cosmic-ray positron fraction
on one or more flights at high altitude using a
long-duration balloon, PEBS will have an unprecedentedly high acceptance of
almost $0.4\,\mathrm{m}^2\,\mathrm{sr}$. A first launch could take
place in~2012. Using a superconducting
magnet to create a mean magnetic field of $0.8\,\mathrm{T}$ and a
scintillating fibre tracker with silicon photomultiplier readout, it will
allow reliable charge-sign and momentum measurements up to at least
$100\,\mathrm{GeV}$. The enormous challenge of reliably identifying positrons in front of
the vast proton background is tackled by a combination of two
independent subdetectors for particle identification. The first one is
an electromagnetic calorimeter which will consist of layers of tungsten
absorber interleaved with scintillator bars, read out by
silicon photomultipliers. The second one is a transition radiation
detector (TRD) similar to the one built for AMS-02, made of an irregular fleece
radiator followed by thin-walled detection tubes.\\
\par
\noindent A detailed Monte Carlo simulation of PEBS, based on Geant4,
was created to study the expected performance of the detector, along with
a reconstruction and analysis suite. The simulation predicts a
momentum resolution of $18\,\%$ for $100\,\mathrm{GeV}$ positrons. For the same energy, the
calorimeter is predicted to have an energy resolution of $6\,\%$ and a proton rejection of 3000 at
$80\,\%$ positron efficiency. The transition radiation detector will
provide an additional rejection factor of~700, again with $80\,\%$ positron
efficiency.
Using testbeam data acquired with a prototype for the
AMS-02-TRD, the accuracy of the simulation of transition radiation and
ionisation losses provided by Geant4 was
studied. Excellent agreement was found between the transition
radiation spectra in data and simulation. Small discrepancies at the
$25\,\%$-level are
present in the tails of the proton energy loss spectra but this makes
the predicted proton rejections uncertain by a factor of two.\\
\par
\noindent In a series of testbeam measurements, the proof of principle
was established for the scintillating fibre tracker with silicon
photomultiplier readout. The intrinsic spatial resolution achieved at the
current level of design is $70\,\mu\mathrm{m}$.\\
\par
\noindent Neutralino dark matter was studied in the minimal
supergravity grand unification (mSUGRA) model. Assuming that
neutralino annihilations are enhanced by boost factors taken from best
fits to the positron fraction data published so far, both PEBS and
AMS-02 will be capable of substantially constraining mSUGRA parameter
space. It is shown that a moderately good fit to the high-energy PAMELA data can
be obtained in the mSUGRA model as well. At the same time, the
low-energy PAMELA data may hint at charge-sign dependent solar
modulation effects.
\end{small}

\cleardoublepage
{\selectlanguage{german}
\begin{center}
{\bf Zusammenfassung}
\end{center}
\begin{small}
In dieser Arbeit werden Detektorkonzepte vorgestellt, die eine
pr\"azise Vermessung des Positronenanteils in der kosmischen Strahlung
bis hin zu bislang unerreichten Energiebereichen zum Ziel haben.\\
\par
\noindent Die Entwicklung und die gro"sr\"aumige Struktur des Universums
werden von dem Ur\-knall\-mo\-dell beschrieben, das durch vielf\"altige
Beobachtungen gest\"utzt wird. In diesem Modell macht die
r\"atselhafte dunkle Materie etwa ein Viertel der gesamten
Energiedichte des Universums aus. Hinweise auf ihre Existenz finden
sich zum Beispiel in galaktischen Rotationskurven, der Dynamik von
Galaxienhaufen und dem Muster der Anisotropien in der kosmischen
Hintergrundstrahlung. Obwohl die Natur der dunklen Materie nach wie
vor unbekannt ist, sind die in supersymmetrischen Erweiterungen des
Standardmodells der Teilchenphysik enthaltenen Neutralinos der
beliebteste Kandidat. Neutralinos, die im galaktischen Halo
zerstrahlen, werden hier als m\"ogliche prim\"are Quellen f\"ur
Positronen in der kosmischen Strahlung betrachtet. F\"ur die
Berechnung des erwarteten sekund\"aren Untergrundes wurde
ein gebr\"auchliches Modell f\"ur die Propagation kosmischer Strahlung
angenommen und seine Unsicherheiten abgesch\"atzt. Die bisher zum
Positronenanteil in der kosmischen Strahlung vorliegenden Daten zeigen
einen \"Uberschuss im Vergleich zu der Erwartung aus rein sekund\"arer Erzeugung, ein
Trend, der k\"urzlich durch Messungen des satellitengest\"utzten
Detektors PAMELA best\"atigt und verst\"arkt wurde. Der Detektor
AMS-02 wird im Jahre~2010 f\"ur die Befestigung an der internationalen
Raumstation bereit stehen. Er wurde f\"ur die
Pr\"azisionsspektroskopie vieler verschiedener Teilchensorten in der
kosmischen Strahlung einschlie"slich Positronen entwickelt.\\
\par
\noindent 
In dieser Arbeit wird ein Entwicklungskonzept f\"ur einen neuen Detektor
vorgestellt, genannt Positron Electron Balloon
Spectrometer~(PEBS). PEBS hat eine Messung des Positronenanteils in der
kosmischen Strahlung bei einem oder mehreren Fl\"ugen mit einem
Langzeit-H\"ohenballon in gro"ser Flugh\"ohe zum Ziel, mit einer
bisher unerreicht gro"sen Apertur von
$0,\!4\,\mathrm{m}^2\,\mathrm{sr}$. Der erste Start k\"onnte im
Jahr~2012 stattfinden. Durch Benutzung eines supraleitenden Magneten,
der ein mittleres Feld von $0,\!8\,\mathrm{T}$ erzeugt, und einem aus
szintillierenden Fasern, die von Si\-li\-zi\-um-Pho\-to\-ver\-viel\-fachern
ausgelesen werden, bestehenden Spurdetektor wird PEBS verl\"assliche
Messungen von Impuls und Ladungsvorzeichen bis mindestens
$100\,\mathrm{GeV}$ erm\"oglichen. Die enorme Herausforderung, die
darin besteht, Positronen vor dem \"uberw\"altigenden Untergrund an Protonen
verl\"asslich zu identifizieren, wird mit einer Kombination zweier
Subdetektoren zur Teilchenidentifikation angegangen. Der erste ist ein
elektromagnetisches Kalorimeter, das aus Wolframlagen bestehen wird,
die von Szintillatorbarren gefolgt werden, die mit
Silizium-Photovervielfachern ausgelesen werden. Der zweite ist ein
\"Ubergangsstrahlungsdetektor (TRD), der dem f\"ur AMS-02 gebauten
\"ahnelt und aus einem irregul\"aren Faserradiator besteht, der von
d\"unnwandigen Detektionsr\"ohrchen gefolgt wird.\\
\par
\noindent 
Eine detaillierte Monte Carlo-Simulation von PEBS wurde
erschaffen, ebenso wie ein Programmpaket f\"ur Rekonstruktion und
Analyse. Die Simulation basiert auf Geant4 und erm\"oglicht es, die zu
erwartende Leistungsf\"ahigkeit von PEBS zu studieren. Sie sagt eine
Impulsaufl\"osung von $18\,\%$ f\"ur $100\,\mathrm{GeV}$ Positronen
vorher. F\"ur das Kalorimeter wird dabei eine Energieaufl\"osung von $6\,\%$
und eine Protonenunterdr\"uckung
um 3000 bei einer Positron-Effizienz von $80\,\%$ vorhergesagt. Der
\"Ubergangsstrahlungsdetektor wird einen weiteren Faktor~700
beitragen, ebenso mit $80\,\%$ Positron-Effizienz. Mit Hilfe von
Teststrahldaten, die mit einem Prototypen f\"ur den AMS-02-TRD
aufgenommen wurden, wurde die Genauigkeit der Simulation von
\"Ubergangsstrahlung und Ionisationsverlusten mit Geant4
studiert. Hervorragende \"Ubereinstimmung zwischen Daten und
Simulation wurde f\"ur die Spektren der \"Ubergangsstrahlung
gefunden. Kleine Abweichungen auf dem Niveau von $25\,\%$ treten in den
Ausl\"aufern der Energieverlustspektren von Protonen zu Tage und
f\"uhren zu einer Unsicherheit der vorhergesagten
Protonunterdr\"uckungsfaktoren von etwa einem Faktor zwei.\\
\par
\noindent 
In einer Reihe von Teststrahlmessungen wurde die Machbarkeit des
Spurdetektors aus szintillierenden Fasern, die von
Si\-li\-zi\-um-Pho\-to\-ver\-viel\-fachern 
ausgelesen werden, gezeigt. Die intrinsische Ortsaufl\"osung, die auf dem
derzeitigen Entwicklungsstand erreicht wird, betr\"agt $70\,\mu\mathrm{m}$.\\
\par
\noindent 
Aus Neutralinos bestehende dunkle Materie wurde im Rahmen des mSUGRA~(minimal
supergravity grand unification)-Modells studiert. Unter der Annahme,
dass die Zerstrahlung von Neutralinos um einen Faktor erh\"oht ist,
der sich aus besten Anpassungen an die bisher ver\"offentlichten
Positrondaten ergibt, werden sowohl AMS-02 als auch PEBS in der Lage
sein, den Parameterraum des mSUGRA-Modells betr\"achtlich
einzuschr\"anken. Es wird gezeigt, dass im mSUGRA-Modell eine einigerma"sen gute
Beschreibung der PAMELA-Daten zu hohen Energien m\"oglich
ist. Gleichzeitig k\"onnten die PAMELA-Daten bei niedrigen Energien
Hinweise auf ladungsabh\"angige solare Modulationseffekte geben.
\end{small}
}

\cleardoublepage
\pagestyle{empty}
\renewcommand*{\chapterpagestyle}{empty}
\setcounter{page}{1}
\tableofcontents
\cleardoublepage
\pagestyle{headings}
\renewcommand*{\chapterpagestyle}{plain}

%~~~~~~~~~~~~~~~~~~~~~~~~~~~~~~~~
\pagenumbering{arabic}

\setlength{\parindent}{0cm}

%~~~~~~~~~~~~~~~~~~~~~~~~~~~~~~~~
\chapter{Introduction}
What are the constituents of matter and how do they interact? Is
the Universe static, or else, how did it evolve into its current
state? These two questions arguably stand out as especially fascinating
among the problems that fundamental science seeks to solve.\\
After a long series of experiments, including those conducted with
enormous detectors used to analyse the debris from collisions of
particles smashed together at breakneck speeds at huge accelerators, the discipline of
particle physics has emerged with a theory -- simply called the
standard model -- describing the interactions of
elementary particles in the framework of quantum field theory. So far,
its predictions have been found to be of exquisite
precision. Electrons and quarks make up the atoms that our planet and
every living being on it are made of, neutrinos help to maintain the
equilibrium inside our star over billions of years, and photons,
gluons, and the $W$ and $Z$ bosons mediate the interactions of these
building blocks of matter, as dictated by a set of simple mathematical
symmetries.\\
At the same time, based on a host of precise astronomical observations
covering many different wavelengths, cosmology has come a long way from
the old picture that saw Earth at the centre of a static
universe. Instead, the evolution of the Universe has been traced
back to within the tiniest fraction of a second from its coming into
being in the so-called Big Bang. It must have started from an initially
hot, dense, and almost perfectly homogeneous state and then cooled
down and expanded over the eons until stars, galaxies, and eventually
planets could form. The expansion of the Universe is
governed by its matter and energy content as
described in the framework of General Relativity.\\
But there are problems. Imprinted on the ubiquitous microwave
radiation filling the cosmos is a snapshot of the acoustic
oscillations in the early Universe, yet their pattern hints at the
presence of additional matter interacting with the plasma only by
gravitational interaction.
The structure of the Universe on large scales could only have formed
in gravitational wells much deeper than calculated from the
distribution of the matter that astronomers can see.
The observed motions of stars and gas around galaxies cannot
be explained by the laws of gravity taking only the attraction of the
luminous matter into account.
These observations, among others, have led to the conclusion that
there must be large amounts of non-luminous, only weakly interacting,
matter out there in the Universe which only betrays itself by its own
mass.
Its energy density today must exceed that
of ordinary baryonic matter by roughly a factor of six. This is
known as the dark matter problem. The existence of dark matter seems
well established by now, but its nature remains elusive.\\
\par
The hunt for dark matter is on. At the Large Hadron Collider now
commissioned near Geneva, physicists hope to create dark matter
particles from energy. In laboratories deep under ground, extremely
sensitive detectors are employed to look for collisions of dark matter
particles with nuclei. In the third approach, the indirect search for
dark matter, one tries to identify
the remnants of pairwise annihilations of dark matter particles taking
place across the Galaxy and its halo. Antiparticles of high energy are expected to be
created in these collisions and as they
have no known primary source in the Galaxy, they can be used as messengers of possible dark matter annihilations.
Positrons are especially promising as
they are stable and thus able to bridge vast distances in the Galaxy and typically
carry away a significant fraction of the dark matter particles' mass.
As a consequence, they stand out from the astrophysical backgrounds more
easily. After diffusing through the Galactic
medium, they may reach Earth as a small admixture to the cosmic
rays.\\
Cosmic rays are a stream of energetic elementary particles and
nuclei reaching our planet from the skies. Apart from maybe shedding
some light on the grand questions outlined above, their study allows us to
draw conclusions about our cosmic surroundings. For example, the
spectra and composition of cosmic rays are sensitive to the structure
of the Galaxy and the interactions with the solar wind and magnetic
fields. In addition, the study of cosmic rays has had a big influence
on particle physics as many elementary particles were first identified
in the cosmic rays. As all but the most high-energetic cosmic rays are
absorbed by Earth's atmosphere long before they can reach the ground,
an instrument trying to measure cosmic rays must be deployed either in
space or on a high-altitude balloon circling at the outskirts of the
atmosphere.\\
\par
This thesis deals with detector concepts aiming at a precise
measurement of the cosmic-ray positron fraction extending to an as yet
unreached range of energy. The first steps have been made by a series
of small detectors flying on high-altitude balloons over the past
decades. Their results have sparked excitement because they indicate
an excess in the relative positron flux over what is expected from
purely secondary production in collisions of protons and nuclei with
interstellar matter. It has been speculated that this excess might be
due to dark matter annihilations. Now, the PAMELA detector is orbiting Earth and
taking data, while the AMS-02 detector is nearing completion and
waiting for a flight to the International Space Station. The PAMELA
positron fraction data have recently become available and confirm the
trend of the earlier measurements.\\
The main focus of this work is a design study for a new experiment,
called Positron-Electron Balloon Spectrometer (PEBS), specifically
designed to measure the cosmic-ray positron fraction in the energy
range from a few hundred MeV up to $100\,\mathrm{GeV}$ and more. The
main motivation is derived from the indirect search for dark matter,
but as briefly touched upon above, the range of topics profiting from
a precise measurement is much wider. As a balloon-borne detector, PEBS
could be realised at a fraction of the cost of a space experiment the
same size. In addition, it could be salvaged after the flight,
allowing for post-flight calibration and checks, and possibly be flown
multiple times. Various space agencies around the globe have been
operating high-altitude balloon programmes reaching flight durations
of up to forty days. The remaining atmosphere would make the data
somewhat harder to interpret compared to a space experiment, but the
necessary corrections turn out to be small and are well understood.\\
\par
Several obstacles have to be overcome before a precise measurement of
the cosmic-ray positron fraction, i.e.~the ratio of the positron flux
to the total electron flux, becomes feasible. First of all, the
fluxes of electrons and positrons are low and drop steeply towards higher energies so that a
precise measurement requires a large geometric acceptance or long
exposures. On the other hand, the size of the detector and its
measurement time are limited by the constraints that the carrier
system imposes on the weight budget
and hence the amount of consumables that can be carried. Moreover, the
proton flux exceeds the electron and positron fluxes by many orders of
magnitude. To reliably identify positrons and electrons in front of
this enormous background, both AMS-02 and PEBS rely on the combination
of two independent subdetectors, a transition radiation detector and
an electromagnetic calorimeter. The separation of electrons and
positrons requires a clean measurement of the particle momentum. To
that end, the two experiments employ a superconducting magnet inside a
helium cryostat and a tracking device. While a conventional silicon
tracker is used in AMS-02, PEBS features an innovative scintillating fibre
tracker with silicon photomultiplier readout that is currently under
development. Silicon photomultipliers are novel devices promising
single-photon detection combined with desirable properties such as
high gain, compactness, auto-calibration and insensitivity to magnetic
fields. The key asset of such a tracking device is its conceptual
simplicity that comes along with low weight and low material
budget.\\
\par
In the design of a new particle detector, Monte Carlo
simulations are of utmost importance allowing one to study the
expected behaviour of the individual components and the entire
detector by simulating it on the computer before even the first
prototype is built. As the design progresses, prototype measurements
in test beams are necessary to verify the simulations and to
demonstrate the proof of principle for the detector design.\\
\par
This thesis is organised as follows. Chapter~\ref{chapter:general}
contains a brief review of Big Bang cosmology and the evidence for the
existence of dark matter. The most popular candidates are presented
and one candidate, the neutralino of the mSUGRA model, is described in
some more detail. After a short description of the physics of cosmic
rays and an explanation of the effects important for their detection
on Earth, the particular propagation model used to study the projected
performance of AMS-02 and PEBS is elaborated on and its uncertainties
are assessed.
In chapter~\ref{chapter:pebs_design}, the designs of PAMELA, AMS-02,
and PEBS are sketched, with a focus on the latter.\\
Chapter~\ref{chapter:pebs_design_study} contains a detailed
description of the design study for PEBS based on Monte Carlo
simulations as well as the reconstruction and analysis techniques used
to extract the projected performance of the detector, presented at the
end of the chapter.
Several module prototypes for the scintillating fibre tracker for PEBS
have been subjected to a proton testbeam at CERN over the past three
years and these measurements are described in
chapter~\ref{chapter:pebs_testbeam}. They provide important input to
the Monte Carlo simulations.
The capability of PEBS and AMS-02 to constrain model parameters in the
case of a discovery is studied in chapter~\ref{chapter:susyscan} for
the example of the mSUGRA model. This is contrasted with the ability
of the presently available cosmic-ray data to constrain this model.
Some concluding remarks and an outlook are given in chapter~\ref{chapter:conclusions}.

\chapter{Cosmic rays and dark matter}
\label{chapter:general}
In this chapter, one of the biggest questions in modern physics, the
nature of dark matter, is explained and it is shown how cosmic-ray
physics might help to solve it. First, the standard model of cosmology
is reviewed. This is vital to understand how certain candidates for
dark matter could have come into existence and populate the Galaxy
today. It is then shown that observations at very different
cosmological scales demonstrate that some form of dark matter must
exist in the Universe today and theories for the most popular
candidates are briefly reviewed. The neutralino is a candidate
predicted by supersymmetric extensions to the standard model of
particle physics. As it was chosen to study the projected performance
of PEBS for the case that the mSUGRA model is realised in nature, this
model and its dark matter candidate are described next. Then, the
cosmic-ray physics needed in the context of the PEBS mission is
introduced and in particular, the propagation model used for the treatment
of the secondary background is examined. Finally, a possible hint at
charge-sign dependent solar modulation of cosmic rays in the recently published
PAMELA data is explored.

\section{Big Bang cosmology}
\label{sec:bb}
From a fruitful interplay of astronomical observations and theoretical
developments over the past decades, a quantitative understanding of
the history of the Universe has
emerged~\cite{ref:kolb,ref:bergstroem_goobar,ref:samtleben}. The
standard model of cosmology is the Hot Big Bang model according to
which the Universe came into existence roughly 14~billion years ago
and started from a very compressed and therefore very hot state. The
expansion of the Universe caused the temperature to decrease.
At first, the energy density was divided among matter -- in
the form of the known elementary particles and dark matter -- and radiation
-- in the form of photons and neutrinos -- and an exotic form of energy called dark energy, which plays the
role of a fluid with negative pressure. Matter and radiation were in
thermal equilibrium, with the radiation dominating the expansion
initially. Around $10^{-6}\,\mathrm{s}$ after the Big Bang, the
temperature dropped to $10^{13}\,\mathrm{K}$, corresponding to an energy scale
of $1\,\mathrm{GeV}$, and thus, protons and neutrons formed from the
quark-gluon plasma. After $1\,\mathrm{s}$ the temperature reached the
level of the nuclear binding energy, $1\,\mathrm{MeV}$. During the era
of nucleosynthesis that began now, the light elements $^2\mathrm{H}$,
$^3\mathrm{He}$, $^4\mathrm{He}$, and $^7\mathrm{Li}$ were
produced. The next important era was reached roughly $300\,000$~years
into the expansion, when the temperature dropped to the level of the
atomic binding energies, around $1\,\mathrm{eV}$. At that point,
protons and electrons combined to form hydrogen atoms and the Universe
became transparent. The radiation emitted from the surface of last
scattering permeates the Universe today and is known as the cosmic
microwave background. It can be observed today to follow an almost
perfect black-body spectrum, at the redshifted temperature of
$2.7\,\mathrm{K}$. The Universe then started to be matter-dominated,
and eventually, the baryonic matter began to form complex structures,
stars and galaxies and clusters of galaxies that dominate the
large-scale structure of the Universe today. It is believed
that dark matter has played an important role in the structure formation, building
the potential wells that the baryonic matter would then collapse
into. Later, emission from the first generation of stars and
supernovae seems to have reionised the Universe temporarily. While the
gravitational pull of matter led to a deceleration of the expansion
during these stages, we are now in an era where the energy density is
dominated by the dark energy, which tends to cause a cosmic acceleration.\\
\par
The observational evidence for the Hot Big Bang Model is
overwhelming. The first hint that the Universe is not static came from
Edwin Hubble's observation that most galaxies are receding away
from us. The velocity $v$ of a given galaxy is on average proportional to
its distance $d$, $v=Hd$. The constant of proportionality is called
Hubble constant and has a value of
$H=100\,h\,\mathrm{km}\,\mathrm{s}^{-1}\,\mathrm{Mpc}^{-1}$ and
$h=0.73\,\pm\,0.03$~\cite{ref:pdg}. The expansion can be understood in
the framework of General Relativity. The Einstein equations relate the
energy-momentum tensor to the properties of space-time. When the
matter content of the Universe is modelled as a perfect fluid, the
Friedmann equation relates the energy density
$\rho$ of the Universe to the time development of its scale factor
$R$:
\begin{equation}
\label{eq:friedmann}
H^2=\left(\frac{\dot{R}}{R}\right)^2=\frac{8\pi{}G\rho}{3}-\frac{k}{R^2}+\frac{\Lambda}{3}
\end{equation}
where $G$ is the gravitational constant, $k=0,\pm{}1$ describes the
curvature of space on large scales, and the cosmological constant
$\Lambda$ is connected to dark energy.\\
The cosmic microwave background was first observed by Penzias and
Wilson in 1965~\cite{ref:penzias}. Its existence had already been predicted for a Hot Big Bang Model
by Gamow more than 20~years in advance. Before recombination, radiation
and matter were coupled via the reaction
$H+\gamma\leftrightarrow{}p+e^-$. When the temperature dropped well
below the ionisation energy of hydrogen, due to the expansion of the
Universe, the Universe became transparent, the photons decoupled from
matter and are travelling on geodesics ever since. These photons
permeate the Universe today.\\
The relative abundance of light elements in the Universe agrees
accurately with what would be synthesised in an initially hot,
expanding universe. While it is generally accepted that the heavier
elements are produced in stars towards the final stages of their
lifetimes, this mechanism cannot explain the fact that the mass
fraction of $^4\mathrm{He}$ is about $24\,\%$. In contrast to this,
taking into account the age of the Galaxy, its luminosity and the
energy yield of hydrogen fusion, it is found that hydrogen burning can
only account for a helium abundance of roughly $1\,\%$. The theory of
Big Bang nucleosynthesis predicts the abundances of the light nuclei $^2\mathrm{H}$,
$^3\mathrm{He}$, $^4\mathrm{He}$, and $^7\mathrm{Li}$, synthesised at
the end of the first three minutes of the Universe, as a function of
the baryon-to-photon ratio $\eta_B$. The observed abundances agree
with the predicted ones, from the number ratio
$^4\mathrm{He}/\mathrm{H}\sim{}0.08$ down to
$^7\mathrm{Li}/\mathrm{H}\sim{}10^{-10}\,$~\cite{ref:pdg}.\\
Lastly, the oldest objects found in the Universe -- globular
clusters of stars and some radioactive isotopes -- do not seem to
exceed an age of around 13~billion years. This indicates that the
Universe is of finite age and is consistent with the age of the
Universe as found in the standard model of cosmology.\\
\par
The contribution of a given component $i$ with energy density $\rho_i$
to the total energy density of the Universe is usually given as a
fraction of the critical density $\rho_c=3H_0^2/8\pi{}G$ that leads to
a Universe whose expansion comes to a halt asymptotically and that is
flat on large scales:
\begin{equation}
\label{eq:omega}
\Omega_i=\frac{\rho_i}{\rho_c}
\end{equation}
From a combination of observations of the anisotropy in the cosmic
microwave background by the WMAP experiment~\cite{ref:wmap_descr}, the luminosity distances to type Ia supernovae~\cite{ref:sn1a},
and the baryon acoustic oscillations in the distribution of galaxies~\cite{ref:bao},
the key parameters of the Hot Big Bang Model are determined to be~\cite{ref:wmap_obs,ref:wmap_cosm}
$\Omega_B=0.0462\,\pm\,0.0015$ for the baryon density,
$\Omega_\mathrm{dm}=0.233\,\pm\,0.013$ for the dark matter density,
$\Omega_\Lambda=0.721\,\pm\,0.015$ for the dark energy density and
$H_0=(70.1\,\pm\,1.3)\,\mathrm{km}\,\mathrm{s}^{-1}\,\mathrm{Mpc}^{-1}$
for the Hubble constant. The total density is found to be
$\Omega_\mathrm{tot}=1.0052\,\pm\,0.0064$, consistent with the
critical value.\\
\par
An important aspect of the Hot Big Bang model is the freeze-out of
heavy particles. Assuming that a massive particle species of mass $m$
never left thermal equilibrium, the number of particles $N$ in a
comoving volume is $N\sim(m/T)^{3/2}\exp(-m/T)$ and therefore
negligible today if $m/T\gg{}1$. However, a so-called relic density of
massive particles can remain if their interaction rate $\Gamma$ drops below the
expansion rate $H$ of the Universe at some point in the expansion
history. As an example, a weakly interacting species
$\chi$ that is annihilated and created in the reaction
$\chi+\bar{\chi}\leftrightarrow{}X+\bar{X}$ is considered. The $X$ particles
are taken to be strongly interacting compared to the $\chi$'s and therefore are
kept in thermal equilibrium. The time evolution of the number density
$n_\chi$ in the comoving frame is then governed by the Boltzmann equation that can be
derived to be
\begin{equation}
\label{eq:boltzmann}
\frac{\mathrm{d}n_\chi}{\mathrm{d}t}=-3Hn_\chi-\langle\sigma|\mathbf{v}|\rangle(n_\chi^2-(n_\chi^\mathrm{EQ})^2)
\end{equation}
The first term on the right-hand side expresses a decrease in number
density that is due to the expansion of the Universe. The second term
describes annihilation of $\chi$ particles, proportional to
$\langle\sigma|\mathbf{v}|\rangle\,n_\chi^2$ and creation in the
back-reaction, proportional to
$\langle\sigma|\mathbf{v}|\rangle\,(n_\chi^\mathrm{EQ})^2$. Here,
$\langle\sigma|\mathbf{v}|\rangle$ denotes the thermally averaged
annihilation cross section for the reaction
$\chi+\bar{\chi}\rightarrow{}X+\bar{X}$ and $n_\chi^\mathrm{EQ}$ is
the equilibrium number density of $\chi$'s. Normalising $n_\chi$ to
the entropy density $s$, which can be shown to be conserved in a
volume of $R^3(t)$, and
introducing $Y_\chi\equiv{}n_\chi/s$ and $x\equiv{}m_\chi/T$,
(\ref{eq:boltzmann}) can be rewritten in the form
\begin{equation}
\label{eq:boltzmann_y}
\frac{x}{Y_\chi^\mathrm{EQ}}\frac{\mathrm{d}Y}{\mathrm{d}x}=-\frac{\Gamma_A}{H}\left(\left(\frac{Y_\chi}{Y_\chi^\mathrm{EQ}}\right)^2-1\right)
\end{equation}
where
$\Gamma_A=n_\chi^\mathrm{EQ}\langle\sigma|\mathbf{v}|\rangle$. This
shows that the evolution is governed by the factor $\Gamma_A/H$. Equation (\ref{eq:boltzmann_y})
can be solved numerically and the solution describes the freeze-out
of a massive particle species. At some value $x_f=m_\chi/T_f$ for the
freeze-out temperature $T_f$, the abundance $Y_\chi$ leaves the
equilibrium curve. The relic abundance will be higher the smaller the
annihilation cross section is. Generically,
an order-of-magnitude estimate is given by~\cite{ref:dmreview}
\begin{equation}
\label{eq:reldens_crosssection}
\Omega_\chi{}h^2\approx\frac{3\cdot{}10^{-27}\,\mathrm{cm}^3\mathrm{s}^{-1}}{\langle\sigma{}v\rangle}
\end{equation}

\section{Evidence for the existence of dark matter}
A variety of observations, both direct and indirect, from galactic to
cosmological scales, lead to the conclusion that a yet unknown form of
matter must exist which contributes significantly to the energy
density in the Universe. Although hints for its existence are
manifold, it has not been identified yet. This is known as the dark
matter problem. Any candidate for dark matter must be only weakly
interacting since it could not have evaded our attention otherwise. It
must be stable on cosmological timescales or it would have disappeared
from the cosmic stage long ago. In the following, the evidence for the existence of dark matter in the
Universe will be reviewed briefly~\cite{ref:pdg,ref:dmreview}.\\
\par
\begin{figure}[htb]
\begin{center}
\begin{tabular}{cc}
\includegraphics[width=0.52\textwidth,angle=0]{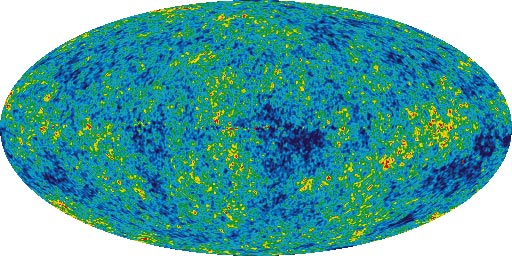}&
\includegraphics[width=0.45\textwidth,angle=0]{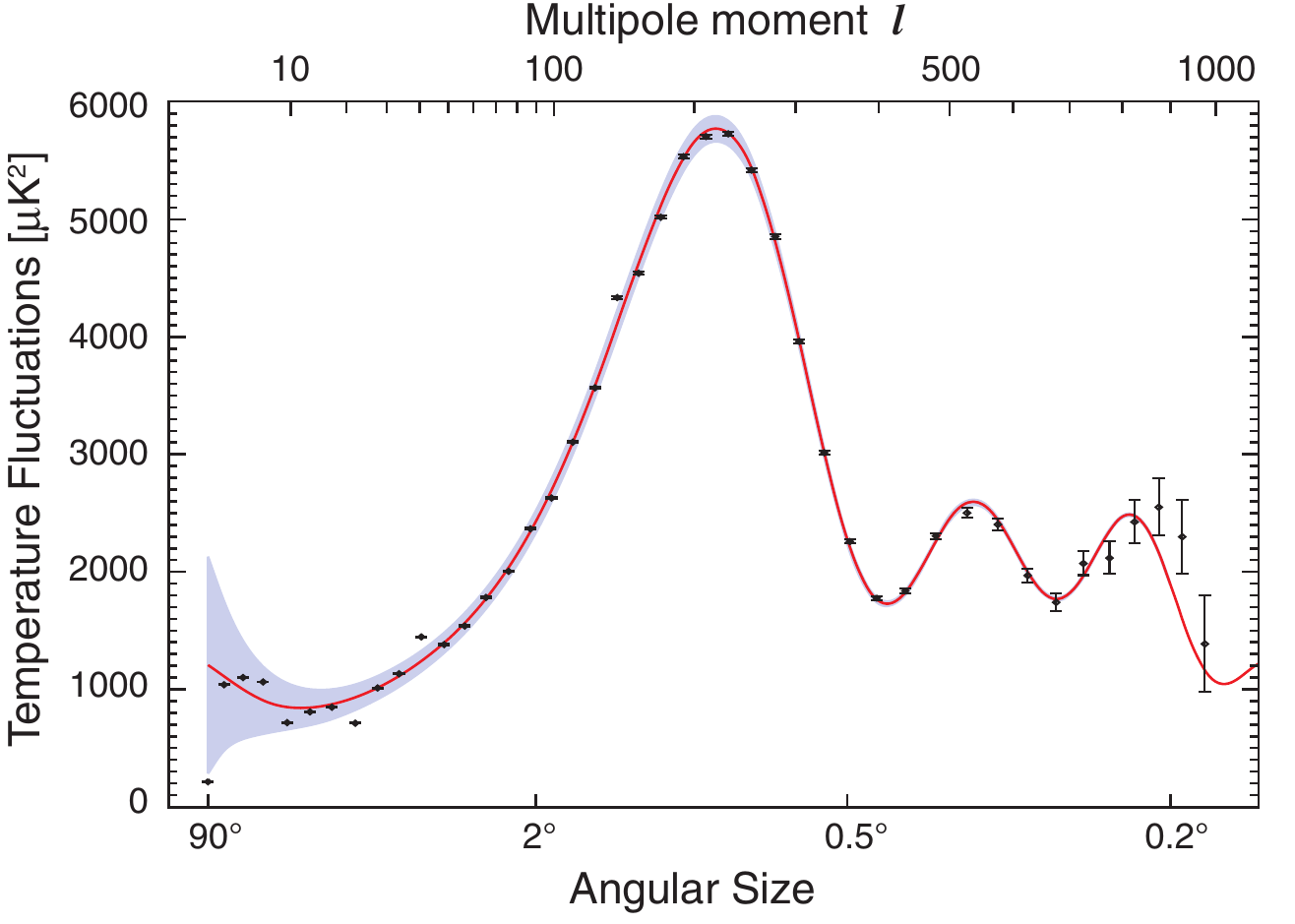}\\
\end{tabular}
\end{center}
\caption{{\it Left:} Sky map of the temperature anisotropies in the cosmic microwave background
measured by the WMAP experiment in five years. Red regions are warmer
and blue regions are colder by about $200\,\mu\mathrm{K}$. {\it
Right:} Corresponding power spectrum. Credit for both pictures:
NASA/WMAP Science Team~\cite{ref:wmap_obs}.}
\label{fig:wmap}
\end{figure}

The most direct evidence for the existence of
dark matter can be obtained at the galactic scale. Dark matter betrays itself in
practically all rotation curves of galaxies measured so far. The
rotation curve describes the rotational velocity $v(r)$ of objects at
a distance $r$ from the galactic centre around the centre. It is based
on measurements of the Doppler shift of suitable emission or
absorption lines. Most importantly, the observations can be extended
beyond the visible disk by looking at the CO and HI($21\,\mathrm{cm}$) line emissions of
gas clouds. If all matter in a galaxy were located in the luminous
disk, one would expect a rotation curve $v\propto{}r^{-1/2}$ outside
the disk according to Kepler's third law. Contrary to this, flat
rotation curves are observed, $v=\mathrm{const}$, consistent
with a mass distribution of $\rho\propto{}r^{-2}$ that extends beyond
the visible disk. It is assumed that this is due to dark matter forming halos
around the disks of galaxies.\\
On larger scales, observations of clusters of galaxies provide hints
for the existence of dark matter. This includes studies of weak
gravitational lensing of background galaxies by a cluster which
depends on the mass distribution inside the cluster. In addition, in a
virialized cluster, the peculiar velocities of the galaxies allow one
to trace the gravitational potential. The same is true for
measurements of the x-ray emissions of the hot gas in the cluster. The
mass-to-light ratios so obtained exceed the one measured in the solar
neighbourhood by more than an order of magnitude which implies the
presence of additional non-luminous matter.\\
On the cosmological scale, the discovery of tiny anisotropies in the
cosmic microwave background opened the door to a new era of precision
cosmology. According to the standard paradigm, small random
density fluctuations in the early Universe caused
gravitational instabilities which formed the seeds of the large-scale structures
observed today. The observed anisotropies then are a combination of a
snapshot of the density distribution at the time of decoupling and the
subsequent gravitational red- and blueshifting of photons leaving
over- or underdense regions. The sky map of the anisotropies in the cosmic microwave background has been
measured, e.g.~by the WMAP experiment over a period of five years (fig.~\ref{fig:wmap} {\it left}).\\
The basic idea for the understanding of the anisotropies at smaller
scales is the realisation that the early Universe must have resounded
with acoustic oscillations. They are created when fluctuations in
density cause photons and matter to fall into potential wells. The
infall will be slowed and eventually reversed by radiation pressure,
and the process will repeat to form an oscillation. The properties of the
oscillations will depend on the fractional contributions of the
various forms of energy to the overall energy content of the Universe,
and hence on the cosmological parameters. The anisotropies can be
characterised by their angular scales. For example, the size of a
causally connected region at the time of decoupling is given by
$H(z=z_\mathrm{rec})^{-1}$, where $z_\mathrm{rec}$ is the redshift at recombination.
The angle subtended by such a region today
is given by $\theta=1/d_AH(z=z_\mathrm{rec})$ and
defines the position of the so-called first acoustic peak. The angular
size distance $d_A$ depends on the expansion history of the Universe
and thus on the cosmological parameters. More specifically, the
anisotropy is
decomposed in spherical harmonics,
\begin{equation}
\label{eq:coeff}
\frac{\delta{}T(\mathbf{\hat{n}})}{T}=\sum\limits_{l=2}^\infty\sum\limits_{m=-l}^la_{lm}Y_{lm}(\theta,\phi)
\end{equation}
A term of index $l$ corresponds to fluctuations with typical
angular scale $\pi/l$. Information about the cosmological parameters
can then be obtained from the power spectrum (fig.~\ref{fig:wmap} {\it right}) given in terms of the
angular averages
\begin{equation}
\label{eq:angularaverage}
C_l=\langle{}a_{lm}a_{lm}^\ast\rangle=\frac{1}{2l+1}\sum\limits_{m=-l}^la_{lm}a_{lm}^\ast
\end{equation}
The cosmological parameters extracted in this way have already been quoted in
section~\ref{sec:bb}. In particular, the discrepancy between the
baryon density $\Omega_B$ and the matter density
$\Omega_M=\Omega_B+\Omega_\mathrm{dm}$ clearly shows the need for a
dark matter component.\\
In addition, Big Bang nucleosynthesis, as discussed in
section~\ref{sec:bb}, predicts the value of $\Omega_Bh^2$ to be in the
range $0.017\leq\Omega_Bh^2\leq{}0.024$ at $95\,\%$ confidence
level~\cite{ref:pdg}, and this is consistent with the value obtained
from the WMAP data.

\section{Dark matter candidates}
Having established the existence of dark matter, the question of its
nature arises. In this section, the most
important candidates will be reviewed briefly~\cite{ref:dmreview}.\\
The most widely studied candidate is provided by supersymmetric
extensions to the standard model of particle physics. This will be
discussed in more detail in section~\ref{sec:dmmsugra}.
\paragraph{Standard model neutrinos}
Standard model neutrinos have been considered as candidates for dark
matter in the past. They only interact by the weak force and they are known to be
massive from the
observation of neutrino oscillations. Among the candidates presented here, they are special because
their existence is not hypothetical but has been well
established. However, assuming that neutrinos do not overclose the
Universe, their relic density can be calculated
to be
\begin{equation}
\label{eq:reldens_neutrino}
\Omega_\nu{}h^2=\frac{\sum{}m_\nu}{93\,\mathrm{eV}}
\end{equation}
However, because of their low mass, neutrinos
are relativistic and therefore a candidate for so-called hot dark
matter. Data on the large-scale structure of the Universe,
combined with anisotropies in the cosmic microwave background and other cosmological
probes can be used to set an upper limit of $0.17\,\mathrm{eV}$ ($95\,\%$ confidence level) on the
neutrino masses~\cite{ref:seljak}, implying a relic density of not more than
$\Omega_\nu{}h^2<0.006$, not enough for neutrinos to be the dominant
form of dark matter.

\paragraph{Kaluza-Klein dark matter}
In theories of universal extra dimensions, it is assumed that there
exist dimensions in addition to the known four-dimensional space-time.
The additional dimensions have not been observed yet, so they have to be compactified which introduces
some characteristic scale $R$. This leads to the appearance of a
so-called tower of new particle states in the effective
four-dimensional theory, with the mass of the $n$-th Kaluza-Klein (KK)
mode given by
\begin{equation}
\label{eq:masskk}
m^{(n)}=\sqrt{(n/R)^2+m^2}
\end{equation}
for a standard-model particle of mass $m$. Assuming a symmetry called KK
parity, the lightest KK state (LKP) can be stable and therefore constitutes
a candidate for dark matter~\cite{ref:kolbkk,ref:kkdm}. It is likely to
be associated with the first excitation of the hypercharge gauge
boson, the $B^{(1)}$. A calculation
of the relic density shows that the LKP can explain the observed relic
density $\Omega_\mathrm{dm}$ of dark matter if its mass is on the order
of $1000\,\mathrm{GeV}$. The $B^{(1)}$ as a dark matter candidate has
the attractive feature of dominantly producing charged leptons in its
annihilation. This would provide a source of hard positrons in the
cosmic rays.

\paragraph{Axions}
The axion~\cite{ref:axionrev} was proposed as a solution to the strong
CP problem that arises in the standard model of particle physics. In
general, the action density of the standard model includes a term
\begin{equation}
\label{eq:axionactsm}
\mathcal{L}_\mathrm{SM}=\ldots+\frac{\theta{}g^2}{32\pi^2}G^a_{\mu\nu}\tilde{G}^{a\mu\nu}
\end{equation}
where $G^a_{\mu\nu}$ are the QCD field strengths, $g$ is the QCD
coupling constant, and $\theta$ is a parameter. The observed physics depends
on the value $\bar{\theta}\equiv\theta-\mathrm{arg}\,\mathrm{det}\,m_q$
where $m_q$ is the quark mass matrix. While the term in
(\ref{eq:axionactsm}) violates the C and CP symmetries, as do the weak
interactions in the standard model, the experimental upper bound on
the electric dipole moment of the neutron limits
$|\bar{\theta}|<10^{-10}$~\cite{ref:axionrev} and the question arises why $\bar{\theta}$
is so small when it can be expected to be an arbitrary number. It was
shown that the introduction of an additional field $A(x)$, called the axion,
can naturally explain why $\bar{\theta}$ is zero. The corresponding
term in the action is
\begin{equation}
\label{eq:withaxionactsm}
\mathcal{L}_\mathrm{axion}=\frac{1}{2}\partial_\mu{}A\partial^\mu{}A+\frac{g^2}{32\pi^2}\frac{A(x)}{f_A}G^a_{\mu\nu}\tilde{G}^{a\mu\nu}
\end{equation}
$f_A$ is a constant with dimension of energy, and the mass and
couplings of the axion can be expressed in terms of this constant,
$m_A,g_{Aii}\propto{}f_A^{-1}$. The allowed axion mass range is
limited from below by cosmological bounds and from above by the
physics of stellar evolution to lie in the range
$10^{-6}\!\sim\!10^{-3}\,\mathrm{eV}$. Nevertheless, the axion is a
viable candidate for cold dark matter, with relic density
$\Omega_A\propto{}m_A^{-7/6}$, because cold, non-thermal axions may
have been produced during the QCD phase transition in the early
Universe. Searches for cosmological and solar axions are underway, but
they have eluded discovery so far.

\paragraph{Other candidates}
Some other candidates~\cite{ref:dmreview} include sterile neutrinos,
gravitinos and axinos. Positrons from annihilation of light scalar
dark matter have been proposed to
cause the $511\,\mathrm{keV}$-line observed in the direction of the
Galactic bulge. Little Higgs models, introduced as an alternative
mechanism to supersymmetry to stabilise the weak scale, may contain a
dark matter candidate, too. Superheavy dark matter particles, so
called wimpzillas, would be interesting also because of their
contribution to the phenomenology of ultra-high energy cosmic
rays. Many more exotic proposals exist. It
should be noted that it is entirely possible that dark matter is made
up of more than one species.

\section{Dark matter in the mSUGRA model}
\label{sec:dmmsugra}
Although the standard model of particle physics has so far been
enormously successful at describing the interactions of matter at the
most fundamental level, it has a number of shortcomings from a
theoretical point of view. Two examples are the hierarchy problem and
the problem of unification of the gauge couplings. The former is
related to the question why the Higgs mass is so small. While the mass
scale of the standard model is set by the vacuum expectation value of
the Higgs $v\approx{}246\,\mathrm{GeV}$, divergent quadratic loop corrections to the Higgs
mass occur, $\delta{}m_H^2\sim\Lambda^2$, where $\Lambda$ is a cut-off
scale at which the standard model must be modified to remain valid. This
is usually associated with the Planck scale,
$M_P=(G_N)^{-1/2}\approx{}1.2\cdot{}10^{19}\,\mathrm{GeV}$, which
means that the mass parameter $\mu$ in the Higgs potential
$V=-\mu^2\phi^\dag\phi+\lambda/4(\phi^\dag\phi)^2$ must be of a
similar amplitude to cancel the divergence. 
This large fine-tuning, where two large mass scales
almost cancel to produce the observed masses of the standard model,
seems unnatural and is known as the hierarchy problem.\\
The second example revolves around the unification of gauge
couplings. The running of the gauge couplings in the standard model as
a function of the energy scale is described by the renormalisation
group equations (RGEs). The inverse gauge couplings $\alpha_1^{-1}(Q^2)$,
$\alpha_2^{-1}(Q^2)$, and $\alpha_3^{-1}(Q^2)$ fail to meet at high
$Q^2\sim(10^{16}\,\mathrm{GeV})^2$ though they come close to doing
so. A unification of the gauge couplings is a highly desirable
property of a fundamental theory.\\
In fact, these and other problems can be overcome in supersymmetric
extensions to the standard
model~\cite{ref:dmreview,ref:kane,ref:aitchison,ref:tata}. The
operators $Q$ of supersymmetry (SUSY) satisfy the algebraic relation
\begin{equation}
\label{eq:susyq}
\{Q_a,\bar{Q}_b\}=2\gamma^\mu_{ab}P_\mu\quad\mathrm{where}\quad{}\bar{Q}_a\equiv(Q^\dag\gamma_0)_a
\end{equation}
and transform bosons into fermions and vice versa,
\begin{equation}
\label{eq:susyj}
Q_a|J\rangle=|J\pm{}1/2\rangle
\end{equation}
Two remarkable features of a supersymmetric theory can be seen from
(\ref{eq:susyq}) and (\ref{eq:susyj}). The link between the SUSY
operators and the 4-momentum operator demonstrates that the concept of
space-time has to be extended to include additional degrees of
freedom, acted upon by the $Q$'s. The second consequence is the
prediction of additional particles, superpartners to the standard
model fields, with identical quantum numbers but different spin. As
none of the superpartners has been discovered yet, their masses must be
different from the standard model ones, meaning that supersymmetry
must be broken. The supersymmetric model that includes the smallest
number of additional particles necessary to give rise to all fields of
the standard model is called the Minimal Supersymmetric Standard Model
\begin{table}[bt]
\small
\begin{center}
\begin{tabular}{lllllll} \hline 
  \multicolumn{2}{c}{Standard Model particles and fields} & \multicolumn{5}{c}{Supersymmetric partners} \\
  & & \multicolumn{3}{l}{Interaction eigenstates} & \multicolumn{2}{l}{Mass 
  eigenstates} \\
  Symbol & Name & Symbol & Name & & Symbol & Name \\ \hline
  $q=d,c,b,u,s,t$ & quark & $\tilde{q}_{L}$, $\tilde{q}_{R}$ & 
  squark & & $\tilde{q}_{1}$, $\tilde{q}_{2}$ & squark \\
  $l=e,\mu,\tau$ & lepton & $\tilde{l}_{L}$, $\tilde{l}_{R}$ & slepton & 
  & $\tilde{l}_{1}$, $\tilde{l}_{2}$ & slepton \\
  $\nu = \nu_{e}, \nu_{\mu}, \nu_{\tau}$ & neutrino & $\tilde{\nu}$ & 
  sneutrino & & $\tilde{\nu}$ & sneutrino \\
  $g$ & gluon & $\tilde{g}$ & gluino & & $\tilde{g}$ & gluino \\
  $W^\pm$ & $W$-boson & $\tilde{W}^\pm$ & wino & & & \\
  $H^-$ & Higgs boson & $\tilde{H}_{1}^-$ & higgsino & 
  \raisebox{-.25ex}[0ex][0ex]{$\left. \raisebox{0ex}[-3.3ex][3.3ex]{}
  \right\}$} &  $\tilde{\chi}_{1,2}^\pm$ & chargino \\
  $H^+$ & Higgs boson & $\tilde{H}_{2}^+$ & higgsino & & & \\
  $B$ & $B$-field & $\tilde{B}$ & bino & & & \\
  $W^3$ & $W^3$-field & $\tilde{W}^3$ & wino & & & \\
  $H_{1}^0$ & Higgs boson & 
  \raisebox{-1.75ex}[0ex][0ex]{$\tilde{H}_{1}^0$} & 
  \raisebox{-1.75ex}[0ex][0ex]{higgsino} & 
  \raisebox{.25ex}[0ex][0ex]{$\left. \raisebox{0ex}[-5.25ex][5.25ex]{}
  \right\}$} & \raisebox{0.5ex}[0ex][0ex]{$\tilde{\chi}_{1,2,3,4}^0$} & 
  \raisebox{.5ex}[0ex][0ex]{neutralino} \\[0.5ex]
  $H_{2}^0$ & Higgs boson & 
  \raisebox{-1.75ex}[0ex][0ex]{$\tilde{H}_{2}^0$} & 
  \raisebox{-1.75ex}[0ex][0ex]{higgsino} & & & \\[0.5ex]
  $H_{3}^0$ & Higgs boson & & & & & \\[0.5ex] \hline
\end{tabular}
\label{tab:susyparticles}
\caption[Standard Model particles and their superpartners in the MSSM]
{\label{tab:susy}Standard Model particles and their superpartners in
  the MSSM~\cite{ref:edsjophd}.}
\end{center}
\end{table}
(MSSM). Its field content is summarised in table~\ref{tab:susy}.\\
The MSSM can be formulated such that a multiplicative quantum number,
called $R$-parity and defined as
\begin{equation}
\label{eq:rparity}
R\equiv{}(-1)^{3B+L+2S}
\end{equation}
is conserved. Standard model particles have $R=1$ and the
superpartners (sparticles) have $R=-1$. Originally introduced to prevent rapid
proton decay, $R$-parity conservation implies that the lightest
sparticle (called the LSP) is stable and can only be destroyed by pair
annihilation. In many MSSM scenarios, the LSP is the lightest neutralino
$\chi\equiv\tilde{\chi}_1^0$ and
this is the best motivated candidate for dark matter known so
far. Because of their properties, neutralinos are among the group of
candidates called
weakly-interacting massive particles (WIMPs).\\
In general, the neutralino is a linear combination of the
superpartners of the $B$ and $W_3$ gauge bosons and of the neutral
Higgses,
\begin{equation}
\label{eq:neutralinomixing}
\chi=N_{11}\tilde{B}+N_{12}\tilde{W}_3+N_{13}\tilde{H}^0_1+N_{14}\tilde{H}^0_2
\end{equation}
The gaugino and higgsino fractions $f_G$ and $f_H$ are then defined as
\begin{equation}
\label{eq:neutrfractions}
f_G=N_{11}^2+N_{12}^2\quad\mathrm{and}\quad{}f_H=N_{13}^2+N_{14}^2
\end{equation}
respectively.\\
The MSSM is based on the same gauge group as the standard model. The
$R$-parity conserving superpotential, in the notation of~\cite{ref:kane}, is given by
\begin{equation}
\label{eq:susypotential}
W=h^U_{ij}\hat{Q}_i\hat{H}_u\hat{u}^c_j+h^D_{ij}\hat{Q}_i\hat{H}_d\hat{d}^c_j+h^E_{ij}\hat{L}_i\hat{H}_d\hat{e}^c_j+\mu\hat{H}_d\hat{H}_u
\end{equation}
Here $\hat{Q}$ and $\hat{L}$ represent the quark and lepton SU(2)
doublet superfields, $\hat{u}^c$, $\hat{d}^c$, $\hat{e}^c$ the
corresponding SU(2) singlets, and $\hat{H}_u$, $\hat{H}_d$ the Higgs
superfields whose scalar components give mass to up- and down-type
quarks and/or leptons, respectively. Generational indices have been
shown explicitly, but group indices have been dropped. The allowed
soft SUSY-breaking terms are given by
\begin{equation}
\label{eq:susybreaking}
\begin{split}
-\mathcal{L}_\mathrm{soft}&=\left(A^U_{(ij)}h^U_{ij}\tilde{Q}_iH_u\tilde{u}^c_j+A^D_{(ij)}h^D_{ij}\tilde{Q}_iH_d\tilde{d}^c_j+
A^E_{(ij)}h^E_{ij}\tilde{L}_iH_d\tilde{e}^c_j+\mathrm{h.c.}\right)+B\mu(H_dH_u+\mathrm{h.c.})\\
&+m^2_{H_d}|H_d|^2+m^2_{H_u}|H_u|^2+m^2_{\tilde{L}}|\tilde{L}|^2+m^2_{\tilde{e}^c}|\tilde{e}^c|^2+m^2_{\tilde{Q}}|\tilde{Q}|^2
+m^2_{\tilde{u}^c}|\tilde{u}^c|^2+m^2_{\tilde{d}^c}|\tilde{d}^c|^2\\
&+\frac{1}{2}\left(M_1\bar{\psi}_B\psi_B+M_2\bar{\psi}_W^a\psi_W^a+m_{\tilde{g}}\bar{\psi}_g^a\psi_g^a+\mathrm{h.c.}\right)
\end{split}
\end{equation}
Here the tilded fields are the scalar partners of the quark and lepton
fields, while the $\psi_i$ are the spin-$1/2$ partners of the
$i=\mathrm{U}(1)_Y,\mathrm{SU}(2)_L,\mathrm{SU}(3)_c$ gauge
bosons. The $A_{(ij)}$ and $B$ are mass parameters.\\
A study of the phenomenology of the MSSM is made difficult by its
large number of free parameters of more than~100, mostly masses and
mixing angles. This number is greatly reduced in the constrained MSSM
-- also called mSUGRA --
model~\cite{ref:chamseddine,ref:hall,ref:barbieri,ref:nath} where the
MSSM is coupled to minimal supergravity from which the following set
of assumptions emerges, inspired by the unification of gauge couplings
at some high unfication scale $M_X$~\cite{ref:kane}:
\begin{itemize}
\item Common gaugino mass $m_{1/2}$. The soft SUSY-breaking gaugino
mass terms are equal to $m_{1/2}$ at $M_X$:
\begin{equation}
\label{eq:m12kane}
M_1(M_X)=M_2(M_X)=m_{\tilde{g}}(M_X)\equiv{}m_{1/2}
\end{equation}
\item Common scalar mass $m_0$. The soft SUSY-breaking scalar mass
terms contributing to the squark, slepton, and Higgs boson masses are
equal to $m_0$ at $M_X$:
\begin{equation}
\label{eq:m0kane}
m^2_{\tilde{Q}}(M_X)=m^2_{\tilde{u}^c}=\ldots{}=m^2_{H_d}(M_X)=m^2_{H_u}(M_X)\equiv{}m_0
\end{equation}
\item Common trilinear scalar coupling $A_0$. The soft trilinear
SUSY-breaking terms are all equal to $A_0$ at $M_X$:
\begin{equation}
\label{eq:akane}
A_t(M_X)=A_b(M_X)=A_\tau(M_X)=\ldots\equiv{}A_0
\end{equation}
\end{itemize}
Requiring radiative electroweak symmetry breaking to occur,
$\mu^2(m_Z)$ can be eliminated as a free parameter in favour of $m_Z$,
but the sign of $\mu$ remains free. Similarly, $B(m_Z)$ can be
eliminated in favour of $\tan\beta(m_Z)$, where
$\tan\beta\equiv{}v_u/v_d$ is the ratio of the Higgs vacuum
expectation values. In the end, the five free parameters
\begin{equation}
\label{eq:msugraparameters}
m_0,\,m_{1/2},\,A_0,\,\tan\beta,\,\mathrm{sgn}\,\mu
\end{equation}
remain. A study of the mSUGRA model in connection
with the indirect search for dark matter will be the subject of chapter~\ref{chapter:susyscan}.

\section{Production and propagation of cosmic rays}
For centuries, mankind has used the light reaching us from distant
celestial objects to gather information about the Universe surrounding
us. Nowadays, in the field of astronomy, the emission in a large part of the electromagnetic
spectrum and from very different sources is examined by a multitude of
observatories.
In 1912, cosmic rays~\cite{ref:ginzburg,ref:gaisser,ref:longair,ref:stanev} were
discovered by Victor Hess during a series of balloon
flights over Austria, Bohemia and Prussia~\cite{ref:hess} who found an increase in the
discharge rate of an electrometer with increasing altitude. They
constitute another source of knowledge about the workings of nature,
and the interplay between their study and conventional particle
physics has created the field of astroparticle physics. A number of
important discoveries in particle physics have been made in cosmic
rays. For example, the positron~\cite{ref:anderson}, the
muon~\cite{ref:anderson_muon,ref:street_muon}, the
pion~\cite{ref:perkins_pion} and the kaon~\cite{ref:kaon}, among others, were first
observed in experiments studying cosmic rays. On the other hand,
particle physics plays a vital role in the understanding of
cosmic-ray production and propagation mechanisms.

\subsection{Basic facts about cosmic rays}
Cosmic rays mainly consist of nuclei. Roughly $90\,\%$ of them are
protons, $9\,\%$ are $\alpha$-particles, and heavier nuclei make up
the rest. Electrons, positrons and antiprotons are found in small
quantities. In general, the relative abundances of the individual
elements follow the abundances found in the solar system. This
indicates that the production mechanism is the same in both cases,
namely fusion of lighter nuclei in the cores of stars near the end of
their life cycle. But there are also important differences. On the one
hand, protons are much more abundant than nuclei with $Z>1$ in the
solar system than in the cosmic rays. This could have something to do
with the fact that hydrogen is relatively hard to ionise for injection
into the acceleration process, or it could reflect a genuine
difference in composition at the source. On the other hand, the
elements Li,~Be,~B and Sc,~Ti,~V,~Cr,~Mn are much more abundant in the
cosmic rays than in solar system material. While they are essentially
absent as end products in stellar nucleosynthesis, they are produced
in spallation processes during interactions of carbon and oxygen or
iron with interstellar matter, respectively. This mechanism can be
used for an understanding of the propagation process.\\
\par
Starting in the GeV-range, the number density of cosmic rays as a
function of energy follows a power law up to very high energies,
\begin{equation}
\label{eq:cr_powerlaw}
\frac{\mathrm{d}N}{\mathrm{d}E}\propto{}E^{-\gamma}
\end{equation}
Up to roughly $10^{15}\,\mathrm{eV}$, one finds
$\gamma\approx{}2.7$. At higher energies, the spectral index $\gamma$
increases, that is the slope of the spectrum becomes steeper, and this
is known as the {\it knee} of the spectrum. This decrease may indicate
that some of the acceleration mechanisms reach a maximal energy
here. Around $10^{19}\,\mathrm{eV}$, the spectral index decreases
again, the particles observed in this range are believed to be of
extragalactic origin, especially as their Larmor radius exceeds the
size of the Galaxy. In fact, the Pierre Auger collaboration has
reported a tentative correlation between the arrival directions of cosmic
rays above $6\cdot{}10^{19}\,\mathrm{eV}$ and the position of active
galactic nuclei within $\sim\!{}75\,\mathrm{Mpc}$~\cite{ref:pa_agn}.
Events with energies above $10^{20}\,\mathrm{eV}$
have even been observed, corresponding to macroscopic energies of a
few Joules.

\subsection{Sources of cosmic rays}
In this section, the widely accepted theory
of cosmic-ray acceleration shall be briefly discussed, limited to the energy range of
interest for this study, below the knee in the cosmic-ray
spectrum~\cite{ref:gaisser}. While it has been observed that particles
are accelerated to GeV energies in solar flares, a different mechanism
is needed to explain acceleration up to the TeV range.\\
From the energy density of cosmic rays,
$\rho_E\approx{}1\,\mathrm{eV}/\mathrm{cm}^3$, a timescale for
diffusion out of the Galaxy of $\tau\sim{}2\cdot{}10^7\,\mathrm{a}$, a
galactic radius of $15\,\mathrm{kpc}$ and a disk height of
$800\,\mathrm{pc}$, the power needed to keep up the cosmic-ray energy
density is roughly estimated by
\begin{equation}
\label{eq:crpower}
P_\mathrm{CR}=\frac{V\rho_E}{\tau}\sim{}4\cdot{}10^{33}\,\mathrm{J}/\mathrm{s}
\end{equation}
Supernovae are the most likely source for that power. Given a typical
energy release of $10^{44}\,\mathrm{J}$ and a mean rate of one
supernova every 30~years, the power released is
$10^{35}\,\mathrm{J}/\mathrm{s}$ so that an efficiency of a few
percent would suffice. First-order Fermi acceleration at a strong
shock describes how a particle diffusing through the turbulent
magnetic fields carried along with a moving plasma gains energy
proportional to $v/c$ during each cycle of passage through the shock
front, where $v$ is the velocity of the shock front. Since there is a
certain possibility in each cycle for the particle to be lost from the
acceleration region, this process naturally leads to the observed
power law of the cosmic-ray spectrum.\\
This picture is now supported by observational evidence. An example is
\begin{figure}[htb]
\begin{center}
\includegraphics[width=0.5\textwidth,angle=0]{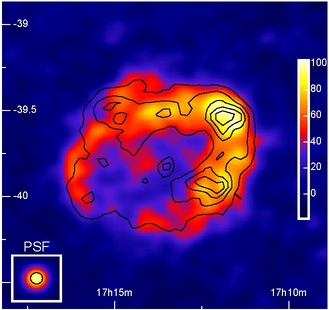}
\end{center}
\caption{$\gamma$-radiation from supernova remnant RX~1713.7-3946 as
measured by HESS~\cite{ref:hesssnr}. The contour lines trace the x-ray emission and the
inset in the lower left corner shows the point spread function of
HESS. Reproduced with permission by the authors.}
\label{fig:hesssnr}
\end{figure}
shown in figure~\ref{fig:hesssnr} depicting the $\gamma$-radiation
emitted by the supernova remnant RX~1713.7-3946, as measured by the
HESS telescope. This shows the internal structure of a source of
TeV-$\gamma$-rays. The $\gamma$-ray spectrum follows a power law with
spectral index of about~2 and extends up to roughly
$10\,\mathrm{TeV}$. This implies the presence of protons accelerated
to even higher energies than that, producing the observed
$\gamma$-rays in collisions with matter present in the vicinity of the
supernova remnant.

\subsection{Propagation of cosmic rays in the Galaxy}
\label{sec:propagation}
On the way of cosmic rays through the Galaxy, their spectra are
altered and their composition is changed by a variety of physical
processes. Hadronic interactions of protons and nuclei with
interstellar matter create secondary charged particles, as well as
$\gamma$-rays by $\pi^0$-production. Electrons lose energy by
bremsstrahlung processes due to the interstellar matter, synchrotron
radiation in the Galactic magnetic field, and inverse Compton
scattering on photons of the cosmic microwave background and of
starlight. Radioactive isotopes decay in flight.\\
As opposed to $\gamma$-rays, charged cosmic-ray particles do not
follow straight lines, but they are scattered on magnetohydrodynamic
waves and discontinuities. As we have no information about the
microscopic structure of the magnetic field in the Galaxy, charged
particles will in effect perform a random walk, and therefore, a
diffusion model is the appropriate description of the propagation
process. This also explains the high level of isotropy found in cosmic
rays and the fact that the Galaxy can store cosmic rays up to high
energies. Intuitively, it is clear that the diffusion coefficient must
have a tendency to increase with energy because the trajectory of a
particle will be more rigid with regard to a magnetic field at higher
energies.\\
\par
The study of cosmic-ray propagation is vital for an indirect search
for dark matter because all the signal particles -- positrons,
antiprotons and $\gamma$-rays -- are produced as secondaries in the
interactions of primary cosmic rays with the interstellar matter. This
means that propagation effects constitute the background that has to
be disentangled from any supposed signal. For example, pions and kaons
produced in collisions of protons with interstellar matter decay to
muons and these subsequently decay to positrons and electrons.\\
\par
While several analytical or semi-analytical approaches to the description of cosmic-ray
propagation exist~\cite{ref:maurinprop}, it was chosen to use the numerical model described
by the Galprop
code~\cite{ref:galpropposelec,ref:galpropnucl,ref:galproppbar,ref:galpropmhd,ref:galproprev},
version 50p,
which was created with the aims of
enabling simultaneous predictions of all relevant observations,
overcoming the limitations of analytical models and incorporating as
much current information as possible, for example on galactic
structure and source distributions. The propagation equation is
written in the form
\begin{equation}
\label{eq:galprop}
\frac{\partial\psi}{\partial{}t}=q(\vec{r},t)+\vec{\nabla}\cdot(D_{xx}\vec{\nabla}\psi-\vec{V}\psi)+\frac{\partial}{\partial{}p}p^2D_{pp}\frac{\partial}{\partial{}p}\frac{1}{p^2}\psi-\frac{\partial}{\partial{}p}\left(\dot{p}\psi-\frac{p}{3}(\vec{\nabla}\cdot\vec{V})\psi\right)-\frac{1}{\tau_f}\psi-\frac{1}{\tau_r}\psi
\end{equation}
Here, $\psi=\psi(\vec{r},p,t)$ is the density per unit of total
particle momentum and $\psi(p)\,\mathrm{d}p=4\pi{}p^2f(\vec{p})$ in
terms of phase-space density $f(\vec{p})$. The terms on the right hand
side of the propagation equation can be understood as follows:
\begin{itemize}
\item $q(\vec{r},t)$ is the source term. The injection spectrum of
nucleons is assumed to be a power law in momentum,
$\mathrm{d}q(p)/\mathrm{d}p\propto{}p^{-\gamma}$, and the distribution
of cosmic-ray sources is chosen as
\begin{equation}
\label{eq:galprop_sources}
q(R,z)=q_0\left(\frac{R}{R_\odot}\right)^\alpha\exp\left(-\beta\frac{R-R_\odot}{R_\odot}-\frac{|z|}{0.2\,\mathrm{kpc}}\right)
\end{equation}
where $q_0$ is a normalisation constant and the parameters here are
chosen to be $\alpha=1/2$ and $\beta=1$. $R_\odot=8.5\,\mathrm{kpc}$ is the distance of the solar system from the Galactic centre.
\item $D_{xx}$ is the spatial diffusion coefficient. It depends on
rigidity $R$ as $D_{xx}=D_0\beta^\lambda(R/R_0)^a$, where
$\beta=v/c$, $D_0$ is a constant and $\lambda$ influences the
behaviour at low rigidities. Typical values of the diffusion
coefficient are
$D_{xx}\sim{}(3-5)\,\cdot\,10^{28}\,\mathrm{cm}^2\,\mathrm{s}^{-1}$ at
$\sim{}1\,\mathrm{GeV}/n$, and $a$ is in the range $0.3$ to $0.6$.
\item Galactic winds lead to a convective transport of cosmic rays and
are described by the convection velocity $V(z)$. $V(z)$ is assumed to
increase linearly with distance from the Galactic plane. This implies
a constant adiabatic energy loss.
\item In addition to spatial diffusion, the scattering of cosmic-ray
particles on randomly moving magnetohydrodynamic waves leads to
stochastic reacceleration, described in the transport equation as
diffusion in momentum space with the diffusion coefficient
$D_{pp}$. It is related to $D_{xx}$ by 
\begin{equation}
\label{eq:galpropDpp}
D_{pp}D_{xx}=\frac{4p^2v_A^2}{3a(4-a^2)(4-a)w}
\end{equation}
where $w$ characterises the level of turbulence. The main free
parameter in this relation is the Alfv\'en speed $v_A$, which is a
characteristic velocity of weak disturbances propagating in a magnetic
field. As only $v_A^2/w$ is relevant here, one can set $w=1$.
\item $\dot{p}$ is the momentum loss rate, and the term involving
$\vec{\nabla}\cdot\vec{V}$ represents adiabatic momentum gain or loss
in the nonuniform flow of gas, with a frozen-in magnetic field whose
inhomogeneities scatter the cosmic rays.
\item $\tau_f$ and $\tau_r$ are the timescales for loss by
fragmentation and radioactive decay, respectively.
\end{itemize}
The structure of the Galaxy is included in the form of the gas
content, which is important for secondary production, and the
interstellar radiation field (ISRF) and magnetic field, which strongly affect electron
energy losses. The distribution of atomic hydrogen is reasonably well known
from 21-cm surveys, but the distribution of molecular hydrogen can
only be estimated using the CO tracer~\cite{ref:galproprev}. The Galactic magnetic field can
be determined from pulsar rotation and dispersion measurements combined
with a model for the distribution of ionised gas. The ISRF comes from
the cosmic microwave background and from stars of all types and is
modified by absorption and reemission by interstellar dust.\\
Interactions and production of secondary particles are governed by the
cross sections for energy losses of nuclei and electrons,
bremsstrahlung and synchrotron radiation, inverse Compton emission and
pion production of $\gamma$-rays, electrons and positrons.\\
Equation (\ref{eq:galprop}) is solved numerically in two dimensions,
assuming cylindrical symmetry, with spatial
boundary conditions assuming free particle escape, and the time
dependence is followed until the steady state is reached. The
propagation equation has to be considered for all relevant particle
species, and the resulting reaction network is solved starting at the
heaviest nucleus, $^{64}\mathrm{Ni}$. Equation (\ref{eq:galprop}) is
solved, computing all the resulting secondary source functions, and
then the process is repeated for the nuclei with $A-1$.\\
Before studying the predictions of the Galprop model in some detail,
one first has to consider how interactions of cosmic-ray particles
with the solar wind and the geomagnetic field affect the particle
spectra measured at the top of the Earth's atmosphere.

\subsection{Solar modulation}
\label{sec:solarmod}
Arriving at the outskirts of the solar system, the fluxes of
cosmic-ray particles are modulated due to interactions with the solar
wind~\cite{ref:longair}. The first hints at this effect came from observations of an
anticorrelation between neutron monitor counts and the sunspot number,
the latter being an indicator of the level of solar
activity~\cite{ref:stanev}. The solar wind consists mostly of protons,
with a typical kinetic energy of $500\,\mathrm{eV}$, velocities of
$350\,\mathrm{km}/\mathrm{s}$, flux of
$1.5\cdot{}10^{12}\mathrm{m}^{-2}\mathrm{s}^{-1}$ and a temperature of
$10^6\,\mathrm{K}$. It originates from the corona of the Sun. A
magnetic field, rooted in the Sun, is frozen into the solar wind
plasma, and the Sun's rotation leads to the creation of the
large-scale structure known as the Archimedes spiral. Cosmic-ray particles
are scattered on the magnetic fields. Gleeson and
Axford~\cite{ref:solarmod} model the solar modulation by taking into
account cosmic-ray diffusion through this magnetic field, convection
by the outward motion of the solar wind, and adiabatic deceleration of
the cosmic rays in this flow. In the force-field approximation, that
is used in the remainder of this thesis, the effect of solar modulation
can be described by a single parameter $\phi$ that depends on the
solar wind speed $V$ and the diffusion coefficient $\kappa$ as
\begin{equation}
\label{eq:solarmodparameter}
\phi=\frac{E+m}{E}\,\frac{T}{3}\int\limits^{r_b}_{r_E}\frac{V(x,t)}{\kappa(x,E,t)}\,\mathrm{d}x
\end{equation}
where $E$, $T$ and $m$ are the total, kinetic and rest energy of a
cosmic-ray particle, respectively. The integral is taken from the
location of the Earth to the boundary of the heliosphere. The interstellar
cosmic-ray flux $J_{IS}$ is then modulated to yield the locally
observed one $J$ as
\begin{equation}
\label{eq:forcefield}
J(E)=\frac{E^2-m^2}{(E+|z|\phi)^2-m^2}\,\cdot\,{}J_{IS}(E+|z|\phi)
\end{equation}
where $z$ is the particle charge. The modulation parameter $\phi$ has
the dimension of a rigidity and is of the order of $500\,\mathrm{MV}$
but it changes with time in accordance with the solar cycle. It must
be stressed that the modulation parameter is not a model-independent
quantity. Because the interstellar flux $J_{IS}$ appears in
(\ref{eq:forcefield}), a value of $\phi$ can only be quoted in the
context of a given propagation model.

\subsection{Effect of the geomagnetic field}
\label{sec:geomag}
The magnetic field of the Earth is the final barrier for cosmic-ray
particles to overcome before they can finally be detected in the
atmosphere or a low orbit. Particles with low rigidities will follow
spiral trajectories around the field lines and eventually lose their
energy. The minimal rigidity that a particle must have in order to
reach Earth through its magnetic field is called the cutoff rigidity $R_S$.
In the vicinity of the Earth, the magnetic
field can be approximated by a dipole field. For this case, a formula
for the cutoff rigidity for vertically approaching particles can be
derived as~\cite{ref:longair}
\begin{equation}
\label{eq:cutoff}
R_S\geq{}14.9\,\mathrm{GV}\cdot\cos^4\lambda
\end{equation}
The cutoff depends on the latitude $\lambda$ of the observer, as
measured with respect to the equatorial plane of the dipole. It is
lowest near the magnetic poles. For particles arriving from any given
direction, the cutoff depends on the azimuthal angle as well. This
leads to the so called east-west effect: For positively charged
particles at the same zenith angle the cutoff is higher from the east
direction and vice versa for negatively charged particles~\cite{ref:stanev}.\\
In reality, the field is not a perfect dipole and interactions with
the solar wind lead to additional distortions. Therefore, an accurate
determination of the geomagnetic cutoff requires a detailed model of
the geomagnetic field and proceeds by backtracing individual particles
with a given position, time, and rigidity through the magnetic field
by integrating the equation of motion, to see if the particle reaches
outer space.\\
The cutoff effect due to the
geomagnetic field leads to a distortion of the spectrum $\Phi(R)$ of a given
cosmic-ray species. For the purposes of this work, it can be described as~\cite{ref:cutoff}
\begin{equation}
\label{eq:geomod}
\Phi^\mathrm{mod,geo}(R)=\Phi(R)\,\cdot\,\frac{1}{1+\left(\frac{R}{R_c}\right)^{-\gamma_c}}
\end{equation}
where $R_c$ is a cutoff rigidity and $\gamma_c$ describes the
steepness of the modulation. Typically, $\gamma_c$ is on the order of unity.

\subsection{Measurements of cosmic rays}
Measurements of cosmic rays can be divided into two groups. Below
$10^{14}\,\mathrm{eV}$, a direct measurement is possible. For
this purpose, instruments known from particle physics, such as emulsion chambers,
scintillators, tracking devices, calorimeters, Cherenkov counters or
transition radiation detectors, are combined in a suitable way to
measure mass, momentum, energy or charge sign of a particle. The PEBS
and AMS-02
detectors belong to this category. Since the thickness of Earth's
atmosphere at sea level amounts to twenty radiation lengths or eight
hadronic interaction lengths, cosmic rays in the energy range of
interest here are absorbed long before reaching the surface of the
Earth. Therefore, a detector measuring primary cosmic rays has to be
flown on a balloon at as high an altitude as possible or in space.\\
Above $10^{14}\,\mathrm{eV}$, another path has to be taken to obtain
acceptable counting rates because of the steep decline in flux as
given in (\ref{eq:cr_powerlaw}). In this region of indirect
measurements, Earth's atmosphere is used as a giant calorimeter to
look for extended air showers created in the wake of an interaction of
the primary particle in the upper layers of the atmosphere. The energy
of the primary can be derived from the size and shape of the shower. Above
primary energies of roughly $10^{15}\,\mathrm{eV}$, the shower
particles can reach the surface of the Earth and be detected
there. For that purpose, an appropriately sized area is instrumented
with detector units, such as scintillators or water Cherenkov
counters. In addition, one can observe the Cherenkov radiation emitted
by the shower particles moving at velocities that exceed the speed of light
in air or the fluorescence light. The latter is emitted after the
excitation of nitrogen molecules in the air by passing shower
particles, and with wavelengths in the range from 300 to
$400\,\mathrm{nm}$. As an example, the Pierre Auger observatory in
Argentina employs a
hybrid detection technique with great success~\cite{ref:auger}.

\subsection{Atmospheric backgrounds}
\label{sec:atmbg}
For balloon-based measurements of cosmic rays, the remaining
atmosphere between the flight altitude and outer space additionally
modifies particle fluxes. Most important for the purposes of this work are
positrons created in the decay chains following hadronic interactions
of protons with the atoms in the upper layers of the atmosphere. A
grammage of approximately $3.7\,\mathrm{g}/\mathrm{cm}^2$ has to be
overcome at an altitude of $40\,\mathrm{km}$ in the polar regions
during summer. Reference~\cite{ref:philip} deals with a Monte Carlo study
of the necessary corrections in great detail. It arrives at the
conclusion that they will be of the order of $10\,\%$ for positrons
above $1\,\mathrm{GeV}$.

\section{Model for cosmic-ray propagation}
\label{sec:galprop}
\begin{figure}
\begin{center}
\begin{tabular}{cc}
\includegraphics[width=0.48\textwidth,angle=0]{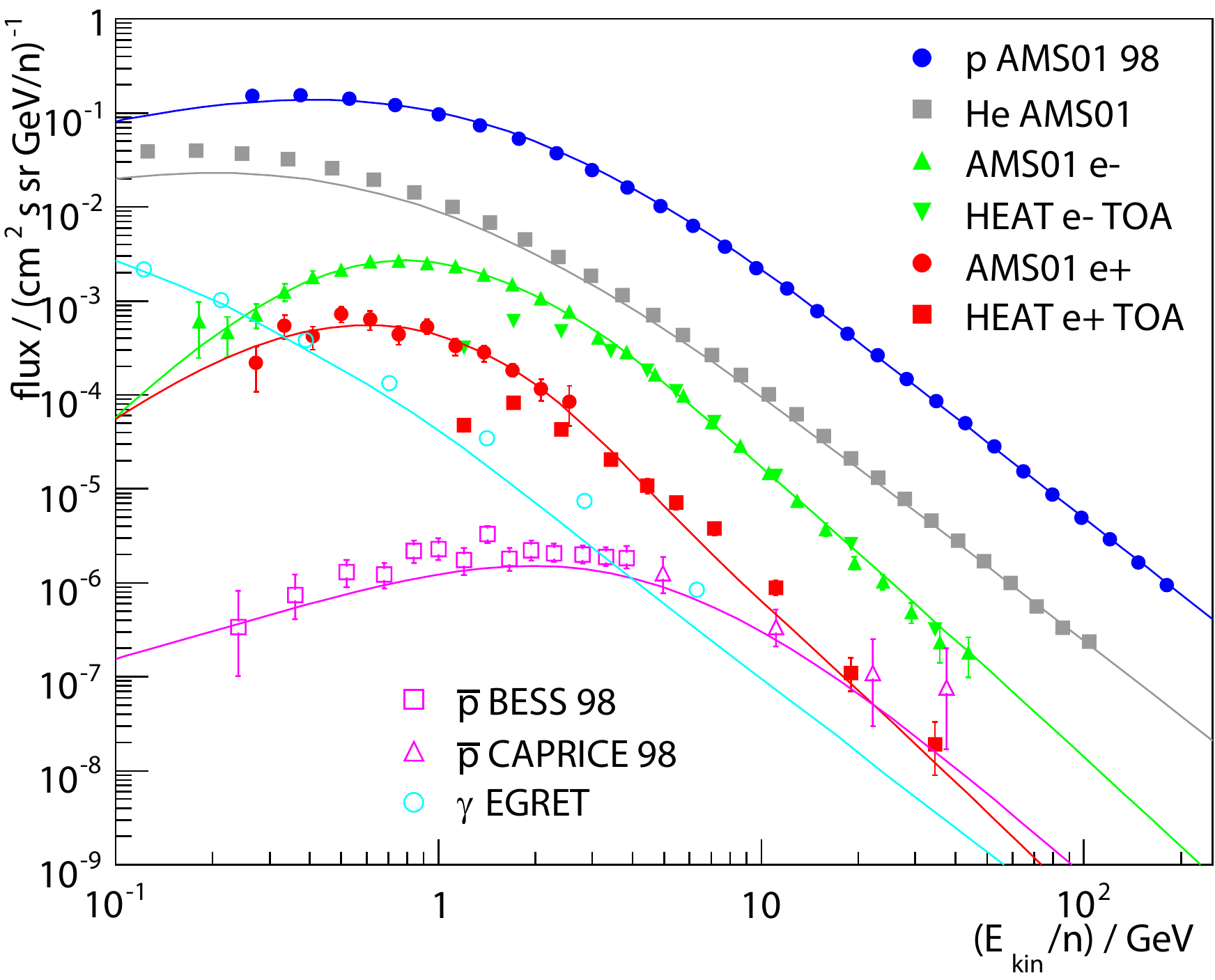}&
\includegraphics[width=0.5\textwidth,angle=0]{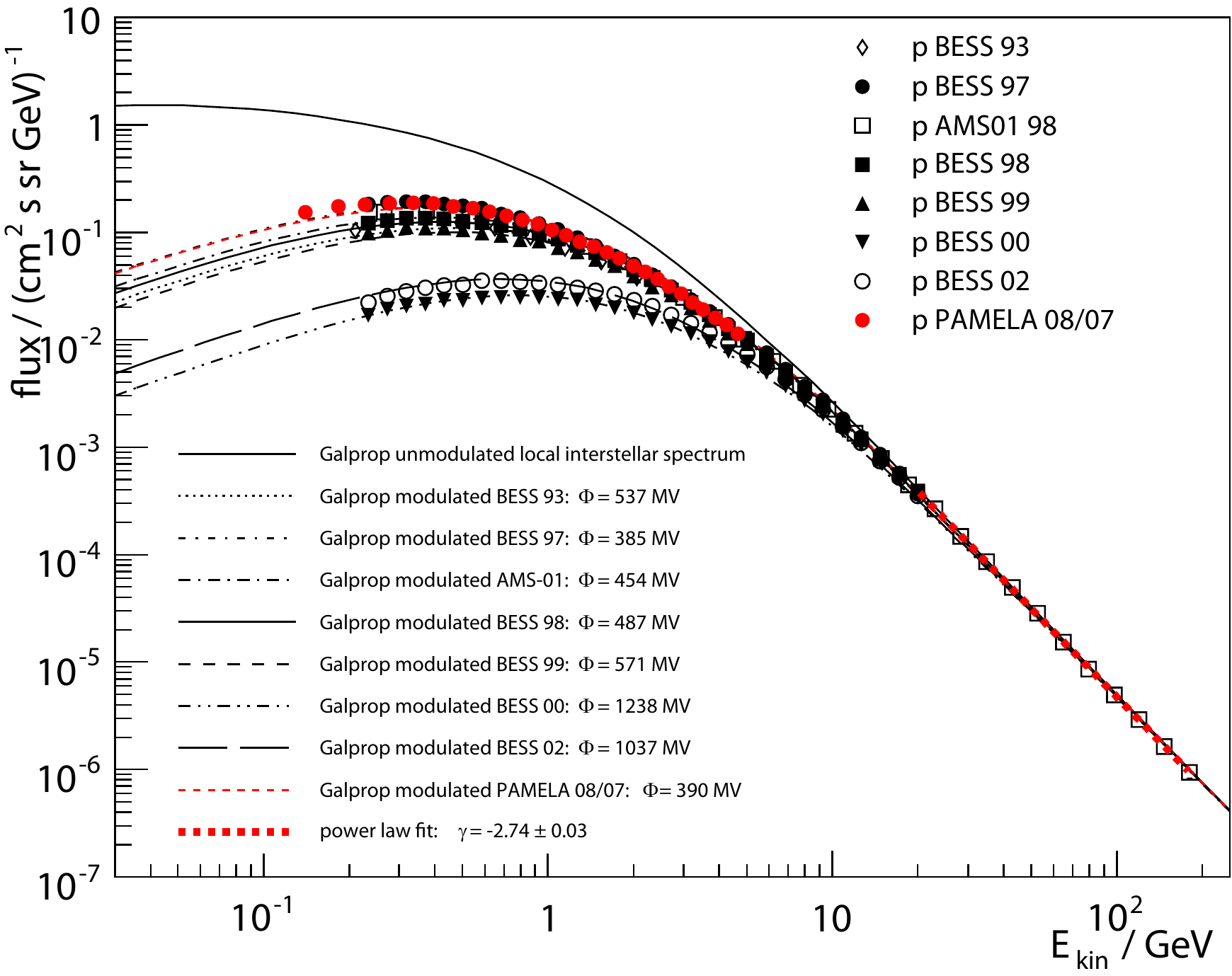}\\
\end{tabular}
\end{center}
\caption{{\it Left:} Measured fluxes of protons~\cite{ref:p_ams},
helium~\cite{ref:he_ams}, electrons and
positrons~\cite{ref:electrons_heat,ref:ams01old},
antiprotons~\cite{ref:bess98,ref:capricepbar}, and diffuse
$\gamma$-rays from the Galactic centre region~\cite{ref:egret}. The
solar modulated predictions by the conventional Galprop model for each
species are included.
{\it Right:} Cosmic-ray proton spectrum as measured by BESS~\cite{ref:p_bess93,ref:p_bess},	
AMS01~\cite{ref:p_ams} and PAMELA~\cite{ref:p_pamela}, together with the prediction by the Galprop conventional
model. The effect of
solar modulation in the data is clearly seen. The unmodulated model
flux is plotted, as well as the modulated ones, using the best-fit
modulation parameter $\phi$ for each data set.}
\label{fig:protons}
\end{figure}
\begin{figure}
\begin{center}
\includegraphics[width=0.5\textwidth,angle=0]{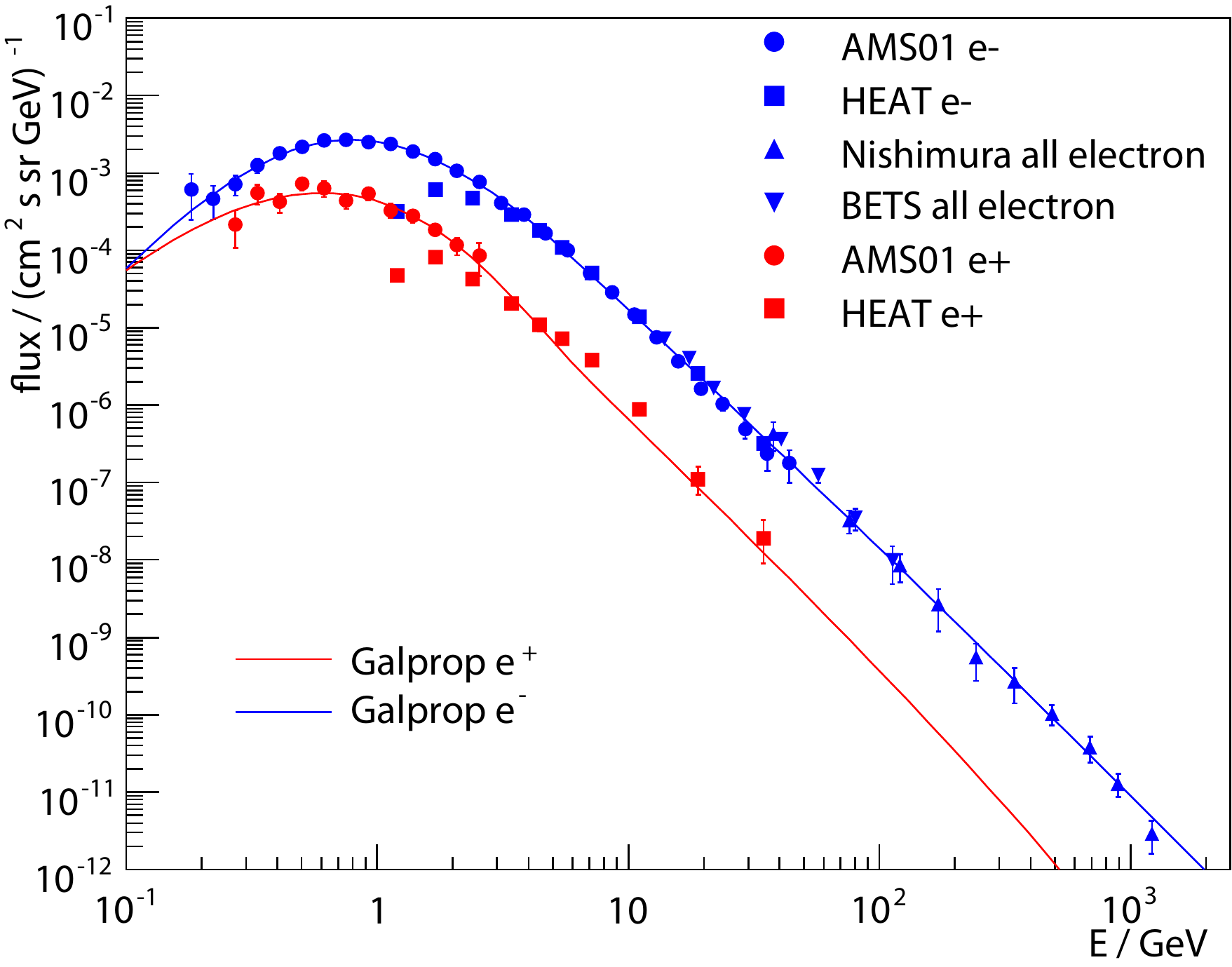}
\end{center}
\caption{Electron spectrum as measured by
HEAT~\cite{ref:electrons_heat}, AMS01~\cite{ref:ams01old}, all electron spectrum as measured by
Nishimura et al.~\cite{ref:electrons_nishimura} and
BETS~\cite{ref:electrons_bets}, and positron spectrum as measured by
HEAT~\cite{ref:electrons_heat} and AMS01~\cite{ref:ams01old}, together
with the respective predictions of the Galprop conventional
model. Modelling of the solar and geomagnetic modulation is performed
for the AMS01 fluxes here. The
same solar modulation parameter of $\phi=442\,\mathrm{MV}$ was used for electrons and
positrons, while the parameters for geomagnetic modulation according
to eq.~(\ref{eq:geomod}) are $R_{c,e^-}=0.72\,\mathrm{GV}$,
$R_{c,e^+}=1.59\,\mathrm{GV}$, $\gamma_{c,e^-}=1.87$, and $\gamma_{c,e^+}=1.02$.}
\label{fig:electrons}
\end{figure}
\begin{figure}
\begin{center}
\includegraphics[width=0.7\textwidth,angle=0]{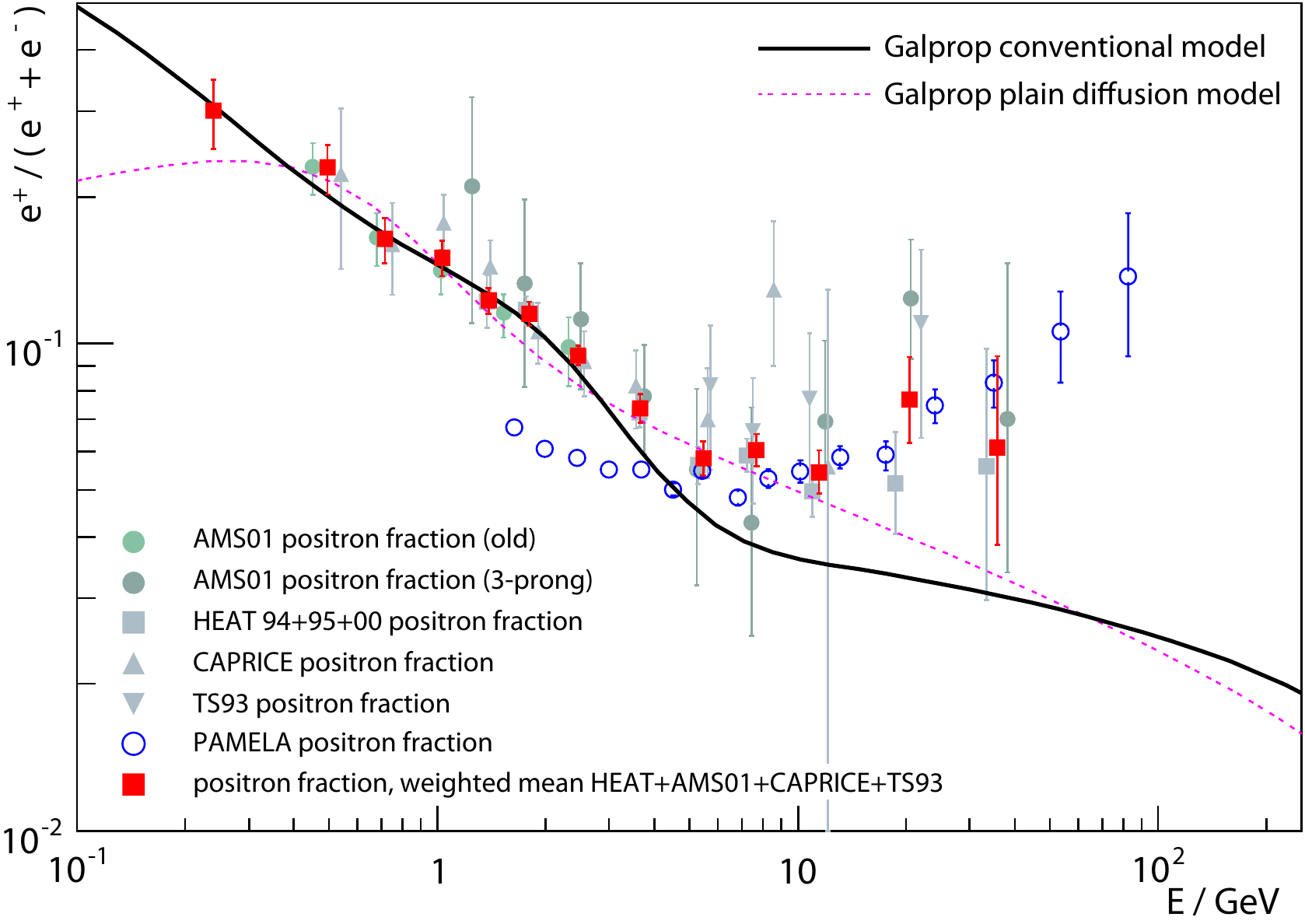}
\end{center}
\caption{Positron fraction data from
  AMS01~\cite{ref:ams01old,ref:ams01new}, HEAT~\cite{ref:heat},
  CAPRICE~\cite{ref:caprice}, and TS93~\cite{ref:ts93}, and the weighted
  mean of these data~\cite{ref:olzem}, together with the secondary background as
  predicted by Galprop's conventional and plain diffusion
  models. Recent data from PAMELA~\cite{ref:pamela_posfrac} are
  included, too.}
\label{fig:positronfraction}
\end{figure}
\begin{figure}
\begin{center}
\includegraphics[width=\textwidth,angle=0]{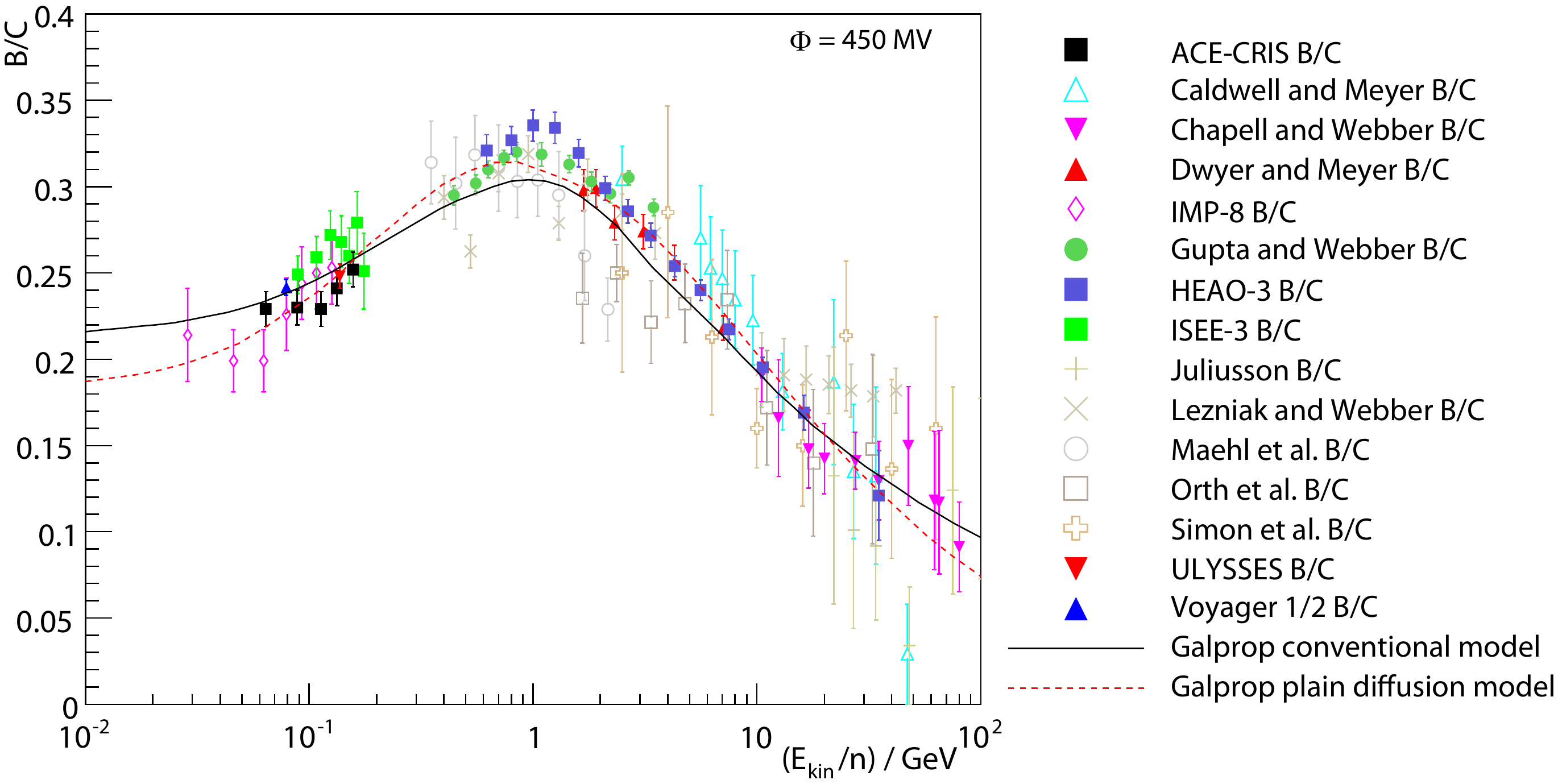}
\end{center}
\caption{Cosmic-ray B/C-ratio as a function of kinetic energy per
  nucleon, compared to Galprop's conventional and plain diffusion
  models. Data are from ACE-CRIS~\cite{ref:bc_ace}, Caldwell and
  Meyer~\cite{ref:bc_cm}, Chapell and Webber~\cite{ref:bc_cw}, Dwyer and
Meyer~\cite{ref:bc_dm}, IMP-8~\cite{ref:bc_imp8}, Gupta and
Webber~\cite{ref:bc_gw}, HEAO-3~\cite{ref:bc_heao3},
ISEE-3~\cite{ref:bc_isee3}, Juliusson~\cite{ref:bc_juliusson}, Lezniak
and Webber~\cite{ref:bc_lw}, Maehl et al.~\cite{ref:bc_maehl}, Orth et
al.~\cite{ref:bc_orth}, Simon et al.~\cite{ref:bc_simon},
ULYSSES~\cite{ref:bc_ulysses}, and Voyager~\cite{ref:bc_voyager}.}
\label{fig:bc}
\end{figure}
\begin{figure}
\begin{center}
\begin{tabular}{cc}
\includegraphics[width=0.5\textwidth,angle=0]{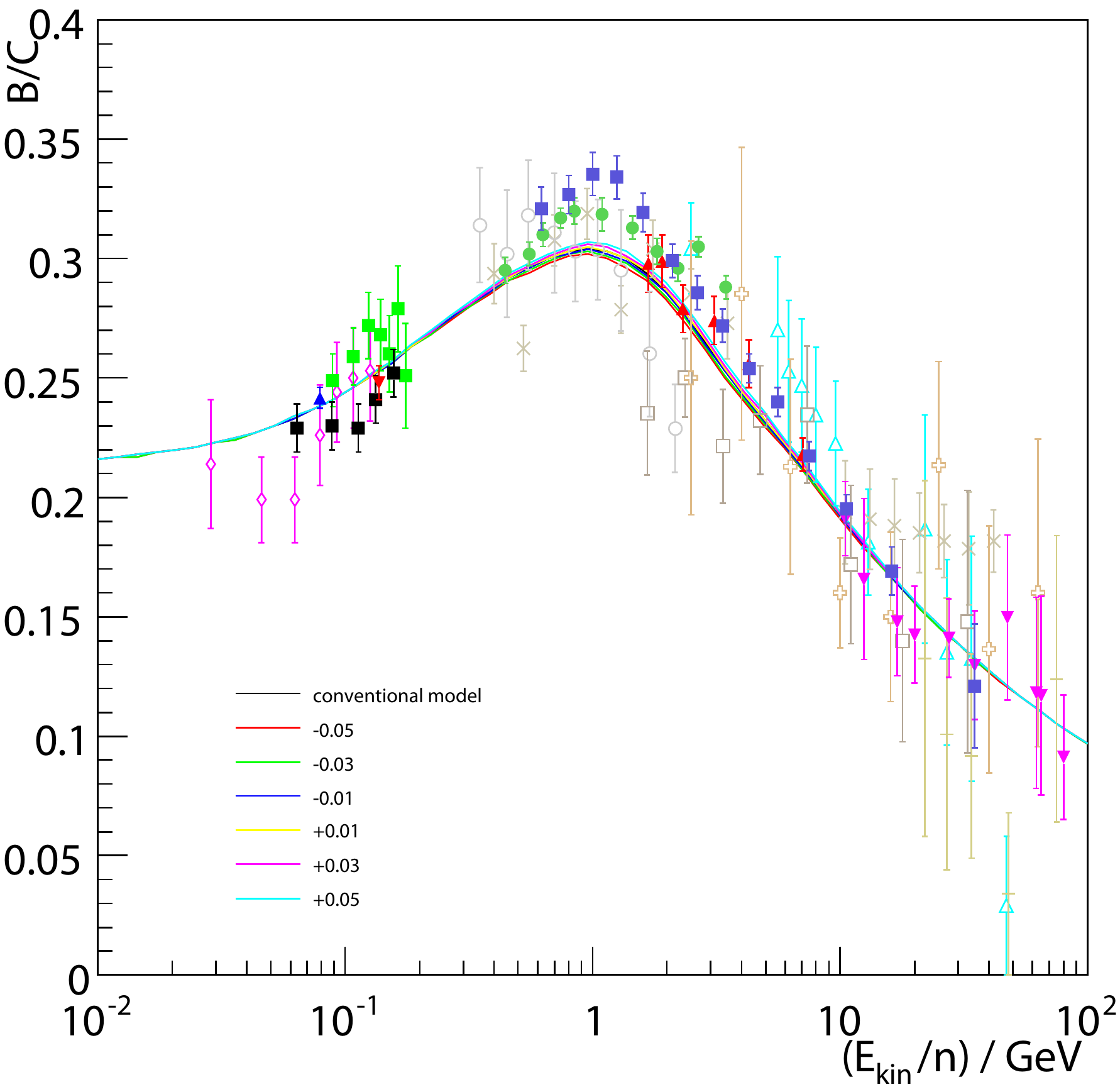}&
\includegraphics[width=0.5\textwidth,angle=0]{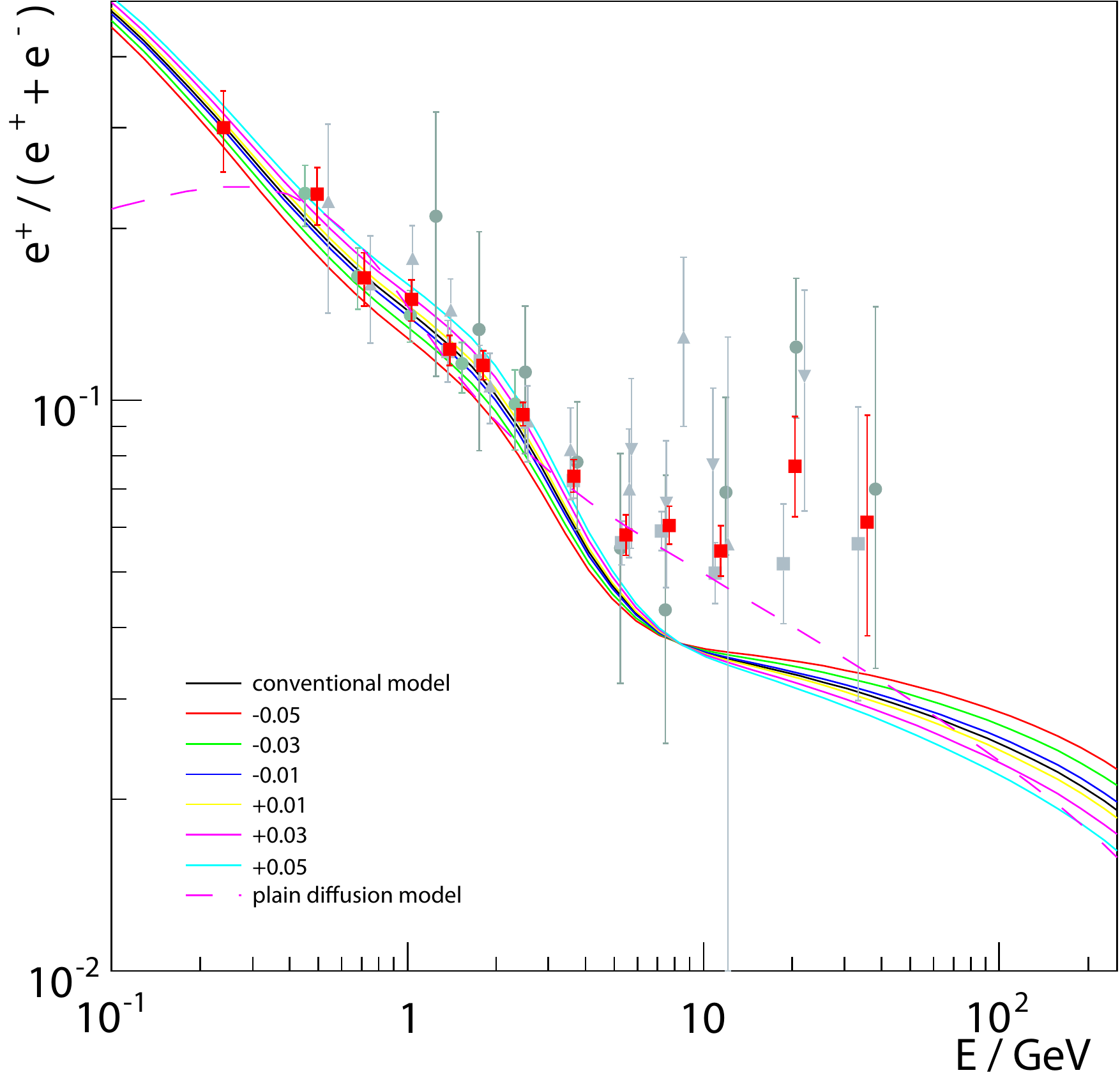}\\
\end{tabular}
\end{center}
\caption{Effect of variation of the nuclear injection spectral indices on B/C ratio ({\it left}) and
positron fraction ({\it right}).}
\label{fig:galprop_index_nuc}
\end{figure}
\begin{figure}
\begin{center}
\includegraphics[width=0.5\textwidth,angle=0]{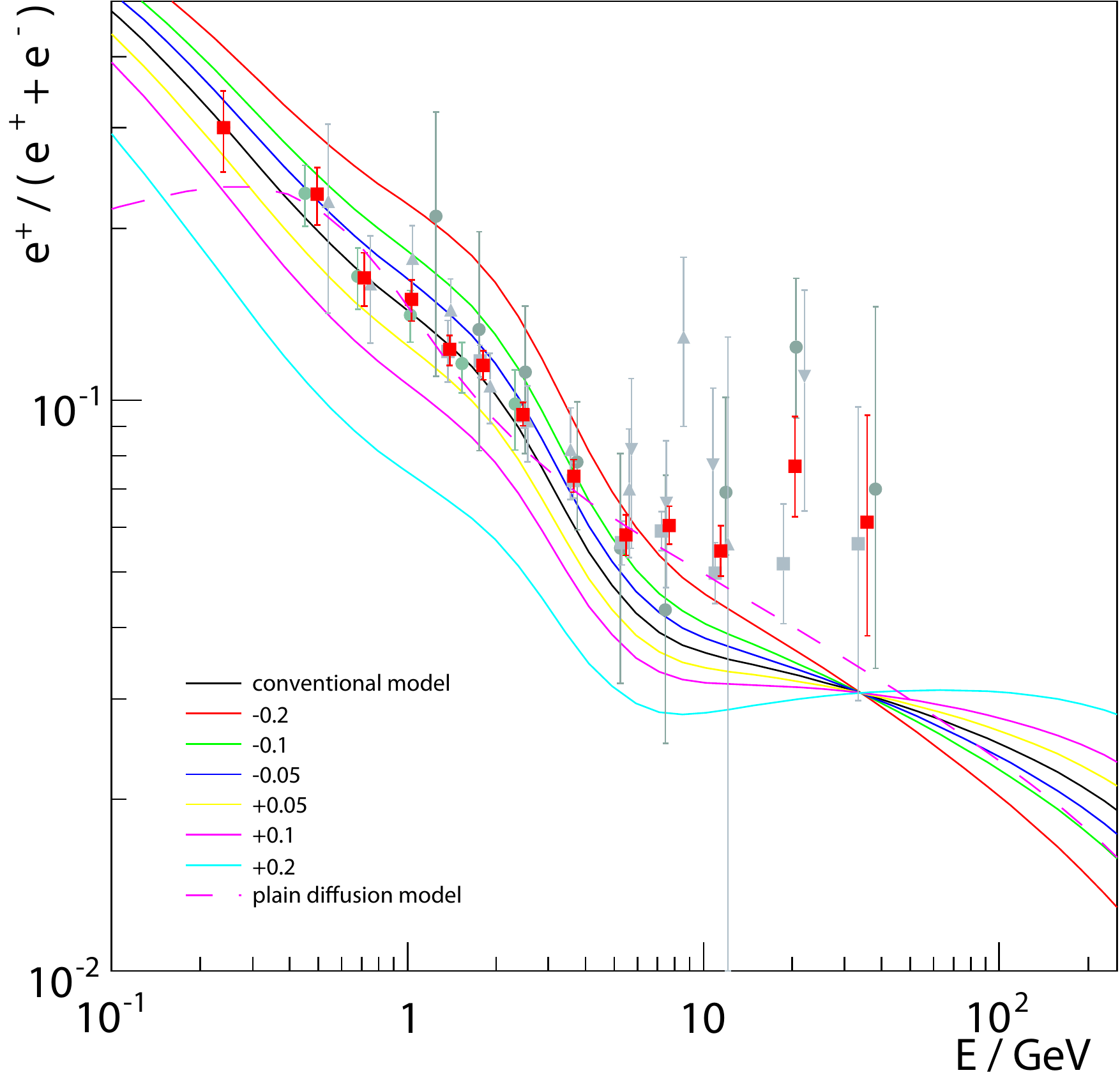}
\end{center}
\caption{Effect of variation of the electron injection spectral
indices on the positron fraction. The B/C ratio remains unchanged.}
\label{fig:galprop_index_elec}
\end{figure}
\begin{figure}
\begin{center}
\begin{tabular}{cc}
\includegraphics[width=0.5\textwidth,angle=0]{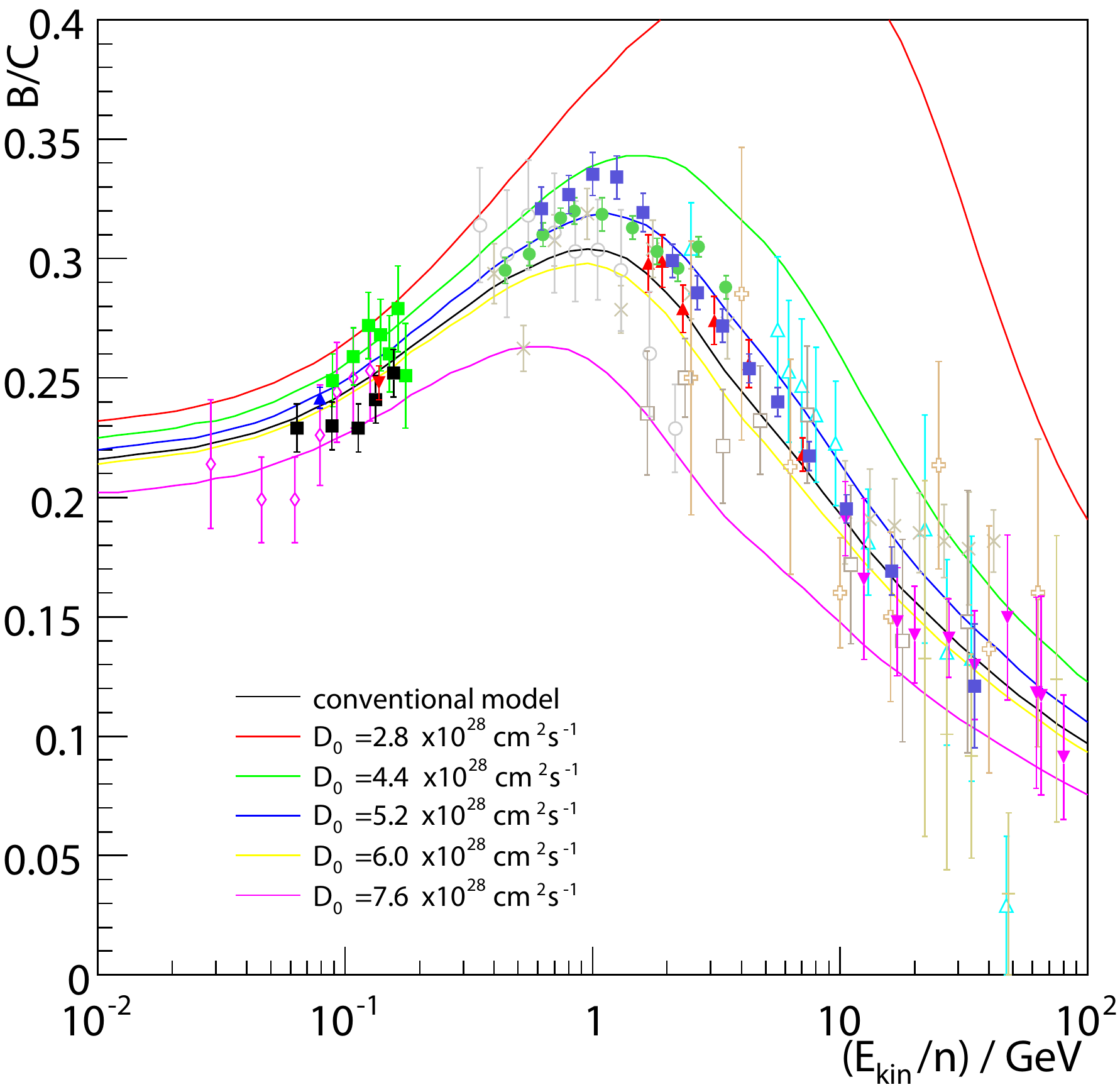}&
\includegraphics[width=0.5\textwidth,angle=0]{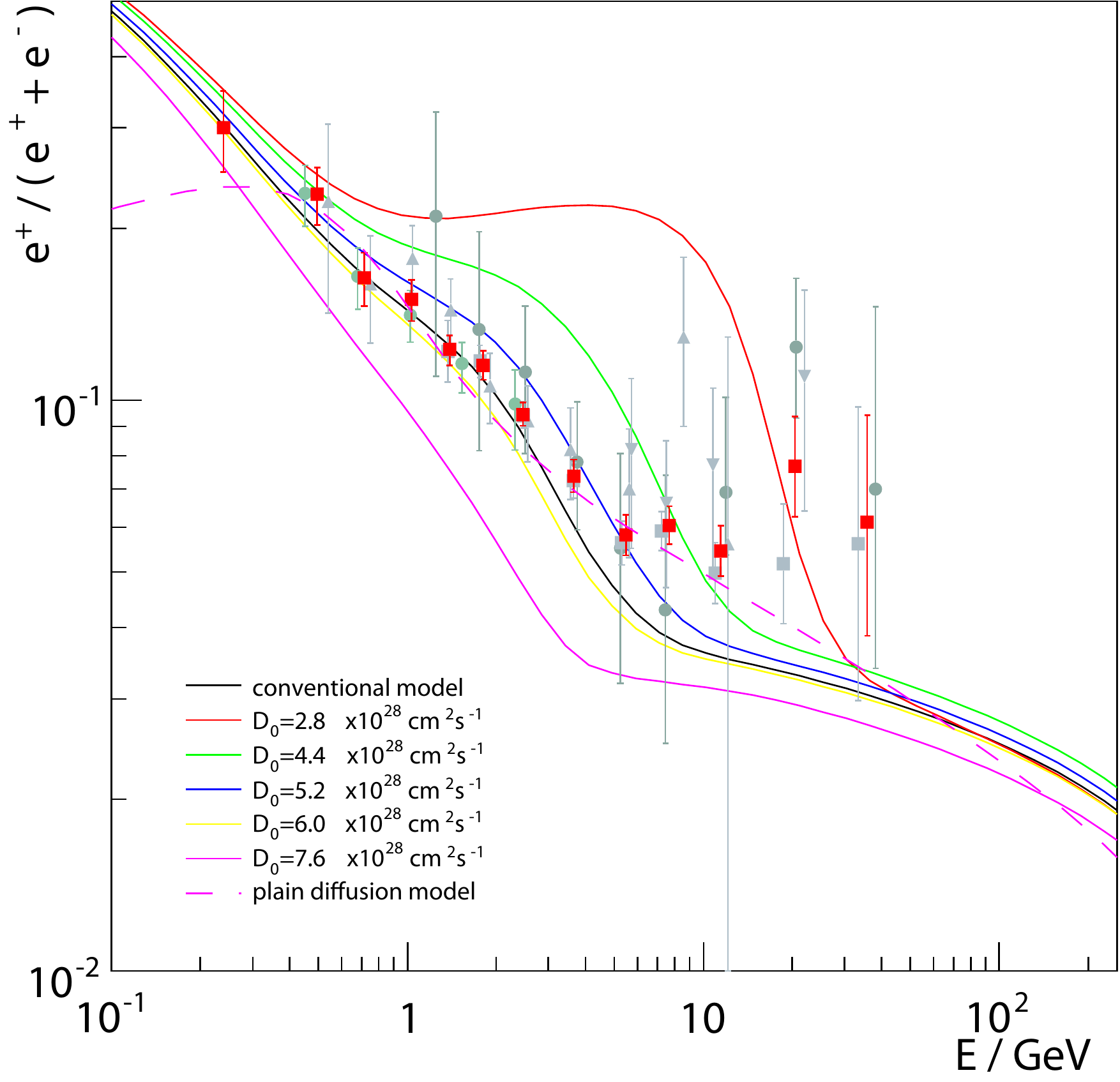}\\
\end{tabular}
\end{center}
\caption{Effect of variation of the diffusion coefficient on B/C ratio ({\it left}) and
positron fraction ({\it right}).}
\label{fig:galprop_d0xx}
\end{figure}
\begin{figure}
\begin{center}
\begin{tabular}{cc}
\includegraphics[width=0.5\textwidth,angle=0]{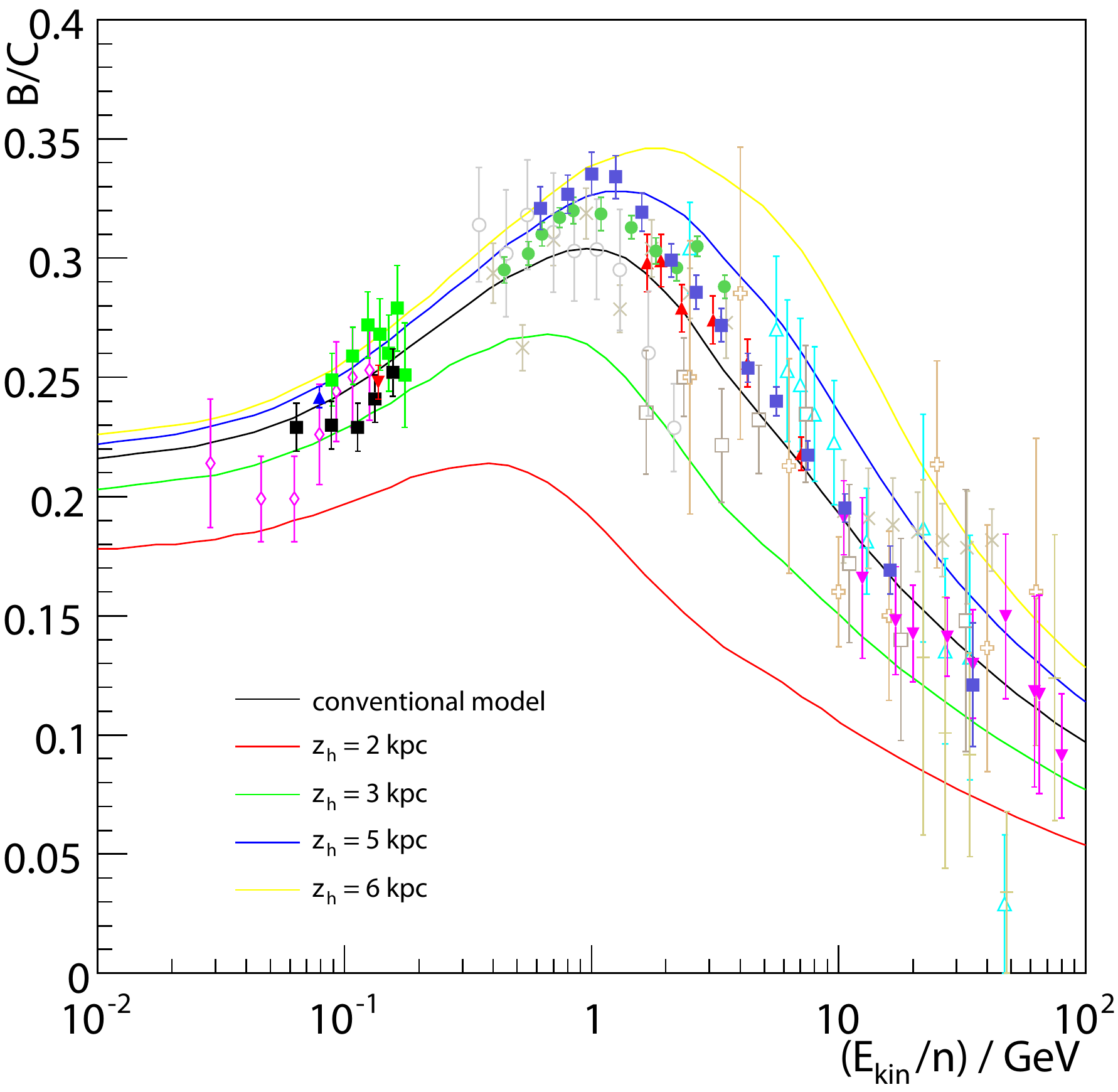}&
\includegraphics[width=0.5\textwidth,angle=0]{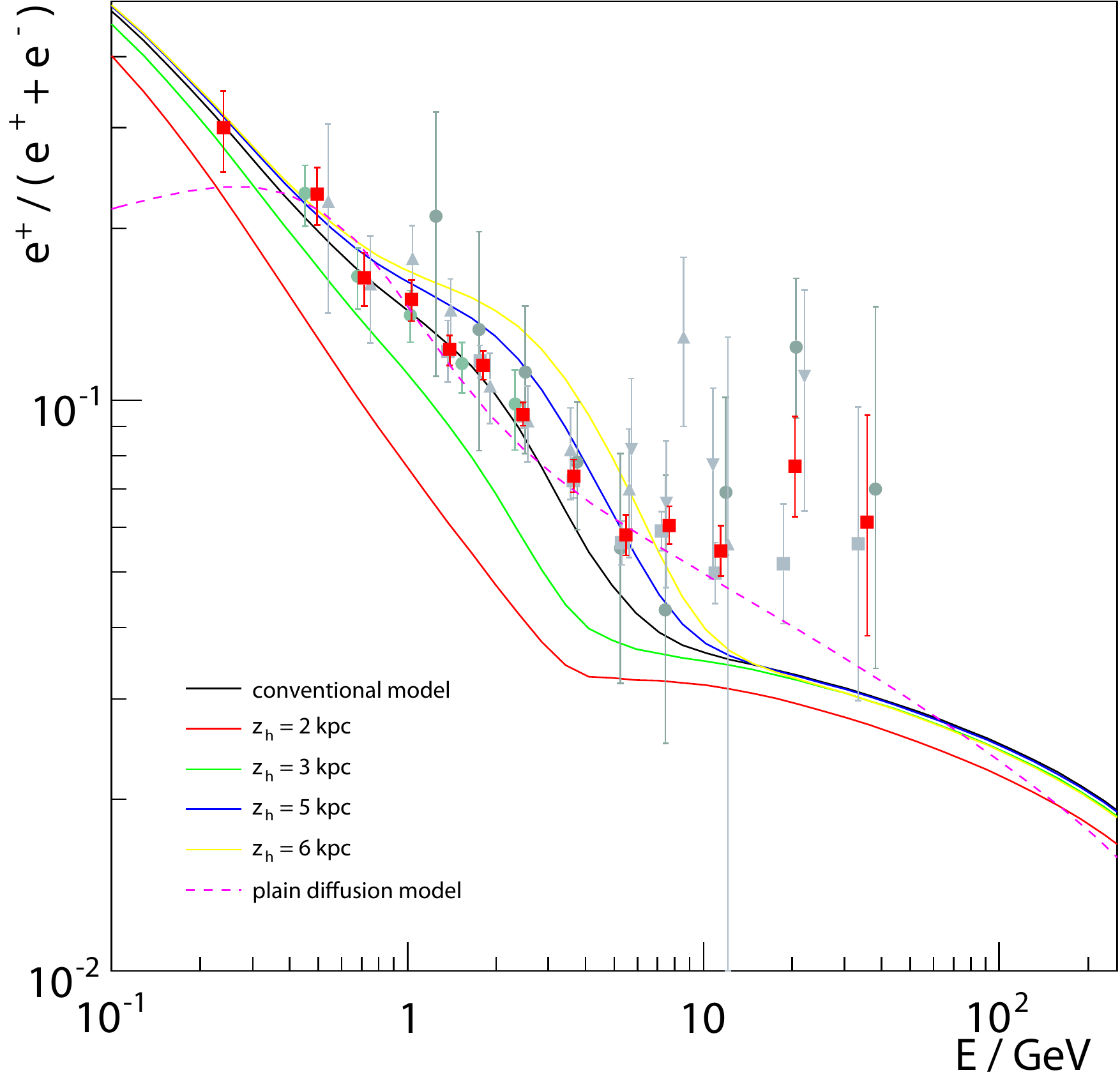}\\
\end{tabular}
\end{center}
\caption{Effect of variation of the halo size on B/C ratio ({\it left}) and
positron fraction ({\it right}).}
\label{fig:galprop_zhalo}
\end{figure}
\begin{figure}
\begin{center}
\begin{tabular}{cc}
\includegraphics[width=0.5\textwidth,angle=0]{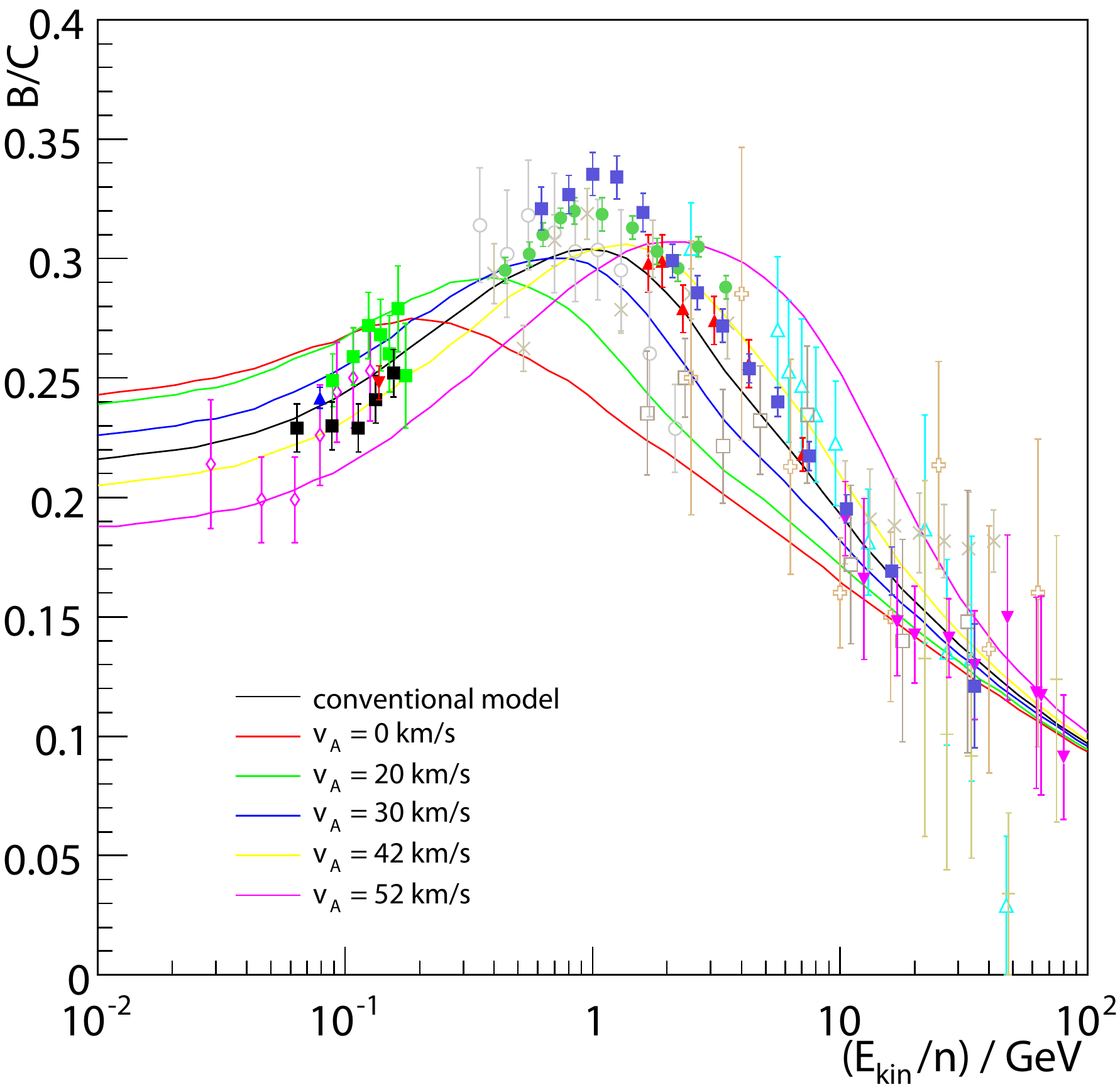}&
\includegraphics[width=0.5\textwidth,angle=0]{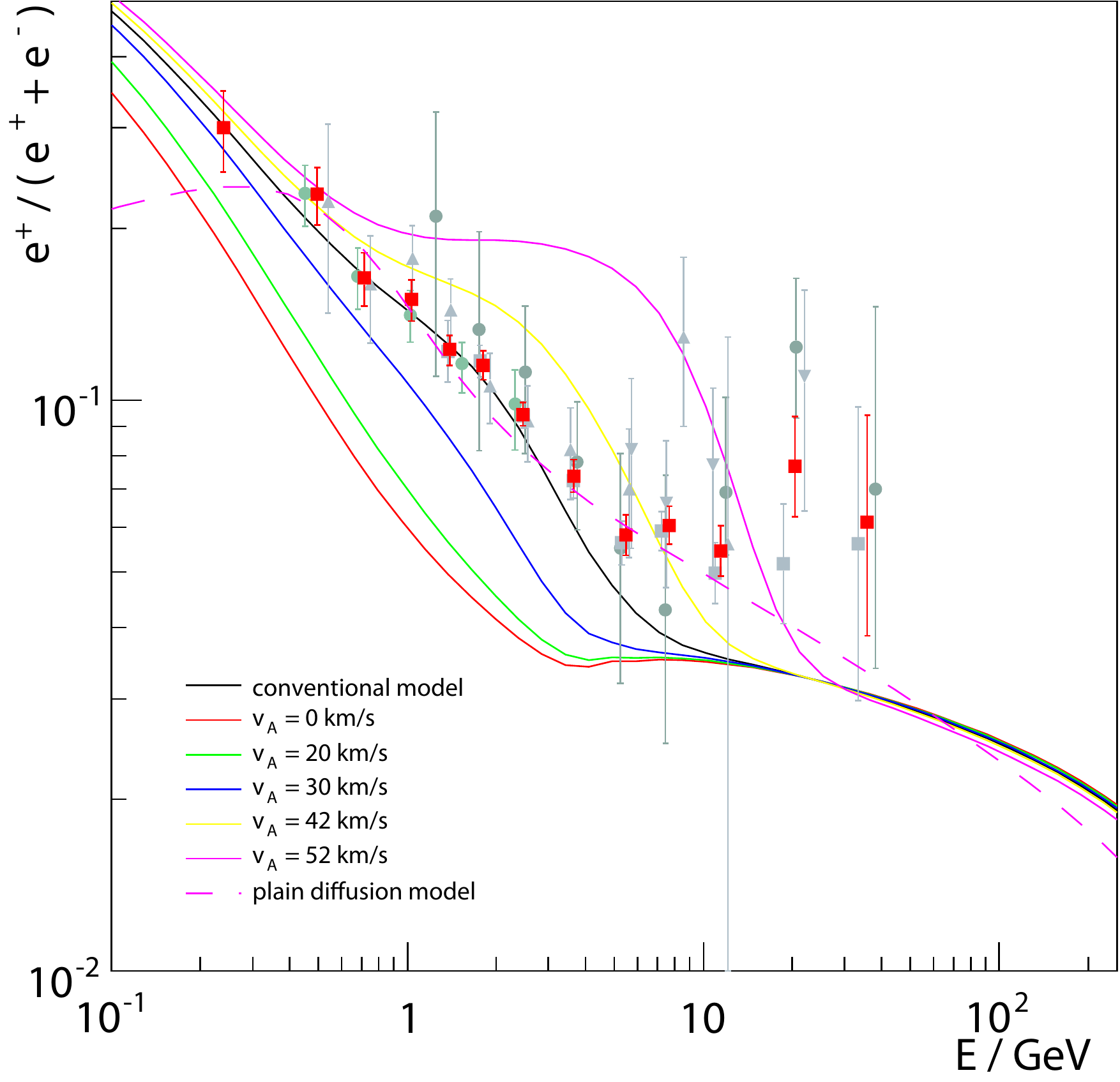}\\
\end{tabular}
\end{center}
\caption{Effect of variation of the Alfv\'en speed on B/C ratio ({\it left}) and
positron fraction ({\it right}).}
\label{fig:galprop_valfven}
\end{figure}
\begin{figure}
\begin{center}
\begin{tabular}{cc}
\includegraphics[width=0.5\textwidth,angle=0]{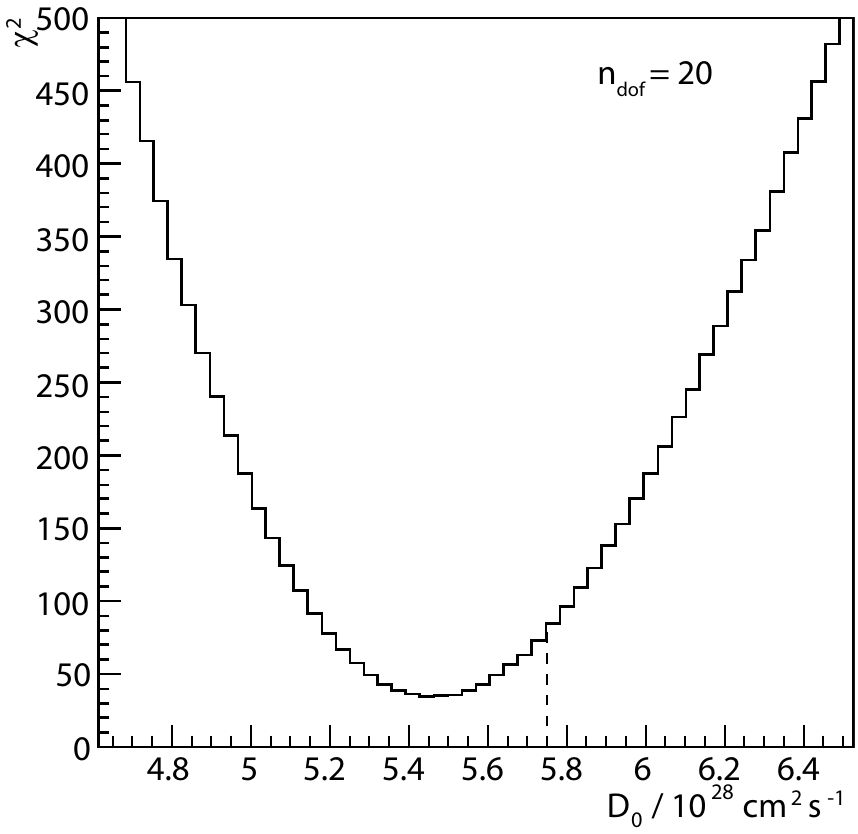}&
\includegraphics[width=0.5\textwidth,angle=0]{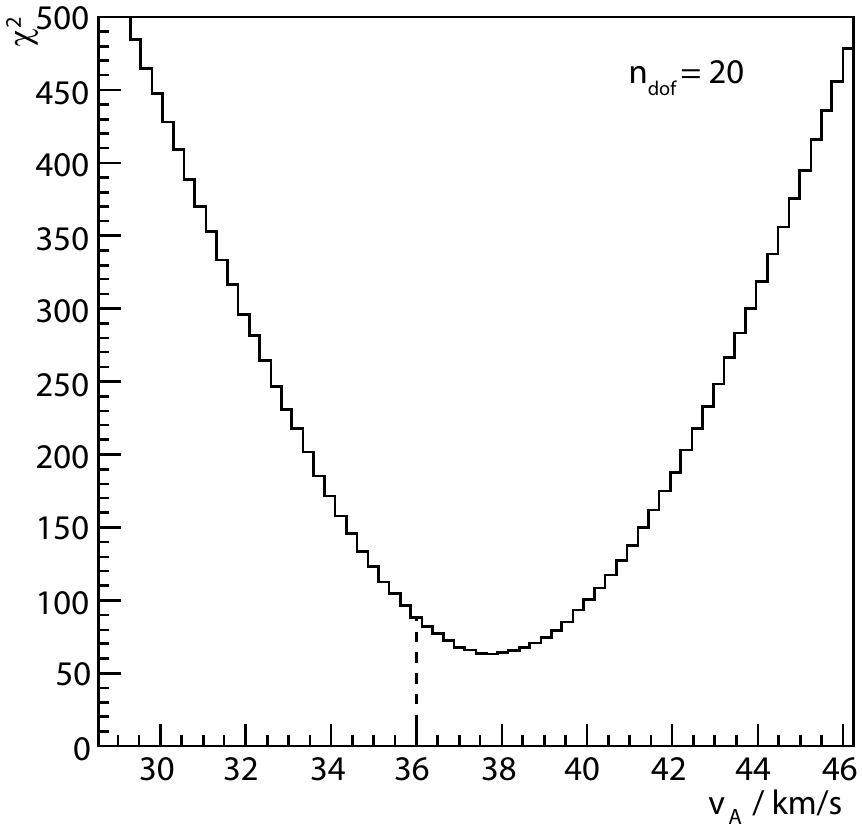}\\
\includegraphics[width=0.5\textwidth,angle=0]{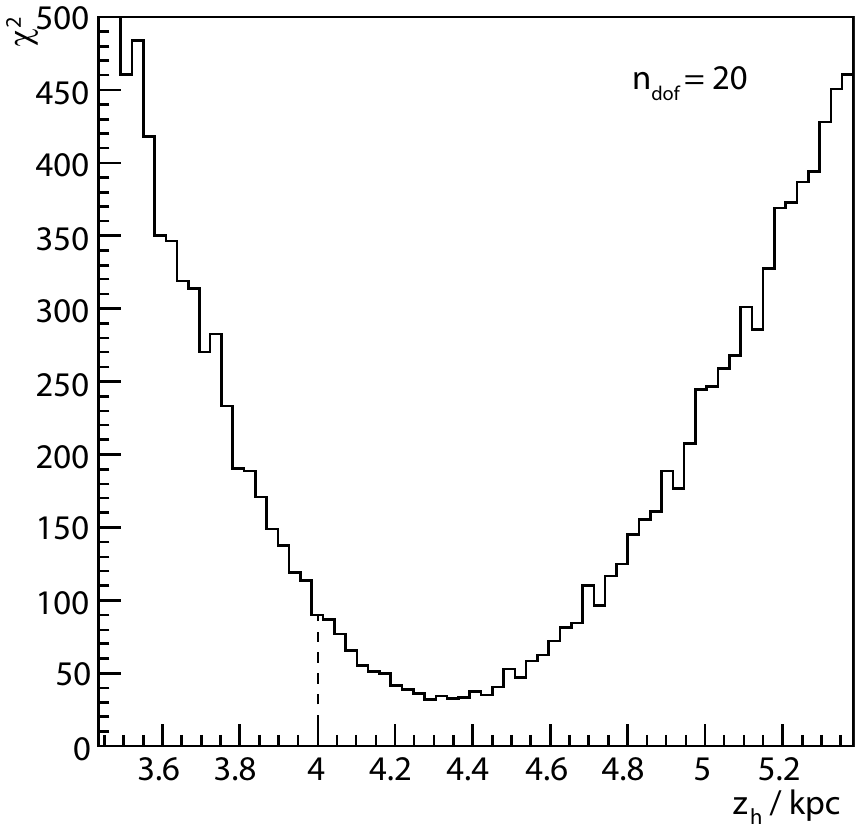}&
\includegraphics[width=0.52\textwidth,angle=0]{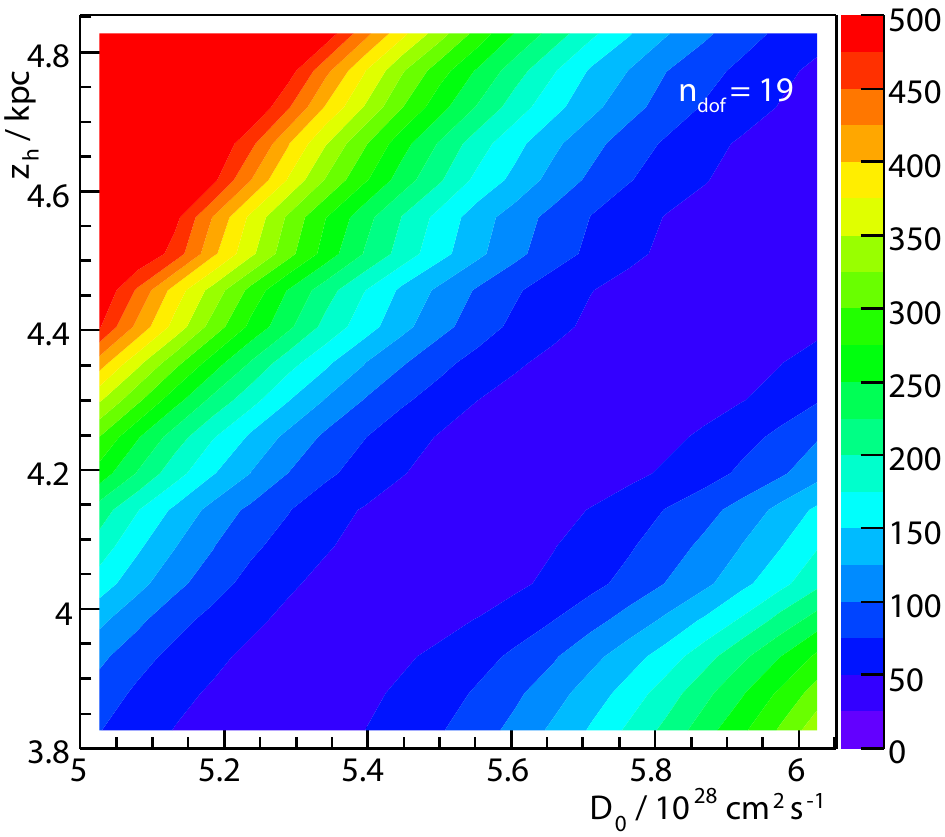}\\
\end{tabular}
\end{center}
\caption{Combined $\chi^2$ values, as defined in the text, for
  variations of the diffusion coefficient, Alfv\'en speed, and halo
  size, and in the $z_h$-$D_0$-plane. The respective nominal parameter
values in the conventional Galprop models are marked by dashed lines.}
\label{fig:galprop_chi2scan}
\end{figure}\begin{figure}[htb]
\begin{center}
\begin{tabular}{cc}
\includegraphics[width=0.5\textwidth,angle=0]{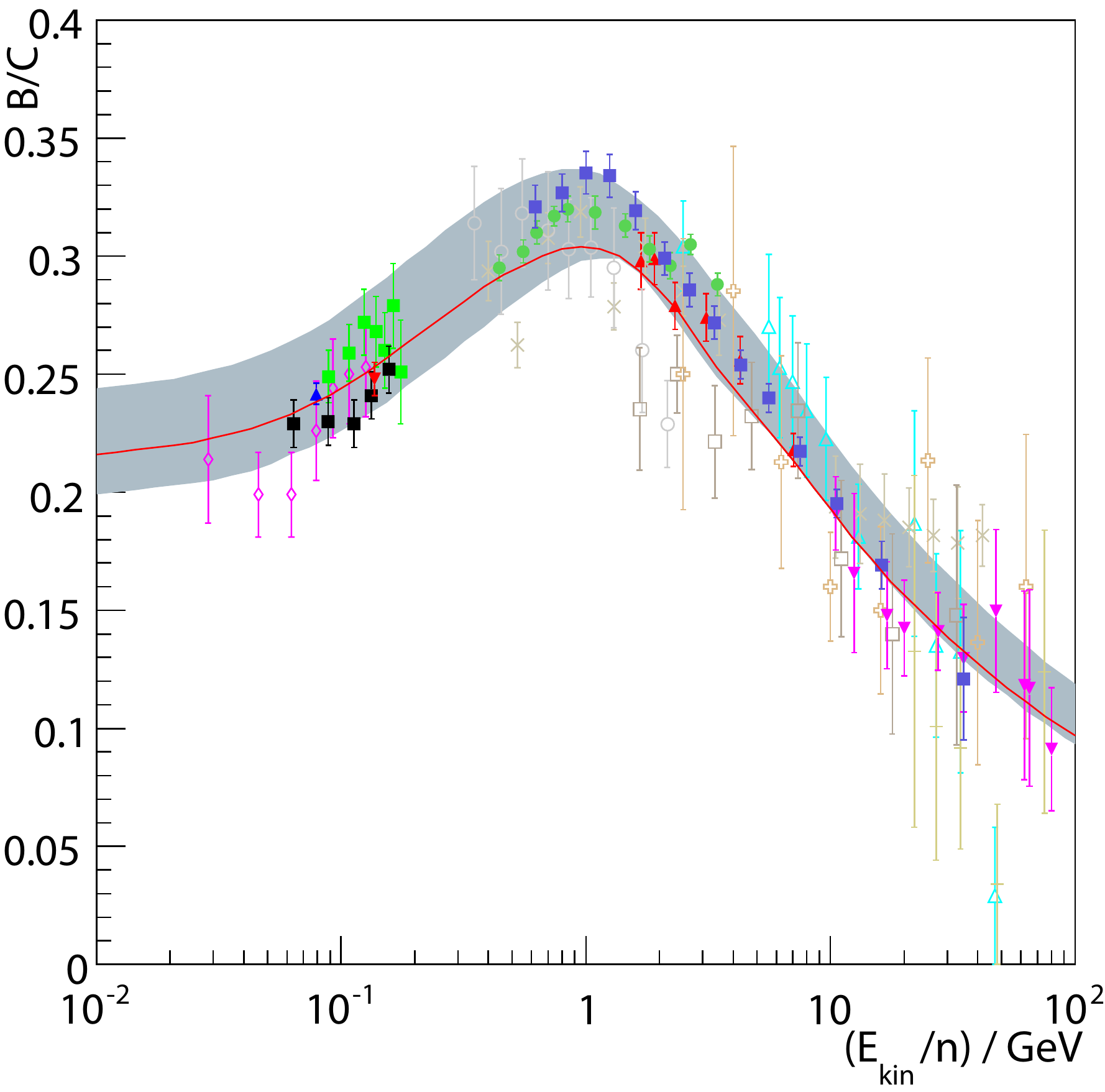}&
\includegraphics[width=0.5\textwidth,angle=0]{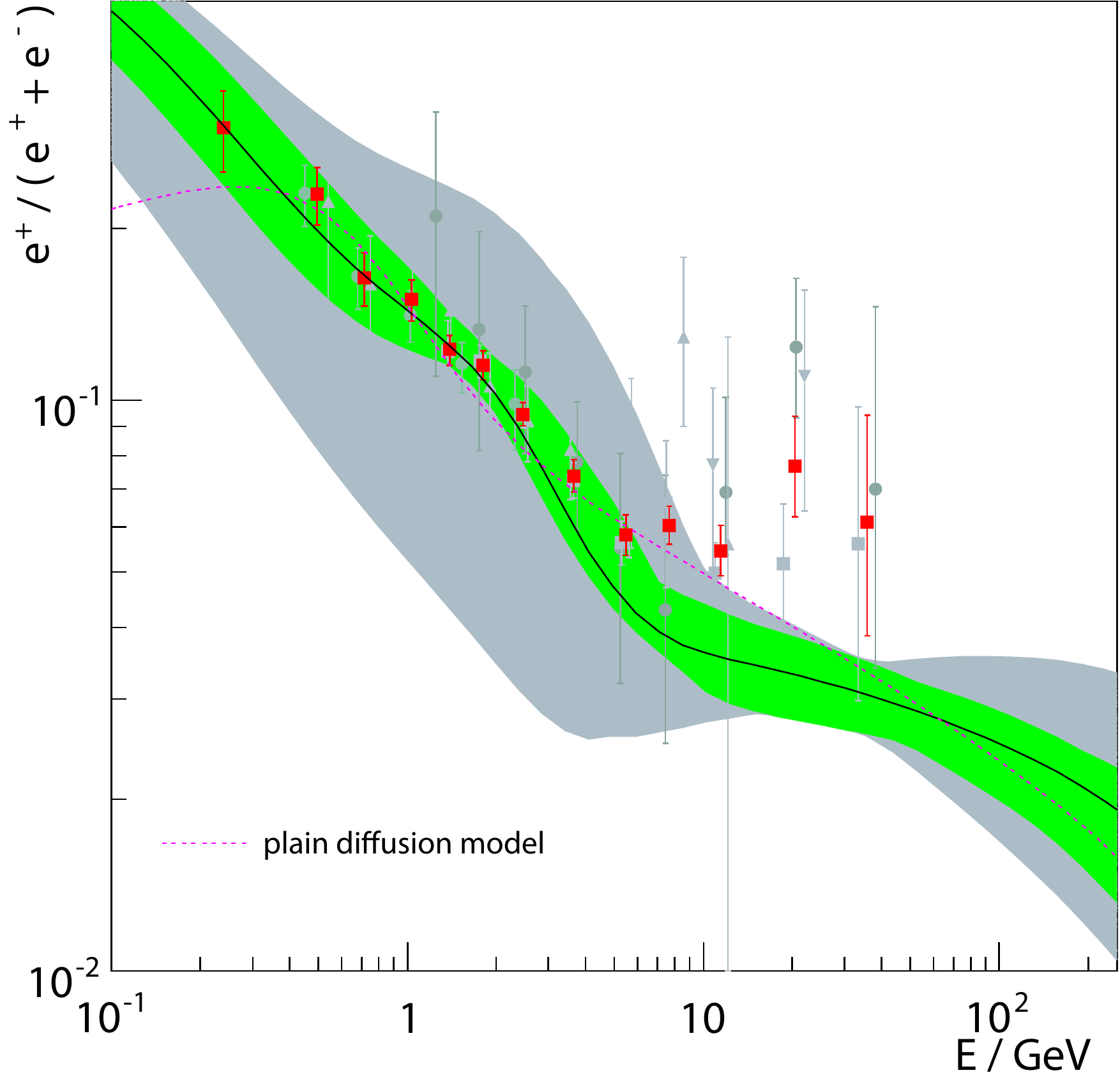}\\
\end{tabular}
\end{center}
\caption{{\it Left:} Wild scan results for models giving a description of the
B/C data with equal or better $\chi^2_\mathrm{HEAO-3}$ than the
conventional Galprop model fall within the grey band. The red line
shows the prediction of the conventional Galprop model itself. {\it Right:} The corresponding predictions for the
positron fraction fall within the grey band. Models additionally giving
$\chi^2_{e^+/(e^++e^-)}<20.8$ for the data points below
$3\,\mathrm{GeV}$, i.e.~outside the signal region, yield the green
uncertainty band.}
\label{fig:galprop_wildscan}
\end{figure}
As already stated in section~\ref{sec:propagation}, Galprop is used as
the propagation model. 
There are two sets of propagation parameters for Galprop that have
been found to give a good description of a wide range of available
cosmic-ray data. They are summarised in~\cite{ref:galpropmhd}.\\
The first standard Galprop model, called plain diffusion model, is based solely on diffusive
transport of cosmic rays, with neither convection nor
reacceleration. The prices to pay for this simplicity are an ad-hoc break in the
rigidity dependence of the diffusion coefficient and an additional
factor of $\beta^{-3}$ that need to be introduced to match the
$\mathrm{B}/\mathrm{C}$-ratio at low energies. Specifically, in this
model $D=\beta^{-2}D_0(R/R_0)^a$ with $a=0$ below and $a=0.6$ above
the reference rigidity $R_0=3\,\mathrm{GV}$ and $D_0=2.2\cdot10^{28}\,\mathrm{cm}^2\,\mathrm{s}^{-1}$.
The spectral index for
the nucleon source term is $\gamma_s=2.3$ below a break rigidity of
$40\,\mathrm{GV}$ and $\gamma_s=2.15$ above. For the electron source
term, the spectral index remains constant at $\gamma_e=2.4$.\\
On the other hand, the diffusive reacceleration -- or conventional --
model includes reacceleration, characterised by an Alfv\'en velocity
of $v_A=36\,\mathrm{km}\,\mathrm{s}^{-1}$. Here, 
the spectral indices for
the nucleon source term are $\gamma_s=1.82$ and $\gamma_s=2.36$ below
and above
a break rigidity of $9\,\mathrm{GV}$ while for the electron source
term, they are $\gamma_e=1.6$ below $4\,\mathrm{GV}$ and
$\gamma_e=2.5$ above. The diffusion coefficient has the well motivated form
$D=\beta{}D_0(R/R_0)^a$ with $a=1/3$ and $D_0=5.2\cdot10^{28}\,\mathrm{cm}^2\,\mathrm{s}^{-1}$.
The halo half-height is $z_h=4\,\mathrm{kpc}$ in both models. An overview of the predictions
of the conventional model for various particle species is given in figure~\ref{fig:protons} ({\it left}).\\
It should be noted that statements about propagation parameters must
always be considered in the context of a given model, for example the
Kolmogorov spectrum ($a=1/3$) is ruled out in the semi-analytic model
studied in~\cite{ref:maurinprop}.\\
\par
The cosmic-ray proton spectrum obtained in the conventional Galprop
model is shown in figure~\ref{fig:protons} ({\it right}). Before it can be compared
to data, solar modulation according to eq.~(\ref{eq:forcefield}) is
applied, the modulation parameter being extracted by a fit to the
data. The data shown were taken by space-borne or balloon
detectors, during several flights across a period of almost a decade,
corresponding to more than half of a solar cycle. The effect of solar
modulation at lower energies is clearly visible. A significant change of its
amplitude is apparent on the timescale of one year.\\
The most precise data on the electron and positron fluxes currently
available are compiled in figure~\ref{fig:electrons}. In the
low-energy regime, the solar modulation effect has
to be modelled to get a good description of the data. While the
electrons used to determine the AMS-01 electron flux have been
selected for low geomagnetic cutoffs, the dip towards the lowest
energies that is evident in fig.~\ref{fig:electrons} can only be
reproduced if a correction for geomagnetic effects, according to
(\ref{eq:geomod}), is applied as well.
In the
high-energy regime, the electron spectrum follows a power law that is
well reproduced by the Galprop models. The overall normalisation of
the spectra is a parameter of the model. The solar modulation
parameter found for the AMS-01 proton flux as
$\phi=(454\,\pm\,7)\,\mathrm{MV}$ is consistent with the value
obtained from the AMS-01 electron spectrum, which is
$\phi=(442\,\pm\,40)\,\mathrm{MV}$.\\
The positron flux has not
been measured reliably above $50\,\mathrm{GeV}$, and there is an
excess over the background expectation which will be discussed in more
detail in chapter~\ref{chapter:susyscan}.\\
Instead of the absolute positron flux, experiments often quote the
positron fraction, that is the flux ratio
$\Phi(e^+)/(\Phi(e^+)+\Phi(e^-))$, because certain systematic
uncertainties, such as uncertainties in the detector acceptance and
trigger live-time, cancel in this case. The effect of solar modulation
is also reduced, but not completely eliminated, as can be seen from
eq.~(\ref{eq:forcefield}). Positron fraction data from space and balloon
experiments, gathered over the past two decades, are assembled in
figure~\ref{fig:positronfraction}. They are shown together with the
prediction of the secondary background as obtained in the conventional
and plain diffusion models of Galprop. For easy comparison, the
weighted mean of the data from HEAT, AMS-01, CAPRICE and TS93 is
included. The
positron fraction data measured by the PAMELA detector have recently
become available in a pre-print~\cite{ref:pamela_posfrac} and are shown in the figure,
too. Towards high energies, the trend of an increasing positron
fraction as seen by the earlier experiments is confirmed. The
PAMELA data extend to higher energies and have smaller statistical
errors than all previous data. At low energies, the PAMELA data points
are lower than those of the earlier measurements and the prediction of
the propagation model. These new data will be put aside for the
moment. Their low-energy behaviour will be investigated
in section~\ref{sec:pamela_modulation}, and a possible exotic signal
will be discussed in section~\ref{sec:msugra_pamela}.\\
It is evident from the figure
that the existing data are still limited both in terms of statistical
uncertainties and energy range. This is caused by the steep decrease
of the positron flux with energy and the increasing difficulty in the experimental
determination of the charge sign for a high-energetic particle.
There is a significant difference in the predicted
secondary positron fraction between the conventional model and the plain
diffusion one, but the change in slope of the positron fraction towards high
energies apparent in the data can be reproduced in neither case.\\
The
figure already hints at the fact that the positron data may not only
be used to look for signals of new physics but is also vital for our
understanding of the propagation mechanism of cosmic rays. 
This statement will be corroborated in the remainder of this section by looking at the
dependence of the positron fraction on the most important parameters
in the propagation model used here. Naturally, this study will also give an idea
about the stability of the positron excess with respect to
uncertainties in the background model.\\
\par
Any comprehensive cosmic-ray model should
describe not only the spectra of primary particles, but also those of as
many secondary particles as possible. For instance, the abundance of
boron in cosmic rays is higher than its abundance in the solar system
by orders of magnitude. It was realised long ago that this is due to
the fact that boron is produced in spallation reactions of carbon on
interstellar matter. The amount of boron produced in this way depends
on the amount of matter traversed. Therefore, the correct description
of the ratio
$\mathrm{B}/\mathrm{C}$ of boron to carbon is a standard test for any
propagation model, although the detailed explanation of the observed
spectral shape remains a challenge. This can be seen in figure~\ref{fig:bc}.
Both Galprop models do a good job at describing the overall shape of the B/C
ratio, but the peak around $1\,\mathrm{GeV}/n$ is not as pronounced as
in the data.\\
In the following, the key parameters of the conventional
model will be varied to see the corresponding changes in the B/C ratio, which is used
to fix the model parameters, and the secondary positron fraction,
which constitutes the background for a possible exotic primary component. The key
parameters are the spectral indices $\gamma_s$ for nuclei and
$\gamma_e$ for electrons at
injection, the diffusion coefficient $D_0$, the Alfv\'en velocity
$v_A$, and the halo size $z_h$. The solar modulation parameters are
kept fixed in the process. During variation of one parameter, the
remaining parameters are fixed at their nominal values (table~\ref{tab:wildscan_table}).\\
The effect of variation of the spectral indices at injection is
illustrated in figures~\ref{fig:galprop_index_nuc}
and~\ref{fig:galprop_index_elec} for the nuclear and electron
injection spectra, respectively. At high energies, where solar
modulation effects do not distort the spectra, a harder proton
injection spectrum leads to a higher number of -- secondarily produced
-- positrons. In the same vein, a harder electron spectrum will
naturally lead
to a higher number of primary electrons at high energies thus reducing
the fraction of positrons.\\
The diffusion coefficient $D_0$ strongly influences the production of
secondaries, positrons and boron among others
(fig.~\ref{fig:galprop_d0xx}). The time needed to cross a distance $x$
in a diffusive process scales as $t\sim{}x^2/2D$. Therefore, particles
spend a longer time in the interstellar matter and thus produce more secondaries for a smaller
diffusion coefficient.\\
A similar effect is caused by a variation of the halo size $z_h$
(fig.~\ref{fig:galprop_zhalo}).
In a larger halo, more secondaries are
produced. The comparison of figures \ref{fig:galprop_d0xx} and
\ref{fig:galprop_zhalo} already shows that a degeneracy exists between
the parameters of the propagation model. This underlines the necessity
of comparing the model to as many different sets of data as possible to
develop a consistent picture of cosmic-ray propagation.\\
Lastly, figure~\ref{fig:galprop_valfven} shows how the Alfv\'en
velocity influences the position of the peak in the B/C-ratio and that
reacceleration must be included in order to describe the B/C-ratio
without an ad-hoc break in the diffusion coefficient. On the other hand,
the effect of variation of $v_A$ on the positron fraction is similar
to what is found when the diffusion coefficient is varied.\\
\par
As a last step, the uncertainty on the secondary
positron fraction as predicted in the conventional Galprop model needs to be estimated. To
that end, the parameters described in this section have been varied
around their nominal values.
In a first approach, $\chi^2$-plots for the parameters primarily affecting the spectra of
secondary species, namely the diffusion coefficient $D_0$, Alfv\'en
velocity $v_A$, and halo size $z_h$, are prepared
(fig.~\ref{fig:galprop_chi2scan}). The $\chi^2$ plotted is a
combined value, defined to quantify the deviation for both B/C data
and the positron fraction as
\begin{equation}
  \label{eq:combinedchi2}
  \chi^2\equiv\chi^2_\mathrm{HEAO-3}+\chi^2_{e^+/(e^++e^-),\leq3\,\mathrm{GeV}}
\end{equation}
An acceptable model is defined as
giving an acceptable $\chi^2$-value with respect to the B/C-data. More
specifically, the HEAO-3 data were chosen as comparison because they
represent a precise measurement that extends over the largest energy
interval. A solar
modulation parameter of $\phi=450\,\mathrm{MV}$ was used.\\
Due to
the rather good precision in the energy range up to a few GeV, the
positron fraction could in principle be used to further constrain the
parameter space of the propagation model. In practice, solar
modulation and geomagnetic effects at low energies complicate the picture, and in
addition, an energy range that is presumed to be free from any primary
signal needs to be defined. In the most conservative approach, only
models yielding too {\it many} positrons can be excluded.
Nevertheless, for a fixed set of modulation parameters, one can use
the low-energy positron fraction data below $3\,\mathrm{GeV}$ to
further constrain propagation models. The corresponding $\chi^2$ is
therefore added to the one from the B/C-data to form the combined
$\chi^2$ defined in~(\ref{eq:combinedchi2}).\\
The $\chi^2$-plots for the one-dimensional lines in the parameter
space must be viewed with caution, however. A somewhat clearer picture
emerges from correlation plots. As an example, the partial degeneracy
of the parameters $z_h$ and $D_0$ is demonstrated in figure~\ref{fig:galprop_chi2scan},
too.\\
Still, the complete picture could only be obtained from a complete
scan of the Galprop parameter space.
However, because the number of parameters is quite
large, a wild scan technique was employed to estimate the uncertainty
coming with the conventional Galprop model, tuned with the intention to
describe as wide a variety of primary and secondary data as possible.
1000~models were randomly generated with parameter values randomly
drawn according to a uniform distribution in the ranges given in
table~\ref{tab:wildscan_table}. For the injection indices, the bounds
\begin{table}[htb]
\begin{center}
\begin{tabular}{cccc}
parameter&nominal value&lower bound&upper bound\\ \hline
$\gamma_s$&1.82/2.36&1.77/2.31&1.87/2.41\\
$\gamma_e$&1.6/2.5&1.4/2.3&1.8/2.7\\
$D_0/\mathrm{cm}^2/\mathrm{s}$&$5.75\cdot{}10^{28}$&$4.4\cdot{}10^{28}$&$7.2\cdot{}10^{28}$\\
$v_A/\mathrm{km}/\mathrm{s}$&36&26&46\\
$z_h/\mathrm{kpc}$&4&3.2&5.5\\
\end{tabular}
\end{center}
\caption{Range of parameter variation for wild scan of the parameter
  space of the conventional Galprop model. Slashes separate values
  below and above the respective break rigidities quoted in the text.}
\label{tab:wildscan_table}
\end{table}
were adapted to the errors obtained from power-law fits to the
high-energy proton and electron spectra. The other parameters were
conservatively bounded such that the deviation from the measured
B/C-ratio became noticeable. The respective ranges can be read off
figures~\ref{fig:galprop_d0xx}~to~\ref{fig:galprop_valfven}.\\
For those
models giving a description of the B/C data equal to or better than
the conventional model ($\chi^2_\mathrm{HEAO-3}\leq79$, $n_\mathrm{dof}=9$), the
corresponding positron fraction was then plotted
(figure~\ref{fig:galprop_wildscan}). Though the resulting band of
models is quite large compared to the measurement errors, the overall
spectral shape remains the same, a power law trend that becomes harder
around $10\,\mathrm{GeV}$. It should be noted that the difference
between the conventional and the plain diffusion models is comparable
to that between two models at the outer ends of the wild
scan uncertainty band. However, none of the models compatible with the
B/C data can explain the excess seen in the positron fraction. 
The final uncertainty band is then constructed from models in
agreement with the positron fraction below $3\,\mathrm{GeV}$,
requiring a
$\chi^2_{e^+/(e^++e^-)}<20.8=5.7+15.1$ with respect to the weighted mean of the
positron fraction data. The conventional model has $\chi^2=5.7$, and
the statistical error at $99\,\%$ confidence level for $n=5$
parameters is then formally obtained by looking for a $\chi^2$
variation of $15.1$~\cite{ref:cowan}. The suitable models are drawn using
green colour in figure~\ref{fig:galprop_wildscan}. The exact value of
the cut seems arbitrary, especially taking the above mentioned
effects into account, but the magnitude of the resulting
uncertainty band remains unaffected. It will therefore be employed again
in chapter~\ref{chapter:susyscan}.

\section{A hint of charge-sign dependent solar modulation in the
  low-energy PAMELA data}
\label{sec:pamela_modulation}
\begin{figure}[htb]
\begin{center}
\begin{tabular}{cc}
\includegraphics[width=0.5\textwidth,angle=0]{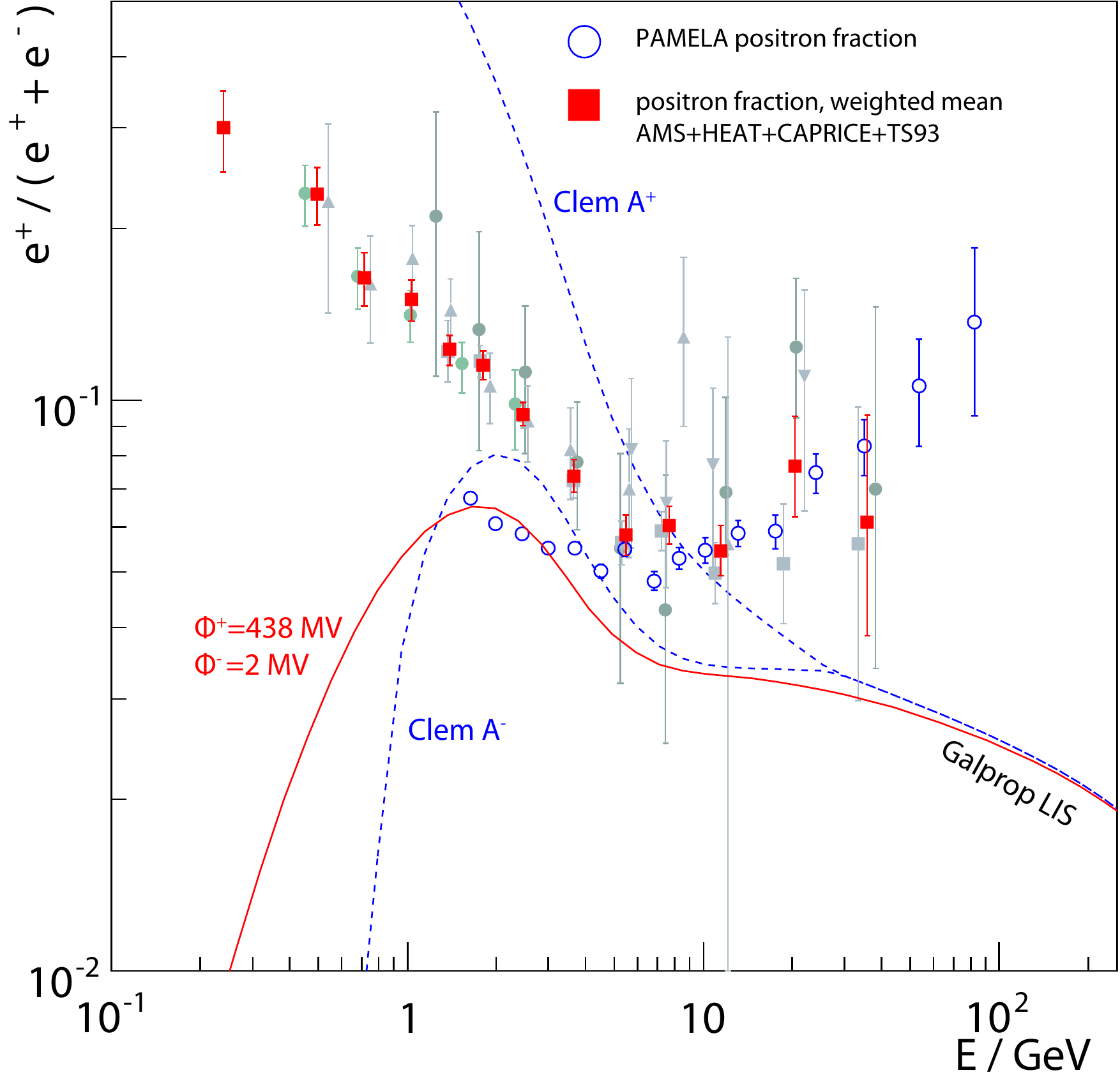}&
\includegraphics[width=0.5\textwidth,angle=0]{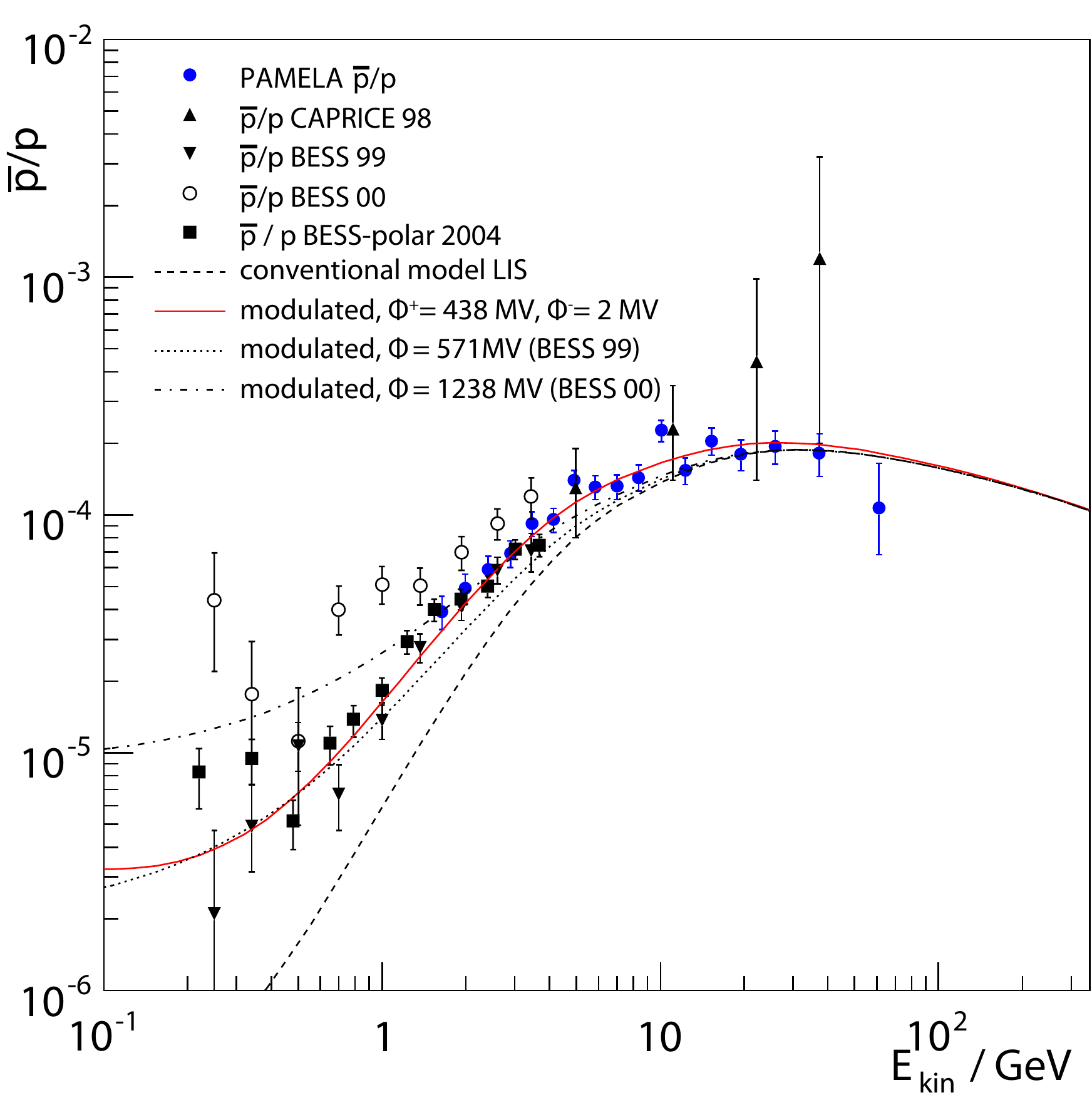}\\
\end{tabular}
\end{center}
\caption{{\it Left:} Positron fraction data compared to predictions
  for the low-energy behaviour, based on the local interstellar
  spectrum (LIS) obtained in the conventional Galprop
  model. The solid red line is based on charge-sign dependent
  modulation parameters in the force-field approximation
  formula~(\ref{eq:forcefield}), the dashed blue lines are obtained in
  the empirical model of Clem et al.~\cite{ref:clem}, as described in the
  text. Data are marked as in fig.~\ref{fig:positronfraction}.
{\it Right:} Antiproton-to-proton ratio measured by
PAMELA~\cite{ref:pamela_pbarp} compared to the prediction obtained
by applying the same charge-sign dependent solar modulation to the
secondary flux predicted by the propagation model as to the positron
fraction. Data from BESS-polar~\cite{ref:pbarp_besspolar}, taken in
the same $A^-$-cycle, are included, too. Some data taken in the
previous $A^+$-cycle, by BESS~\cite{ref:bess99} and
CAPRICE~\cite{ref:capricepbar}, are compared to predictions in the
charge-symmetric case, with parameters obtained from fits to the
corresponding proton spectra.}
\label{fig:chargesign_solarmod}
\end{figure}
At first glance, the positron fraction measured by PAMELA at low
energies, below $10\,\mathrm{GeV}$, seems puzzling. While the other
measurements taken in recent years agree well in this energy range,
the PAMELA data points indicate significantly fewer positrons
(fig.~\ref{fig:positronfraction}). If the pronounced rise at high energies
apparent in the PAMELA data is to be
taken seriously, it must first be shown that this observation does not
point to a systematic error in the response of the PAMELA apparatus
nor the data analysis. In this section, it is argued that charge-sign
dependent solar modulation~\cite{ref:clem} is a possible way to explain all
measurements quoted above.\\
\par
The amplitude of solar modulation varies along with the solar cycle,
with its well known half-period of eleven years. A good measure for
the activity of the sun is the sunspot number, which has been observed
almost continuously for the last centuries. Although the magnetic
field of the sun is complex, it is nearly always dominated by the
dipole term. The projection of this dipole on the solar rotation axis
can be either positive or negative and these two states are referred
to as $A^+$ and $A^-$, respectively. The dipole reverses direction at
each sunspot maximum, leading to alternating magnetic polarity in
successive solar cycles~\cite{ref:babcock}.\\
\par
It can be expected that solar modulation depends on the charge sign of
a particle, affecting positrons and electrons differently. As a simple
extension of the force-field approximation used so far, it can be
assumed that the modulation parameter $\phi$ in (\ref{eq:forcefield})
is charge-dependent and takes different values $\phi^+$ and $\phi^-$
for positively and negatively charged particles, respectively.\\
Allowing for different values of $\phi^+$ and $\phi^-$, a fit of the
local interstellar positron fraction calculated in the conventional
Galprop model to the PAMELA positron fraction data below
$4\,\mathrm{GeV}$ yields values of $\phi^+=438\,\mathrm{MV}$ and
$\phi^-=2\,\mathrm{MV}$ with statistical uncertainties of $4\,\mathrm{MV}$
(fig.~\ref{fig:chargesign_solarmod} {\it left}). These
values mean that electrons can reach the Earth almost unhindered by
the solar wind while positrons are moderately suppressed so that the
fraction of positrons is reduced with respect to the charge-symmetric
case.\\
In an empirical model of charge-dependent solar modulation, Clem et
al.~\cite{ref:clem} assumed that the flux $J_E$ of a given species
with charge sign $q$ measured at Earth is related to the interstellar
flux $J_{IS}$ by
\begin{equation}
  \label{eq:clemmodel}
  J_E(R,\hat{\phi},qA)=C(qA,R)\,\cdot\,M(R,\hat{\phi})\,\cdot\,J_{IS}(R)
\end{equation}
where $R$ is rigidity, $A$ is the solar magnetic polarity, and
$\hat{\phi}$ is the phase of the solar cycle. $C$ and $M$ are two
modulation factors, and $\hat{\phi}$ is associated to the modulation
parameter considered before. This simple model neglects the adiabatic
deceleration present in (\ref{eq:forcefield}), but it has the
advantage that $J_E$ can be expressed in terms of just $J_{IS}$ and
$\varrho(R)$, the ratio of the total electron fluxes in the $A^+$-cycle to
the total electron fluxes in the $A^-$-cycle. The empirical data on
$\varrho(R)$ as assembled by Clem et al.~can be parameterised as
\begin{equation}
  \label{eq:clemparam}
  \varrho(R)=0.166\,\log(R/\mathrm{GeV})+0.452
\end{equation}
It can then be shown that $J_E$ is given by
\begin{equation}
  \label{eq:clemformula_aplus}
  J_E=\frac{J_{IS}^2(R+1)-RJ_{IS}}{2J_{IS}-1}\qquad\mathrm{for}\:A^-
\end{equation}
and
\begin{equation}
  \label{eq:clemformula_aminus}
  J_E=\frac{J_{IS}^2(R+1)-J_{IS}}{R(2J_{IS}-1)}\qquad\mathrm{for}\:A^+
\end{equation}
The resulting curves are included in
fig.~\ref{fig:chargesign_solarmod}. It can be seen that the trend
predicted for the $A^-$ cycle, during which the PAMELA data were
taken, is rather similar to the curve predicted in the
$\phi^\pm$-model presented above.\\
\par
The antiproton-to-proton ratio measured by PAMELA~\cite{ref:pamela_pbarp} can be used to
cross-check the $\phi^\pm$-model. In fact, using the same $\phi^\pm$-values
as for the positron fraction, the $\bar{p}/p$-ratio can be reproduced
well (\ref{fig:chargesign_solarmod} {\it right}). To put this result
into perspective, some data taken in the
previous $A^+$-cycle are compared to predictions in the
charge-symmetric case, with parameters obtained from fits to the
corresponding proton spectra. The BESS data taken at solar maximum
show some discrepancy with respect to the prediction obtained in the
charge-symmetric case. It is unclear however if this points to a
problem in the modulation model, the propagation model providing the
interstellar fluxes, or both.\\
\par
\begin{figure}[htb]
\begin{center}
\includegraphics[width=0.9\textwidth,angle=0]{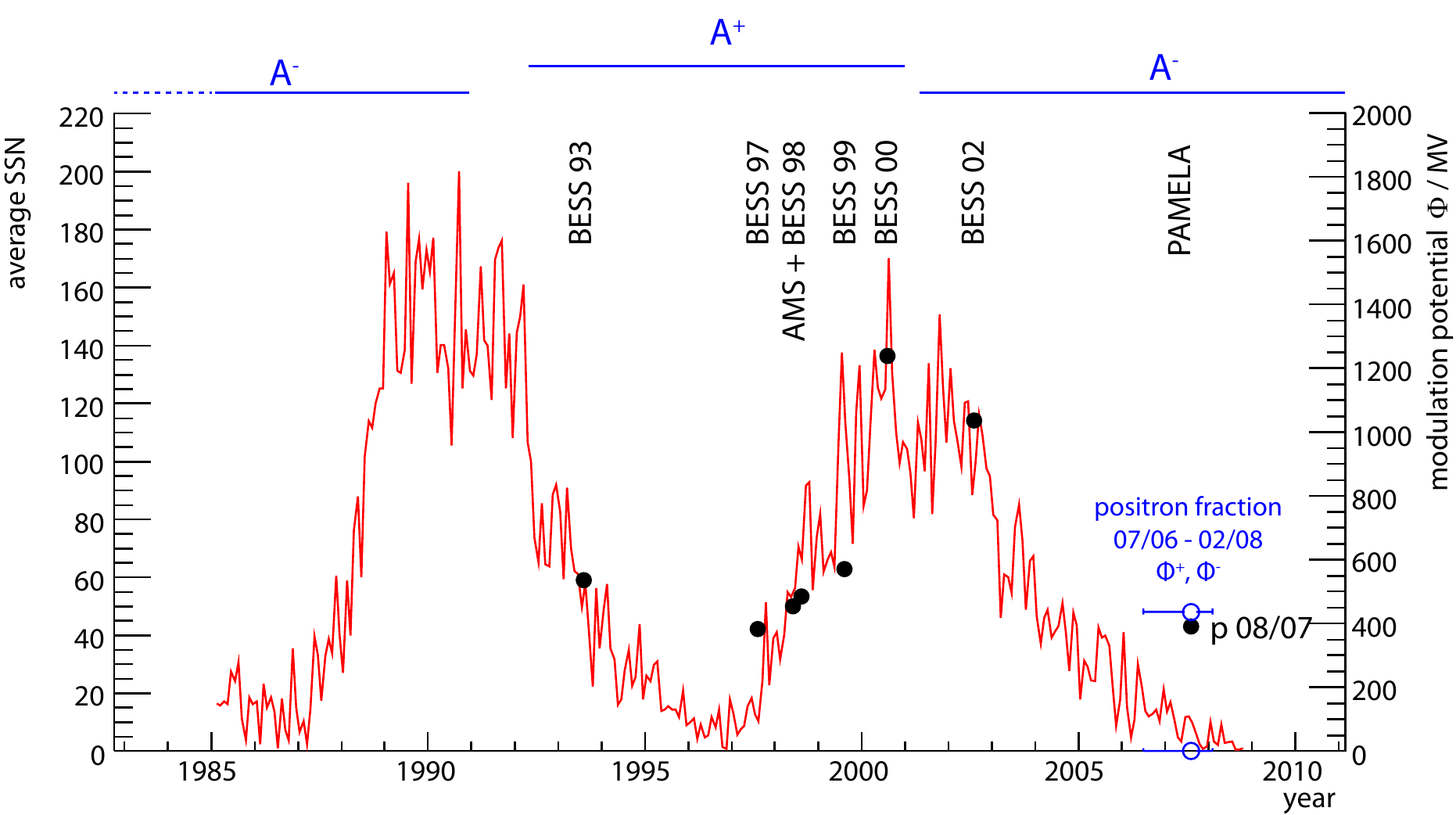}
\end{center}
\caption{Monthly average of the observed sunspot
  numbers~\cite{ref:sunspotnumber} since 1985, compared to the values of the
  modulation parameter extracted from the fits to the proton spectra
  included in fig.~\ref{fig:protons}. The charge-sign dependent parameters
  used to describe the low-energy positron fraction measured by PAMELA
  are depicted, too. The cycle of the solar magnetic polarity is
  indicated by the bars at the top of the figure, with the approximate
start and end dates taken from~\cite{ref:snodgrass} and~\cite{ref:durrant}.}
\label{fig:sunspots}
\end{figure}
Looking at the correlation of the solar activity, expressed in terms
of the sunspot number, with the modulation potential obtained from
fits to the proton spectra taken in recent years (\ref{fig:sunspots}),
a good match is found in general. The value for the PAMELA protons is
somewhat higher than expected from the trend implied by the solar
data, by a margin comparable to the difference in the values of
$\phi^+$ and $\phi^-$ quoted above.\\
\par
A prediction of this model of charge-dependent solar modulation is an
unexpected and rather drastic decrease of
the positron fraction at the lowest energies, below
$1\,\mathrm{GeV}$ (fig.~\ref{fig:chargesign_solarmod}). A new measurement
at these energies during the current solar cycle is therefore highly
desirable.\\
\par
\begin{figure}[htb]
\begin{center}
\includegraphics[width=0.5\textwidth,angle=0]{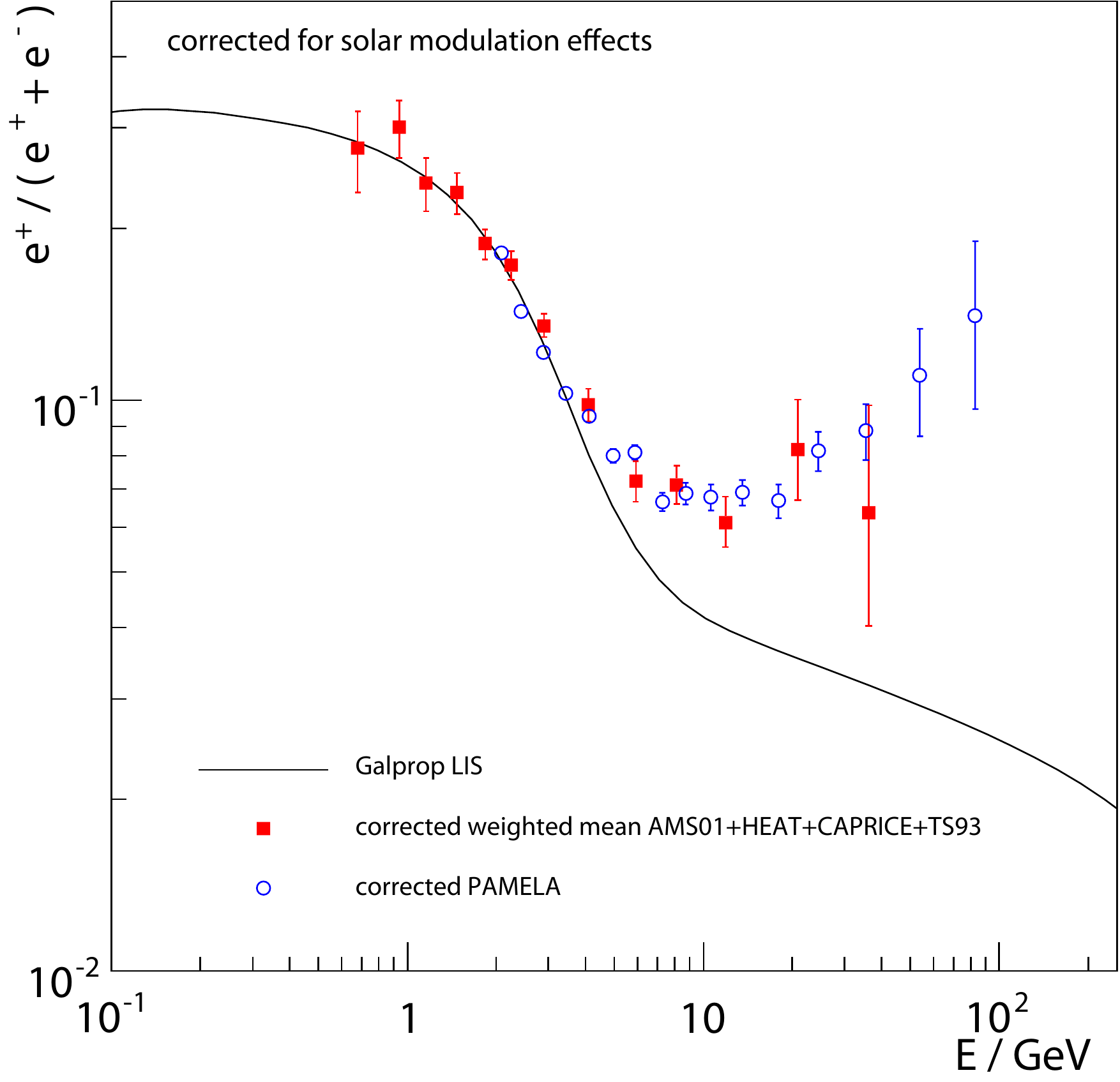}
\end{center}
\caption{Positron fraction data corrected for solar modulation effects
according to the Galprop conventional model. PAMELA data have been
corrected based on the charge-sign dependent model, the weighted mean
of the previously published data has been corrected based on a
charge-symmetric model using $\phi=442\,\mathrm{MV}$.}
\label{fig:corrposfrac}
\end{figure}

Using the model of solar modulation presented in this section and the
local interstellar positron and electron spectra calculated in the
Galprop conventional model, both the previously published data and the new
PAMELA positron fraction data can be corrected for the solar
modulation effects to obtain estimates of the interstellar amplitudes (fig.~\ref{fig:corrposfrac}).
The result shows that in this model, the PAMELA data are in very good
agreement with the weighted mean of the data from AMS-01, HEAT,
CAPRICE, and TS93.

\chapter{Detectors to measure the cosmic-ray positron fraction}
\label{chapter:pebs_design}
In this chapter, first the two
contemporary players in the field of indirect dark matter search using
the cosmic-ray positron fraction, AMS-02 and PAMELA, will be presented
briefly. Then, an overview of the design concept for the new PEBS detector~\cite{ref:vcihenning,ref:icrchenning,ref:vcigregorio}
will be given. It is shown which detector components are needed in order to
fulfil the mission objectives and how these components will work
together. Details of their working principles, implementation in the
simulation software, and the projected performance will be treated in
chapter~\ref{chapter:pebs_design_study}.

\begin{figure}[htb]
\begin{center}
\begin{tabular}{cc}
\includegraphics[width=0.4\textwidth,angle=0]{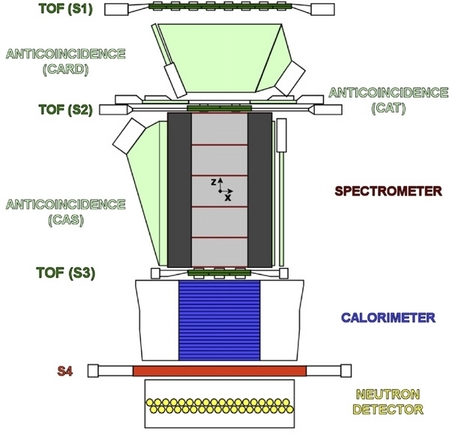}&
\includegraphics[width=0.5\textwidth,angle=0]{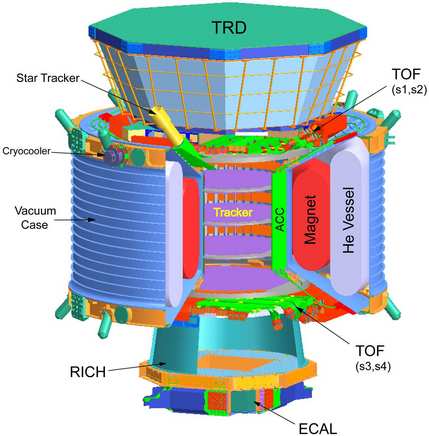}\\
\end{tabular}
\end{center}
\caption{Schematic drawings of the PAMELA {\it (left)} and
AMS-02 detectors {\it (right)}. PAMELA drawing reprinted from~\cite{ref:pamelaoverview}, Copyright 2006, with
permission from Elsevier.}
\label{fig:pamela_ams02}
\end{figure}

\section{PAMELA}
The PAMELA experiment~\cite{ref:pamelaoverview}
(fig.~\ref{fig:pamela_ams02} {\it left}) is a satellite-borne
detector designed to study charged particles in the cosmic radiation
with a particular focus on antiparticles. It was launched in
June~2006~\cite{ref:pamelalaunch} and some preliminary results on
in-flight performance~\cite{ref:pamelainflight} and the abundances of
light nuclei and isotopes~\cite{ref:pamelaabundances} have been
published. First results on positron~\cite{ref:pamela_posfrac} and antiproton~\cite{ref:pamela_pbarp} data have also
been made available. The PAMELA apparatus measures roughly $1.3\,\mathrm{m}$ in
height, weighs $470\,\mathrm{kg}$ and has a power consumption of
$355\,\mathrm{W}$. The overall acceptance is
$21.5\,\mathrm{cm}^2\,\mathrm{sr}$. Its magnetic spectrometer consists
of a $0.43\,\mathrm{T}$ permanent magnet and six $300\,\mu\mathrm{m}$
thick silicon detector planes yielding a spatial resolution of
$3\,\mu\mathrm{m}$ in the bending plane and $11.5\,\mu\mathrm{m}$ in
the non-bending plane~\cite{ref:pamelaspatialres}. The sampling
electromagnetic calorimeter~\cite{ref:pamelaecal} comprises
44~single-sided silicon sensor planes interleaved with 22~plates of
tungsten absorber, each having a thickness of $0.26\,\mathrm{cm}$,
giving a total depth of $16.3\,X_0$. It reaches a proton rejection
factor of about $10^5$ at $90\,\%$ positron efficiency. A neutron
detector complements the electron-proton discrimination capabilities
of the calorimeter. The detector is completed by the time-of-flight
and anticoincidence systems.

\section{AMS-02}
AMS-02~\cite{ref:amsbattiston} (fig.~\ref{fig:pamela_ams02} {\it right})
is a particle detector designed for installation on the
International Space Station and a mission duration of three years. 
It contains eight main components. A superconducting magnet, cooled by
superfluid helium, creates a strong magnetic field, for a total
bending power of $BL^2=0.8\,\mathrm{Tm}^2$. A silicon tracker
consisting of eight layers of double-sided sensors measures the
intersection points of particle tracks with a precision of
$10\,\mu\mathrm{m}$ and $30\,\mu\mathrm{m}$ in the bending and
non-bending planes, respectively. It provides a proton rigidity
resolution of $20\,\%$ at $500\,\mathrm{GeV}$ and is able to determine
the charge of nuclei up to iron ($Z=26$). A laser alignment system is
used to measure the positions of the tracker ladders with a precision
of $5\,\mu\mathrm{m}$. A 20-layer transition radiation detector
identifies electrons and positrons with a rejection factor against
protons of $10^3$ at $1.5\,\mathrm{GeV}$ to $10^2$ at
$300\,\mathrm{GeV}$. A sandwich of lead and scintillating fibres
serves as the electromagnetic calorimeter. It allows three-dimensional
shower reconstruction and has a thickness of $16.7\,X_0$. A proton
rejection of $10^4$ is reached in the range between
$1.5\,\mathrm{GeV}$ and $1\,\mathrm{TeV}$ and the energy resolution is
parameterised as
$\sigma(E)/E=10.2\,\%/\sqrt{E/\mathrm{GeV}}\oplus{}2.3\,\%$. Four
layers of time-of-flight hodoscopes provide precision time
measurements with a resolution of $120\,\mathrm{ps}$ and
$\mathrm{d}E/\mathrm{d}x$ measurements. A ring-imaging Cherenkov
counter measures the velocity, to a precision of $0.1\,\%$, and the charge
of nuclei. A system of anticoincidence counters rejects particles
passing outside the magnet aperture. A system of two star trackers
allows the precise reconstruction of the origin of high-energy $\gamma$-rays.
The overall acceptance of AMS-02 was calculated to be roughly
$875\,\mathrm{cm}^2\,\mathrm{sr}$ by
a Monte Carlo integration (sec.~\ref{sec:acceptance}).

\section{PEBS}
\subsection{Design overview}
\label{sec:pebs_design_overview}
\begin{figure}[htb]
\begin{center}
\includegraphics[width=0.5\textwidth,angle=0]{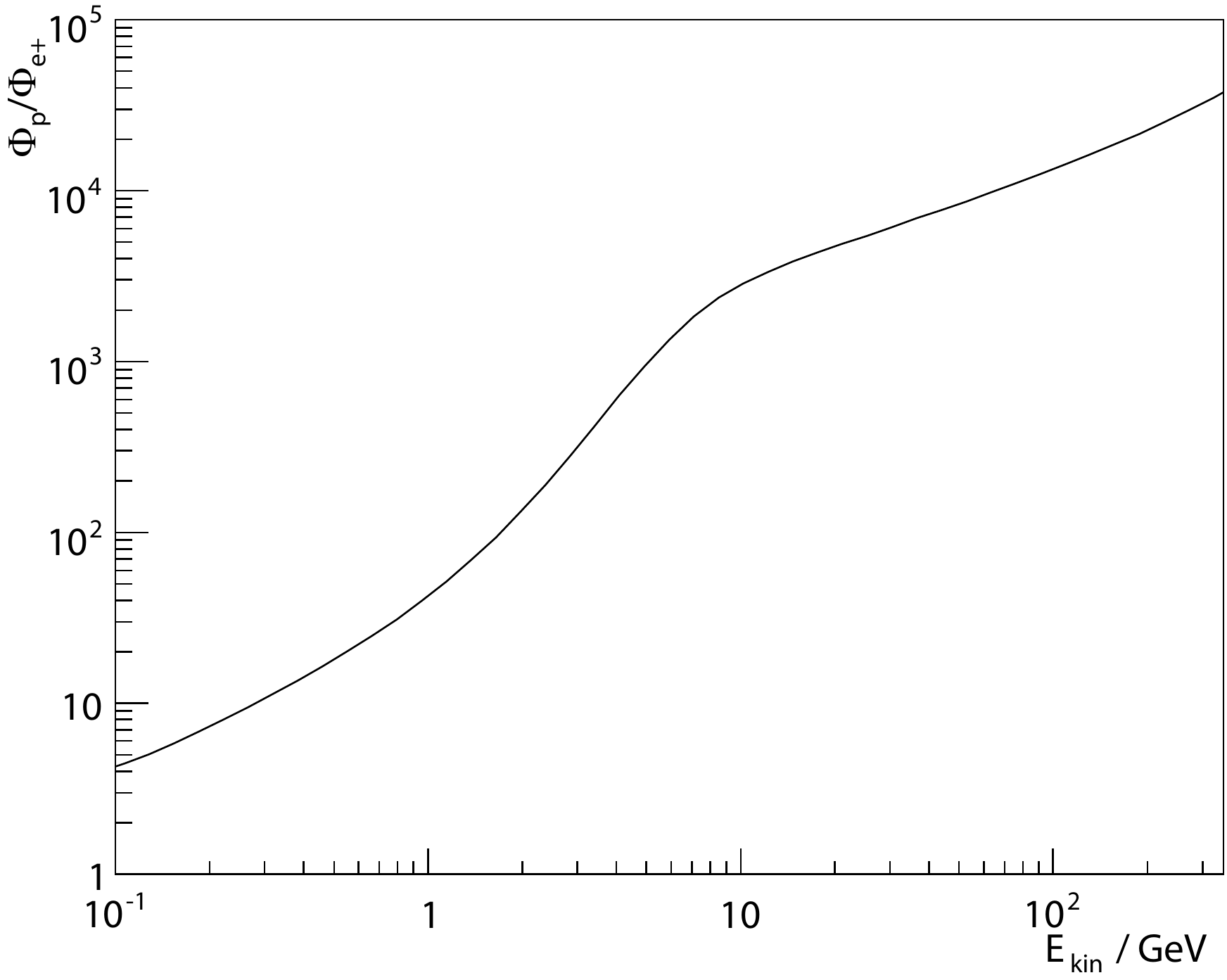}
\end{center}
\caption{Local interstellar ratio of proton to positron flux, as a function of energy, calculated in the conventional Galprop model.}
\label{fig:protontopositron}
\end{figure}
\begin{figure}
\begin{center}
\includegraphics[width=\textwidth,angle=0]{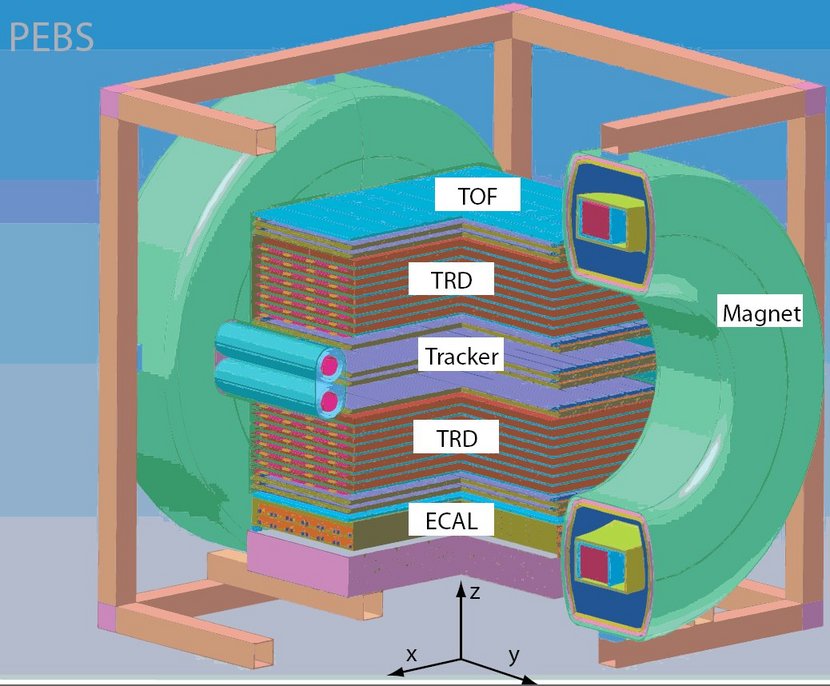}
\end{center}
\caption{Cutaway three-dimensional view of the PEBS detector.
The magnet coils and support are visible
enclosing the inner detector consisting of the time-of-flight system (TOF),
tracker, transition radiation detector (TRD) and electromagnetic
calorimeter (ECAL). A cylinder of solar panels enclosing the structure shown here will be used for power
generation. Carbon fibre and aluminium support structures are shown, too.}
\label{fig:isatec}
\end{figure}
\begin{figure}
\begin{center}
\includegraphics[height=0.7\textwidth,angle=90]{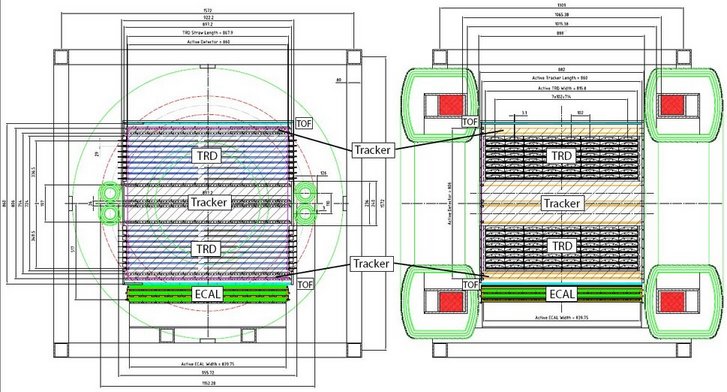}
\end{center}
\caption{Schematic drawing of PEBS, including the dimensions (in mm).
The bending plane ($xz$) and non-bending plane ($yz$) views
are shown in the lower and upper parts of the figure, respectively.}
\label{fig:pebs2d}
\end{figure}

The primary mission goal of the PEBS detector is a precise and
reliable measurement of the cosmic-ray positron fraction, i.e.~the
ratio of the positron flux $\Phi_{e^+}$ to the total electron flux
$\Phi_{e^+}+\Phi_{e^-}$, in an energy range extending from the
$\mathrm{GeV}$-region up to $100\,\mathrm{GeV}$ and above.
The main motivation for this endeavour is
drawn from the observation of an excess in the positron fraction
data available so far which seems to be hard to explain from secondary
production alone (sec.~\ref{sec:galprop}) but may arise from the annihilation
of long-sought dark matter particles with masses of the order of
$100\,\mathrm{GeV}$, for example in the scenario treated in
chapter~\ref{chapter:susyscan}.
For such a measurement to become feasible,
four main obstacles need to be overcome:
\begin{enumerate}
\item The positron flux in the cosmic rays decreases steeply with
energy (fig.~\ref{fig:protons}). Therefore a detector needs a very
high geometric acceptance in order to obtain sufficient statistics in
the high-energy regime.
\item Positrons constitute only a small fraction of the overall
composition of cosmic rays. They are vastly outnumbered by protons
which therefore are the main background for this
measurement. The flux ratio
$\Phi_p/\Phi_{e^+}$ as a function of energy
(fig.~\ref{fig:protontopositron}) reaches a value of $10^4$ and
even more at energies above $100\,\mathrm{GeV}$. This means that one
may not misidentify more than 1~in~1,000,000 protons for a clean
positron sample with a remaining proton contamination of $1\,\%$.
As outlined below, this will be achieved by the
combination of an electromagnetic calorimeter (ECAL, sec.~\ref{sec:ecal}) and a
transition radiation detector (TRD, sec.~\ref{sec:trd}), each
providing roughly a factor~1000 to the proton rejection.
\item A reliable determination of the charge sign is needed in order
to separate electrons from positrons. As the positron flux is expected
to be of the order of 1~to~$10\,\%$ of the electron flux, the fraction
of events with incorrectly reconstructed charge sign must not be above
the per mille level. The magnet and tracker (sec.~\ref{sec:tracker}) are designed
with this goal in mind.
\item Earth's atmosphere with its thickness of roughly twenty
radiation lengths prohibits a measurement of cosmic rays within the
energy range of interest here from the ground. An attractive
alternative to a space-based measurement is found in high-altitude
balloons as explained in section~\ref{sec:balloons}. Such a platform
entails strong constraints on weight and power consumption that have
to be taken into account during the design of the detector. As briefly
mentioned in section~\ref{sec:atmbg}, the remaining atmosphere at
flight altitude will require small corrections to extrapolate to the
top-of-atmosphere fluxes, too.
\end{enumerate}
Figures~\ref{fig:isatec}~and~\ref{fig:pebs2d} show drawings of the
PEBS design. The detector resides between two Helmholtz coils. The
coils are made of superconducting Al-stabilised Niobium filaments and are located inside a
helium cryostat. This superconducting magnet (sec.~\ref{sec:magnet}) will be able to create a
magnetic field with a mean flux density of $B=0.8\,\mathrm{T}$ and mean $BL^2=0.63\,\mathrm{Tm}^2$ for a
period of 20-40~days until the helium is
evaporated.
The field created by the superconducting magnet will be used by the
tracker~\cite{ref:gregorio,ref:gregoriophd} to measure the curvature of the tracks of charged cosmic rays
and hence their momentum and charge sign. The tracker (sec.~\ref{sec:tracker})
consists of eight rows of double-layered modules of scintillating fibres of $250\,\mu\mathrm{m}$ diameter which are read
out by linear silicon photomultiplier (SiPM, sec.~\ref{sec:sipms})
arrays. A momentum resolution of $18\,\%$ is expected for positrons of
$100\,\mathrm{GeV}$.\\
For proton suppression, two independent subdetectors,
an electromagnetic calorimeter (sec.~\ref{sec:ecal}) and a transition
radiation detector (sec.~\ref{sec:trd}), are used. The calorimeter (ECAL) employs a sandwich of
tungsten absorber plates and scintillator bars, again read out by SiPMs, to
determine the energy and the shape of an electromagnetic shower
induced by electrons and positrons. At an energy of
$100\,\mathrm{GeV}$, an energy resolution of $6\,\%$ is predicted by
the Monte Carlo simulation. The corresponding proton rejection is
$3\times10^3$ at $80\,\%$ positron efficiency.\\
The design of the transition
radiation detector (TRD) is based on the one used in the AMS-02 experiment. The x-ray transition
radiation photons created in radiator fleece layers are detected by straw-tube
proportional counters filled with a mixture of xenon and CO$_2$. The
expected proton rejection at $100\,\mathrm{GeV}$ is 700 at $80\,\%$ positron efficiency.\\
The time-of-flight (ToF) system will consist of scintillator panels of
$5\,\mathrm{mm}$ thickness mounted above and below the tracker. Both
upper and lower ToF will have one or two layers of scintillator panels
rotated by 90~degrees with respect to each other. Each layer is built
out of nine scintillator panels with SiPM readout on both ends.

\subsection{Carrier system}
\label{sec:balloons}
While space experiments have the undisputed virtue of being able to
measure the spectra of cosmic rays completely undisturbed by
Earth's atmosphere, scientific high-altitude balloons constitute an
interesting alternative for several reasons. The experiment can be salvaged after the flight
and be recalibrated, refitted and eventually repeated for the gradual
improvement of the statistical accuracy, and it can be
conducted at a much lower cost.\\
High-altitude scientific balloons reach an altitude of
$40\,\mathrm{km}$ and carry a payload of up to 3~tons. Their skin has
a thickness of $20\,\mu\mathrm{m}$ and expands to
a volume of $10^6\,\mathrm{m}^2$. During summer stable weather
conditions develop at Earth's poles including circumpolar winds
that can carry a balloon on a more or less predictable trajectory
around the North or South Pole. The record in flight duration is held
by the CREAM~\cite{ref:cream} experiment which was flown above
Antarctica circumnavigating the South Pole three times for a total of
42~days in December~2004. Launch sites in the Northern hemisphere
include Kiruna, Sweden, and the Svalbard archipelago. Other
experiments successfully flown with balloons include
ATIC~\cite{ref:atic}, BESS~\cite{ref:bessprogram}, CAPRICE~\cite{ref:capriceinstr}, HEAT~\cite{ref:heatinstr}
TRACER~\cite{ref:tracer}, and ISOMAX~\cite{ref:isomax}, which was
equipped with a superconducting magnet similar to the one foreseen for
PEBS.\\
Table~\ref{tab:powerweightpebs} contains a power and weight budget for
the PEBS gondola, partially based on the figures for CREAM.
\begin{table}[htb]
\begin{center}
\begin{tabular}{lrr}
&weight$\,/\,\mathrm{kg}$&power$\,/\,\mathrm{W}$\\ \hline
flight control and flight train&50&0\\
control and data handling&38&60\\
telecommunications&66&84\\
electrical power&250&107\\
mechanical structure&110&0\\
altitude control&75&4\\
system cabling&85&2\\ \hline
sub total&674&257\\
science&1600&343\\ \hline
total&2274&600 \\ \hline
\end{tabular}
\end{center}
\caption{Power and weight budget for the PEBS gondola.}
\label{tab:powerweightpebs}
\end{table}

\subsection{Secondary targets}
\label{sec:secondaries}
In addition to a measurement of the cosmic-ray positron fraction,
other interesting questions may be tackled with the PEBS
detector. Low-energetic antiprotons are of interest because they might
also contain clues to additional primary sources of cosmic rays. In
addition, as with other low-energetic particles, their continued
measurement over a series of years promises a better understanding of
heliospheric physics including solar modulation, as well as
geomagnetic effects. A measurement of the $^3\mathrm{He}/^4\mathrm{He}$ and B/C ratios is of great
interest for an improved understanding of cosmic-ray propagation.
Detailed studies of the feasibility of these ideas need to
be conducted in the future.

\chapter{Design study for PEBS based on Monte Carlo simulations}
\label{chapter:pebs_design_study}
The design and construction of PEBS are complex tasks. An essential
tool in this process is a simulation of the detector that can be used
to study its performance in a fast and inexpensive
way, without having to resort to a lengthy series of prototype
building.\\
The passage of a particle through the detector is governed by the
principle of chance at the most fundamental level. This is completely
counter-intuitive and demonstrates the quantum nature of the physical
processes involved. For example, a proton passing the calorimeter may
do so almost completely unhindered, leaving only a few clouds of
ionisation along its track. The next time, though flying on exactly the
same path, an interaction with a nucleus of the calorimeter material
may occur leading to the disintegration of the two particles and the
creation of a bunch of new ones. A neutral pion might be among those,
and its decay to two photons will lead to an electromagnetic shower in
the calorimeter and thus a radically different event topology.\\
When a particle crosses a detector element, made of a certain
material, every interaction process occurs with a certain
probability. This is encoded in the cross section $\sigma$ of a
process relating the mean number $\mathrm{d}N$ of interactions to the
number density $n$ of targets and thickness $\mathrm{d}x$ of the material
traversed by
\begin{equation}
\label{eq:def_cs}
\mathrm{d}N = \sigma{}n\,\mathrm{d}x
\end{equation}
From the discussion above, it is clear that the correct approach to
predicting the detector performance is a Monte Carlo simulation. The
toolkit best suited and most widely used for this task is
GEANT4~\cite{ref:g4}. It has widespread applications in high-energy
physics, space and radiation physics and medical physics, among
others.\\
\par
A full simulation of PEBS based on GEANT4 has been created. This
simulation can be used to predict the overall behaviour of
the detector, vary design parameters and study
the change in detector performance, test reconstruction algorithms,
develop analysis techniques, and many more. In
section~\ref{sec:mcprinciple}, the working principle of the simulation is explained.
The detector design and its
implementation in the simulation are described in
section~\ref{sec:design_study_pebs}. This section also contains
information about the various process activated for the simulation as
well as the modelling of the response of the different subdetectors to
passing particles. Part of these models use input from testbeam data,
and the extraction of the required information is detailed in
section~\ref{sec:trdtestbeam} as well as in chapter~\ref{chapter:pebs_testbeam}.\\
The simulation alone is not sufficient to get a realistic picture of
the detector performance. The events it produces contain raw
information that is arranged in a similar way as is expected from the
real detector. The task of the reconstruction program presented in
section~\ref{sec:reconstruction} is to transform the raw data of the
individual subdetectors into higher-level objects. This includes track
finding and fitting in the tracker and the TRD, and reconstruction of
a possible electromagnetic shower in the calorimeter.\\
Results for the
most important figures of merit for PEBS will be presented in the final
section~\ref{sec:performance} of this chapter.

\section{Working principle of the Monte Carlo simulation}
\label{sec:mcprinciple}
For every GEANT4 application, the user must provide a
description of the detector geometry, a list of the interaction
processes to be used, and instructions for generating the primary
particle entering the detector. The program will then calculate the
energy deposition in those elements of the detector marked as
sensitive by the user. This information can subsequently be used to
model the response of the detector and create so called simulated
events - a picture of the passing particle as taken by the detector.\\
For a given particle, the simulation proceeds by transporting it
through its current material in small steps. First, a proposed step
length is randomly generated
for every process activated for the particle
species, based on the process cross section for the current
material. If the shortest step length is shorter than the distance to
the boundary to the next material, the corresponding process will be set to
be invoked after the step. Before that, all continuous processes, such
as ionisation energy loss, are invoked and the energy, position and
time of the particle are updated according to the step taken. During
evaluation of the continuous processes or the chosen process, the
particle may be destroyed and/or additional particles may be
created. If the range of a particle created in an interaction inside a
given material exceeds a certain cut value, the particle will be added
to the list of secondaries to be propagated. Otherwise, the
corresponding energy loss is added at the place of the
interaction. Increasing this range cut will reduce the computing time needed
at the possible expense of simulation accuracy.\\
Transportation during the step takes the given magnetic field
into account. After completion of a step, the procedure starts anew,
and when the particle reaches the boundary of the simulated volume or
is destroyed, any secondary particles are tracked in the same way,
possibly creating additional new particles. The event is finished when
no secondaries are left to be tracked.\\
\par
Particles can be generated from a fixed point or uniformly generated
on a surface, with fixed or random momentum, and with perpendicular
incidence or from an angular distribution corresponding to an isotropic
flux, projected onto a plane, i.e.~drawn from uniformly distributed
$\cos^2\theta$ and $\phi$.
To increase efficiency in the early test stages, initial trajectories
can be forced to approximately lie inside the detector acceptance, so
that the number of events with
particles traversing the magnet is minimised and CPU time is saved.\\
\par
The PEBS simulation was designed with two main goals in mind. First,
the performance to be expected from PEBS should be projected as
accurately as possible. This requires that the key detector components
be modelled to a high degree of detail. At the same time, the user
should be as flexible as possible to vary the main design
parameters of the detector which is important for the design
optimisation process. The latter is accomplished by using Geant4's
command language enabling the user to control many different aspects
of the simulation with macro files. For example, the tracker fibre
diameter or the number of layers in the ECAL and its absorber material can be modified in this
way.

\section{Design study for PEBS and implementation in the Monte Carlo simulation}
\label{sec:design_study_pebs}
\begin{figure}[htb]
\begin{center}
\includegraphics[width=0.7\textwidth,angle=0]{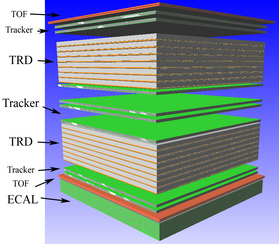}
\end{center}
\caption{View of the inner detector as modelled in the Geant4
simulation. The scintillating fibres of the tracker, not individually
resolved on this scale, are painted in light green, and the foam in
between is dark grey. The TRD radiator is drawn
in grey, with the TRD straw tubes in brown sitting in between. The TOF
panels are dark orange and the ECAL at the bottom has absorber plates
in grey and fibre bars in green.}
\label{fig:g4innerdetector}
\end{figure}
In the following, the working principles of the individual
subdetectors and their current designs are presented. The
implementation of their geometry in the simulation
(fig.~\ref{fig:g4innerdetector}) is shown, too.

\subsection{Magnet}
\label{sec:magnet}
\begin{figure}[htb]
\begin{center}
\begin{tabular}{cc}
\includegraphics[width=0.5\textwidth,angle=0]{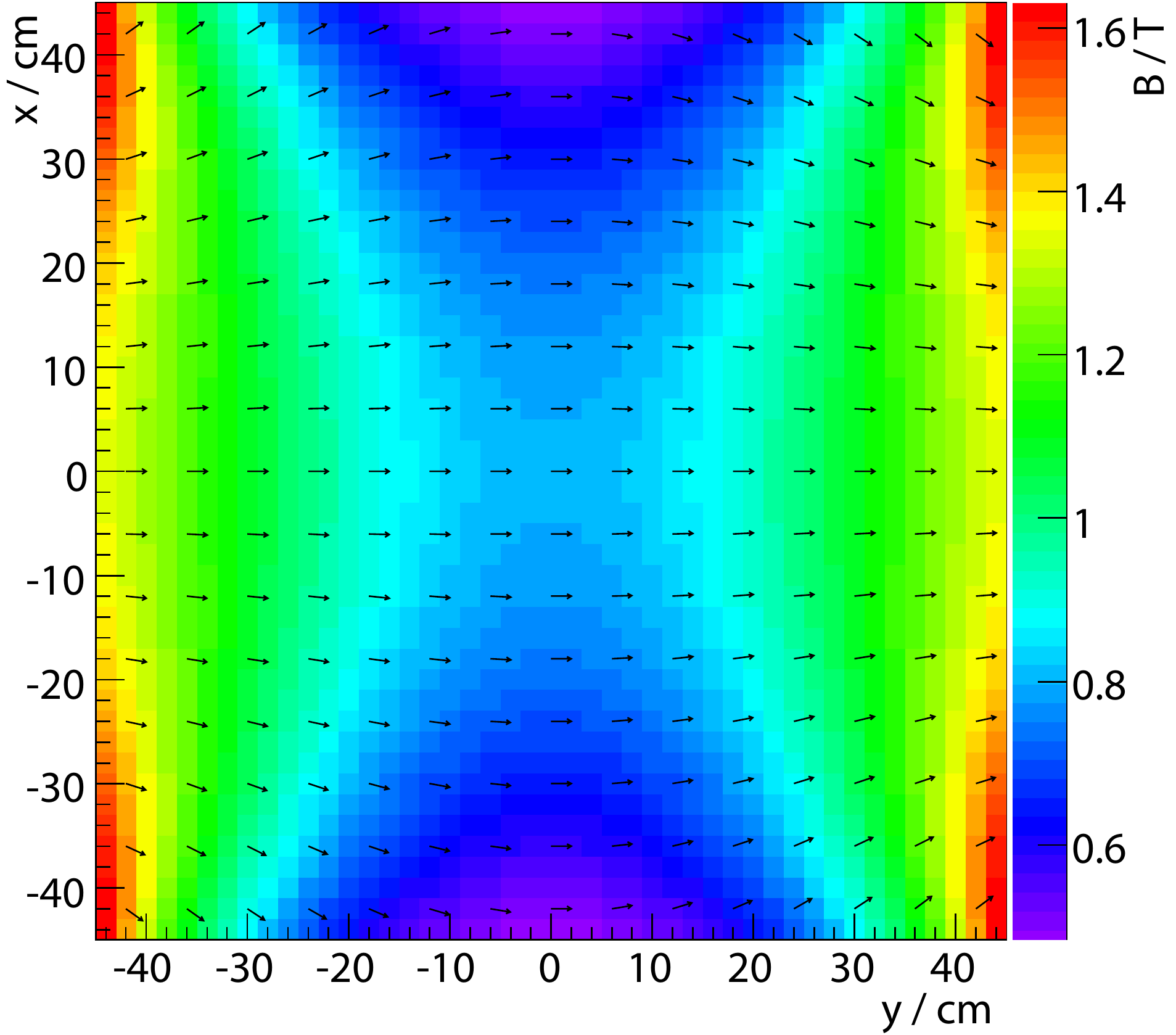}&
\includegraphics[width=0.48\textwidth,angle=0]{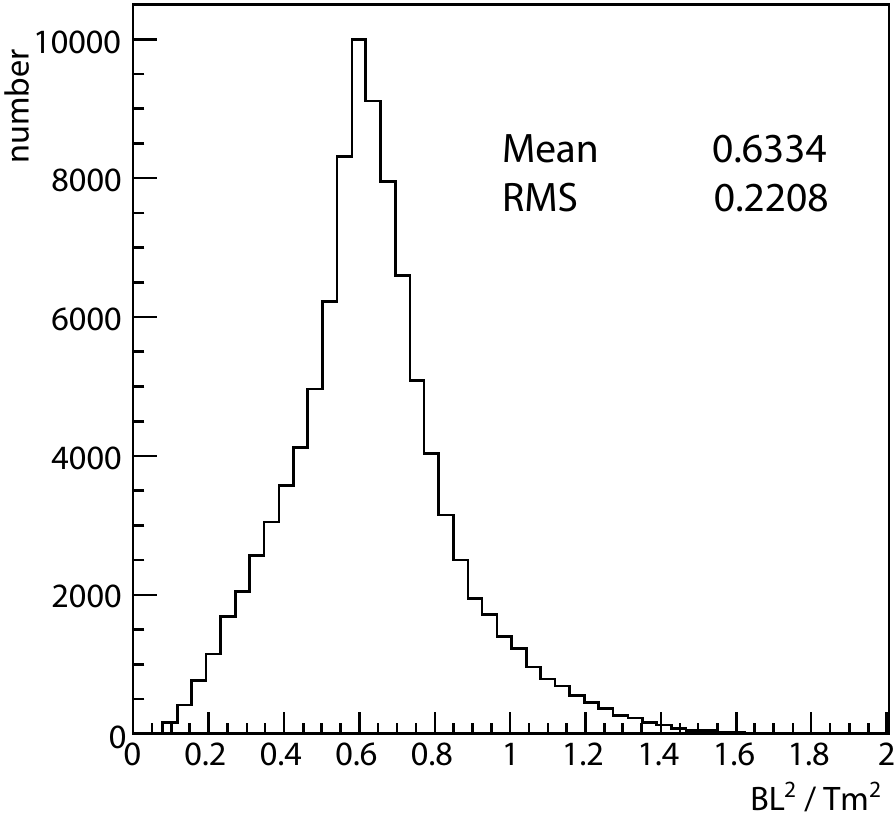}\\
\end{tabular}
\end{center}
\caption{{\it Left:} Magnetic field amplitude $|\vec{B}|$ at $z=0$. The magnet
coils are located to the left and right. {\it Right:} Distribution of
$BL^2$ as defined in the text, for tracks within the acceptance of PEBS.}
\label{fig:magfield}
\end{figure}
The magnet design foresees a pair of superconducting Helmholtz coils of $1150\,\mathrm{mm}$ diameter
creating a mean flux density of $0.8\,\mathrm{T}$ inside the tracker
volume. The coils will be located inside a helium cryostat and the
amount of helium to be carried will allow flight durations of
20-40~days. The opening aperture of the magnet measures $882\times882\,\mathrm{mm}^2$.
Figure~\ref{fig:magfield} shows the magnetic
field map in a plane containing the centre of the detector.
The mean bending power of the magnet is found to be
$BL^2=0.63\,\mathrm{Tm}^2$. Here, $BL^2$ is defined by
\begin{equation}
\label{eq:bl2}
BL^2\equiv\int{}B_\perp(\ell)\,\mathrm{d}\ell\,\cdot\,L
\end{equation}
so that two field configurations with the same $BL^2$ will lead to the same
angular deflection of a charged particle.
\\
In the simulation, a homogeneous magnetic field of $0.8\,\mathrm{T}$
is used inside the
tracker volume, to simplify the tracking and reconstruction
algorithms for the time being and because the magnet design is still
in a state of flux anyway. The corresponding
$BL^2=0.57\,\mathrm{Tm}^2$ is somewhat lower than in the current
design and the resulting
estimates for the detector performance will therefore be conservative.\\
The insensitive material parts surrounding the tracker, such as
the magnet coils and
cryostat, solar panels, electronics crates, and so on, are not
included in the simulation yet.

\subsection{Tracker}
\label{sec:tracker}
\subsubsection{Working principle}
The purpose of the tracker is to measure the momentum and thus the
charge sign of an incoming particle. The measurement exploits the fact
that the path followed by a particle of momentum $p$ and charge $z$ in a constant magnetic
field $\vec{B}$ is a helix. While the TRD is used to measure its pitch
angle $\lambda$, the tracker is used to determine the radius $R$ of
curvature of the helix which is related to the momentum
by~\cite{ref:pdg}
\begin{equation}
  \label{eq:rigidity}
  \frac{p}{\mathrm{GeV}}\cos\lambda=0.3\,z\,\frac{B}{\mathrm{T}}\,\frac{R}{\mathrm{m}}
\end{equation}
The uncertainties in a measurement of momentum are best discussed in
terms of the errors of the curvature $k\equiv{}1/R$ because the
distribution of this quantity is approximately Gaussian. The overall
error $\sigma_k$ on the curvature measurement has contributions from
the position error as well as from multiple scattering and may be approximated by
\begin{equation}
\label{eq:curverror}
\sigma_k^2=\sigma_{k,\mathrm{res}}^2+\sigma_{k,\mathrm{msc}}^2
\end{equation}
The term due to multiple scattering is only important for low momenta
and is approximately given by
\begin{equation}
\label{eq:msc}
\sigma_{k,\mathrm{msc}}=\frac{0.016\,\mathrm{GeV}/c\,\cdot\,|z|}{L/\mathrm{m}\,p\beta\cos^2\lambda}
\sqrt{\frac{L}{X_0}}
\end{equation}
where $L$ is the total track length and $X_0$ is the radiation length
defined in section~\ref{sec:ecalprinciple}.\\
The contribution from the limited spatial resolution of the individual
tracker elements is given by~\cite{ref:gluckstern}
\begin{equation}
\label{eq:res_uniform}
\sigma_{k,\mathrm{res}}=\frac{\sigma}{L^{\prime{}2}}\,\sqrt{\frac{720}{N+4}}
\end{equation}
if $N$ independent measurements are taken at equal distances with
resolution $\sigma$ perpendicular to the trajectory for a total length
$L^\prime$ measured in the projection to the bending plane.\\
This uncertainty can be reduced in the so-called optimised spacing of
detection planes, where $N/4$ measurements are taken at the entry and
exit of the tracker each, and $N/2$ at the centre. The curvature error
then becomes
\begin{equation}
\label{eq:res_opt}
\sigma_{k,\mathrm{res}}=\frac{\sigma}{L^{\prime{}2}}\,\sqrt{\frac{256}{N}}
\end{equation}

\subsubsection{Design and implementation in the simulation}
\label{sec:tracker_design}
\begin{figure}[htb]
\begin{center}
\includegraphics[width=0.75\textwidth,angle=0]{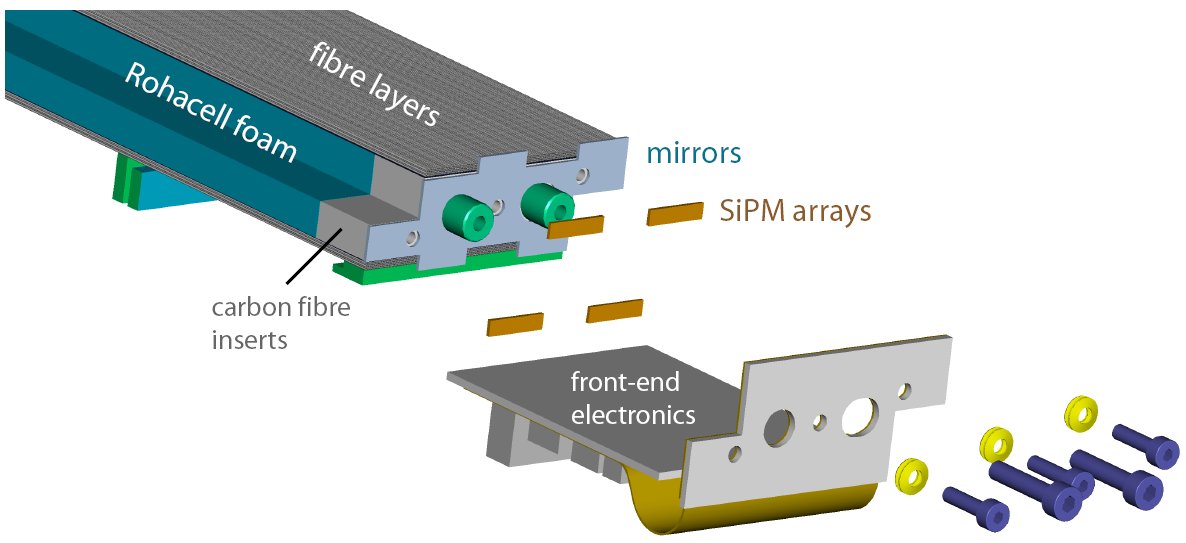}
\end{center}
\caption{Exploded view of a PEBS tracker module. Two stacks of
scintillating fibres are carried by Rohacell foam with carbon fibre
inserts at the ends. Plates equipped with reflective foils and
silicon photomultiplier arrays are screwed to the sides of the module. The large screws
will be used to mount the module to the tracker walls.}
\label{fig:trackermodule}
\end{figure}
\begin{figure}[htb]
\begin{center}
\includegraphics[width=0.8\textwidth,angle=0]{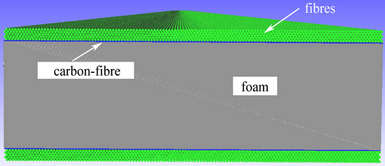}
\end{center}
\caption{Visualisation of a tracker module in the Geant4
  simulation. The fibres, carbon-fibre spacers and Rohacell foam are visible.}
\label{fig:trackermodulevis}
\end{figure}

The tracker is constructed
from modules like the one depicted in
figure~\ref{fig:trackermodule}. Particles traversing the module
deposit energy in two stacks of scintillating fibres, creating scintillation photons.
The scintillation light is then guided by total internal reflection to the
fibre ends where it is detected by
novel silicon photomultiplier (SiPM) arrays~\cite{ref:roman}
(sec.~\ref{sec:sipms}).
Each fibre stack consists of five layers of 128~round fibres of
$250\,\mu\mathrm{m}$ diameter and $860\,\mathrm{mm}$ length, positioned with a horizontal pitch of
$275\,\mu\mathrm{m}$ and glued together. The distance between the individual channels on an SiPM array matches this
fibre pitch. The vertical fibre spacing is taken
to be $5\,\mu\mathrm{m}$. The stacks are produced by threading the
fibres on a rotating barrel
with premachined grooves. These grooves have to be somewhat further
apart than the diameter of the fibres to allow for inhomogeneities in
the fibre diameter. The stacks are supported by spacers
consisting of two carbon-fibre skins of $100\,\mu\mathrm{m}$ thickness
with $10\,\mathrm{mm}$ of Rohacell foam in between. The performance of such
a fibre module in a proton testbeam is studied in section~\ref{sec:testbeam2008}.\\
The SiPM arrays are located on ceramic
plates mounted to both fibre ends which also contain highly reflective
mirroring foils. The SiPM arrays and mirrors are arranged in such a
way that each fibre is covered by a mirror on one end and by an SiPM
array channel on the other. This compact module design allows the
construction of a tracker sensitive to particles almost across its
entire aperture.
The modules will be grouped into eight layers, with two layers each
located at the top and bottom of the tracker and four layers around the
centre. In this configuration, the momentum resolution is optimised
(eq.~(\ref{eq:res_opt})) and space is created for the TRD. For the
simulation, a somewhat simpler module shape is used at the moment (fig.~\ref{fig:trackermodulevis}).

\subsection{Silicon photomultipliers for scintillating fibre readout}
\label{sec:sipms}
\begin{figure}[htb]
\begin{center}
\begin{tabular}{cc}
\includegraphics[width=0.74\textwidth,angle=0]{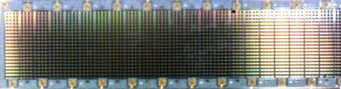}&
\includegraphics[width=0.26\textwidth,angle=0]{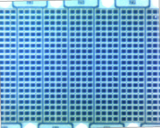}\\
\end{tabular}
\end{center}
\caption{{\it Left:} Close-up view of a Hamamatsu MPPC~5883 SiPM array, suitable for use in a scintillating fibre tracker.
{\it Right:} Mircroscope picture of several channels of the Hamamatsu MPPC~5883 SiPM array revealing the
individual pixels. Each SiPM on the array has $4\times20=80$~pixels.}
\label{fig:tb08_arrays}
\end{figure}
A silicon photomultiplier (SiPM, fig.~\ref{fig:tb08_arrays})~\cite{ref:sipm1,ref:sipm2,ref:sipm3,ref:sipm_renker} is a
novel solid state photodetector. In principle, it consists of a matrix
of avalanche photo diodes (APDs) that are operated in the Geiger
mode, i.e.~above the breakdown voltage. Simply put, an APD is based on
a p-n junction between two semiconductors with a reverse bias voltage
applied to it so that photo-electrons are accelerated in the resulting
electric field creating an avalanche of secondary ionisation. The
single matrix elements are
called pixels and the output signal is then
proportional to the number of pixels that have fired. The mode of
operation leads to a high intrinsic gain of the SiPM on the order of
$10^5$ to $10^6$. In
addition, they have the virtues of insensitivity to
magnetic fields, high quantum efficiency, as well as
compactness and auto-calibration. The latter is due to the fact that
the single photo-electron peaks are visible in the spectrum (fig.~\ref{fig:tracker_amps} {\it left}).
Dark noise rates of SiPMs typically are on the order of a few hundred $\mathrm{kHz}$.\\
Silicon photomultipliers will be
used for the readout of both the scintillating fibres in the tracker
and the scintillator layers of the electromagnetic
calorimeter and possibly also for the time-of-flight system.
Their modelling is therefore an important part of the
digitisation chain in the simulation and their key features, present for
both tracker and calorimeter, are summarised in the following. A detailed discussion and
measurements of the properties of SiPMs in the context of the PEBS
project can be found in~\cite{ref:gregorio}. In addition to individual
SiPMs suitable for use in the ECAL, SiPM arrays have recently become
available that combine a number of SiPMs next to each other on the
same wafer. These compact devices are well suited for the tracker
readout.\\
\par
Since the actual avalanches created in the SiPM pixels are not
simulated by Geant4, a model for the response of the SiPMs has to be
adopted. The ionisation energy loss $\mathrm{d}E/\mathrm{d}x$ of a
particle over a simulated step of length $\Delta{}x$ is provided by
the simulation. This will be turned into photons by the scintillating
fibre material with an efficiency $\epsilon_\mathrm{scint}$ (with
units of $\mathrm{keV}^{-1}$). The fibre will trap a fraction
$\epsilon_\mathrm{trap}$ of these photons due to total internal
reflection, and the mirrored surface at one end of the fibre will lead
to an increase by a factor of $f_\mathrm{mirror}$. Finally, a fraction
$\epsilon_\mathrm{opt}$ of the photons arriving at the SiPM end of the
fibre will be optically coupled into the SiPM, which in turn has a
limited geometrical efficiency $\epsilon_\mathrm{geom}$, given by the
ratio of active pixel area to the total area subtended by the fibre
end. The product $\epsilon_\mathrm{qe}\,\epsilon_\mathrm{abe}$ of the quantum efficiency and the avalanche breakdown efficiency of a single SiPM
pixel will then determine if an incoming photon will actually trigger
a breakdown. The mean number of photons in an SiPM is then given by
\begin{equation}
\label{eq:meanphotons}
N_\mathrm{ph}=\frac{\mathrm{d}E}{\mathrm{d}x}\,\Delta{}x\,\epsilon_\mathrm{scint}\,\epsilon_\mathrm{trap}\,f_\mathrm{mirror}\,\epsilon_\mathrm{opt}\,\epsilon_\mathrm{geom}\,\epsilon_\mathrm{qe}\,\epsilon_\mathrm{abe}
\end{equation}
At the moment, cross-talk, arising from UV photons created
during the avalanche process and subsequently triggering a
neighbouring pixel, is not considered. Attenuation in the fibre, i.e. a
position-dependent $\epsilon_\mathrm{trap}$ is neglected at the moment
as the attenuation length is large compared to the fibre length in our
case.\\
For all simulations presented here, the manufacturer's
values of $\epsilon_\mathrm{scint}=8\,\mathrm{keV}^{-1}$ and
$\epsilon_\mathrm{trap}=0.056$ for round fibres and
$\epsilon_\mathrm{trap}=0.073$ for square fibres were adopted, for the Bicron fibres
employed during the testbeam campaign presented in chapter~\ref{chapter:pebs_testbeam}~\cite{ref:bicronfibres}.
It is assumed that $f_\mathrm{mirror}=1.6$. The
quantity $\epsilon_\mathrm{opt}$ is difficult to determine and it is
practical to combine the last four factors in
equation~(\ref{eq:meanphotons}) in a single overall coupling
efficiency
\begin{equation}
  \label{eq:effcoup}
  \epsilon_\mathrm{coup}\equiv\epsilon_\mathrm{opt}\,\epsilon_\mathrm{geom}\,\epsilon_\mathrm{qe}\,\epsilon_\mathrm{abe}
\end{equation}
This quantity determines the overall light yield measured with the SiPMs. It contains the photo-detection efficiency
\begin{equation}
\label{eq:pde}
\epsilon_\mathrm{pde}=\epsilon_\mathrm{geom}\,\epsilon_\mathrm{qe}\,\epsilon_\mathrm{abe}
\end{equation}
As presented in detail in
chapter~\ref{chapter:pebs_testbeam}, a typical light yield to be
expected from the PEBS tracker modules is 10~photo-electrons per
cluster for the standard configuration. The value of $\epsilon_\mathrm{coup}=0.1$ has been chosen to
match this result.
\begin{figure}[htb]
\begin{center}
\begin{tabular}{cc}
\includegraphics[width=0.5\textwidth,angle=0]{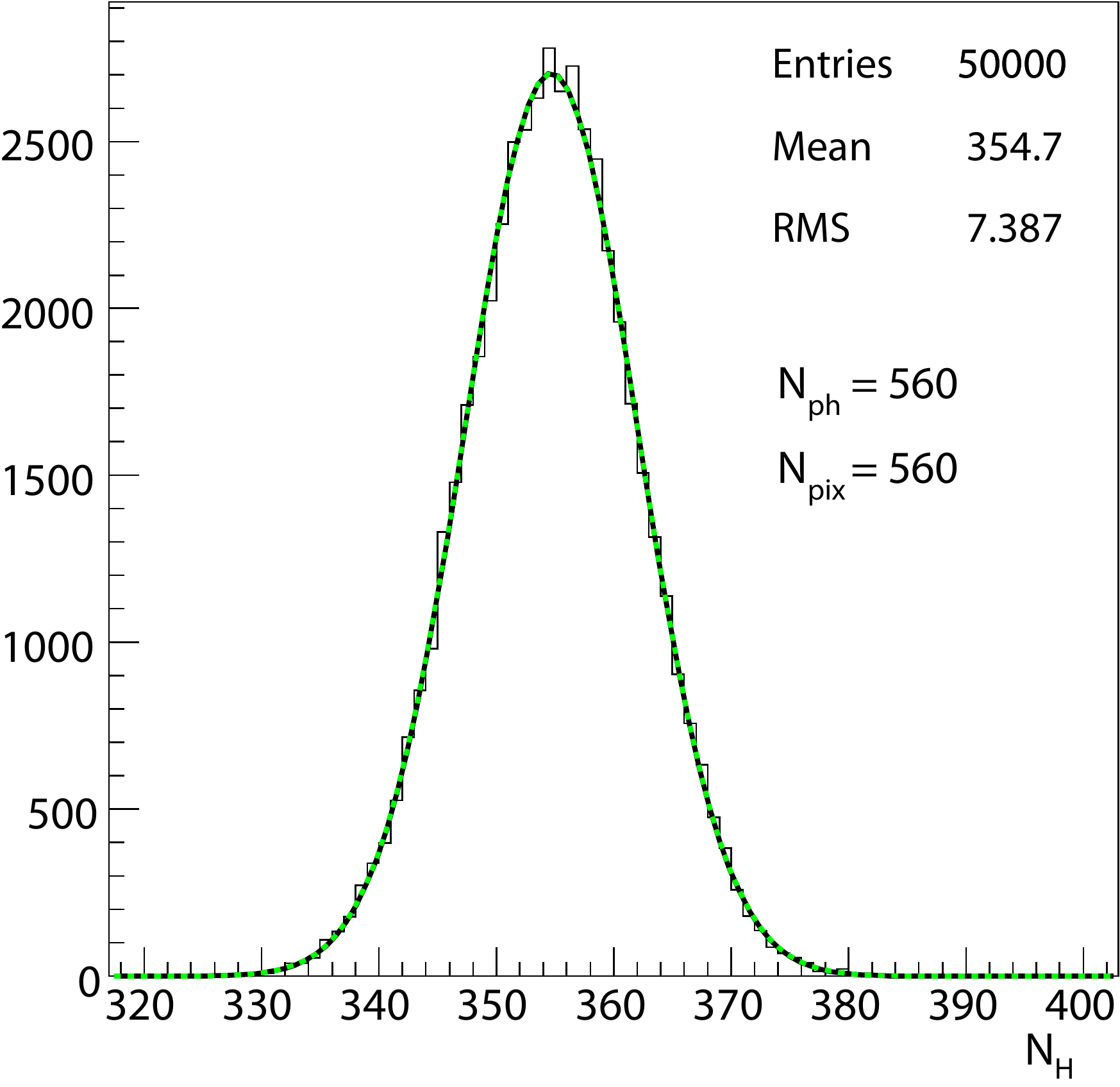}&
\includegraphics[width=0.5\textwidth,angle=0]{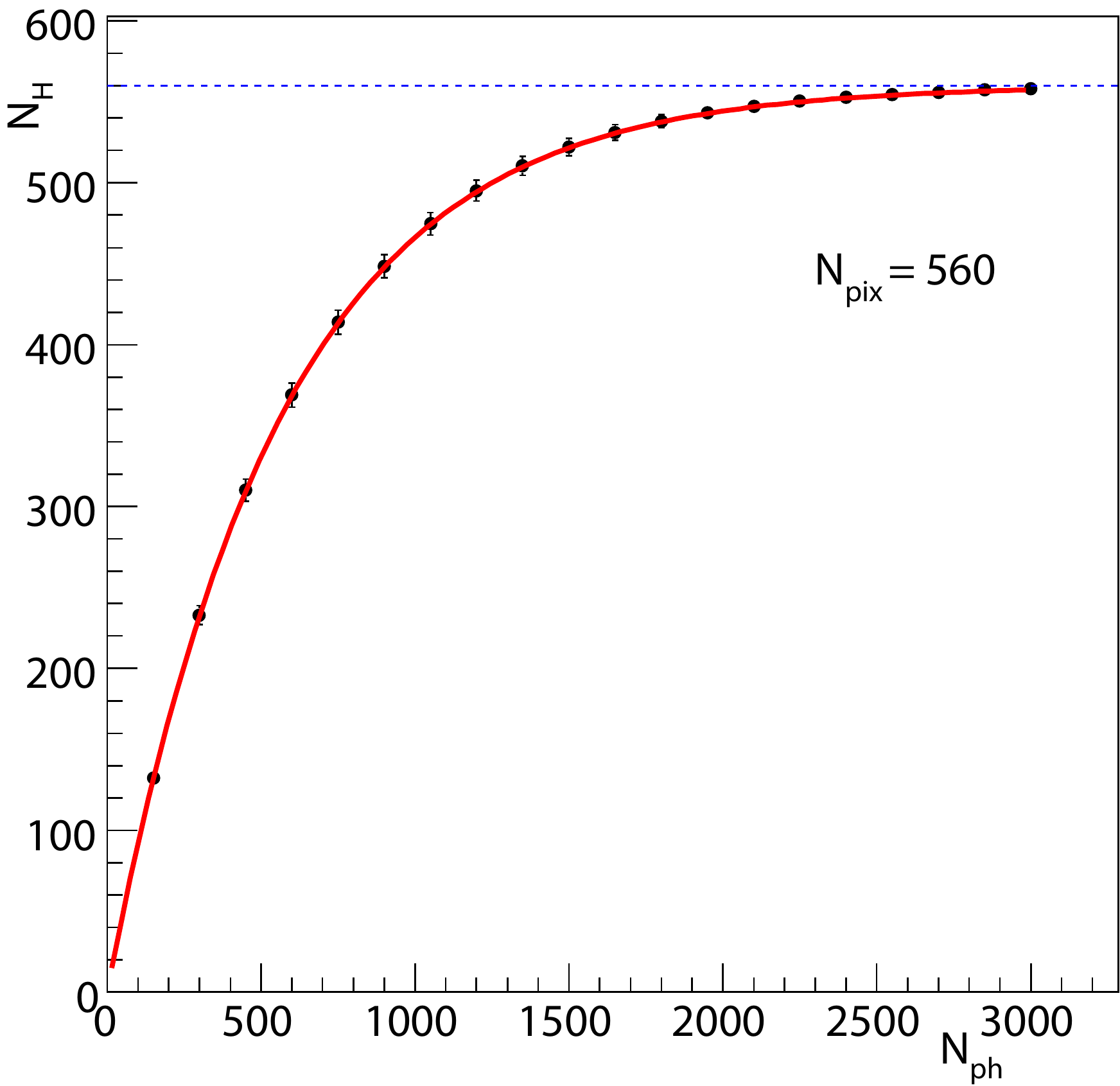}\\
\end{tabular}
\end{center}
\caption{{\it Left:} Simulated distribution of raw amplitude $N_H$ (fired pixels) for a
  Geiger-mode array of 560~SiPM pixels, with $N_\mathrm{ph}=560$ incoming photons,
  assuming perfect geometric and quantum efficiency, together with a
  fit of a Gaussian distribution (black line) and a Gaussian with mean
and variance according to (\ref{eq:urnmodel}) and
(\ref{eq:urnmodelvariance}) (green dotted line). {\it Right:} Raw
  amplitudes obtained for an array of 560~pixels, as a function of
  the number of incoming photons. The error bars correspond to the
  standard deviation obtained from a fit of a Gaussian to the
  distribution found at each $N_\mathrm{ph}$. Also included is the
  analytic curve according to equation (\ref{eq:pixeldigi}).}
\label{fig:sipmarraysim}
\end{figure}
\\
In reality, the number of pixels fired is subject to random
fluctuations. Therefore, the digitisation for each SiPM proceeds as
follows:
\begin{itemize}
\item From all steps during the simulated particle propagation in the given fibre, the accumulated energy
  deposition $\Delta{}E$ is calculated.
\item The number $N_\mathrm{gen}$ of generated photons is drawn from a Poissonian
  distribution with mean $\Delta{}E\cdot\epsilon_\mathrm{scint}$.
\item The number of photons coupled to the available SiPM pixels is drawn
  according to a binomial distribution with the number of trials
  $N_\mathrm{gen}$ and probability
  $\epsilon_\mathrm{trap}\cdot{}f_\mathrm{mirror}\cdot\epsilon_\mathrm{coup}$.
\item In the case of the tracker, certain fibres deposit their photons
  on two adjacent SiPMs. This is simulated by drawing the number of
  photons on the left SiPM from a binomial distribution with the
  number of photons obtained from the previous step and probability
  $1/2$. The right SiPM then receives the remaining photons.
\item According to the noise level $N_\mathrm{noise}$ set by the user,
  and given as a mean number of photons, a number of noise photons is drawn
  for each SiPM from a Poissonian distribution with mean
  $N_\mathrm{noise}$.
\item The photons reaching the sensitive area of an SiPM are
distributed on the pixel matrix. For the simulation of the pixel matrix, a random pixel number is drawn
for each photon from a uniform distribution and the corresponding
pixel is then activated. The total number of activated pixels then
yields the raw SiPM amplitude, given in units of pixels.
\item The number $n$ of pixels is smeared by a
  Gaussian with a width of $\sigma_n=\sqrt{n+1}\sigma$~\cite{ref:vladikecal}. The parameter
  $\sigma$ determines the width of the photo-electron peaks in
  the SiPM spectrum. The purpose is to model the electronic noise
  of the SiPMs and thus to obtain the
  smeared-out peak structure characteristic of an SiPM
  (cf.~fig.~\ref{fig:tracker_amps}). For the tracker, a value of
  $\sigma=0.1$ is adopted, while the measured value of
  $\sigma=0.15$~\cite{ref:vladikprivate} is taken for the ECAL SiPMs.
\item Hit information is stored for each SiPM whose amplitude exceeds a given
  threshold.
\end{itemize}
The response
of an SiPM to a given number of incident photons is
illustrated in figure~\ref{fig:sipmarraysim}.
It shows the distribution of the number $N_H$ of fired pixels for a SiPM
matrix of $N_\mathrm{pix}=560$~pixels hit by 560~incident photons. As some pixels are
hit by more than one photon, the mean number of $N_H$ is smaller than
$N_\mathrm{pix}$.\\
In order to
get a reliable determination of the energy of an ECAL shower, the
departure of the SiPM response from linearity, as is evident in the
figure, has to be corrected for. To this end, the
number of fired pixels has to be translated to the true amplitude,
i.e.~the number of incident photons. This requires an analytic
expression for the mean number of pixels $N_H$ fired by a number of
photons $N_\mathrm{ph}$ incident on an array of $N_\mathrm{pix}$
pixels.
In the case of $N_\mathrm{pix}\gg 1$, such an expression can be
derived as follows:\\
A small group $\mathrm{d}N_\mathrm{ph}^\prime$ of photons will
increase the number of fired pixels by $\mathrm{d}N_H^\prime$. If
$N_H^\prime$ pixels have already fired at that point, one has
\[
\mathrm{d}N_H^\prime=\mathrm{d}N_\mathrm{ph}^\prime\cdot\left(1-\frac{N_H^\prime}{N_\mathrm{pix}}\right)
\]
After rearrangement, one can take the continuous limit and integrate
both sides:
\[
\int\limits_0^{N_H}\frac{\mathrm{d}N_H^\prime}{1-\frac{N_H^\prime}{N_\mathrm{pix}}}=\int\limits_0^{N_\mathrm{ph}}\mathrm{d}N_\mathrm{ph}^\prime
\]
This leads to
\begin{equation}
  \label{eq:pixeldigi}
N_H=N_\mathrm{pix}\left(1-\exp\left(-\frac{N_\mathrm{ph}}{N_\mathrm{pix}}\right)\right)  
\end{equation}
From the inverse of the above, the desired amplitude $N_\mathrm{ph}$ can be
calculated.\\
From a combinatorial point of view, the problem of the SiPM response
can be treated by considering an urn model where $N_\mathrm{ph}$ balls
are distributed into $N_\mathrm{pix}$ urns and asking for the number
of urns with at least one ball. This approach is more difficult but it
yields the exact
result for the mean of this number, and also its variance~\cite{ref:urnmodel}:
\begin{equation}
\label{eq:urnmodel}
N_H=N_\mathrm{pix}\left(1-\left(1-\frac{1}{N_\mathrm{pix}}\right)^{N_\mathrm{ph}}\right)
\end{equation}
\begin{equation}
\label{eq:urnmodelvariance}
\sigma^2(N_H)=N_\mathrm{pix}(N_\mathrm{pix}-1)\,\cdot\,\left(1-\frac{2}{N_\mathrm{pix}}\right)^{N_\mathrm{ph}}+
N_\mathrm{pix}\,\cdot\,\left(1-\frac{1}{N_\mathrm{pix}}\right)^{N_\mathrm{ph}}-
N_\mathrm{pix}^2\,\cdot\,\left(1-\frac{1}{N_\mathrm{pix}}\right)^{2N_\mathrm{ph}}
\end{equation}
(\ref{eq:urnmodel}) becomes (\ref{eq:pixeldigi}) in the limit $N_\mathrm{ph}\rightarrow\infty$.

\subsection{Time-of-flight system}
\label{sec:tof}
In the simulation, the time-of-flight system is modelled by two layers of scintillator
modules above and below the tracker, each. Each layer consists of
modules made of polystyrene with dimensions
$100\times{}5\,\mathrm{mm}$.

\subsection{Electromagnetic calorimeter}
\label{sec:ecal}
\subsubsection{Principle of operation}
\label{sec:ecalprinciple}
Electromagnetic calorimeters~\cite{ref:calorimetry_fabjan,ref:wigmans} are widely used in high energy physics to
measure the energies of electrons and photons. They are based on the
physics of the electromagnetic
cascade. For electrons with energies
larger than $\sim\!10\,\mathrm{MeV}$, the main energy loss mechanism
is bremsstrahlung. On the other hand, photons in this energy range
lose their energy predominantly by pair creation. This means that an
impinging electron will radiate bremsstrahlung photons, and the
ones with sufficient energy will in turn create secondary electrons and positrons. These
processes will be repeated, thus giving rise to a cascade of particles
with lower and lower energies. Below a  critical energy $\epsilon$,
the dominant energy loss mechanisms are ionisation and excitation for
electrons and Compton scattering and photoelectric effect for
photons. Therefore, towards the end of the cascade, energy will be
dissipated instead of being used for the creation of additional particles. These
processes thus lead to the development of an {\it electromagnetic shower}.\\
The main features of a shower can be characterised by a single
parameter, the radiation length $X_0$ that depends on the absorber
material. It represents the average distance $z$ that an electron
needs to travel in a material to reduce its energy to $1/\mathrm{e}$
of its initial energy $E_0$. Similarly, the intensity $I_0$ of a
photon beam will be reduced by a factor of $1/\mathrm{e}$ after a path
length of $9/7\,X_0$. For the common absorber materials lead and
tungsten, the radiation lengths are $X_0^\mathrm{Pb}=5.612\,\mathrm{mm}$ and
$X_0^\mathrm{W}=3.504\,\mathrm{mm}$, respectively.\\
The longitudinal shower profile can then be described as a function of
$t\equiv{}z/X_0$, where $z$ is the coordinate along the shower axis, and is parameterised to a good approximation by the formula
\begin{equation}
\label{eq:showershape}
\frac{\mathrm{d}E}{\mathrm{d}t}=E_0b\frac{(bt)^{a-1}\mathrm{e}^{-bt}}{\Gamma(a)}
\end{equation}
where $E_0$ is the total energy and $a$ and $b$ are parameters. $b$
sets the scale of the shower and $a$ in turn gives the position of the
shower maximum which is calculated to be
\begin{equation}
\label{eq:showermax}
t_\mathrm{max}=\frac{a-1}{b}
\end{equation}
from (\ref{eq:showershape}). It depends logarithmically on energy,
\begin{equation}
\label{eq:showermaxvse}
t_\mathrm{max}\approx\log\frac{E_0}{\epsilon}+t_0
\end{equation}
where $t_0=-1/2$ for electrons and $t_0=+1/2$ for photons and $\epsilon$ is the critical energy,
at which losses by bremsstrahlung and ionisation are equally large. In
lead, $\epsilon\approx{}7\,\mathrm{MeV}$.\\
The transverse size of an electromagnetic shower is determined by
multiple scattering and, to a lesser extent, by bremsstrahlung photons
emitted away from the shower axis. The typical scale is the Moli\`ere
radius $R_M$ which is given in terms of the radiation length and the
critical energy, and it can be approximated by
\begin{equation}
\label{eq:moliereradius}
R_M\approx 21\,\mathrm{MeV}\frac{X_0}{\epsilon}
\end{equation}
It represents the average lateral deflection of electrons at the
critical energy after traversing one radiation length. The values of
the Moli\`ere radii for lead and tungsten are
$R_M^\mathrm{Pb}=16\,\mathrm{mm}$ and
$R_M^\mathrm{W}=9.3\,\mathrm{mm}$, respectively.\\
A calorimeter can be used for an energy measurement because the total energy
deposition by all charged particles in a shower is proportional to
the energy of the incident particle. The total track length $T_0$ of
the shower, defined as the sum of all ionisation tracks, is
$T_0\propto{}X_0E_0/\epsilon$. The intrinsic energy resolution of an
ideal calorimeter will then be due to statistical fluctuations in the
track length: $\sigma(E)\propto\sqrt{T_0}$. In reality, imperfections
in the calorimeter response will lead to an additional constant term
in the relative energy resolution, which becomes
\begin{equation}
\label{eq:ecalres}
\frac{\sigma(E)}{E}=\frac{a}{\sqrt{E}}\oplus{}c
\end{equation}
where $\oplus$ denotes a quadratic sum. As will be seen, leakage
effects due to the finite size of the calorimeter and the limited dynamic range of the SiPMs used for the
readout give the most
important contribution to $c$ and might even cause it to rise
with energy.\\
In general, calorimeters can be divided into two groups. In the first
group, called homogeneous calorimeters, the same material serves as
absorber and detection medium. For example, PbWO$_4$ combines high
charge number $Z$ and thus small radiation length with the properties
of a scintillator so that the light yield obtained in a crystal of
such a material is a measure of the energy of an incident electron. In
the other group, called inhomogeneous calorimeters, layers of absorber
material, such as lead or tungsten, are followed by active detection
layers, for example in the form of a plastic scintillator. The
drawback of inhomogeneous calorimeters is their lower energy
resolution due to the fact that only a part of the shower is visible
in the active material and that part is subject to statistical
fluctuations. On the other hand, the sandwich design of an
inhomogeneous calorimeter allows one to not only measure the {\it
energy}, but also the {\it shape} of the shower, i.e.~determine the
shower profile $\mathrm{d}E/\mathrm{d}t$. This means that an
inhomogeneous calorimeter can be used for particle identification and
it is therefore the design of choice for PEBS, where the calorimeter
plays a vital role in the suppression of the proton background. In
addition, the degradation of the energy resolution due to the leakage
effect can be reduced by fitting the shower profile with
(\ref{eq:showershape}) and thus extrapolating the energy deposition on
the missing layers.\\
The particle identification capability is based on the principle that
the shower induced by a proton traversing the calorimeter looks different than
that of an electron. First, protons create secondaries by hadronic
interactions, for which the hadronic interaction length $\lambda_I$
sets the scale. For a good absorber, the radiation length is much smaller
than the interaction length, $X_0\ll\lambda_I$. 
Second, the angles of the trajectories of secondaries with respect to
the shower axis are large, compared to the bremsstrahlung case. This
can be seen from the distribution of transverse momenta in inelastic
hadron-hadron scattering, which has a cross section
$\mathrm{d}\sigma/\mathrm{d}p_T^2\propto\exp(-bp_T)$ with
$b\approx{}6\,\mathrm{GeV}^{-1}$~\cite{ref:berger}. Hence, the mean
transverse momentum is given by
$\langle{}p_T\rangle=0.33\,\mathrm{GeV}$.
To summarise, proton showers start deeper in the calorimeter, are
broader than electromagnetic ones, and have less regular shape.

\subsubsection{Design}
\begin{figure}[htb]
\begin{center}
\includegraphics[width=\textwidth,angle=0]{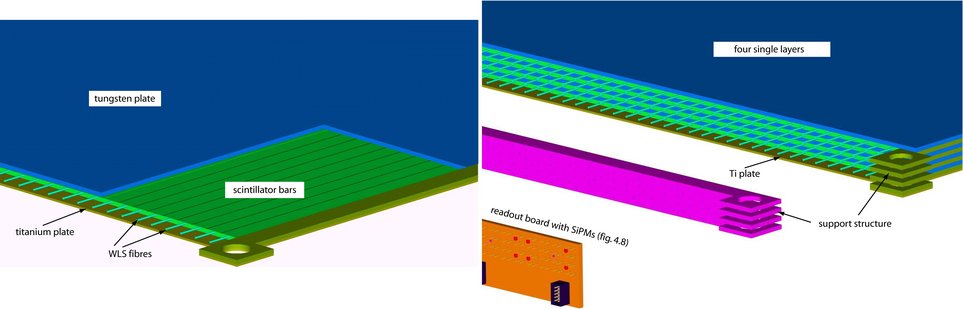}
\end{center}
\caption{Exploded drawing of a single ECAL layer at the bottom of a
superlayer ({\it left}) and one superlayer ({\it right}).}
\label{fig:ecalexploded}
\end{figure}
\begin{figure}[htb]
\begin{center}
\begin{tabular}{cc}
\includegraphics[width=0.45\textwidth,angle=0]{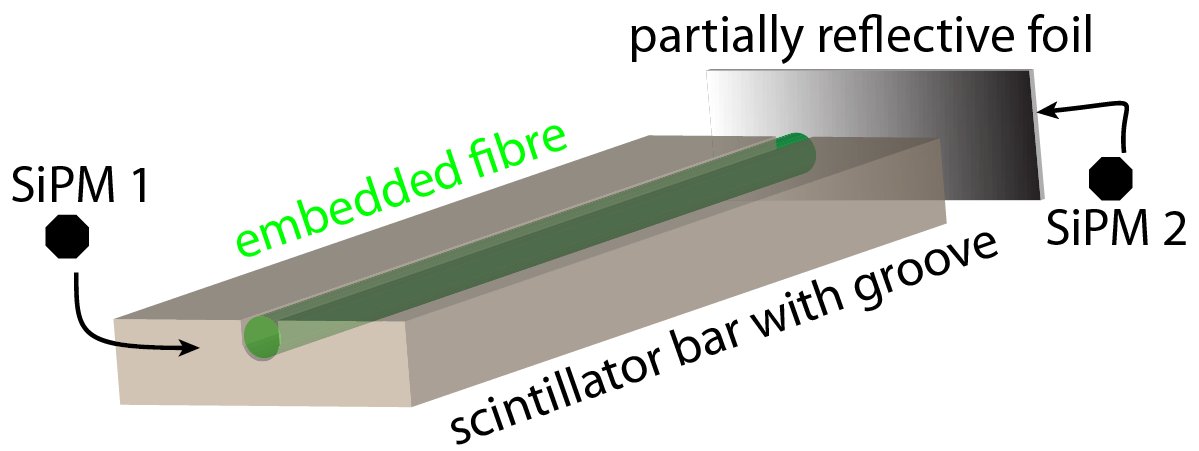}&
\includegraphics[width=0.55\textwidth,angle=0]{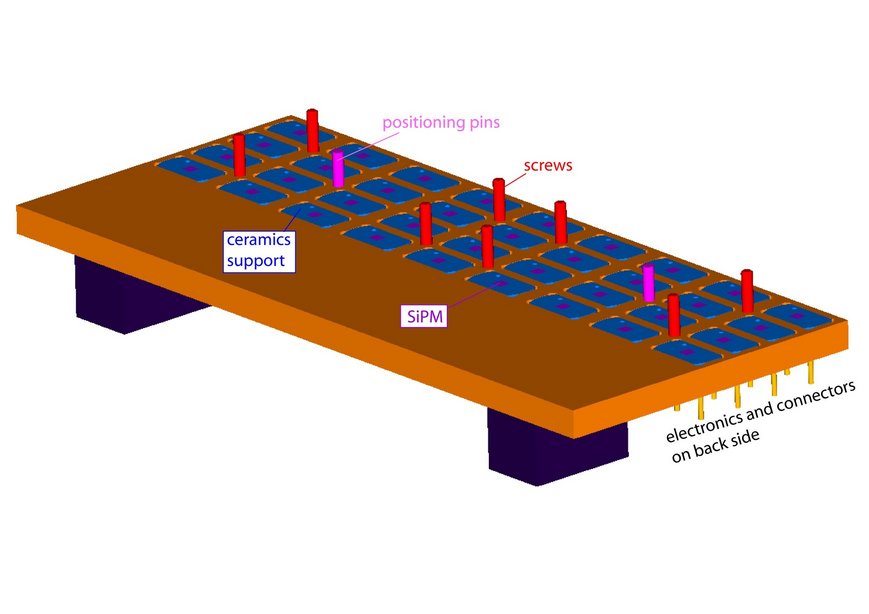}\\
\end{tabular}
\end{center}
\caption{{\it Left:} ECAL fibre readout sketch. {\it Right:} Mechanical drawing of the ECAL board foreseen to position the
silicon photomultipliers at the sides of the calorimeter sandwich.}
\label{fig:ecalboard}
\end{figure}
The electromagnetic calorimeter is located at the bottom of the
detector.
It performs a destructive measurement of the energy of
electrons and can be used for proton suppression.
From the considerations above, the design parameters of the
electromagnetic calorimeter can
already be narrowed down considerably.
A high-$Z$ absorber material is needed, and lead or tungsten are the
prime candidates. Both have an excellent ratio of the radiation and
hadronic interaction lengths
$X_0/\lambda_I\approx{}0.035$ leading to good proton suppression, as
well as
short $X_0$ allowing the construction of a thin calorimeter which in
turn means large geometric acceptance. The overall thickness of the
calorimeter should allow a measurement that includes at least the
shower maximum. According to (\ref{eq:showermaxvse}) this translates
into a minimum thickness of roughly $10X_0$. The longitudinal and
transverse readout granularities must be somewhat smaller than $X_0$
and $R_M$, respectively.\\
\par
The design chosen features twenty
layers of $2\,\mathrm{mm}$ tungsten absorber sheets followed by a
layer of scintillator bars of $2\,\mathrm{mm}$ thickness and
$7.75\,\mathrm{mm}$ width each. Embedded in each scintillator bar is a
wavelength shifting (WLS) fibre used to transport the scintillation light to
the fibre ends where it is detected by SiPMs.\\
Four layers each will be grouped into a superlayer, supported by a
$1\,\mathrm{mm}$ plate of titanium (fig.~\ref{fig:ecalexploded}). Five superlayers are placed with
alternating directions to allow a three-dimensional image of the
shower to be taken. The total thickness of the calorimeter will be as
small as $8.4\,\mathrm{cm}$ and be equivalent to 11.4~radiation
lengths.\\
A challenge in the construction of the calorimeter is introduced by the
fact that its dynamic range must be sufficient to cover the range
extending from minimally ionising particles to the light produced in
the shower maximum of a $200\,\mathrm{GeV}$~electron. This problem is
overcome by putting a partially reflective foil with a reflectivity of
roughly $90\,\%$ in front of one of the two SiPMs
(fig.~\ref{fig:ecalboard} ({\it left})), thereby reducing the light incident on
one SiPM and increasing it on the other. The precise positioning of
the SiPMs in front of the WLS fibres will be done by embedding them in
an aluminium plate like the one sketched in figure~\ref{fig:ecalboard} ({\it right})
which can be screwed to the side of the absorber-fibre sandwich and
will also contain the necessary front-end electronics.

\subsubsection{Simulation of fibre and SiPM response}
The electromagnetic calorimeter is made out of alternating layers of tungsten
absorber and scintillator bars. Layers are grouped into
multilayers and these are rotated to alternating orientations so that a
three-dimensional picture of the shower development can be taken.\\
\par
The response of the detection chain in the electromagnetic
calorimeter, consisting of the scintillator bars, partially reflective
foil and silicon photomultipliers (fig.~\ref{fig:ecalboard}), is modelled such that it matches the
results of test measurements performed at
ITEP Moscow~\cite{ref:vladikecal}. There, the light yield of a scintillator
bar with an embedded wavelength-shifting fibre that was equipped with a mirror on one end and
read out by a SiPM on the other end was measured for minimally
ionising particles (MIPs) crossing the fibre bar over its entire length. An
average yield of $\bar{n}_\gamma=10$ photons was found. Without the
reflective end, a ratio of $A=1.5$ was found for the light yields at
the near and far ends of the fibre bars. The average number
$n_{\gamma,\mathrm{mip}}$ of trapped photons {\it per hemisphere} for
a MIP in the ECAL fibre bars will therefore be
\begin{equation}
\label{eq:ngammamip}
n_{\gamma,\mathrm{mip}} =
\bar{n}_\gamma\,\cdot\,\frac{1}{\int_0^1\left(A^{-x}+A^{-2+x}\right)\,\mathrm{d}x}
= \bar{n}_\gamma\,\cdot\,\frac{\log{}A}{1-\frac{1}{A^2}}
\end{equation}
For an actual simulated energy deposition $\Delta{}E$ in the
calorimeter, the mean number $n_\gamma$ of trapped photons per
hemisphere is then obtained by using the energy loss
$\mathrm{d}E/\mathrm{d}x=2.052\,\mathrm{MeV}/\mathrm{cm}$ of the
scintillator material and the thickness $d$ of a scintillator bar:
\begin{equation}
\label{eq:ecalngamma}
n_\gamma=\frac{\Delta{}E}{d\,\frac{\mathrm{d}E}{\mathrm{d}x}}\,\cdot\,n_{\gamma,\mathrm{mip}}
\end{equation}
Let $x_s\in[0,1]$ be the relative coordinate along the fibre bar. The
light attenuation in the fibre as well as the reflectivity $R$ and
transmissivity $T=1-R$ of the partially reflective foil at $x_s=1$ are then taken
into account by calculating the average number of photons arriving at
SiPM~1~($x_s=0$)
and SiPM~2~($x_s=1$) as
\begin{equation}
\label{eq:sipm12}
n_1=n_\gamma\,\cdot\,\left(\mathrm{e}^{-\log{}A\cdot{}x_s}+\mathrm{e}^{-\log{}A\cdot{}(1-x_s)}\,\cdot\,R\,\cdot\,\frac{1}{A}\right)\quad\mathrm{and}\quad
n_2=n_\gamma\,\cdot\,\left(\mathrm{e}^{-\log{}A\cdot{}(1-x_s)}\,\cdot\,T\right)
\end{equation}
The number of fired pixels $n_\mathrm{pix}^1$ and $n_\mathrm{pix}^2$
on the two SiPMs is generated by drawing
two numbers of incident photons from Poisson distributions with mean
$n_1$ and $n_2$, respectively, and applying the procedure outlined in
section~\ref{sec:sipms}. It was found~\cite{ref:vladikprivate} that a
systematic uncertainty from the non-uniformity of the SiPM response has
to be taken into account by randomly smearing the number of pixels
used in the simulation of the SiPM response by applying a Gaussian
smearing with a standard deviation of
$\sigma_\mathrm{pix}^\mathrm{syst}=10$ pixels.\\
Finally, the output signals $N_1$ and $N_2$
of the two SiPMs are drawn from two Gaussian distributions of mean
$n_\mathrm{pix}^{1,2}$ and standard deviation
$\sqrt{n_\mathrm{pix}^{1,2}}\sigma_\mathrm{peak}$ to take into account
the smearing of the photo peaks as described by the smearing parameter
$\sigma_\mathrm{peak}=0.15$.\\
For a good energy resolution and linearity of the calorimeter, the
reconstructed number of incident photons
$n_{\gamma,\mathrm{rec}}^{1,2}$ must now be extracted from the number of
fired pixels. To this end, formula~(\ref{eq:urnmodel}) is
inverted. Attenuation is corrected for using the same factors as in
eq.~(\ref{eq:sipm12}).\\
To finally exploit the beneficial effect of the partially reflective foil, a
weighted average of the two SiPM signals is calculated as the last
step. Assuming Gaussian likelihoods, the weighted average is
calculated as~\cite{ref:barlowbook}
\begin{equation}
\label{eq:weightedaverage}
n_{\gamma,\mathrm{rec}}=\frac{\frac{n_{\gamma,\mathrm{rec}}^1}{(\sigma_{\gamma,\mathrm{rec}}^1)^2}+
\frac{n_{\gamma,\mathrm{rec}}^2}{(\sigma_{\gamma,\mathrm{rec}}^2)^2}}{\frac{1}{(\sigma_{\gamma,\mathrm{rec}}^1)^2}+\frac{1}{(\sigma_{\gamma,\mathrm{rec}}^2)^2}}
\end{equation}
For the evaluation of~(\ref{eq:weightedaverage}), the errors on the
reconstructed numbers of photons are needed. They are obtained from a
toy Monte Carlo study incorporating the effects outlined in this
\begin{figure}[htb]
\begin{center}
\includegraphics[width=0.5\textwidth,angle=0]{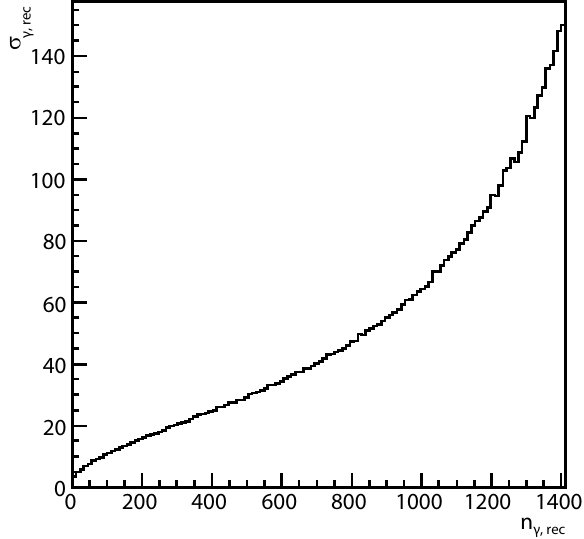}
\end{center}
\caption{Results of the toy Monte Carlo used for the determination of
the error on the number of reconstructed photons in an ECAL fibre bar.}
\label{fig:ecal_nreco_error}
\end{figure}
paragraph minus attenuation. Figure~\ref{fig:ecal_nreco_error}
contains the resulting histogram which is used for the interpolation
of the errors.
In the case that the reconstructed number of photons for SiPM~1
(without the reflective foil in front) exceeds a given threshold,
only the SiPM behind the reflective foil is used for the signal
reconstruction.

\subsection{Transition radiation detector}
\label{sec:trd}
\subsubsection{Principle of operation}
\begin{figure}[htb]
\begin{center}
\includegraphics[width=8cm,angle=0]{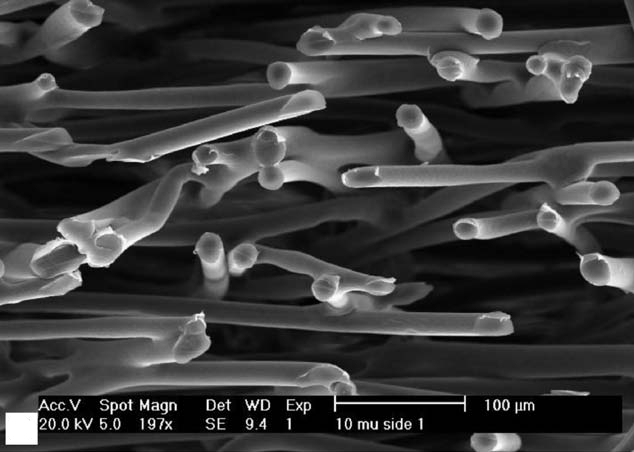}
\end{center}
\caption{Scanning electron microscope picture of an irregular radiator material: polypropylene
fibres (Freudenberg LRP375BK). Reprinted from~\cite{ref:tr_spectra}, Copyright 2006, with
permission from Elsevier.}
\label{fig:radiator_pic}
\end{figure}

A moving charged particle crossing the boundary between two media of
different refractive index emits transition radiation. Qualitatively,
this occurs because the field configuration created by the moving
charge must be different deep inside the second medium as compared to
the one deep inside the first. The necessary change in configuration
leads to some pieces of the field being shaken off as radiation. We
will briefly review the physics of transition radiation here to show
how this effect can be used for particle
identification~\cite{ref:jackson,ref:dolgoshein_trd,ref:egorytchev,ref:struczinski}.\\
A calculation of the spectral and angular dependence of the radiated
energy for a charge entering a medium with electric permittivity
$\epsilon$ begins with the observation that for frequencies above the
optical resonance region, the index of refraction is not far from
unity. This means that the Fourier component of the induced
polarisation $\mathbf{P}(\mathbf{x},\omega)$ can be evaluated
approximately by
\begin{equation}
\label{eq:tr_pol}
\mathbf{P}(\mathbf{x},\omega)=\left(\frac{\epsilon(\omega)}{\epsilon_0}-1\right)\mathbf{E}_i(\mathbf{x},\omega)
\end{equation}
where $\mathbf{E}_i$ is the Fourier transform of the electric field of
the incident particle in vacuum. This can be used to calculate the
dipole radiation field caused by the polarisation
$\mathbf{P}(\mathbf{x},\omega)$.\\
For a particle with charge $ze$ crossing the
boundary between media with electric permittivities $\epsilon_1$ and
$\epsilon_2$, the resulting angular and spectral dependence of
the transition radiation energy is
\begin{equation}
\label{eq:tr_theta_omega}
\frac{\mathrm{d}^2I}{\mathrm{d}\omega\mathrm{d}\theta}=\frac{2z^2\alpha\hbar\theta^3}{\pi}\left(\frac{1}{\gamma^{-2}+\theta^2+\omega_1^2/\omega^2}-
\frac{1}{\gamma^{-2}+\theta^2+\omega_2^2/\omega^2}\right)^2
\end{equation}
Here, $\omega_{1,2}$ are the plasma frequencies of the media which
determine the electric permittivity,
\begin{equation}
\epsilon_i(\omega)\simeq{}1-\frac{\omega_i^2}{\omega^2}
\end{equation}
The plasma frequency $\omega_p$ depends on the number density $n_e$ of electrons in the
medium, $\omega_p^2=n_ee^2/\epsilon_0m_e$. For a typical radiator,
polyethylene, $\hbar\omega_p^{\mathrm{CH}_2}=20.9\,\mathrm{eV}$, while for
air $\hbar\omega_p^\mathrm{air}=0.7\,\mathrm{eV}$.\\
Integrating over the emission angle $\theta$ gives the differential
energy spectrum
\begin{equation}
\frac{\mathrm{d}I}{\mathrm{d}\omega}=\frac{z^2\alpha\hbar}{\pi}\left(\left(\frac{\omega_1^2+\omega_2^2+2\omega^2/\gamma^2}{\omega_1^2-\omega_2^2}\right)\cdot\log\left(\frac{1/\gamma^2+\omega_1^2/\omega^2}{1/\gamma^2+\omega_2^2/\omega^2}\right)-2\right)
\end{equation}
The total energy radiated at a single boundary then becomes
\begin{equation}
\label{eq:tr_int}
I=\frac{z^2\alpha\hbar}{3}\,\gamma\,\frac{(\omega_1-\omega_2)^2}{\omega_1+\omega_2}
\end{equation}
The proportionality of the energy $I$ to $\gamma=E/mc^2$ is the key to
particle identification. For electrons in the GeV-range,
$\gamma\gtrsim{}1000$, while $\gamma\sim{}1$ for protons of the same
energy due to their much higher mass. The angular and spectral
distributions can be shown to be such that most of the emission is in
the forward direction, in a cone with half-angle $1/\gamma$, and is
emitted in the form of x-ray photons for highly relativistic
particles, with the energy scale of the photons given by
$\gamma\omega_p$.
A proportional chamber filled with a gas with a high x-ray
absorption coefficient, such as xenon, is commonly used for detection
of transition radiation for that reason. As the x-ray photons are
emitted almost along the direction of the charged particle, the
detected signal will be the sum of the energy deposition due to
ionisation and the absorbed x-ray photons.\\
As the probability of emission of a transition radiation photon at a
single boundary is of the order of $\alpha$ according to
(\ref{eq:tr_int}), radiators consisting of many interfaces have to be
employed in practice. An important quantity is the formation zone of a material, given by
\begin{equation}
\label{eq:formationzone}
Z=\frac{4c}{\omega}\left(\frac{1}{\gamma^2}+\theta^2+\frac{\omega_p^2}{\omega^2}\right)^{-1}
\end{equation}
If the thickness of a radiator element is smaller than $Z$, the yield
will be strongly suppressed. The
calculation of the transition radiation yield behind a stack of
radiators then takes additional effects into account. First, transition radiation
is now emitted at both sides of a single radiator. Then, if the spacing
between the individual radiators is larger than the formation zone in
air the total flux of x-ray photons will be the incoherent sum of the
individual fluxes. In addition, absorption of x-ray photons inside the
radiator stack has to be considered.\\
The radiator that was chosen for
PEBS, based on the experience with the AMS-02-TRD, is of irregular
form. It consists of a fleece made of
polypropylene-polyethylene fibres of nominal thickness
$10\,\mu\mathrm{m}$ that are revealed in a picture taken with an
electron microscope (fig.~\ref{fig:radiator_pic}).
Here, transition radiation is emitted at the numerous fibre-air interfaces.
The calculation of the transition radiation yield of such a radiator
is further complicated by the fact that the thicknesses of the fibres
and corresponding air gaps are not constant but follow a given random
distribution. This is handled by the Geant4 simulation, as outlined in
more detail in section~\ref{sec:trdtestbeam}.

\subsubsection{Design and implementation in the simulation}
\begin{figure}[htb]
\begin{center}
\includegraphics[width=0.7\textwidth,angle=0]{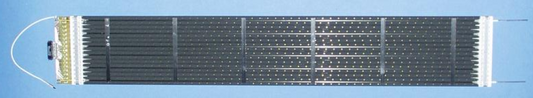}
\end{center}
\caption{A picture of a TRD straw module used for the AMS-02 experiment.}
\label{fig:strawmodule}
\end{figure}
The transition radiation detector
is based on the one constructed for the AMS-02
experiment~\cite{ref:ams02,ref:fopp}.
It consists of two parts of eight layers each, located between the
tracker layers. Each TRD layer contains a mat of irregular radiator fleece
of $20\,\mathrm{mm}$ thickness and eight modules of sixteen straw tubes
each (figure~\ref{fig:strawmodule}) of $868\,\mathrm{mm}$ length. The straw tubes contain a
thin tungsten wire operated at
high voltage and an 80/20
mixture of Xe/CO$_2$ and they are used to detect both ionisation
losses of charged particles and the x-ray transition radiation photons
created by light particles such as positrons in the fleece
radiator. For
\begin{figure}[htb]
\begin{center}
\includegraphics[width=0.5\textwidth,angle=0]{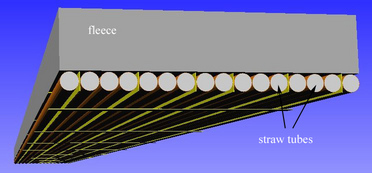}
\end{center}
\caption{Visualisation of a TRD module in the Geant4
  simulation. The radiator fleece, straw tubes, and carbon-fibre
  stiffeners are visible.}
\label{fig:trdmodulevis}
\end{figure}
mechanical stability, longitudinal and lateral carbon-fibre
stiffeners are added. The tasks of the TRD are proton suppression and
determination of the slope angle $\lambda$ of tracks.\\
In the simulation (fig.~\ref{fig:trdmodulevis}), the straw tube walls are modelled as tubes of $72\,\mu\mathrm{m}$ Kapton, with a
tungsten wire of $30\,\mu\mathrm{m}$ diameter in the middle. The gas
mixture has a density of $4.885\,\mathrm{mg}/\mathrm{cm}^3$. The
radiator fleece
consists of polypropylene and polyethylene with number ratio
$85/15$ and has a density of $0.06\,\mathrm{g}/\mathrm{cm}^3$. The performance of the TRD depends on the microscopic
properties of the radiator, and section~\ref{sec:trdtestbeam} was
devoted to their determination and modelling.\\
The overall material budgets of the tracker and the TRD are
calculated to be only $6\,\%$ and $5.5\,\%$ of a radiation length, respectively.

\subsection{Physics processes}
Geant4 offers a wide variety of physics models for the interactions of the
different particle species with the available materials, covering an energy range from
$250\,\mathrm{eV}$ up to several $\mathrm{PeV}$~\cite{ref:g4}. In many
cases, concurrent or alternative models are available, and it is the
user's responsibility to choose the most appropriate set of physics
implementations suitable for her application. The choice of models
describing hadronic processes is especially large and it ranges from
data-driven models to parameterised and theory-driven models, some of
them optimised for speed, others for accuracy. In the following, the
choice of models for the simulation of the PEBS detector will be
outlined briefly and the Geant4 class names employed will be given for
reference in those cases where they are not apparent.\\
For the interactions of $\gamma$-rays, the standard photoelectric
effect, Compton scattering, and pair conversion are used. Electrons
and positrons undergo multiple scattering, ionisation losses, and
bremsstrahlung losses. In addition, positrons will annihilate at
rest. The standard processes are used throughout, with the important
exception of the ionisation losses in the TRD gas mixture. These are
described by the photo-absorption ionisation model as implemented in
the class {\tt G4PAIModel}. This is important for a proper
determination of the TRD proton rejection and is elaborated on in
section~\ref{sec:trdtestbeam}. The same electromagnetic processes are
implemented using the appropriate standard classes for muons, protons,
deuterons, tritons, pions, kaons, $\alpha$-particles, heavier ions, as well as for some
more exotic mesons and baryons and their respective antiparticles. A
default cut value for the range of secondary particles of
$0.5\,\mathrm{mm}$ was used throughout the simulation, with the
exception addressed in section~\ref{sec:trdtestbeam}. For
completeness, the more exotic effects of photo-, electro-, and
muo-production of hadrons as well as synchrotron radiation are
included as implemented in the {\tt G4EmExtraPhysics}
class. Particular care has to be taken with the simulation of
transition radiation, which remains a challenging
task. Section~\ref{sec:trdtestbeam} contains details about the
procedure.\\
The physics of inelastic hadronic interactions is given by the {\tt
HadronPhysicsQGSP\_BIC} class~\cite{ref:g4physics,ref:g4hadr}.
It applies a quark-gluon string model for the
modelling of interactions of high energy hadrons above
$25\,\mathrm{GeV}$. Interactions at intermediate energies are modelled by the low
energy parameterised (LEP) model, and below $10\,\mathrm{GeV}$, the Geant4
binary cascade~\cite{ref:g4binarycascade} for primary protons and
neutrons replaces the LEP
model. The binary cascade is used because it promises a better
description of the production of secondary particles produced in
interactions of protons and neutrons with nuclei.
The high energy interaction creates an exited nucleus, which is passed
to the Geant4 precompound model modelling the nuclear de-excitation.
Nuclear capture of negative
particles and neutrons at rest is modelled by the
Chiral Invariant Phase Space (CHIPS) model.\\
Also included are the standard decay physics, hadron elastic
scattering, and stopping physics. Ion physics is handled by the {\tt
G4IonBinaryCascadePhysics} class.

\subsection{Adjustment of parameters for the simulation of transition
radiation with testbeam data}
\label{sec:trdtestbeam}
This section is devoted to a comparison of the transition radiation (TR) and ionisation energy loss
distributions obtained from the Monte Carlo simulation based on Geant4 to
electron and proton testbeam data~\cite{ref:trdstudy}. The data were taken with a prototype consisting of 20~layers of fleece
radiator and straw tube proportional chambers, filled with Xe/CO$_2$,
during a testbeam conducted in the context of the AMS-02 experiment at
CERN in the year 2000. Impressive agreement is found for the
simulation of TR for electrons and very good agreement for the ionisation loss
distributions for protons over a wide range of incident energies. The
implications of slight deviations in the tails of the proton energy loss
spectra on calculated proton suppression factors are studied briefly
to assess the accuracy of the predictions obtained in this design
study. The testbeam data are also used to adjust various parameters of
the Geant4 simulation governing the microscopic structure of the
radiator fleece.
Only few validation studies for Geant4 concerned with
transition radiation and ionisation energy losses in thin absorbers
have been published so far~\cite{ref:g4pai,ref:grichine_nim_02,ref:g4trpackage,ref:grichinetrd}.

\subsubsection{Description of the testbeam prototype and simulation}
\label{sec:descr}
\begin{figure}
\begin{center}
\begin{tabular}{cc}
\includegraphics[width=0.3\textwidth,angle=0]{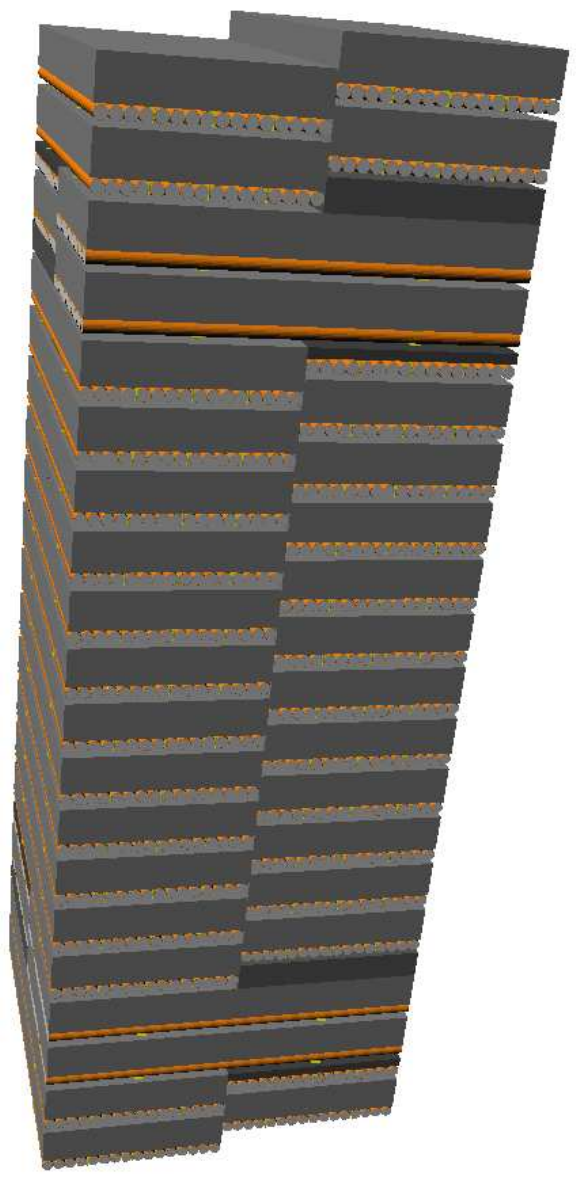}&
\includegraphics[width=0.5\textwidth,angle=0]{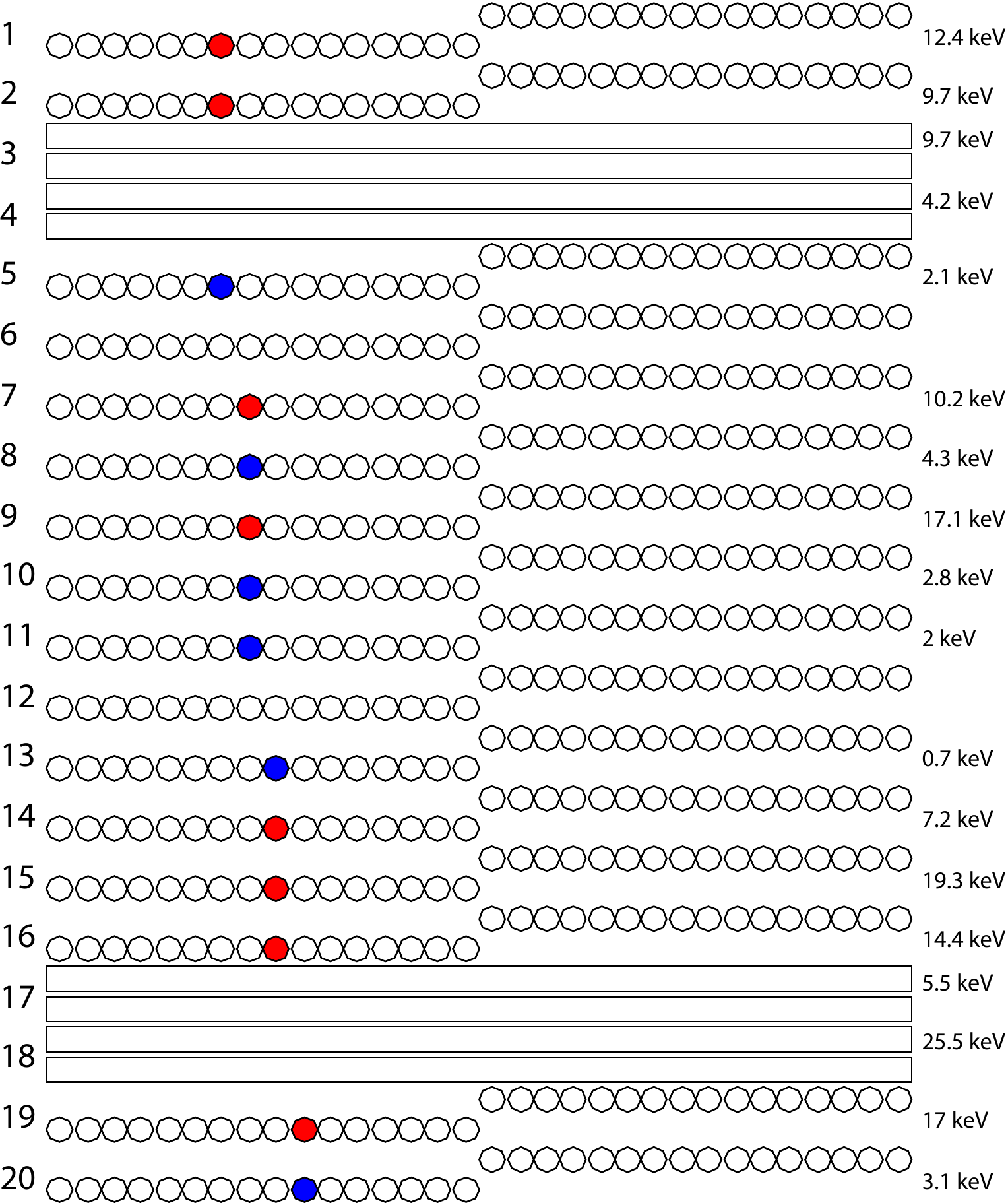}\\
\end{tabular}
\end{center}
\caption{{\it (left)} Visualisation of the testbeam prototype in the Geant4
simulation. The staggered modules and rotated layers are visible, complete with radiator
fleece stacks, gas-filled straw tubes and carbon stiffeners.
{\it(right)} Schematic drawing of an electron event in a front view of
the prototype. Several high energy depositions ($>6.5\,\mathrm{keV}$) along the track are
characteristic of an electron.}
\label{fig:g4vis}
\end{figure}

The setup used in the testbeam consisted of prototype modules for the
AMS-02-TRD which were grouped in 20~layers (see fig.~\ref{fig:g4vis}).
In each module, the TR x-ray photons are generated in a
$2\,\mathrm{cm}$ thick irregular fleece radiator made of polyethylene and
polypropylene, with a density of
$\rho=0.06\,\mathrm{g}/\mathrm{cm}^3$. They are subsequently detected
in proportional wire
chambers in the form of straw tubes made of
aluminised kapton foils which have an inner diameter of $6\,\mathrm{mm}$ and
are filled with an $80:20$ mixture of Xe/CO$_2$ at a pressure of
$p=1\,\mathrm{atm}$. The $30\,\mu{}\mathrm{m}$ gold-plated tungsten wires in the straw
tubes were operated at a voltage of $1480\,\mathrm{V}$ for a gas-gain of
6000. Temperature and pressure were continuously monitored during the
testbeam.\newline
Layers 3,~4,~17~and~18 were rotated by $90^\circ$ with respect to the
others to provide track coordinates in all spatial dimensions. Two
towers of modules were put next to each other and staggered by half
the thickness of a module.\newline
Protons and electrons were recorded at several energies at the CERN X7
and H6 beamlines in 2000. An extensive description of the testbeam setup is found in~\cite{ref:ams02,ref:orboeck}. 

The prototype setup was implemented in a Geant4.8.2 simulation
(fig.~\ref{fig:g4vis}). This includes the
radiator mats, kapton straw tubes with gas mixture and tungsten wires,
and carbon fibre stiffeners, as well as the trigger scintillators and
the acryl glass entry window that the beam particles had to traverse. A
gas density of $\rho_{\mathrm{Xe/CO}_2}=4.885\,\mathrm{mg}/\mathrm{cm}^3$ was employed.
Primary protons and electrons were generated according to the energies chosen and angular distribution
measured in the testbeam.

Geant4 offers several physics
process classes for the simulated generation of transition radiation
photons. They have in common that they are based mainly on the
approach of calculating the work done by a relativistic charge
crossing the boundary between two media against the electric fields
induced by the charge in the vicinity of the
boundary~\cite{ref:g4trpackage}.
The classes available include models with fixed and gamma-distributed radiator foil and gas
thicknesses, as well as a model for the simulation of additional TR
photons generated in straw tube walls. In the first case, both a model
with TR photon absorption taken into account and a
simplified transparent version are available.

The appropriate model for the irregular fleece radiator employed in the testbeam setup is the {\tt
G4GammaXTRadiator}~\cite{ref:grichine_nima_04}. In this model, the fleece fibre and gas
thicknesses $z$ are assumed to vary according to the distribution~\cite{ref:grichinetrd}
\begin{equation}
p_j(z) =
\left(\frac{\alpha_j}{\bar{z}_j}\right)^{\alpha_j}\,\frac{z^{\alpha_j-1}}{\Gamma(\alpha_j)}\,\exp\left(-\frac{\alpha_j
z}{\bar{z}_j}\right)
\end{equation}
where $\bar{z}_j$ is the mean thickness of the $j$-th medium in the
radiator and the $\alpha_j$ parameters describe the relative thickness
fluctuations $\delta_j$ which are found to be
\begin{equation}
\delta_j =
\frac{\sqrt{\langle(z_j-\bar{z}_j)^2\rangle}}{\bar{z}_j}=\alpha_j^{-1/2}\quad
(j=1,2)
\end{equation}
The signal detected in the straw tubes is the sum of those from
absorbed TR x-ray photons and the ionisation losses of the primary
particle. The standard electromagnetic processes in Geant4 fail to
reproduce the energy loss spectra in thin absorbers, such as the straw
tube gas mixture. Instead, the photo absorption ionisation (PAI) model~\cite{ref:allison_pai}
as implemented in Geant4~\cite{ref:g4pai} is used.\newline
It should be stressed that the Geant4 processes were applied as-is,
i.e.~no modifications (``tuning'') of any kind were made. Only the simulation parameters related
to the microscopic structure of the radiator need to be chosen.

A description of the raw data processing procedure is given
in~\cite{ref:ams02}. Once energy calibration and gas density dependent
gas gain corrections have been performed, a
track fit is done to select clean single-track events, and the energy
depositions for each layer along the track are stored for further analysis.

\subsubsection{Adjustment of parameters for the simulation of
transition radiation}
\label{sec:parameters}
The simulation results depend on a number of parameters, some of which
are known, e.g.~the gas density $\rho$ and number
fraction $n_\mathrm{Xe}$ of xenon, while others have to be determined from a fit
to the data. In particular, the structure of the radiator fleece mats
is for a given density determined by the mean fibre diameter
$\bar{z}_1$ and the fluctuation parameters $\alpha_{1,2}$ of the
fibres and gas in the radiator, respectively.

The parameters are obtained from a $\chi^2$-minimisation procedure as
follows. First of all, MC and data electron spectra are normalised
with respect to each other, taking only the energy interval
$[E_\mathrm{min}=0.5\,\mathrm{keV};E_\mathrm{max}=20\,\mathrm{keV}]$ into account. This
eliminates the influence of artefacts in the data, namely
inefficiencies of the straw tubes (at low energies) and cutoff values
introduced by the ADCs employed (at high energies). The $\chi^2$ is
then calculated from the error-weighted quadratic sum of the MC-data
differences over all layers and all energy bins within the interval
quoted above:
\begin{equation}
\chi^2 = \sum\limits_{i \atop E_j\in [E_\mathrm{min},E_\mathrm{max}]} \frac{(F^\mathrm{MC}_{i,j}-F^\mathrm{data}_{i,j})^2}{\sigma^2_{\mathrm{data},i,j}+\sigma^2_{\mathrm{MC},i,j}}
\end{equation}
where $i$ numbers layers, $E_j$ is the $j$-th energy bin,
$F_{i,j}\equiv(\mathrm{d}N/\mathrm{d}E)_{i,j}$ and the errors are approximated by
$\sigma_{i,j}=N_i^{-1}\cdot\sqrt{\Delta{}N_{i,j}}$ in the usual
way. Here, $N_i$ is the normalisation factor for layer $i$ and the
$\Delta{}N_{i,j}$ are the raw counts.

\begin{figure}[htb]
\begin{center}
\begin{tabular}{cc}
\includegraphics[width=0.58\textwidth,angle=0]{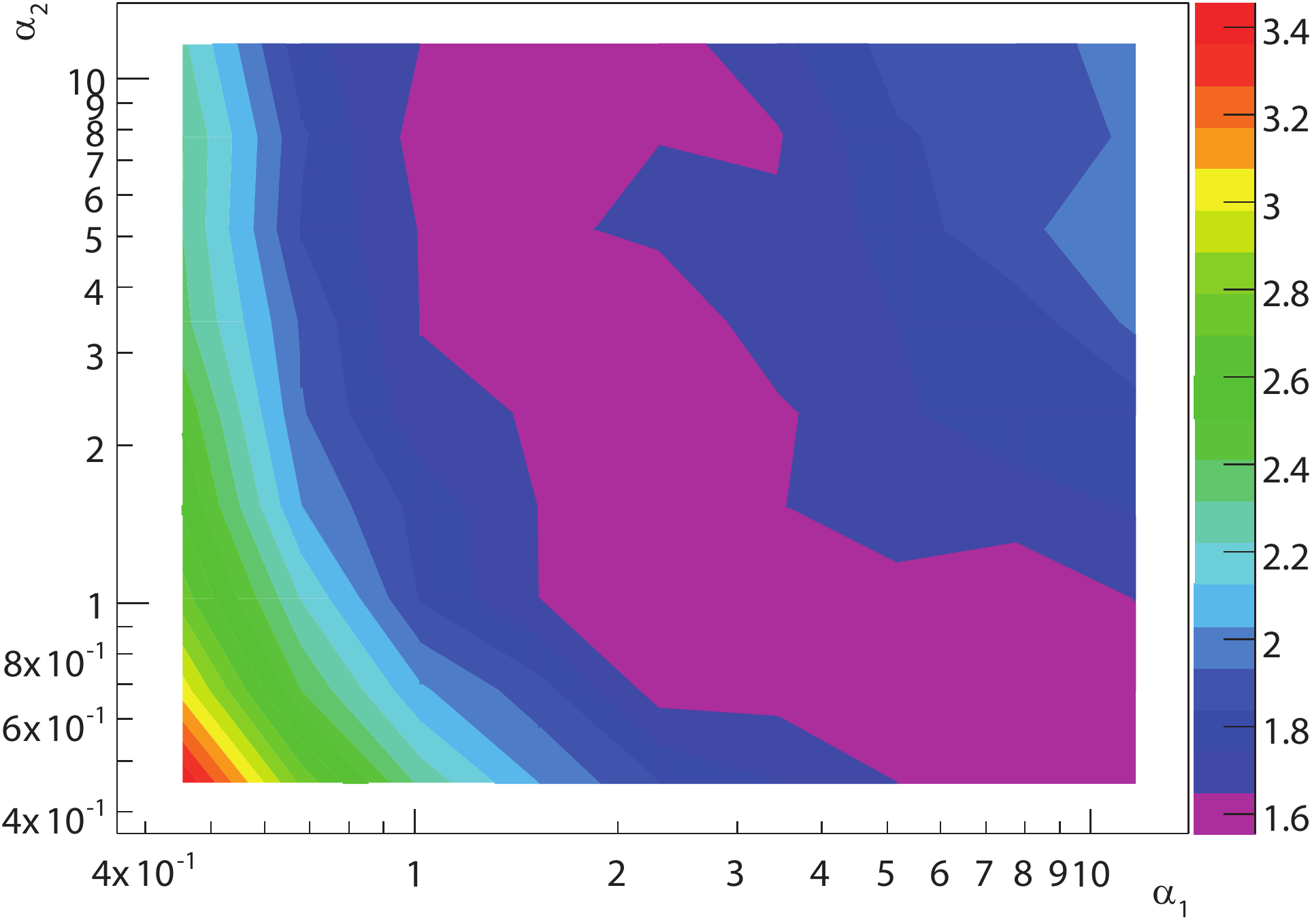}&
\includegraphics[width=0.42\textwidth,angle=0]{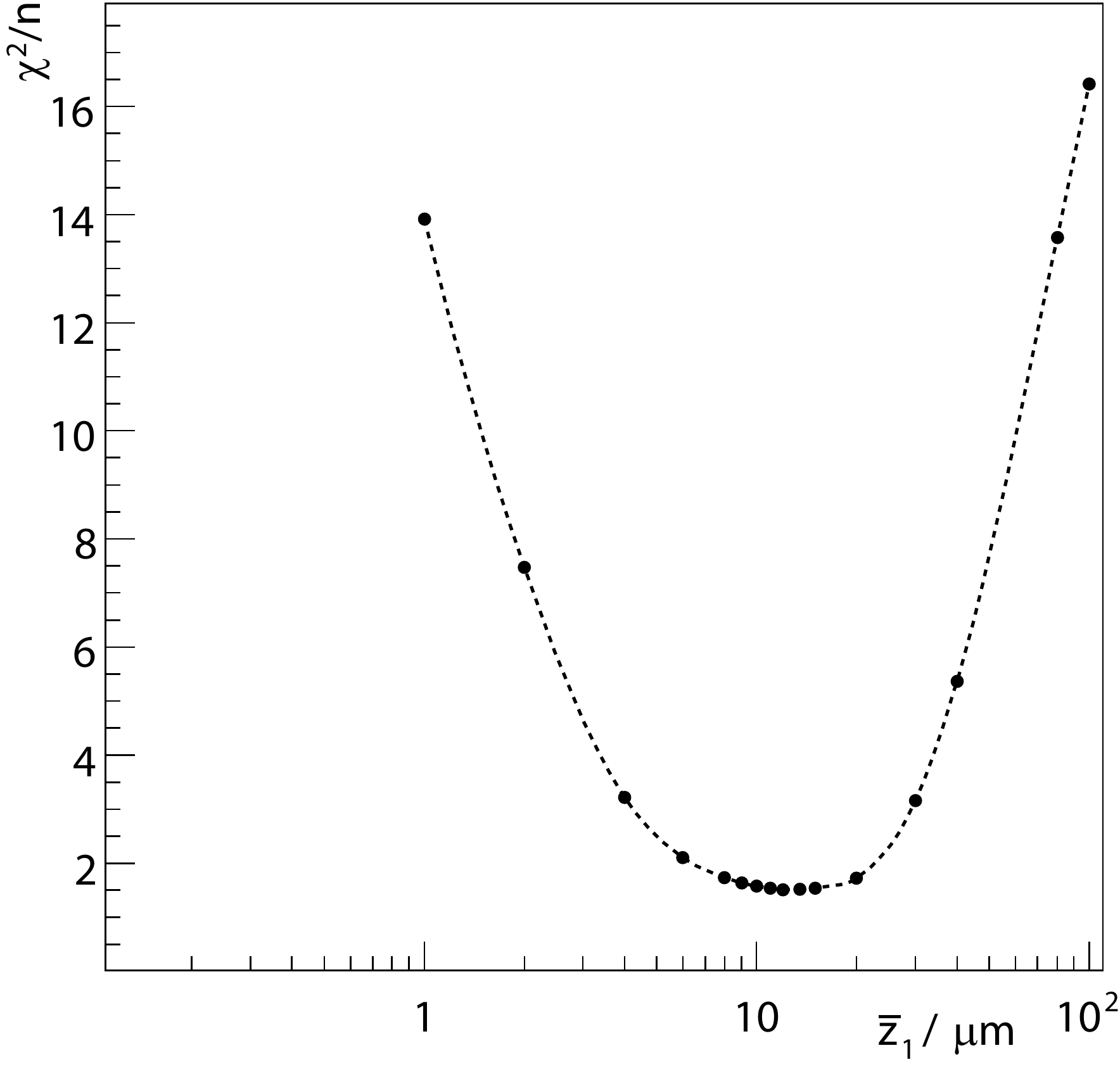}\\
\end{tabular}
\end{center}
\caption{{\it Left:} $\chi^2/n$ values from variation of $\alpha_1$ and
$\alpha_2$. A minimum is found at $(5,\frac{2}{3})$. {\it Right:} $\chi^2/n$ values from variation of $\bar{z}_1$. The nominal
value is $\bar{z}_1=10\,\mu\mathrm{m}$.}
\label{fig:chi2alpha_var}
\end{figure}

Because the fibre thickness fluctuation parameters $\alpha_{1,2}$ are
unknown, first the nominal value of $\bar{z}_1=10\,\mu{}\mathrm{m}$ is
assumed and $\chi^2$ is calculated for simulation
results in the $(\alpha_1,\alpha_2)$-plane. Figure~\ref{fig:chi2alpha_var}
shows the result as a contour plot. Plotted is the value $\chi^2/n$
where $n$ is the total number of energy bins used. A minimum is found at roughly
$\alpha_1=5$ and $\alpha_2=\frac{2}{3}$. As can be seen from the
figure, the quality of the match does not depend too sensitively on
the chosen values as the region with values close to the minimum is
quite large, spanning almost a decade.
The quoted values for
$\alpha_{1,2}$ are fixed, and $\bar{z}_1$ is subsequently varied. The resulting $\chi^2/n$-plot is depicted in
figure~\ref{fig:chi2alpha_var}.
A minimum is found at $\bar{z}_1=12\,\mu{}\mathrm{m}$ which is in good
agreement with the nominal value.

\subsubsection{Comparison of TR spectra}
\label{sec:comparison}
\begin{figure}[htb]
\begin{center}
\includegraphics[width=0.8\textwidth,angle=0]{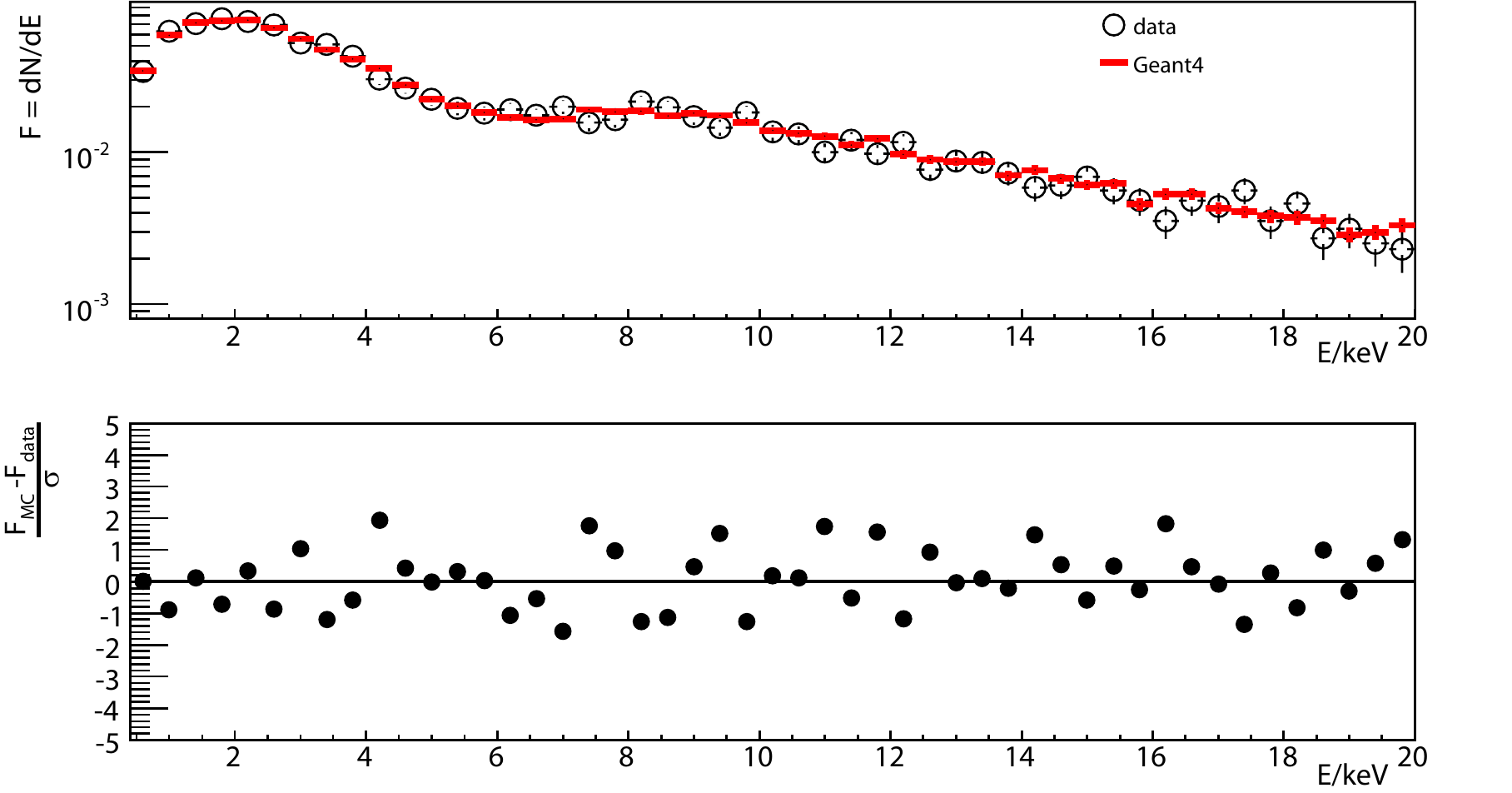}
\end{center}
\caption{Ionisation loss and transition radiation spectra from data
and Geant4 simulation, for $20\,\mathrm{GeV}$ electrons, in layer~1 of the
prototype {\it (top)} and corresponding residual plot {\it
(bottom)}. Here,
$\sigma=\sqrt{\sigma^2_\mathrm{data}+\sigma^2_\mathrm{MC}}$.}
\label{fig:layer1}
\end{figure}
\begin{figure}[htb]
\begin{center}
\includegraphics[width=0.8\textwidth,angle=0]{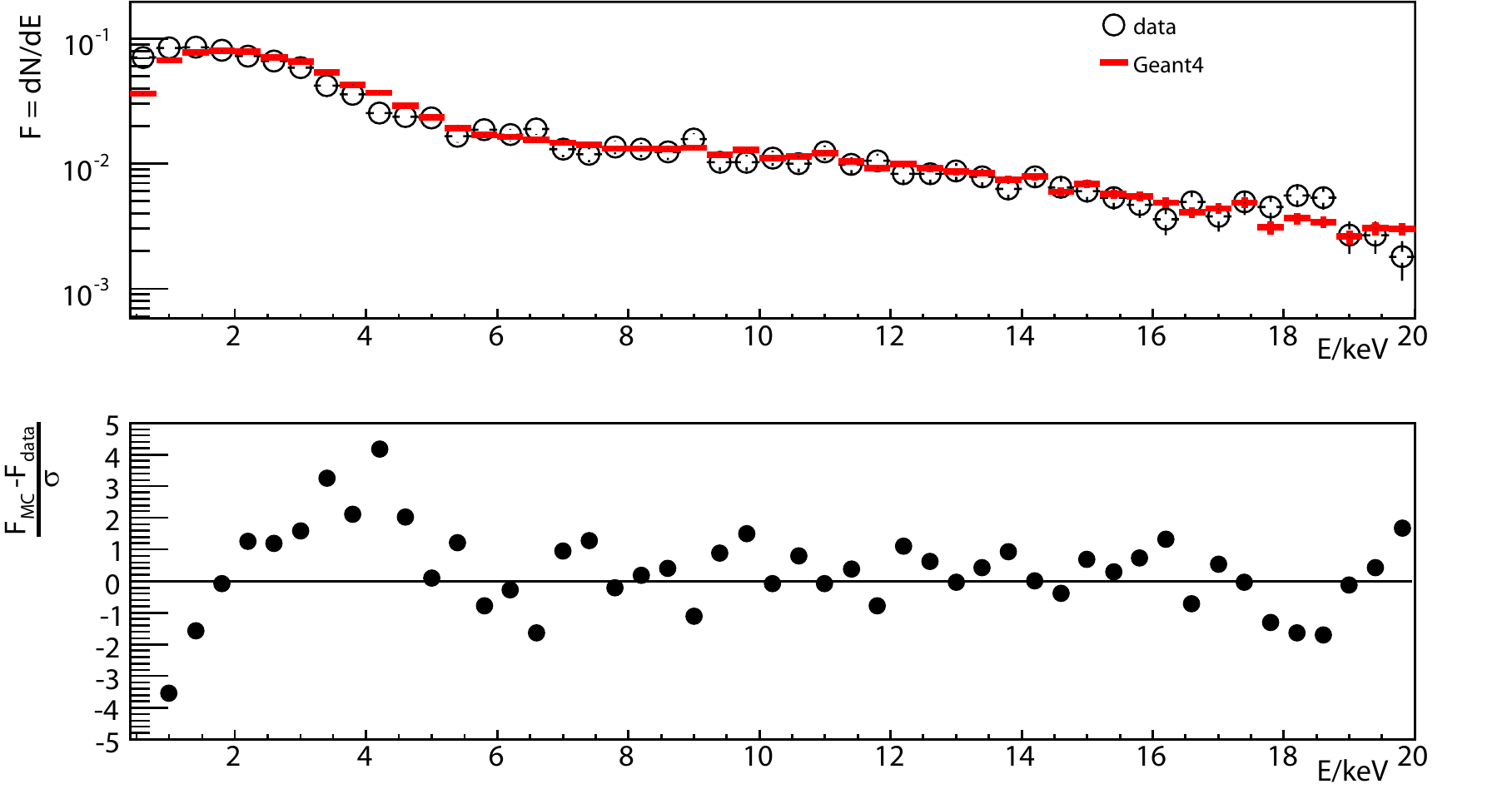}
\end{center}
\caption{Ionisation loss and transition radiation spectra from data
and Geant4 simulation, for $20\,\mathrm{GeV}$ electrons, in layer~3 of the
prototype {\it (top)} and corresponding residual plot {\it
(bottom)}.}
\label{fig:layer3}
\end{figure}
\begin{figure}[htb]
\begin{center}
\includegraphics[width=0.8\textwidth,angle=0]{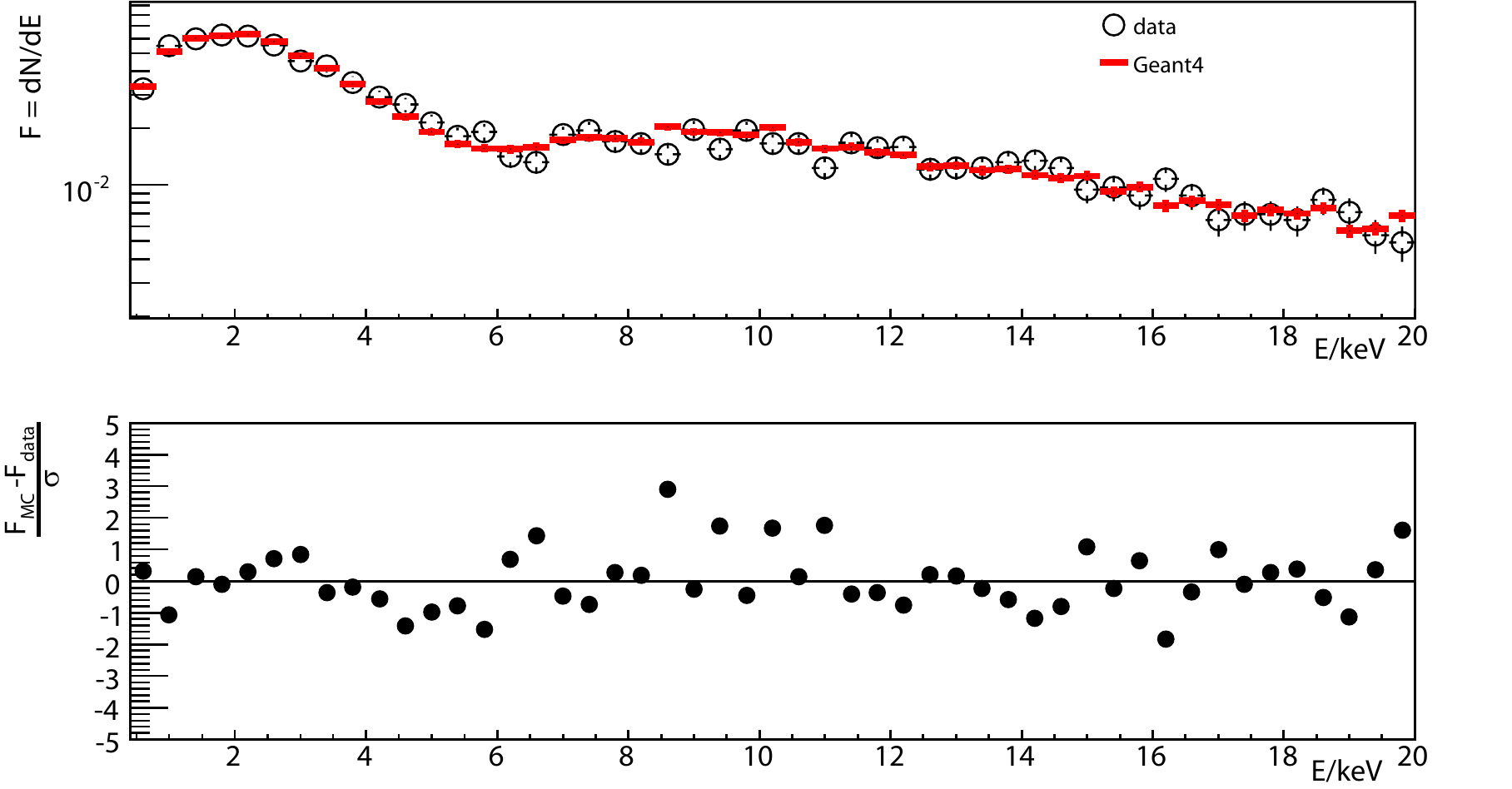}
\end{center}
\caption{Ionisation loss and transition radiation spectra from data
and Geant4 simulation, for $20\,\mathrm{GeV}$ electrons, in layer~5 of the
prototype {\it (top)} and corresponding residual plot {\it (bottom)}.}
\label{fig:layer5}
\end{figure}
\begin{figure}[htb]
\begin{center}
\includegraphics[width=0.8\textwidth,angle=0]{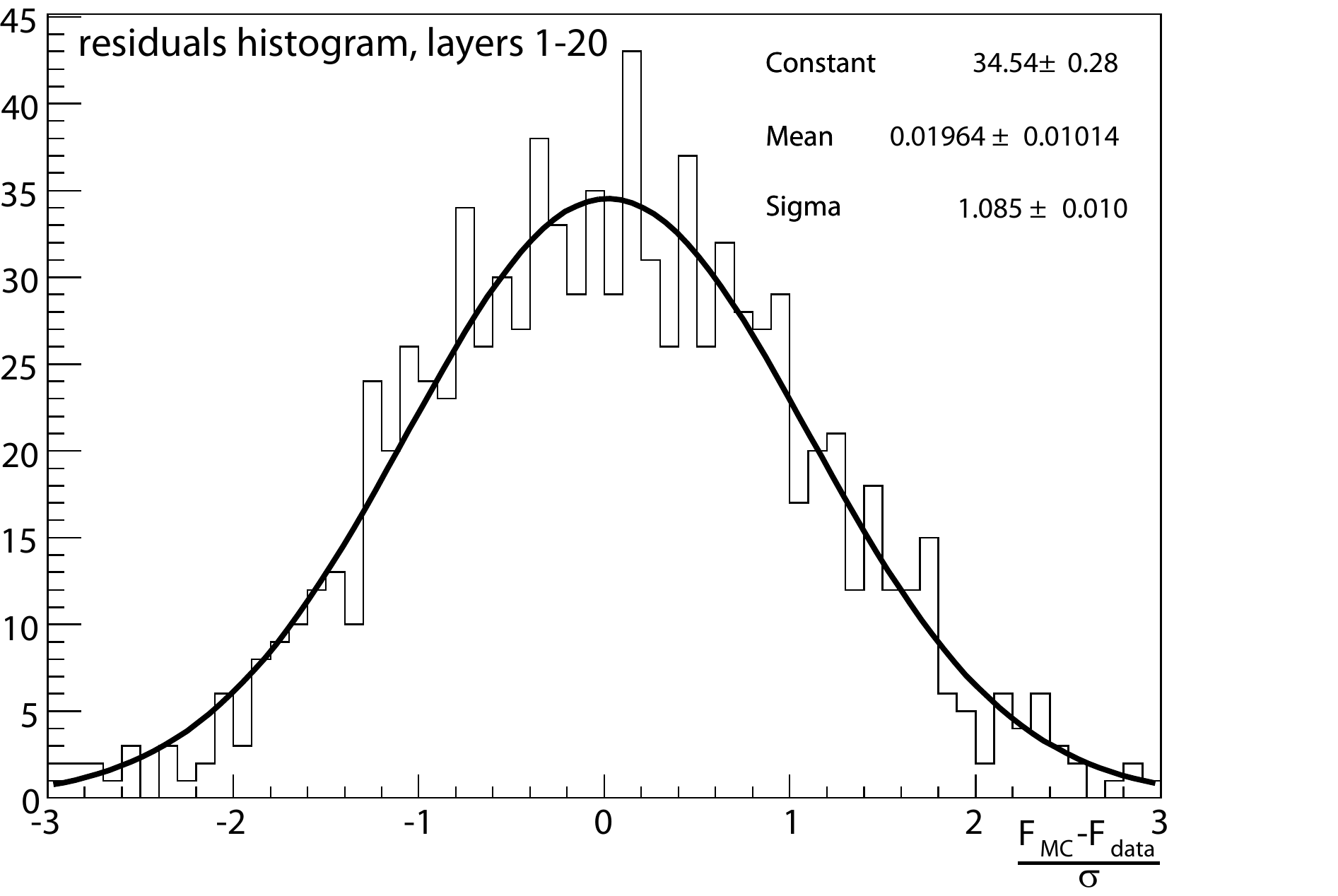}
\end{center}
\caption{Histogram of residuals for electron spectra, containing energy bins
from all layers.}
\label{fig:residuals_electrons}
\end{figure}
\begin{figure}[htb]
\begin{center}
\includegraphics[width=0.7\textwidth,angle=0]{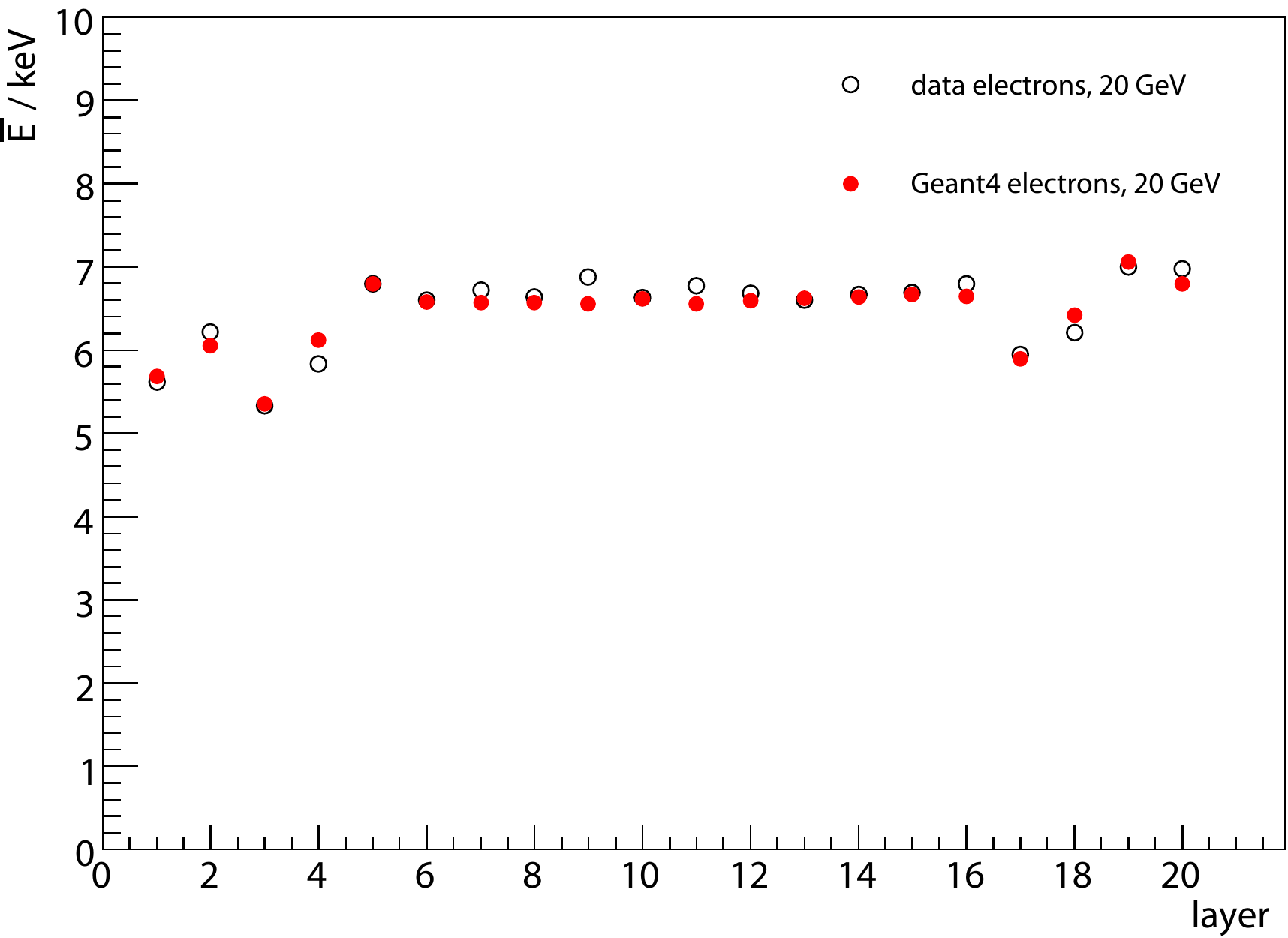}
\end{center}
\caption{Mean electron energy deposition as a function of layer
number, for data and simulation.}
\label{fig:layerspectrum}
\end{figure}

With all parameters set to the optimised values derived above,
the electron spectra obtained from data
and simulation can be
compared.
Figures~\ref{fig:layer1},~\ref{fig:layer3} and~\ref{fig:layer5} show
data and simulation spectra for layers 1,~3~and~5, respectively. These
layers were chosen because an incoming particle had to traverse
different amounts of radiator fleece material before entering each of
these (cf.~fig.~\ref{fig:g4vis}). Only half a layer of fleece was
present in front of layer~3 for the events considered here, while the
fleece in front of layer~5 had $1.5$ times the usual thickness. The
agreement between data and simulation spectra is very good. This can also be seen from the residual
plot in figure~\ref{fig:residuals_electrons} showing a histogram of
all values of
$\frac{F^\mathrm{MC}_{i,j}-F^\mathrm{data}_{i,j}}{\sigma_{i,j}}$,
where $\sigma_{i,j}=\sqrt{\sigma^2_{\mathrm{data},i,j}+\sigma^2_{\mathrm{MC},i,j}}$,
from all layers and bins within the threshold interval. They follow a
Gaussian distribution with mean and RMS values of $0.02$ and $1.1$.
Figure~\ref{fig:layerspectrum} shows the mean energy deposition for
data and simulated electrons of $20\,\mathrm{GeV}$. The good description of
the TR yield in each layer is illustrated here. The deviation of the
simulation from the data is below the layer-to-layer
variation in the data. Both the effect of the slope towards
saturation in the first layers as the effect of varying radiator
thickness around the rotated layers 3,~4,~17,~18 are well reproduced.

\subsubsection{Comparison of proton energy loss spectra}
\label{sec:protons}
\begin{figure}[htb]
\begin{center}
\begin{tabular}{cc}
\includegraphics[width=0.5\textwidth,angle=0]{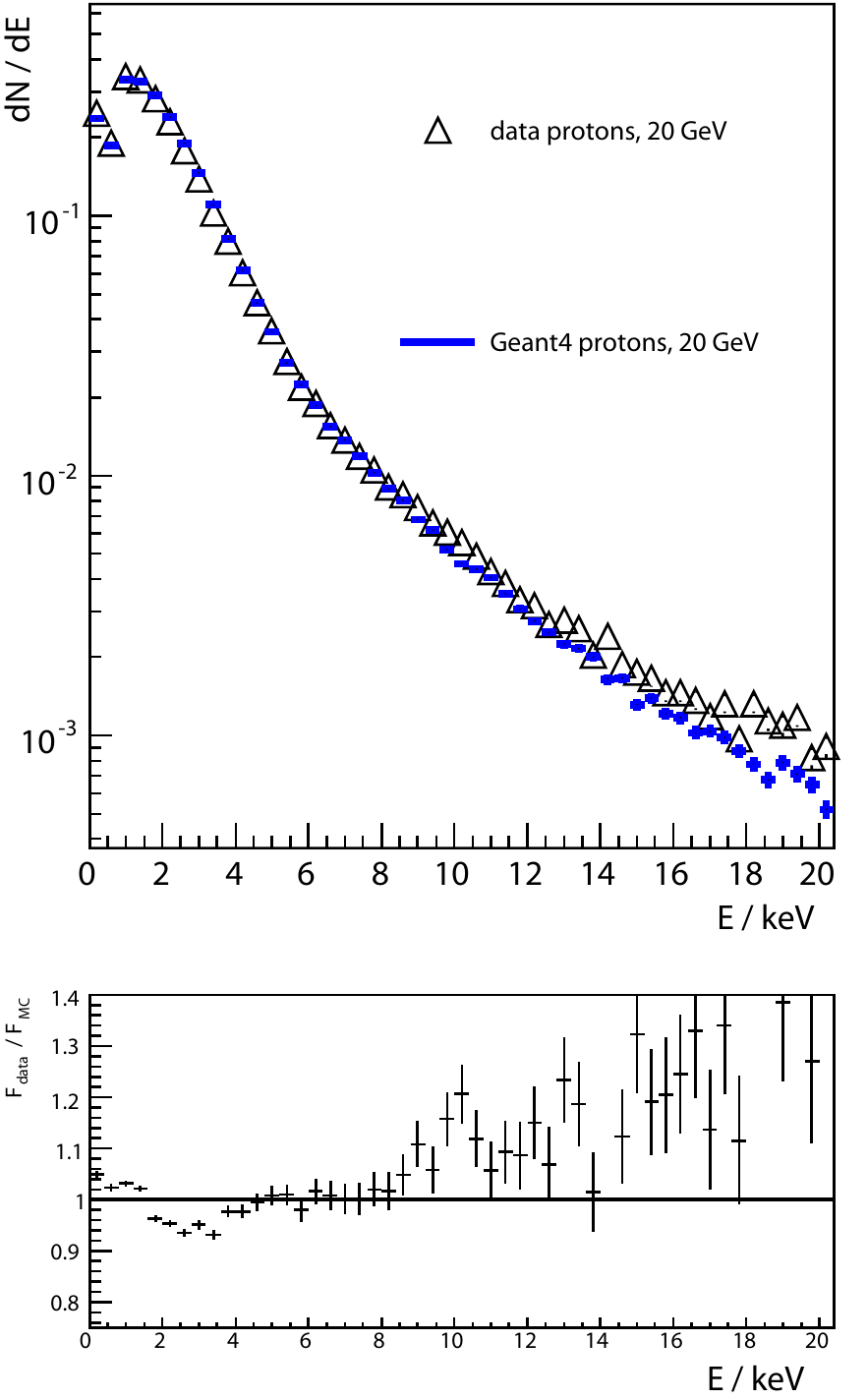}&
\includegraphics[width=0.5\textwidth,angle=0]{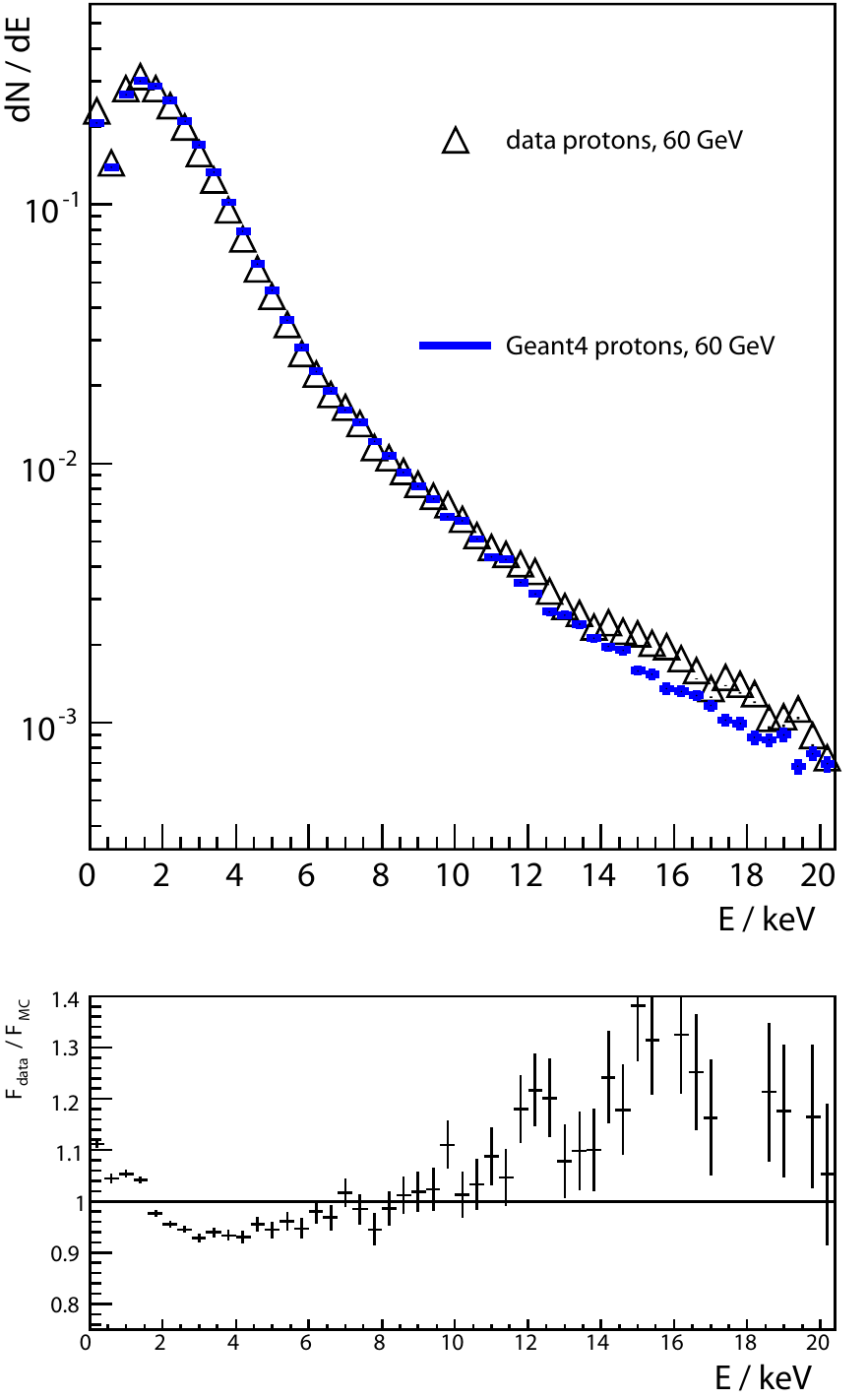}\\
\end{tabular}
\end{center}
\caption{{\it Top:} Ionisation energy loss spectrum from all layers
for $20\,\mathrm{GeV}$ {\it (left)} and $60\,\mathrm{GeV}$ {\it (right)}.
{\it Bottom:} Corresponding ratio of spectrum from testbeam
data to Geant4 spectrum.}
\label{fig:protons20_60}
\end{figure}

\begin{figure}[htb]
\begin{center}
\includegraphics[width=0.5\textwidth,angle=0]{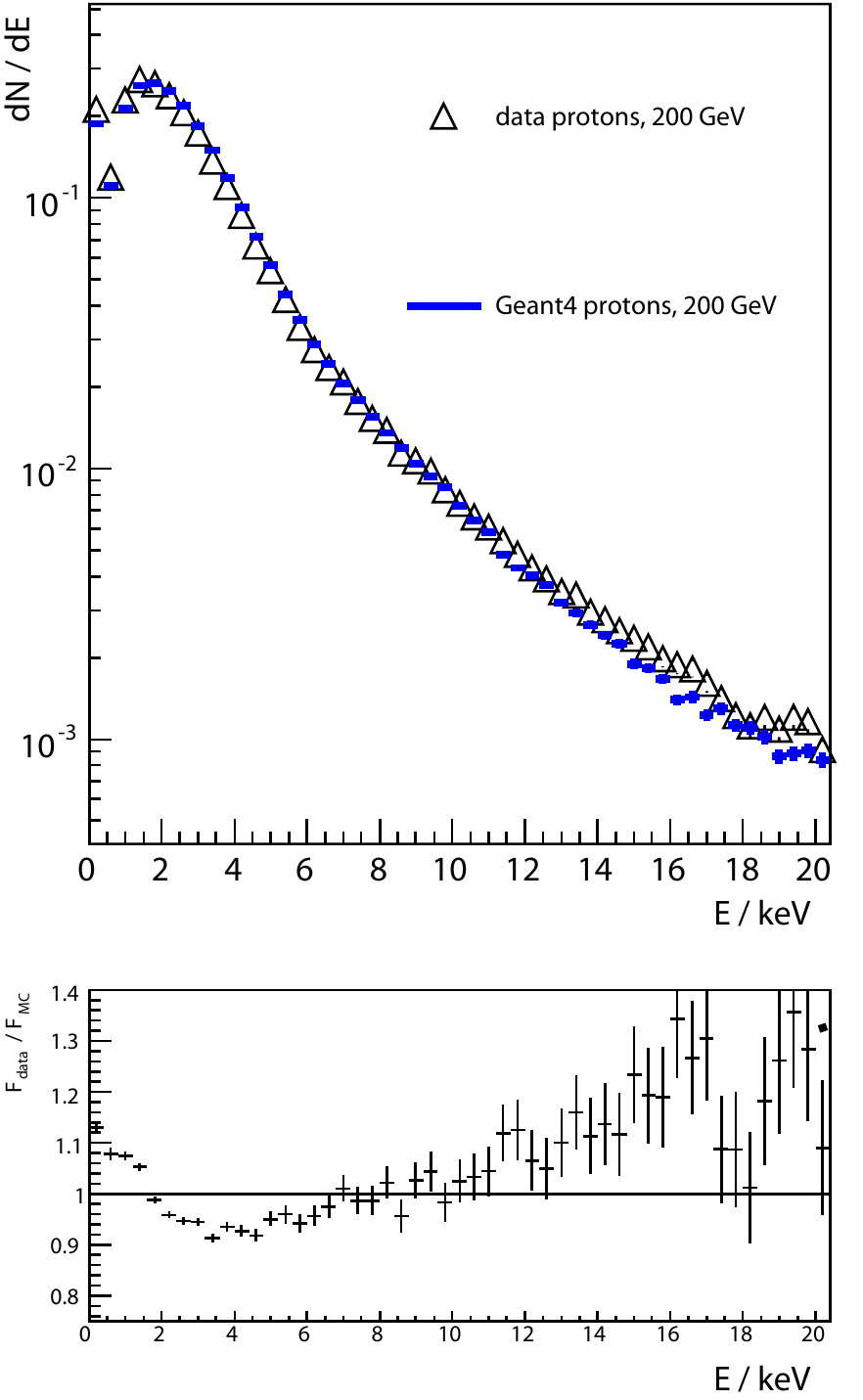}
\end{center}
\caption{{\it Top:} Ionisation energy loss spectrum from all layers for $200\,\mathrm{GeV}$
protons. {\it Bottom:} Corresponding ratio of spectrum from testbeam
data to Geant4 spectrum.}
\label{fig:protons200}
\end{figure}

For studies of the proton rejection power of a TRD, the ionisation
energy loss spectra for protons must be reproduced accurately by the
simulation. Figures~\ref{fig:protons20_60}
and~\ref{fig:protons200} show the spectra for data and
for simulated protons of $20$, $60$ and $200\,\mathrm{GeV}$, respectively. All layers have been included
because the energy deposition is independent of the layer number for
protons. The range cut, employed in Geant4 to determine the production
of secondary particles, was set to $10\,\mathrm{mm}$, larger than the diameter of the straw
tubes. Reducing this value unexpectedly worsens the agreement of
simulation and data. This might point to a problem in the handling of
knock-on electrons in the PAI model employed.
While the overall agreement for lower tube energies is good, the figures
show an underestimation of energy
depositions above roughly $12\,\mathrm{keV}$ in the Geant4 simulation, on the
order of $25\,\%$. 

\subsubsection{Proton rejection study}
\label{sec:rejection}

An important figure of merit of a transition radiation detector is its
proton rejection $R_p$ for a certain electron efficiency
$\epsilon_e$. Here, $R_p\equiv{}1/P(e|p)$ where $P(e|p)$ is the
probability for a proton to be misidentified as an electron, when
satisfying the selection criteria that a fraction $\epsilon_e$ of the
single-track electrons survive.

The likelihood method is a simple way to determine the proton
rejection. From the single-layer energy depositions $(E_k)_{k=1}^{20}$, the
likelihoods for an event to be proton- or electron-like are calculated
as the geometric
means $\bar{P}_{p,e}=\sqrt[20]{\prod_{k=1}^{20} P_{p,e}(E_k)}$,
where the single-layer probabilities $P_{p,e}$ are taken from the
normalised mean respective spectra. The event is then classified as an
electron if $-\log(L)\equiv{}-\log(\bar{P}_e/(\bar{P}_e+\bar{P}_p))$
is below a certain cut.
\begin{figure}[htb]
\begin{center}
\includegraphics[width=0.7\textwidth,angle=0]{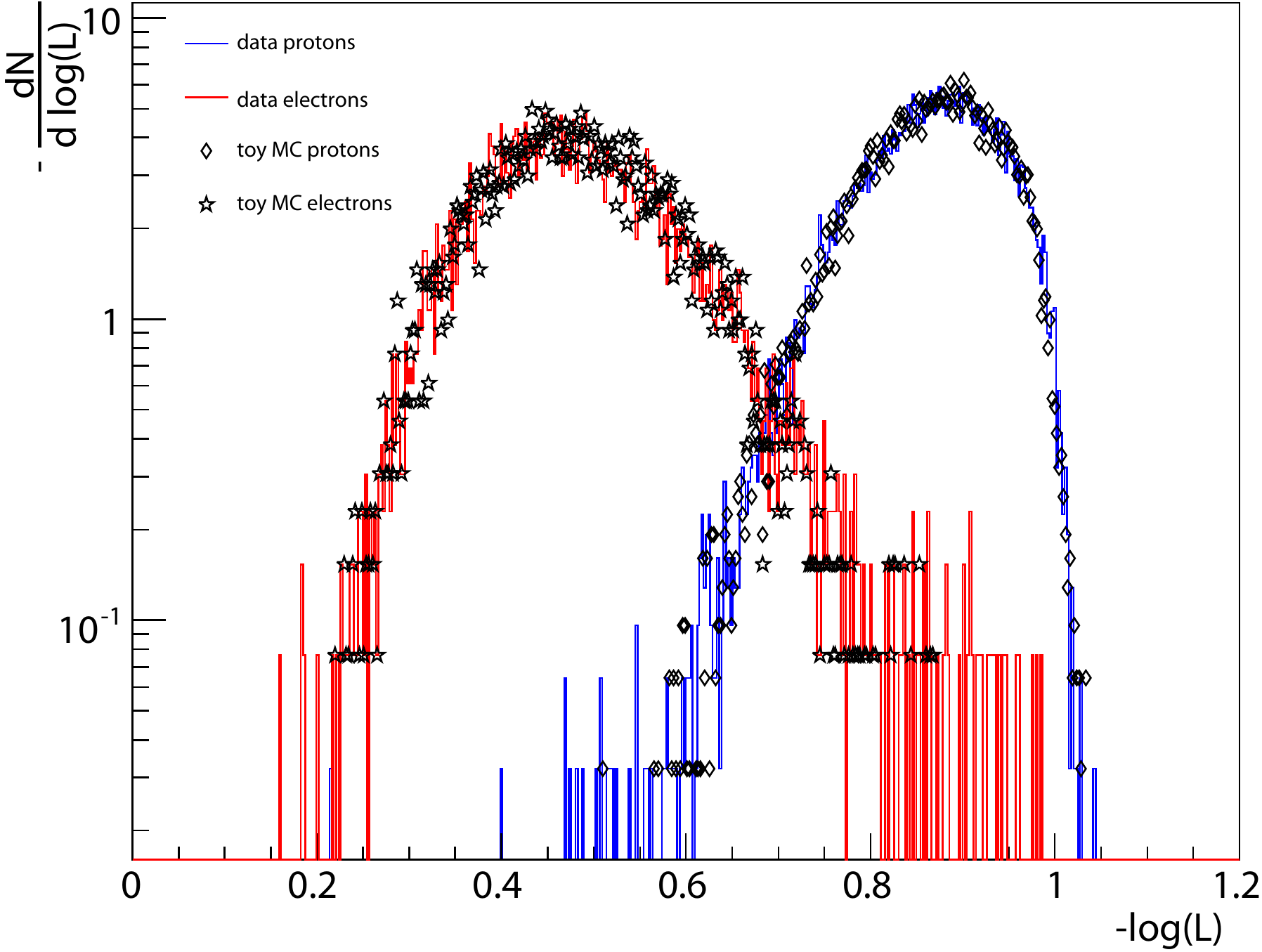}
\end{center}
\caption{Log-Likelihood distributions for $200\,\mathrm{GeV}$ protons and
$20\,\mathrm{GeV}$ electrons. The results of the toy MC study described in the
text are added to estimate the distributions expected for purely
statistical fluctuations of the energy loss.}
\label{fig:likelihoods}
\end{figure}
Protons that are misidentified as electrons by this method can be divided into two
categories. This is illustrated in figure~\ref{fig:likelihoods}
showing as an example the $-\log(L)$-distributions obtained for
testbeam data of $200\,\mathrm{GeV}$ protons
and $20\,\mathrm{GeV}$ electrons. The
first and most important category of misidentified protons comprises protons with several high
energy depositions on their track which are due to the statistical
fluctuations in the energy loss. This effect can be evaluated by a
simple toy Monte Carlo (``toy MC'') study as follows: Energy
depositions $(E_k)_{k=1}^{20}$ are
generated randomly according to
the mean normalised spectra obtained from testbeam data, as shown in
the preceding sections. This procedure is repeated $N$ times.
The $-\log(L)$-values obtained in this way are
also drawn in figure~\ref{fig:likelihoods}. Comparing their
distribution to the data reveals the second category of
misidentified protons: events with a small $-\log(L)$-value ($-\log(L)\lesssim0.55$) that is clearly
in the electron-like region. Such events are extremely unlikely to arise
from statistical fluctuations. Instead, they indicate the presence of
particles other than protons, e.g. secondaries arising from interactions in the
material traversed or from beam contamination. For example, an incident proton can transfer a high
amount of energy to an atomic electron which in turn flies almost
parallel to the path of the primary. This leads to an event with high
energy depositions in almost every layer and subsequently an electron-like
value of $-\log(L)$. The electron contamination of the proton beam depends strongly on
energy but remains below the order of $10\,\%$~\cite{ref:beamx7}. This
number was further reduced by using Cherenkov counters to veto on
fast particles like electrons in the testbeam. Their inefficiency has
been estimated to be $1.15$ \textperthousand~$\,$~\cite{ref:orboeck}.
Therefore, the contamination of the proton
sample will only play a minor role. \newline
On the other hand, there is a number of electron
events in the $-\log(L)$-region characteristic of protons ($-\log(L)\gtrsim0.9$) that is not
accounted for by statistical fluctuations in the energy deposition. As
the corresponding energy depositions along the track are {\it lower}
than expected in this case, it is clear that this component cannot be
due to additional particles. Instead, it must be caused by contamination with 
relativistic particles with low TR yield, such as protons, muons, pions or kaons. From
fig.~\ref{fig:likelihoods} this contamination is estimated to be of
the order of a few \textperthousand.

In the context of this thesis, it is interesting to quantify the
effect of the observed slight deviations in the tails of the Geant4 proton
energy loss distributions on calculated proton rejections. To that
end, two toy MC studies have been conducted as described above, one
using the spectra obtained in the testbeam, the other using the Geant4
spectra. This was repeated for various incident proton energies. Spectra for
$20\,\mathrm{GeV}$ electrons, where the TR yield is in saturation, were used as
\begin{figure}[htb]
\begin{center}
\begin{tabular}{cc}
\includegraphics[width=0.5\textwidth,angle=0]{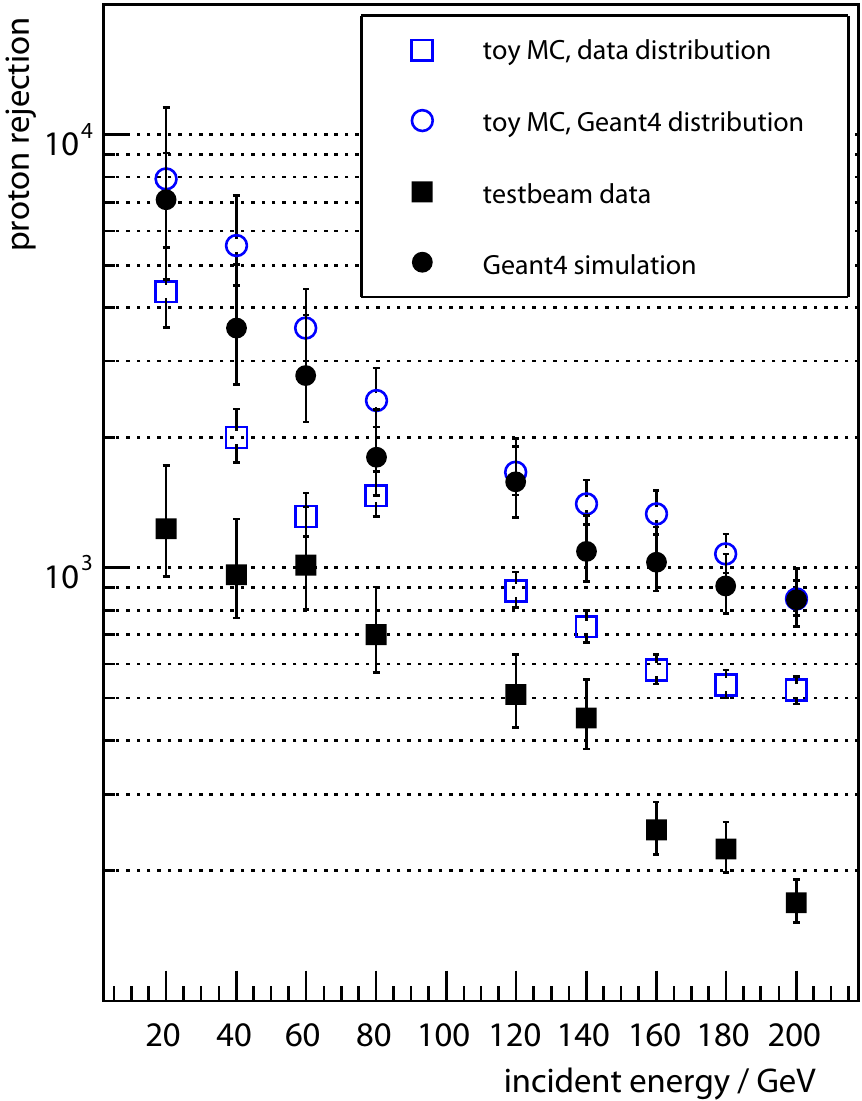}&
\includegraphics[width=0.48\textwidth,angle=0]{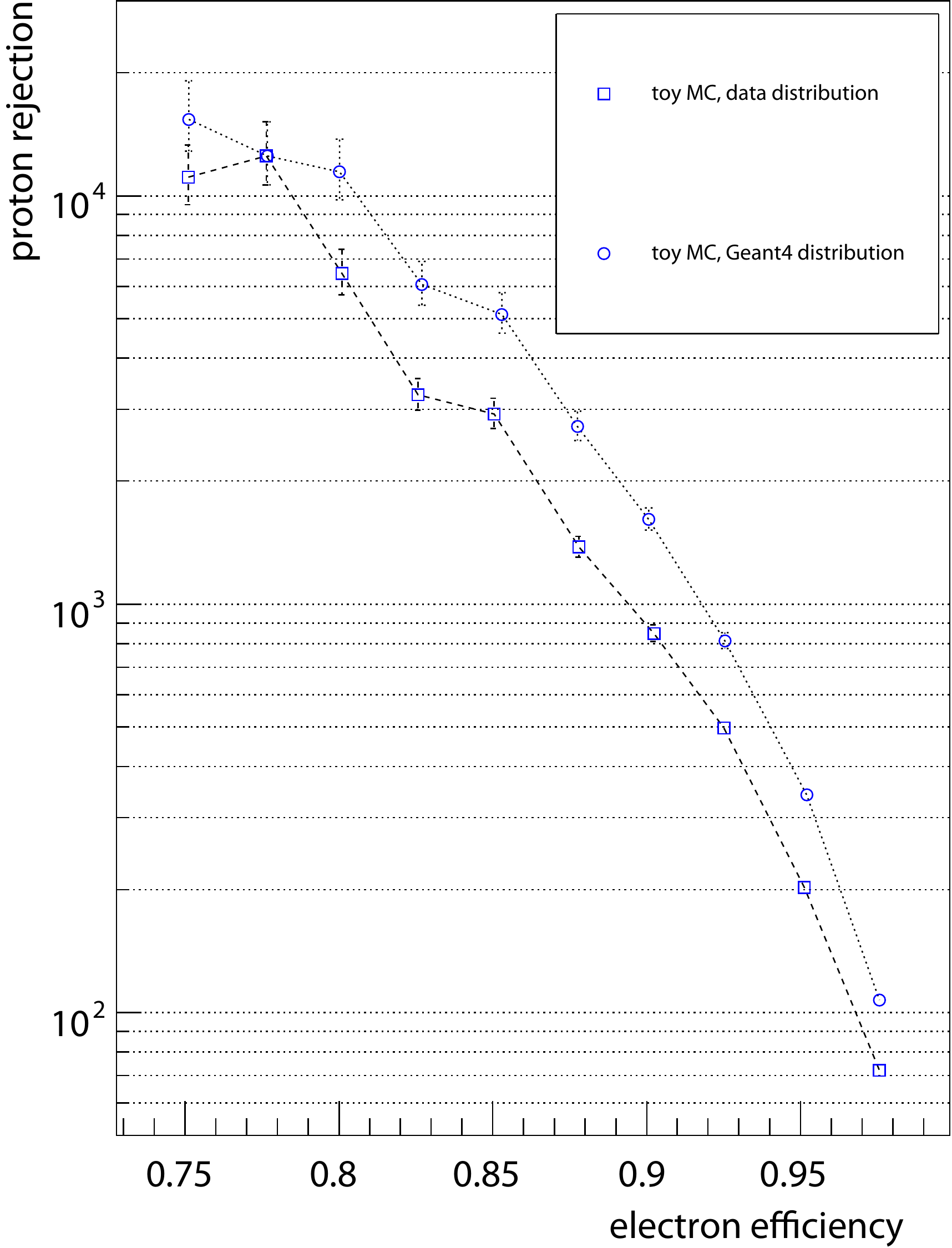}\\
\end{tabular}
\end{center}
\caption{{\it Left:} Proton rejection for various energies of the incident
protons, at a corresponding electron efficiency of $90\,\%$. The
results of the toy MC study are used to evaluate the
effect of the observed slight deviations in the tails of the Geant4 proton
energy loss distributions with respect to the testbeam data, reducing
the sensitivity to beam contaminations and other systematic effects as
far as possible. The
proton rejections obtained when taking the
actual $E_k$ values from testbeam data or simulated events are shown,
too.
{\it Right:} Proton rejection as a function of electron efficiency, for
incident protons of $120\,\mathrm{GeV}$, in the context of the toy MC study
described in the text, for the distributions obtained from testbeam
data and Geant4 simulation.}
\label{fig:rejection_energy_efficiency}
\end{figure}
reference. The resulting proton rejections are depicted in the
left-hand side of
figure~\ref{fig:rejection_energy_efficiency}, each for an electron efficiency of
$90\,\%$. The rejections obtained in the two cases differ by roughly a
factor of two. However, the right-hand side of the figure
demonstrates the very steep dependence of the proton rejection on the
corresponding electron efficiency. The observed difference in
rejection translates into a difference in electron efficiency of
$\sim\!2.5\,\%$ at $\epsilon_e\sim{}0.9$. These curves offer a more complete
picture of the discrepancy between testbeam data and Geant4 proton
spectra than the rejection value at a fixed electron efficiency.

Figure~\ref{fig:rejection_energy_efficiency} also shows the
proton rejections obtained when taking the
actual $E_k$ values from testbeam data or simulated events. They are
lower than the rejections calculated for the respective toy MCs. This
is due to a contribution from events belonging to the second category
introduced above. The difference is larger in the case of the data
than the Geant4 simulation. While this might be due to inaccurate
cross sections in the simulation, the beam contaminations mentioned
above prevent us from drawing such a conclusion.\\
To summarise, while slight deviations in the tails towards higher
energy depositions in the proton spectra prevent us from accurately
predicting the behaviour of the TRD with respect to proton
suppression, a good picture of the overall performance can be
obtained. The parameters describing the microscopic structure of the
fleece radiator that were found to give the best description of the
electron spectra in this section are used throughout this design study.

\section{Event reconstruction}
\label{sec:reconstruction}
The events produced by the simulation only contain raw information,
such as amplitude and position, for the hit channels in the various
subdetectors. In addition, information about the primary and secondary
tracks exceeding a momentum of $100\,\mathrm{MeV}$ and created above
the ECAL, is included. Before one can use these events to study the
behaviour of the detector, higher-level objects have to be created
that represent pictures of the particles as seen by the
\begin{figure}[htb]
\begin{center}
\begin{tabular}{cc}
\includegraphics[width=0.5\textwidth,angle=0]{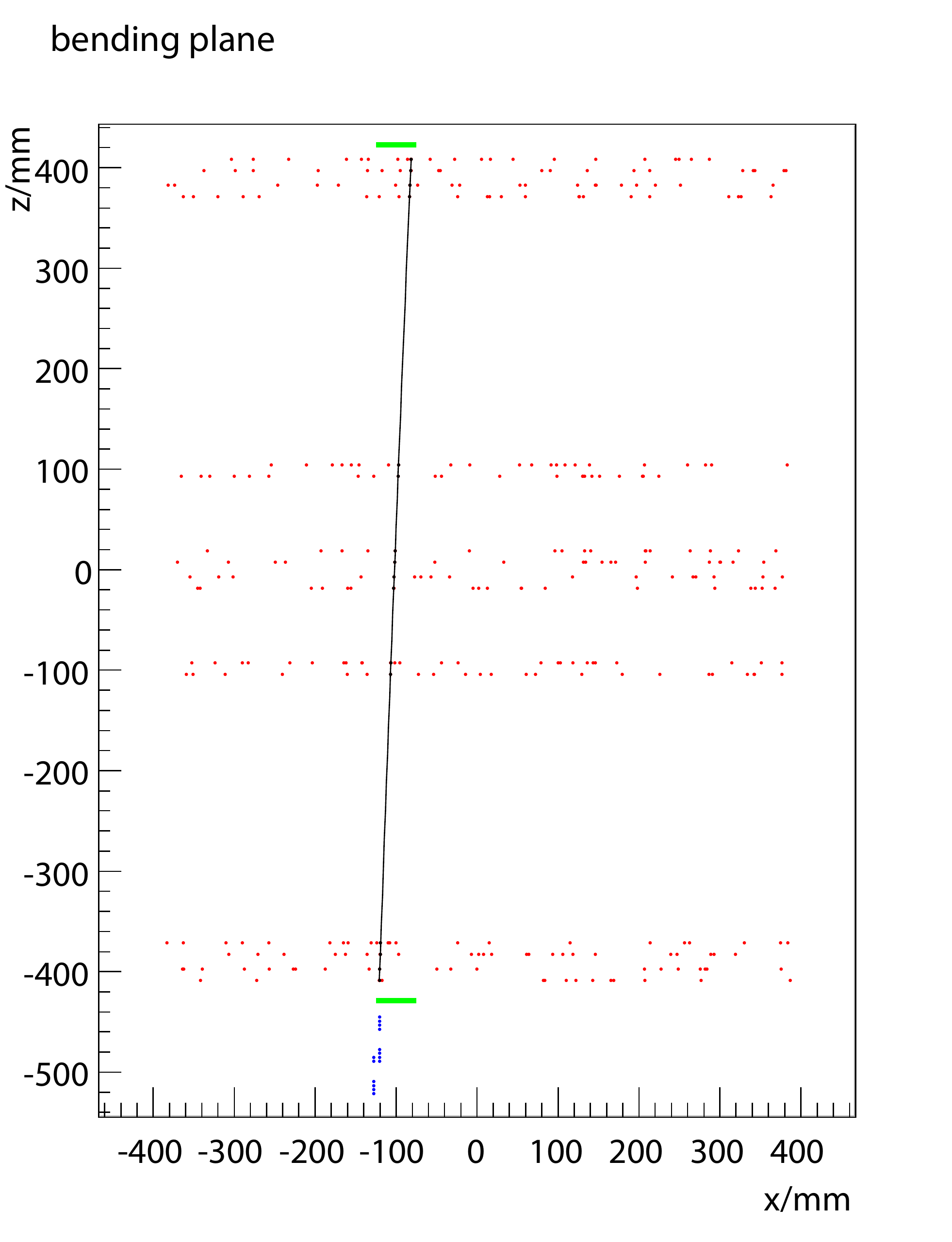}&
\includegraphics[width=0.5\textwidth,angle=0]{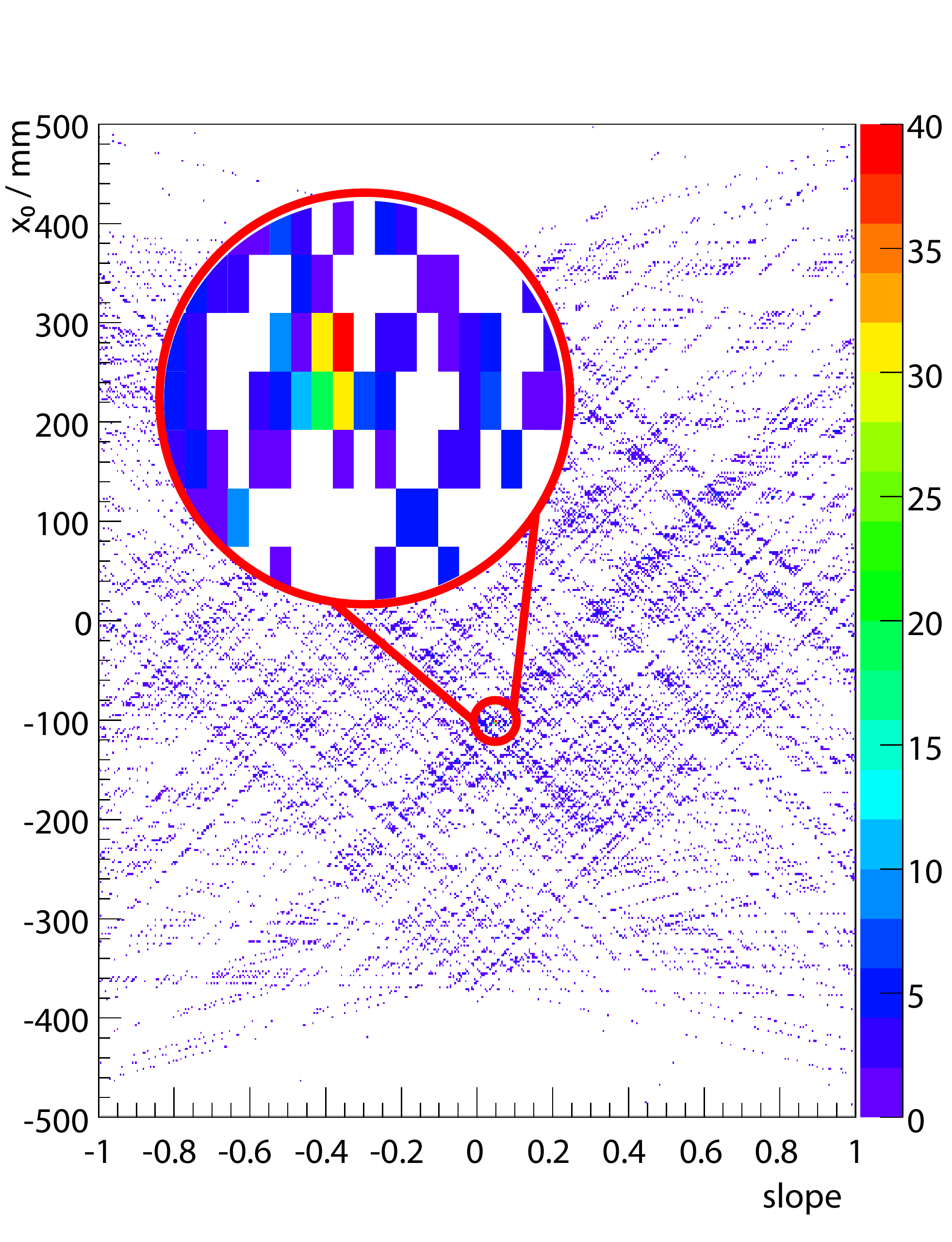}\\
\end{tabular}
\end{center}
\caption{{\it Left:} Example of a $20\,\mathrm{GeV}$ proton track, for
an average noise level of $0.24$ photons in the tracker. Tracker hits
are marked in red, ECAL hits in blue, and TOF hits in green.
{\it Right:} Histogram used for track finding. The clearly discernible
maximum in the $x_0$-slope-plane corresponds to the track corridor.}
\label{fig:track_finding}
\end{figure}
subdetectors. For the tracker and the TRD, tracks have to be found and
their curvature and orientation have to be determined to yield the
particle's momentum. In the calorimeter, a shower has to be identified
and the shower orientation and parameters can then be retrieved.
In general, one is interested in clean single-track events, which offer the best
prospects for unambiguous particle identification.\\
\par
In a first step, a track in the TRD is identified as follows. First, a
seed track is created from the digis on those layers with exactly one
digi. This seed track is then interpolated and additional digis within
one tube diameter from the seed track are added. This simple algorithm
works because the noise level in the TRD is expected to be low.\\
\par
In the second step, clusters of neighbouring SiPM channels are identified in
the tracker and for the subsequent track finding and fitting, the
position is calculated from a weighted mean of the amplitudes $A_i$:
\begin{equation}
  \label{eq:weightedmean}
  \vec{x}=\frac{\sum_iA_i\vec{x}_i}{\sum_iA_i}
\end{equation}
The SiPMs to be employed in the tracker are noisy to a certain
extent. Therefore, a track finding algorithm had to be developed
to reliably find tracks even in the presence of noise hits. The idea
for the approach adopted here is inspired by the Hough
transform~\cite{ref:hough}, and the algorithm proceeds as follows:
\begin{itemize}
\item For each {\it pair} of tracker clusters, calculate the intercept
  and slope for the straight line defined unambiguously by the two
  clusters, and fill the resulting values into a two-dimensional
  histogram. The idea here is that a track can be approximated by a
  straight line to first order. This means that all pairs of clusters
  lying on the track will give the same slope and intercept. On the
  other hand, a pair of clusters containing a noise cluster will only
  rarely lead to a given value of slope and intercept.
\item Find the maximum bin in the histogram obtained in this way and
  calculate the corresponding straight line. Loop the tracker layers
  adding the cluster nearest to the straight line, provided the
  distance does not exceed a maximum value. An example of the
  histogram in the intercept-slope-plane obtained for an event simulated
  with a mean noise level of $0.24$ photons is shown in
  figure~\ref{fig:track_finding}. 
\item Find outliers in the candidate track. This is done both by using
  a robust circle fit provided by Blobel~\cite{ref:blobel_robustfit}
  to identify outliers, and by looking for clusters whose removal from
  the candidate track leads to a significantly reduced $\chi^2$ in a
  circle fit.
\item Because a $\chi^2$ fit prefers a low number of clusters on the
  track, a single noise cluster will sometimes be chosen instead of a
  segment of the real track on the outer tracker superlayers. This
  situation is remedied by looking for a segment of clusters on the
  uppermost and lowermost layers, respectively, with
  similar slope as the candidate track's. The single cluster is
  replaced by the clusters of the segment so obtained if the
  inclination of each pair that can be formed from its clusters
  matches the one of the candidate track.
\item A final track fit is then performed. A circle fit is done for
  the tracker clusters in the bending plane to obtain the curvature, while a line fit is
  performed for the TRD hits to get the inclination angle
  $\lambda$. The rigidity is then calculated according to
  eq.~(\ref{eq:rigidity}).
\end{itemize}
\begin{figure}[htb]
\begin{center}
\includegraphics[width=0.5\textwidth,angle=0]{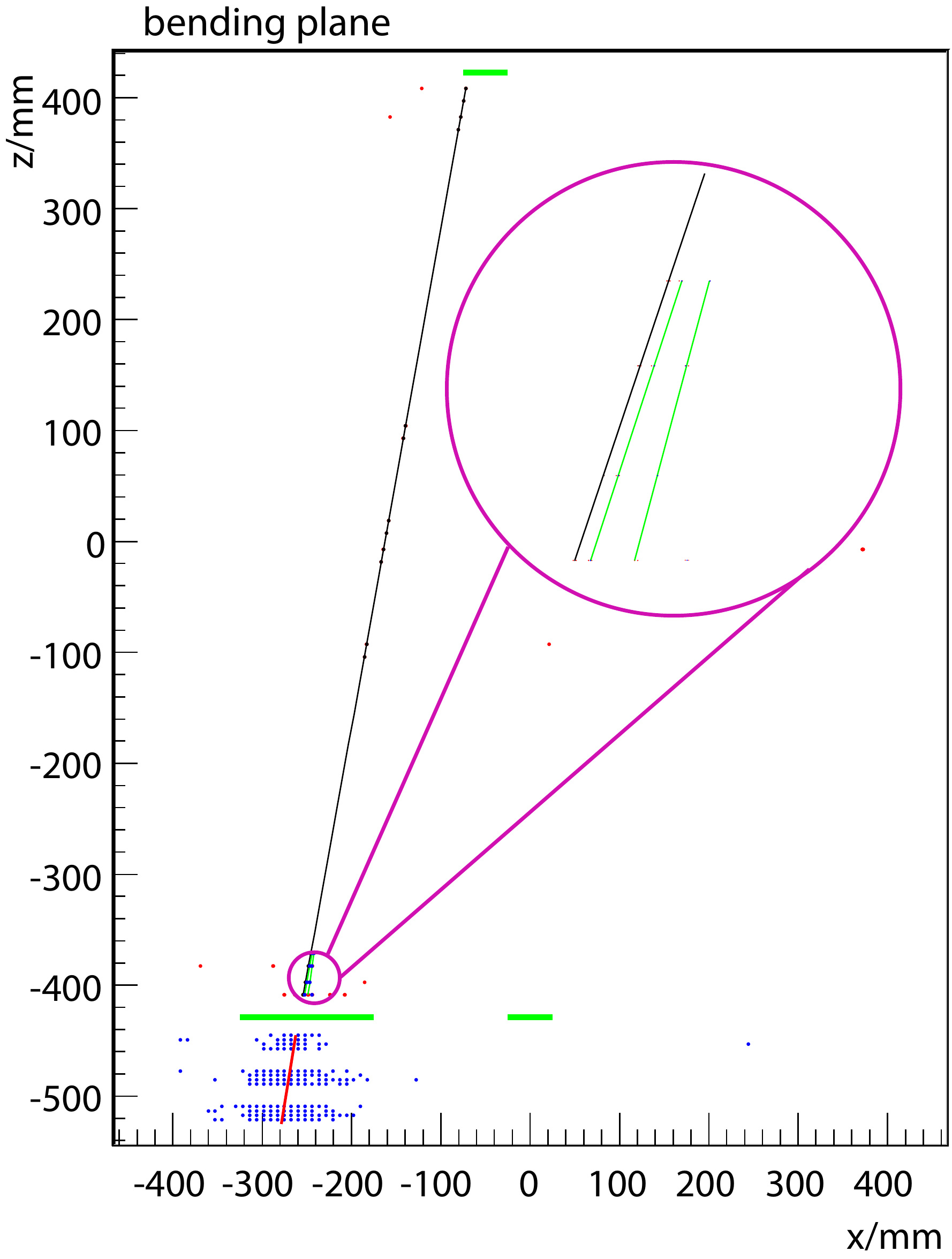}
\end{center}
\caption{Example of an event with additional track segments caused by
secondaries created by a hard bremsstrahlung photon. The event
information is drawn in the same way as in
fig.~\ref{fig:track_finding}. Here, a $100\,\mathrm{GeV}$ positron
radiates a photon of $19.9\,\mathrm{GeV}$ inside the upper TOF system
and the photon in turn converts to an $e^+$-$e^-$-pair inside the
lower TRD. The resulting three-prong signature on the lowest tracker
layers is
magnified. Additional track segments identified by the algorithm
described in the text are shown in green. The primary positron was
misreconstructed as having a momentum of $-45.9\,\mathrm{GeV}$ in this
event.}
\label{fig:examplesegment}
\end{figure}
It turns out that a good determination of the charge sign of a
particle (sec.~\ref{sec:momres}) can only be achieved if care
is taken to identify events in which an incoming electron or positron
radiates a hard bremsstrahlung photon which subsequently undergoes
pair creation inside or above the tracker volume. Typically,
additional track segments close to the primary track will be present in the lower quarter of the
tracker in this case. If the track finding algorithm chooses
clusters from a segment belonging to one of the secondaries, the
ensuing determination of the momentum will be incorrect
(fig.~\ref{fig:examplesegment}).
The algorithm
for finding additional track segments is similar to the one used for
track finding. After the clusters belonging to the primary track have
been removed from the sample, line segments on the lowest four tracker
layers are searched in the vicinity of the track, by looking for peaks
in a slope-intercept-histogram calculated from each pair of clusters.\\
\par
\begin{figure}[htb]
\begin{center}
\includegraphics[width=0.7\textwidth,angle=0]{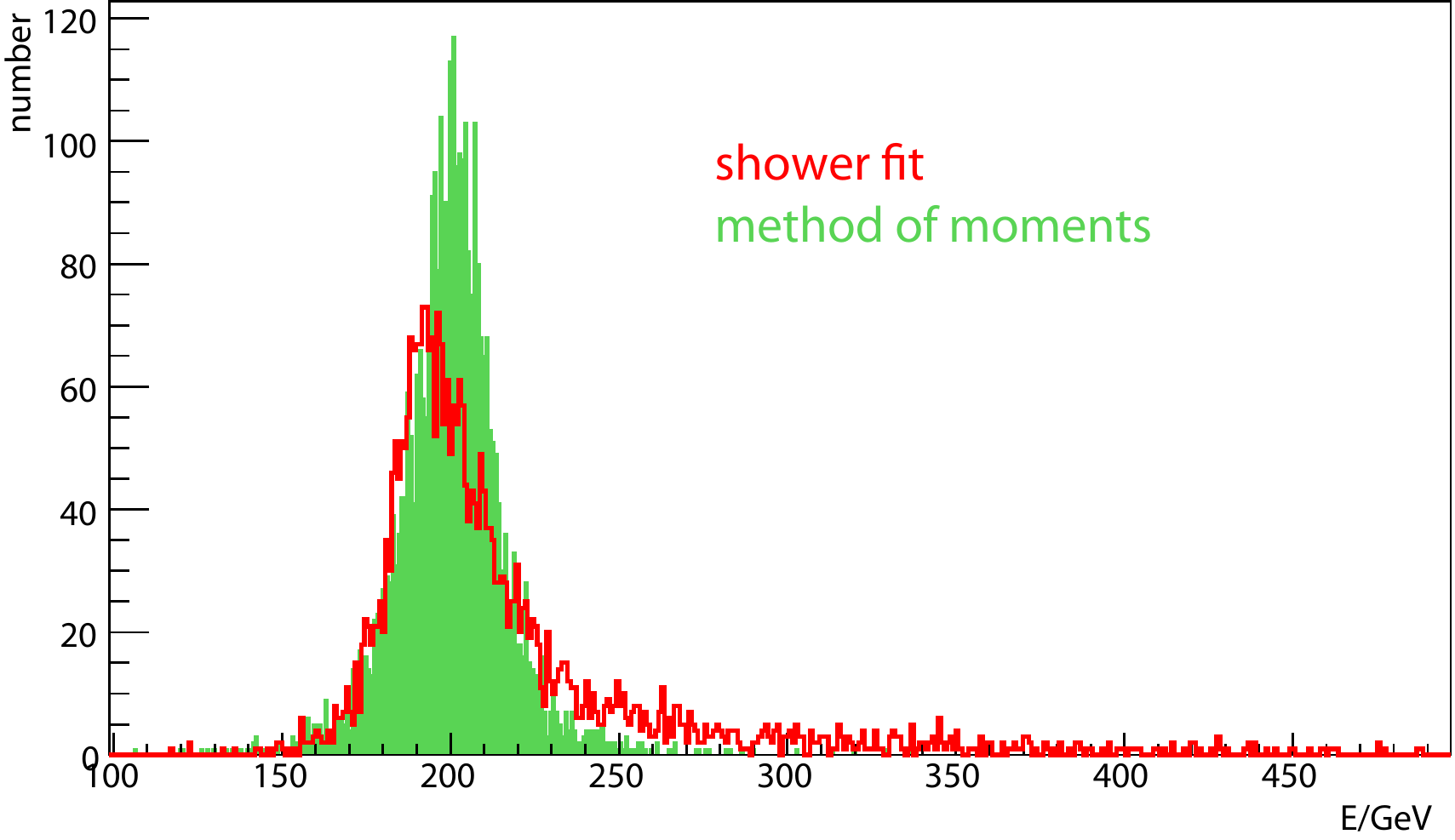}
\end{center}
\caption{Comparison of reconstructed ECAL energies for
$200\,\mathrm{GeV}$ positrons, calculated by the method of moments and
obtained from a fit of the shower profile (\ref{eq:showershape}).}
\label{fig:ecalcomparison}
\end{figure}

In the third step, a possible shower in the electromagnetic
calorimeter is identified. First, the shower axis is reconstructed. To
this end, the total amplitude in each ECAL
layer is calculated. Then, two robust straight line fits with equal
weights are performed separately for the $xz$- and $yz$-projections of the shower
to determine the shower axis.
The vertical shower profile
$\mathrm{d}E/\mathrm{d}z$ is then calculated to determine the shower
shape and energy variables as follows.\\
Starting from (\ref{eq:showershape}) and for perpendicular incidence, the energy deposition
$\Delta{}E(z)$ in a layer at position $z$ is given by
\begin{equation}
\label{eq:showershapez}
\Delta{}E(z)=E_0\frac{(bz)^{a-1}\mathrm{e}^{-bz}}{\Gamma(a)}\,b\,\Delta{}z
\end{equation}
The goal of the following is to extract the shower energy $E_0$ and
the shower shape parameters $a$ and $b$. The scale of
the shower development is given by $b$, and $a$ then determines the position of the
shower maximum (eq.~(\ref{eq:showermax})).\\
For an ideal calorimeter of infinite length, $E_0$ is given by
\[
\int\limits_0^\infty\frac{\Delta{}E}{\Delta{}z}(z)\,\mathrm{d}z=E_0
\]
For the PEBS calorimeter with its limited depth $z_m$, leakage effects have
to be taken into account. In fact,
\begin{equation}
\label{eq:leakagecorr}
\frac{1}{E_0}\int\limits_0^{z_m}\frac{\Delta{}E}{\Delta{}z}(z)\,\mathrm{d}z=\int\limits_0^{bz_m}\frac{x^{a-1}\mathrm{e}^{-x}}{\Gamma(a)}\,\mathrm{d}x=\gamma_p(a,bz_m)
\end{equation}
Here, $\gamma_p$ denotes the incomplete Gamma function. It can be
calculated numerically and is included in standard computer libraries
such as the GNU scientific library~\cite{ref:gsl}.
From (\ref{eq:leakagecorr}), $E_0$ can be
calculated if $a$ and $b$ are known. For their determination, the
first two moments $m_{1,2}$ of the vertical shower distribution are
calculated,
\begin{equation}
\label{eq:moments}
m_i=\frac{\int_0^{z_m}z^i\frac{\Delta{}E}{\Delta{}z}(z)\,\mathrm{d}z}{\int_0^{z_m}\frac{\Delta{}E}{\Delta{}z}(z)\,\mathrm{d}z}
\end{equation}
The calculation gives
\begin{equation}
\label{eq:momentsresults}
m_1=\frac{a}{b}\,\frac{\gamma_p(a+1,bz_m)}{\gamma_p(a,bz_m)}\quad\mathrm{and}\quad{}m_2=\frac{a(a+1)}{b^2}\,\frac{\gamma_p(a+2,bz_m)}{\gamma_p(a,bz_m)}
\end{equation}
For each event, the moments $m_{1,2}$ are calculated using the
definition (\ref{eq:moments}) and $a$ and $b$ can then be extracted
using an iterative solution of (\ref{eq:momentsresults}), starting
from the zeroth-order equations
\begin{equation}
\label{eq:iterzero}
m_1=\frac{a}{b}\;\wedge\;m_2=\frac{(a+1)a}{b^2}\quad\Rightarrow\quad{}b=\frac{m_1}{m_2-m_1^2}\;\wedge\;a=bm_1
\end{equation}
Using these starting values for $a$ and $b$, the incomplete Gamma
functions in (\ref{eq:momentsresults}) can be calculated and better
values of $a$ and $b$ can be obtained. The iteration is continued
until convergence is reached.\\
In order to compare showers of different angles of inclination, the
shower development is better described using the radiation length as the
scale. From inspection of eq.~(\ref{eq:showershape}), this can simply
be achieved by introducing a new scale parameter $\hat{b}$ and
requiring scale invariance
\begin{equation}
\label{eq:ecal_si}
\hat{b}t=bz
\end{equation}
Introducing the effective radiation length per absorber layer
$w\equiv\Delta{}t/\Delta{}z=(d/X_0)/\cos\lambda$,
where $d$ is the thickness of an absorber layer, one gets
\begin{equation}
\label{eq:ecal_bhat}
\hat{b}=\frac{b}{w}=\frac{b}{d/X_0}\,\cos\lambda
\end{equation}
$a$ and $\hat{b}$ are stored for further analysis. It turns out that
the determination of the shower parameters using this method of
moments is more stable and reliable than a fit of the longitudinal
shower profile (fig.~\ref{fig:ecalcomparison}). To give an example,
the energy resolution for $200\,\mathrm{GeV}$ positrons is found to be
$6.5\,\%$ for the method of moments, but only $8.3\,\%$ for the shower
fit. The method of moments has the additional advantage that no statistical
errors of the individual energy depositions are required as for a fit
based on $\chi^2$ minimisation.

\section{Projected performance}
\label{sec:performance}
This section deals with the performance of the tracker,
electromagnetic calorimeter, and transition radiation detector, as
predicted by the Geant4 Monte Carlo simulation described above. The
results presented here were obtained using a dedicated analysis suite
which works with the reconstructed objects that were found as outlined
in section~\ref{sec:reconstruction}.

\subsection{Momentum resolution}
\label{sec:momres}
\begin{figure}[htb]
\begin{center}
\begin{tabular}{cc}
\includegraphics[width=0.5\textwidth,angle=0]{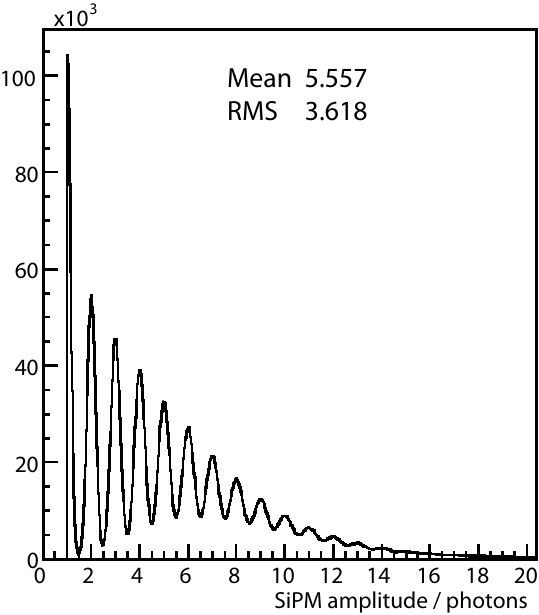}&
\includegraphics[width=0.5\textwidth,angle=0]{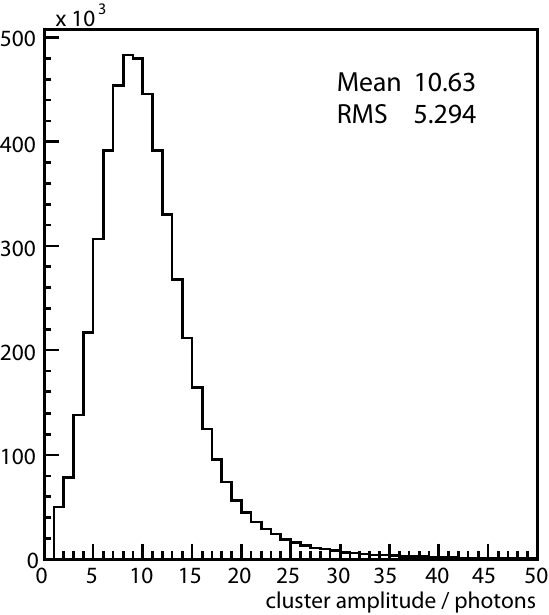}\\
\end{tabular}
\end{center}
\caption{{\it Left:} Distribution of amplitudes (in photons) on all
SiPM channels belonging to reconstructed proton tracks.
{\it Right:} Distribution of cluster amplitudes (in photons) for
reconstructed proton tracks.}
\label{fig:tracker_amps}
\end{figure}
\begin{figure}[htb]
\begin{center}
\begin{tabular}{cc}
\includegraphics[width=0.46\textwidth,angle=0]{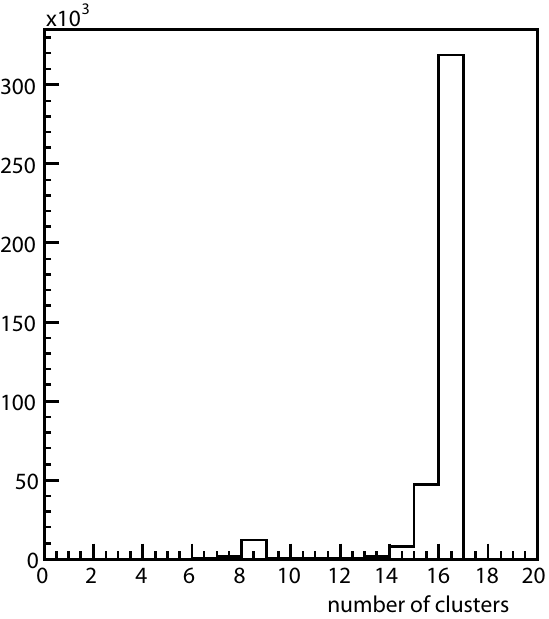}&
\includegraphics[width=0.5\textwidth,angle=0]{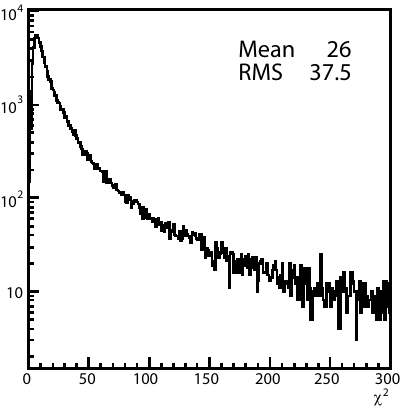}\\
\end{tabular}
\end{center}
\caption{{\it Left:} Distribution of number of cluster in tracker track. The
number of layers is~16. Occasionally, tracks with clusters located
only on the eight central tracker layers
are found for low-energetic events ($\lesssim{}3\,\mathrm{GeV}$)
because the track-finding algorithm is optimised for high-momentum
tracks. {\it Right:} Distribution of $\chi^2$-values for track fits, for positrons
of all energies. A quality cut is done at $\chi^2_\mathrm{cut}=150$.}
\label{fig:nclusters_tracker}
\label{fig:trackchi2}
\end{figure}
\begin{figure}[htb]
\begin{center}
\begin{tabular}{cc}
\includegraphics[width=0.5\textwidth,angle=0]{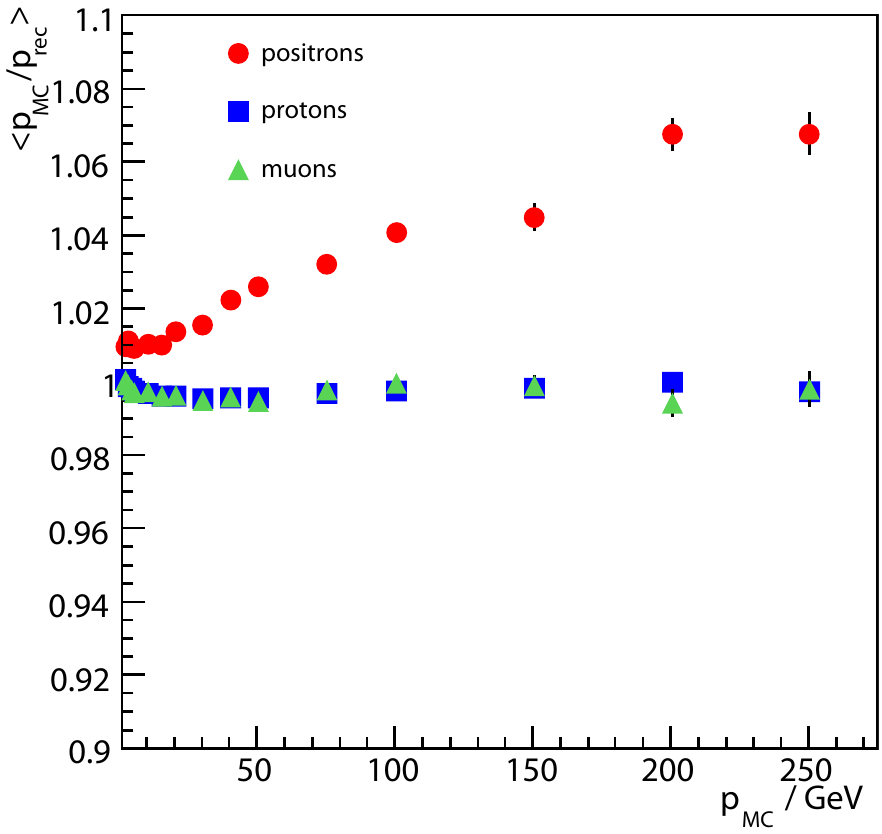}&
\includegraphics[width=0.5\textwidth,angle=0]{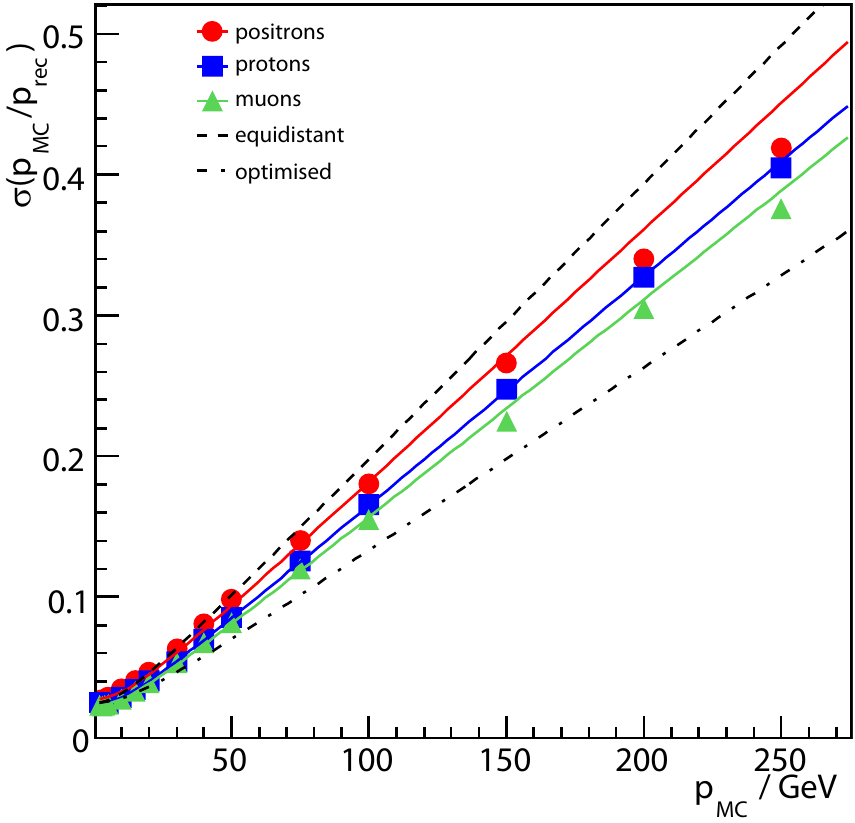}\\
\end{tabular}
\end{center}
\caption{Tracker momentum resolution as a function of incident momentum
for positrons, protons, and muons. Mean value of
$p_\mathrm{MC}/p_\mathrm{rec}$ ({\it left}) and
$\sigma(p_\mathrm{MC}/p_\mathrm{rec})$ of an individual measurement ({\it right}), together with
fits according to equation (\ref{eq:momres}) and the expectations for
the case of the equidistant and optimised geometry according to
equations (\ref{eq:res_uniform}) and (\ref{eq:res_opt}), respectively.}
\label{fig:momres}
\end{figure}

The key figure of merit for the tracker is its momentum
resolution. The momentum resolution is directly related to the error
in the curvature measurement by
equations~(\ref{eq:rigidity}) and~(\ref{eq:curverror}). 
The results in this section were obtained using a noise level of zero
for the tracker. 
The effective coupling efficiency (eq.~(\ref{eq:effcoup})) was adjusted
such that the mean cluster amplitude is close to 10~photons, in
accordance with the preliminary testbeam results found in
section~\ref{sec:testbeam2008}. This is illustrated in
figure~\ref{fig:tracker_amps} showing both the amplitude distribution
of the SiPM array channels individually and the summed cluster
amplitudes in the tracker. In the former, the auto-calibrating
capability of SiPMs is apparent from the individually resolved first
photo-peaks. The Gaussian smearing of the photo-peaks mentioned in
section~\ref{sec:sipms} is visible, too. In the latter, the mean cluster
amplitude of roughly 10~photons is demonstrated.\\
A number of track quality cuts was applied before considering an event
for the analysis of the momentum resolution. A minimum number of eleven
clusters (figure~\ref{fig:nclusters_tracker}) and a minimum track length of $750\,\mathrm{mm}$ were
required for the track to ensure a sufficient lever arm for the
momentum measurement. Events containing additional track segments in the
lower quarter of the tracker, possibly caused by secondaries, were
discarded. Furthermore, tracks with $\chi^2>150$ for the
track fit were discarded. The $\chi^2$ distribution is shown in
figure~\ref{fig:trackchi2} before the cut was applied. A high value of
the $\chi^2$ hints at the inclusion of a spurious hit in the track
which usually leads to a wrong momentum determination.\\
For the determination of the momentum resolution, the quantity
$p_\mathrm{MC}/p_\mathrm{rec}$ was then histogrammed versus
$p_\mathrm{MC}$ for simulated events. Here, $p_\mathrm{MC}$ is the
generated momentum of a particle and $p_\mathrm{rec}$ is its
reconstructed momentum. $p_\mathrm{MC}/p_\mathrm{rec}$ is the
appropriate quantity because the distribution of the curvature is
approximately Gaussian. For each bin in $p_\mathrm{MC}$, the
resulting distribution is then fit by a Gaussian to obtain the mean
and standard deviation. These two values are shown in
figure~\ref{fig:momres} as a function of $p_\mathrm{MC}$ and
separately for three particle species, positrons, protons, and muons.
The mean $\langle{}p_\mathrm{MC}/p_\mathrm{rec}\rangle$ is close to one for
muons and protons while it deviates from one by a few percent in the
case of positrons showing that $p_\mathrm{rec}$ serves as a good,
unbiased estimator for the true momentum. The momentum resolution
$\sigma(p_\mathrm{MC}/p_\mathrm{rec})$ is slightly different for the
three species. Muons have the cleanest signature and for all practical
purposes do not interact inside the tracker. In general, protons also
produce clean signatures but occasional interactions lead to a small
deterioration of the momentum resolution. For positrons, both
$\langle{}p_\mathrm{MC}/p_\mathrm{rec}\rangle$ and
$\sigma(p_\mathrm{MC}/p_\mathrm{rec})$ suffer from bremsstrahlung
losses inside the tracker volume. This effect is illustrated in
\begin{figure}[htb]
\begin{center}
\begin{tabular}{cc}
\includegraphics[width=0.5\textwidth,angle=0]{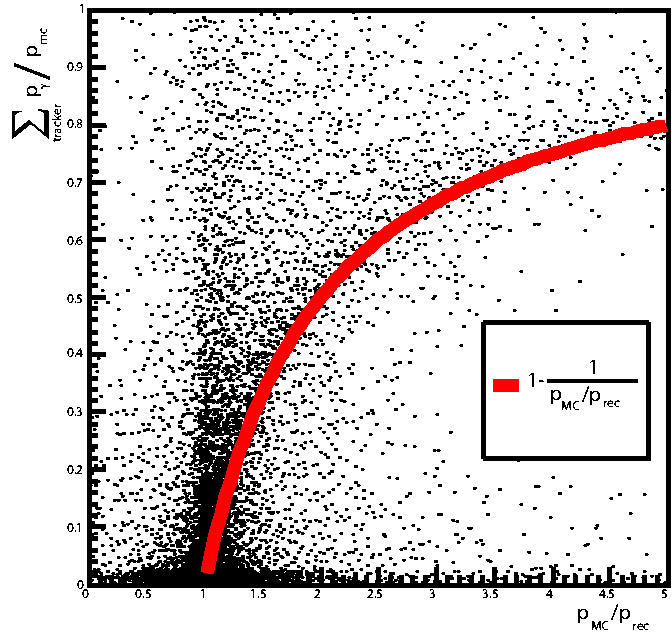}&
\includegraphics[width=0.5\textwidth,angle=0]{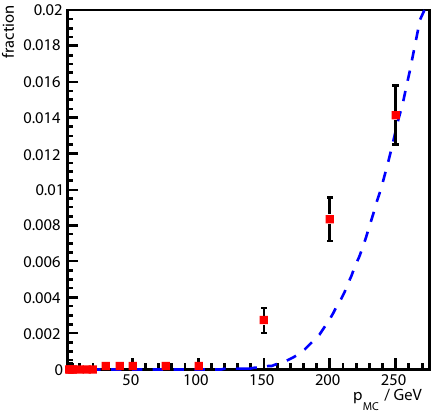}\\
\end{tabular}
\end{center}
\caption{{\it Left:} Relative amount of bremsstrahlung losses of
incident positrons in the tracker as a function of
$p_\mathrm{MC}/p_\mathrm{rec}$. The expected curve
(eq.~(\ref{eq:bremsstrahlungloss})) for the extreme
case, where the error in the reconstructed momentum equals the
bremsstrahlung loss, is included, too.
{\it Right:} Fraction of positron events reconstructed with wrong
charge sign by the tracker as a function of the incident momentum,
together with the expectation according to eq.~(\ref{eq:wrongchargesign}).}
\label{fig:bremsstrahlung_chargesigneff}
\end{figure}
figure~\ref{fig:bremsstrahlung_chargesigneff}. There, the quantity
\[
y\equiv\frac{\sum\limits_\mathrm{tracker}p_\gamma}{p_\mathrm{MC}}
\]
i.e.~the fraction of momentum radiated away as bremsstrahlung, is
plotted as a function  of $x\equiv{}p_\mathrm{MC}/p_\mathrm{rec}$. For
the extreme case that the error in reconstructed momentum is given by
the bremsstrahlung losses, one obtains
\begin{equation}
\label{eq:bremsstrahlungloss}
x=\frac{p_\mathrm{MC}}{p_\mathrm{rec}}=\frac{p_\mathrm{MC}}{p_\mathrm{MC}-\sum{}p_\gamma}=\frac{p_\mathrm{MC}}{p_\mathrm{MC}-yp_\mathrm{MC}}\quad\Rightarrow\quad
y=1-\frac{1}{x}
\end{equation}
The concentration of events between the line
$p_\mathrm{MC}/p_\mathrm{rec}=1$ and this curve in
figure~\ref{fig:bremsstrahlung_chargesigneff} demonstrates the
importance of the bremsstrahlung losses.\\
Bremsstrahlung losses also cause non-Gaussian tails in the
distribution of the reconstructed momenta. These are not taken into
account by the Gaussian fits which is justifiable by the fact that the
energy measurement in the calorimeter will be used to filter events
with large bremsstrahlung losses.\\
According to equations~(\ref{eq:rigidity})~and~(\ref{eq:curverror}),
the momentum resolution can be parameterised as
\begin{equation}
\label{eq:momres}
\sigma\left(\frac{p_\mathrm{MC}}{p_\mathrm{rec}}\right)=a_\mathrm{msc}\oplus{}b_\mathrm{res}\cdot{}p_\mathrm{MC}
\end{equation}
where $a\oplus{}b\equiv\sqrt{a^2+b^2}$. The parameter $a_\mathrm{msc}$ reflects the
contribution from multiple scattering and dominates at low
energies. The parameter $b_\mathrm{res}$ describes the uncertainty arising from the
limited position resolution and is the dominating term at high
energies. The momentum resolution shown in figure~\ref{fig:momres} is
fitted to eq.~(\ref{eq:momres}), separately for the three species.
Values of $a_\mathrm{msc}^\mu=2.3\,\%$ and
$b_\mathrm{res}^\mu=0.16\,\%/\mathrm{GeV}$ for muons, $a_\mathrm{msc}^p=2.4\,\%$ and
$b_\mathrm{res}^p=0.16\,\%/\mathrm{GeV}$ for protons, and $a_\mathrm{msc}^e=2.7\,\%$ and
$b_\mathrm{res}^e=0.18\,\%/\mathrm{GeV}$ for positrons were
obtained. As expected, these values lie between the theoretical expectations for
the equidistant (eq.~(\ref{eq:res_uniform})) and optimised
(eq.~(\ref{eq:res_opt})) geometries, calculated for 16~tracker layers
with a mean $BL^2=0.57\,\mathrm{Tm}^2$ and a spatial resolution of
$56\,\mu\mathrm{m}$ that was extracted from the simulation for
isotropically incident muons. As the testbeam results presented in section~\ref{sec:testbeam2008} will show,
this number is on the optimistic side as compared to what is achieved at the current level of design. An
intrinsic resolution of $70\,\mu\mathrm{m}$ has been achieved for perpendicularly incident particles whereas the
simulation predicts $43\,\mu\mathrm{m}$ for this case.\\
\par
For a clean measurement of positrons on top of a background of
electrons, a good efficiency for the determination of the charge sign
is mandatory. The fraction of positron events reconstructed with wrong
charge sign by the tracker as a function of the incident momentum is
given in figure~\ref{fig:bremsstrahlung_chargesigneff}. It is
\begin{figure}[htb]
\begin{center}
\includegraphics[width=\textwidth,angle=0]{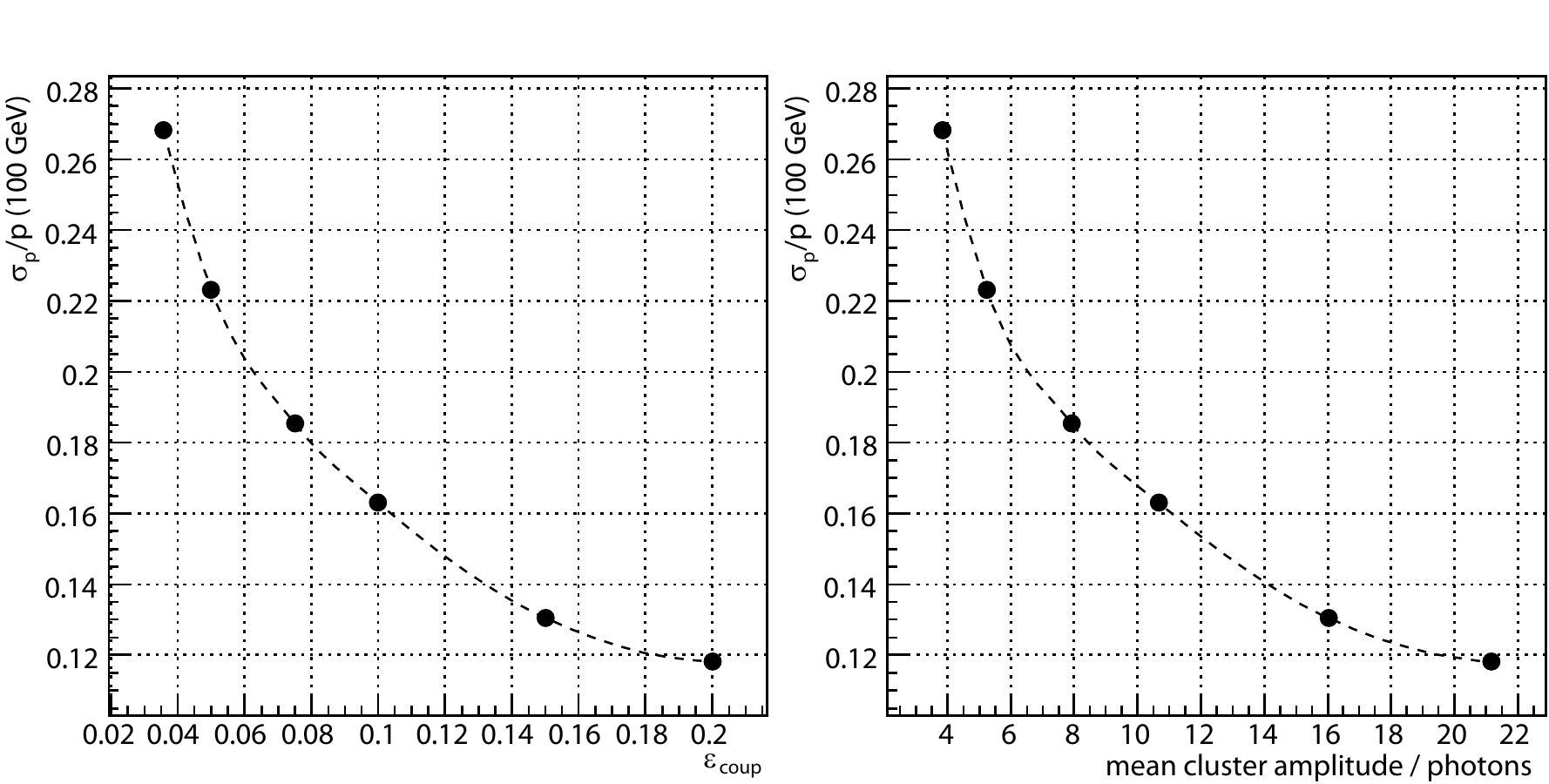}
\end{center}
\caption{Momentum resolution for $100\,\mathrm{GeV}$ protons according to a fit
of (\ref{eq:momres}), as a function of the overall coupling efficiency
$\epsilon_\mathrm{coup}$ ({\it left}) or the corresponding mean
cluster amplitude in photons ({\it right}).}
\label{fig:momres_effcoup}
\end{figure}
negligible below $100\,\mathrm{GeV}$ and gradually rises due to the
deteriorating momentum resolution towards higher energies. If the
rigidity resolution follows a Gaussian distribution with standard
deviation $\sigma_p(p)$, the expected fraction $f$ of events with
misreconstructed charge sign can be calculated to be
\begin{equation}
\label{eq:wrongchargesign}
f=\frac{1}{2}\left(1-\mathrm{erf}\left(\frac{1}{2}\,\frac{\sqrt{2}}{\sigma_p(p)}\right)\right)
\end{equation}
This is the trend followed by the simulation results in
figure~\ref{fig:bremsstrahlung_chargesigneff}.\\
\par
The track positions that enter the momentum determination are
calculated as weighted means of the individual SiPM channel amplitudes
according to eq.~(\ref{eq:weightedmean}). The higher the light yield
of the scintillating fibre and SiPM chain
per particle the more accurate this weighted mean can be
calculated. Figure~\ref{fig:momres_effcoup} contains the momentum
resolution of $100\,\mathrm{GeV}$ protons, again according to a fit to
eq.~(\ref{eq:momres}), as a function of the light yield. The curve is
shown both as a function of the overall coupling efficiency
$\epsilon_\mathrm{coup}$, as defined in (\ref{eq:effcoup}), which is
actually used as an input to the simulation, and the resulting mean
total cluster amplitude of the tracker clusters so obtained. A rather
strong dependence of the momentum resolution on the light yield is observed.

\subsection{Angular resolution}
\begin{figure}[htb]
\begin{center}
\begin{tabular}{cc}
\includegraphics[width=0.5\textwidth,angle=0]{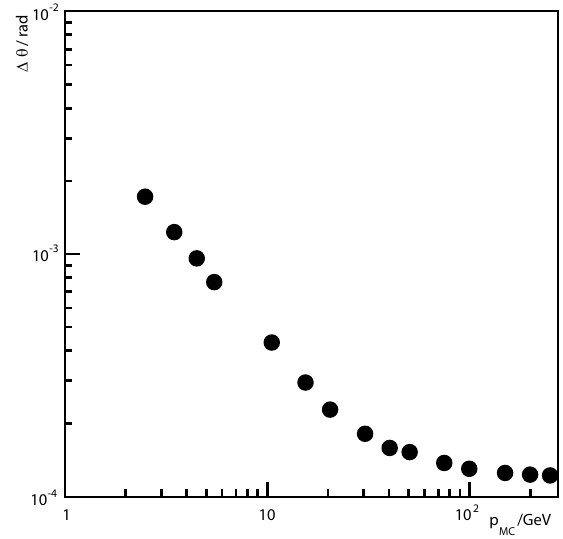}&
\includegraphics[width=0.5\textwidth,angle=0]{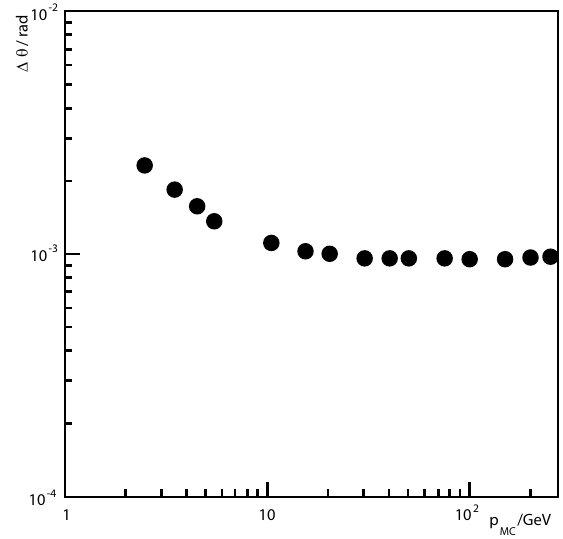}\\
\end{tabular}
\end{center}
\caption{Median difference of reconstructed and generated angles, separately
for the projection to the bending plane as measured by the tracker
({\it left}) and to the non-bending plane as measured by the TRD ({\it
right}), as a function of the momentum of the incident positrons.}
\label{fig:angular_resolution}
\end{figure}
The combined angular resolution of the tracker and the TRD is
important. For example, the reconstructed track direction is used together with the reconstructed shower axis
in the calorimeter to separate positrons from protons. The angular
resolution obtained for positrons in the simulation is shown in
figure~\ref{fig:angular_resolution}. The difference of the reconstructed and
generated angles is $\Delta\theta$. It is measured by the tracker in the bending plane and by
the TRD in the non-bending plane. For the former, a resolution on the order of
$0.1\,\mathrm{mrad}$ is found at high energies, and
$1\,\mathrm{mrad}$ for the latter.

\subsection{Impact of SiPM noise}
\begin{figure}[htb]
\begin{center}
\begin{tabular}{cc}
\includegraphics[width=0.48\textwidth,angle=0]{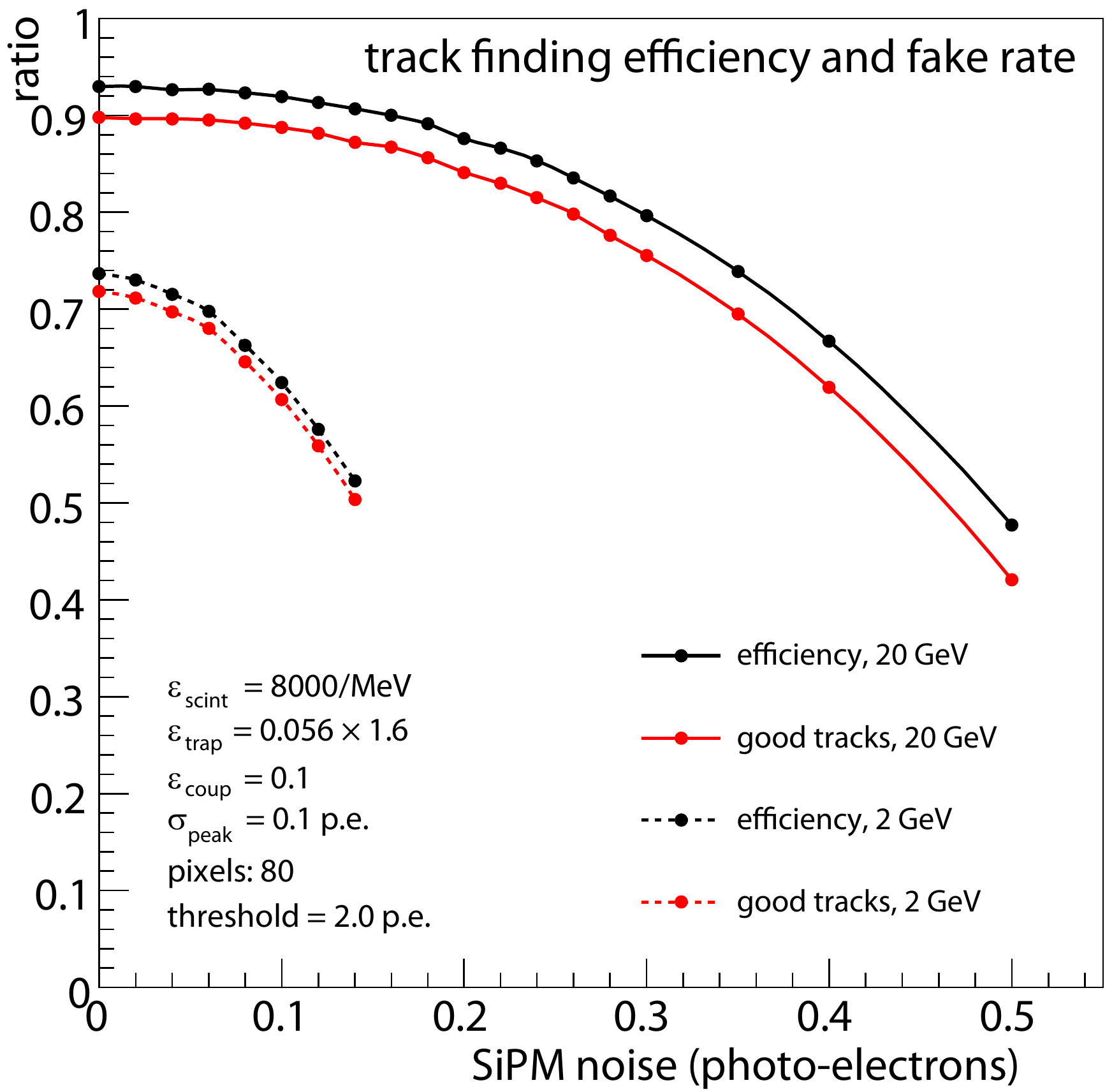}&
\includegraphics[width=0.5\textwidth,angle=0]{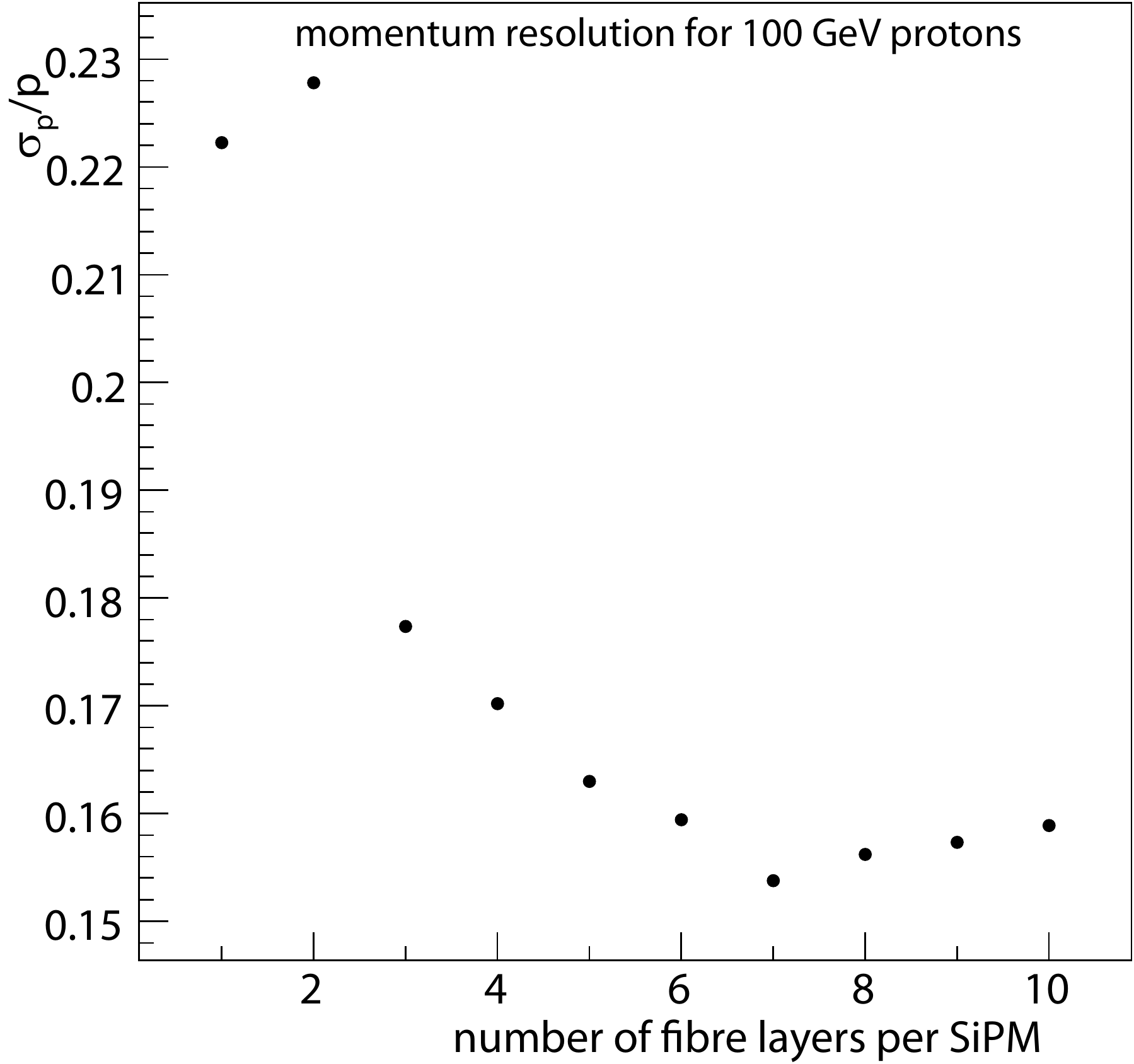}\\
\end{tabular}
\end{center}
\caption{{\it Left:} Track finding efficiency as a function of the mean noise level
in the tracker SiPMs. The plot shows the fraction of proton events for which a track
surviving the quality cuts described in the text is found. An event is
further labelled as containing a good track if both momentum and track
direction are reconstructed correctly. Protons of
$2\,\mathrm{GeV}$ and $20\,\mathrm{GeV}$ momentum are examined, respectively.
{\it Right:} Momentum resolution for $100\,\mathrm{GeV}$ protons as a
function of the number of fibre layers in front of each SiPM array
channel in the tracker.}
\label{fig:trackfinding_noise}
\end{figure}
In reality, fake hits will be present in the tracker that are caused
by the unavoidable noise of the SiPM arrays.
The track finding algorithm presented in
section~\ref{sec:reconstruction} was designed to overcome the
difficulties caused by SiPM noise as far as possible. To study the
effectiveness of the algorithm and thus the tracker performance for
different noise levels, the mean noise level of the SiPM channels,
given in photo-electrons, was varied. Using a threshold of 2~photo-electrons
per SiPM channel, the track finding efficiency was extracted for each
setting of the noise level. Protons of $2\,\mathrm{GeV}$ and
$20\,\mathrm{GeV}$ momentum were examined. The result is shown in
figure~\ref{fig:trackfinding_noise}. The curves labelled ``efficiency''
include all events including a track passing the tracker quality cuts
described in section~\ref{sec:momres}. In addition, a ``good track''
is defined as having $|p_\mathrm{rec}-p_\mathrm{MC}|/\sigma_p<4$ and a
reconstructed direction within $2^\circ$ of the true direction. For
high-momentum tracks, the tracker performance begins to deteriorate at
a noise level of around $0.1$. For low-momentum tracks, the track
finding algorithm is not that appropriate as it is optimised for
tracks with small curvature, and the track finding efficiency is
therefore lower.\\
Although the mean noise level of an SiPM depends strongly on the type
of SiPM and the
conditions under which it is operated, especially the temperature, the
testbeam results presented in section~\ref{sec:testbeam2008} show that
the noise level will be around $0.06$. This number is valid for the
Hamamatsu SiPM arrays studied in the testbeam at a temperature of
$T\sim16^\circ\mathrm{C}$. Typically, the noise level is reduced in half when the
temperature of the device is lowered by
$\sim8^\circ\mathrm{C}$~\cite{ref:gregoriophd}.

\subsection{Impact of module thickness}
An obvious way to increase the light output of the scintillating fibre
stacks used in the tracker is to increase their thickness in terms of
the number of fibre layers per
module. Figure~\ref{fig:trackfinding_noise} shows the momentum
resolution obtained for $100\,\mathrm{GeV}$ protons as a function of
the number of fibre layers per SiPM. The height of the SiPM channels
was scaled accordingly but the number of pixels was kept fixed. The
momentum resolution reaches a flat minimum at seven layers of
fibres. It slowly rises again beyond this point which can be
understood from the increasing uncertainty in the $z$ coordinate for a
thick fibre stack. In practice, the increasing difficulty of
producing a homogeneous fibre stack with many layers has to be taken
into account, as small deviations from the nominal positions of fibres
in the lower layers tend to cause larger displacements in the upper
layers~\cite{ref:gregoriophd}. If this difficulty can be overcome in the
production of the fibre stacks, a configuration using seven fibre layers
can be expected to give the best spatial resolution.

\subsection{ECAL energy resolution}
\begin{figure}[htb]
\begin{center}
\includegraphics[width=0.5\textwidth,angle=0]{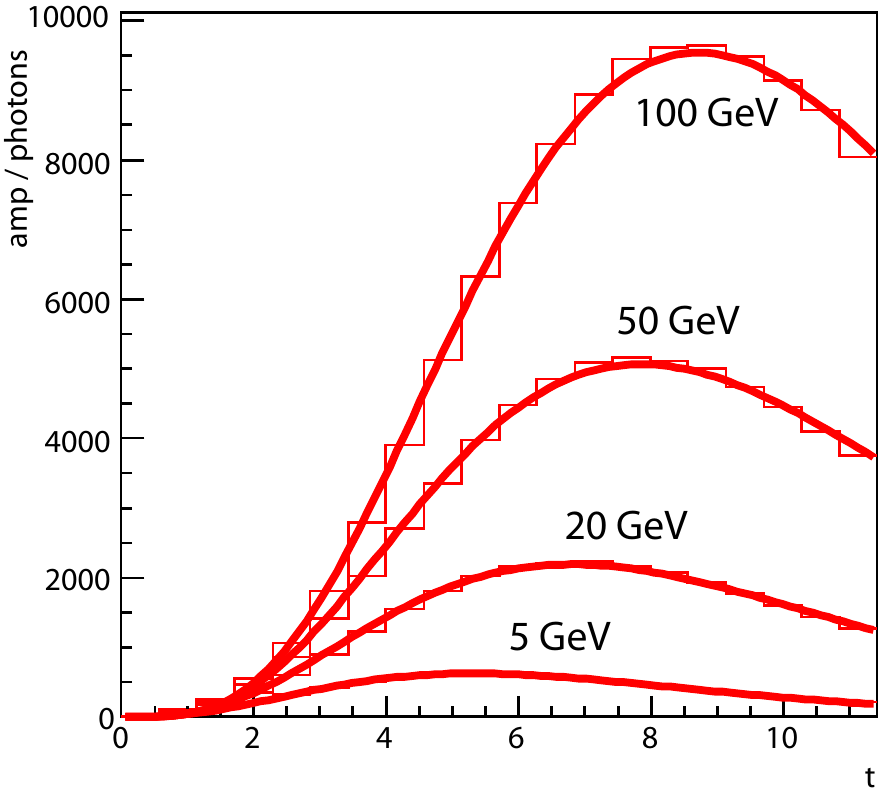}
\end{center}
\caption{Average longitudinal ECAL shower shapes for various energies of the
incident positrons. The mean amplitude (in reconstructed photons) is
plotted as a function of the depth $t=x/X_0$ in the calorimeter. One
bin corresponds to one layer of absorber. A fit of the
parameterisation according to (\ref{eq:showershape}) is included, too.}
\label{fig:longitudinal_shower_shapes}
\end{figure}
\begin{figure}[htb]
\begin{center}
\includegraphics[width=0.5\textwidth,angle=0]{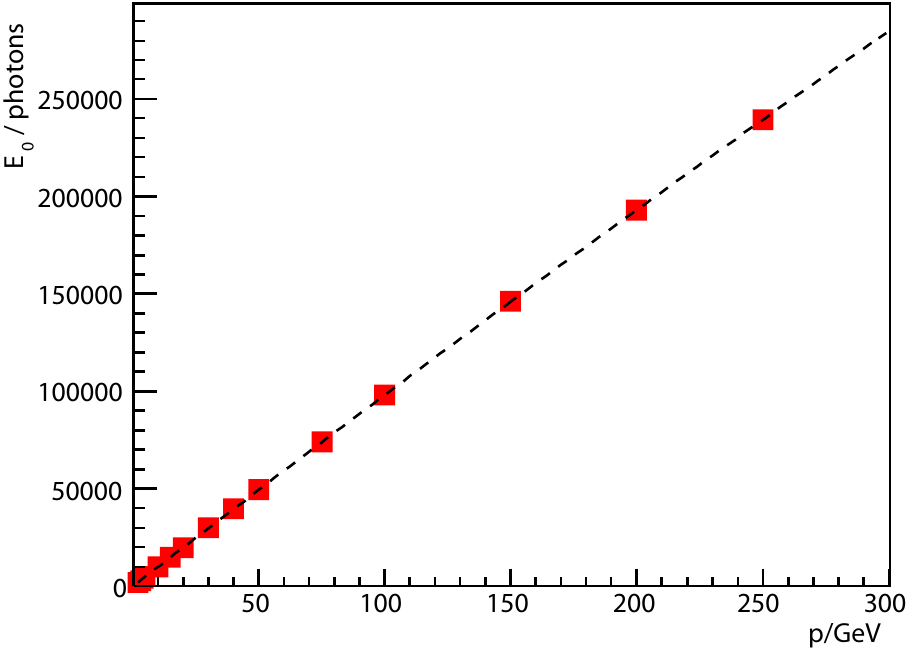}
\end{center}
\caption{Linearity of ECAL response: Mean reconstructed shower energy
$E_0$ (in photons), as a function of the incident momentum. The
response function has been fitted by a parabola.}
\label{fig:ecal_linearity}
\end{figure}
The key figures of merit for the electromagnetic calorimeter are its
energy resolution and its rejection power against protons
(sec.~\ref{sec:ecalrejection}).
Figure~\ref{fig:longitudinal_shower_shapes} illustrates the
calorimetric energy measurements of electrons and positrons. It shows
the mean longitudinal shower shapes obtained from the simulation. The
response to positrons of different energies is depicted in the
figure. Both the increase in integrated amplitude according to
(\ref{eq:showershape}) and the logarithmic dependence of the location of the
shower maximum on the incident energy according to (\ref{eq:showermaxvse}) are
visible. The figure also demonstrates that the shower maximum is well
contained in the ECAL at all energies of interest here.\\
The energy resolution of the ECAL is derived from distributions of the
reconstructed shower energy $E_0$ (in photons), separately for various incident
energies. As a first step, the calibration curve of the ECAL is
obtained (fig.~\ref{fig:ecal_linearity}). It shows the
mean value of $E_0$ found from Gaussian fits to the
$E_0$-distributions of the individual measurements as a function of the incident momentum. The response
shows a slight deviation from linearity that can be attributed to the
imperfect correction of the SiPM response function. A parabolic fit
has been applied to the ECAL response curve. This fit was then used to
calculate the reconstructed energy (in GeV) on an event-by-event
basis. The resulting distributions were fit by Gaussians to yield the
mean and standard deviation and thus the relative energy resolution
$\sigma(E)/E$. Only showers completely contained within the ECAL
acceptance were used for this study. 
\begin{figure}[htb]
\begin{center}
\includegraphics[width=0.5\textwidth,angle=0]{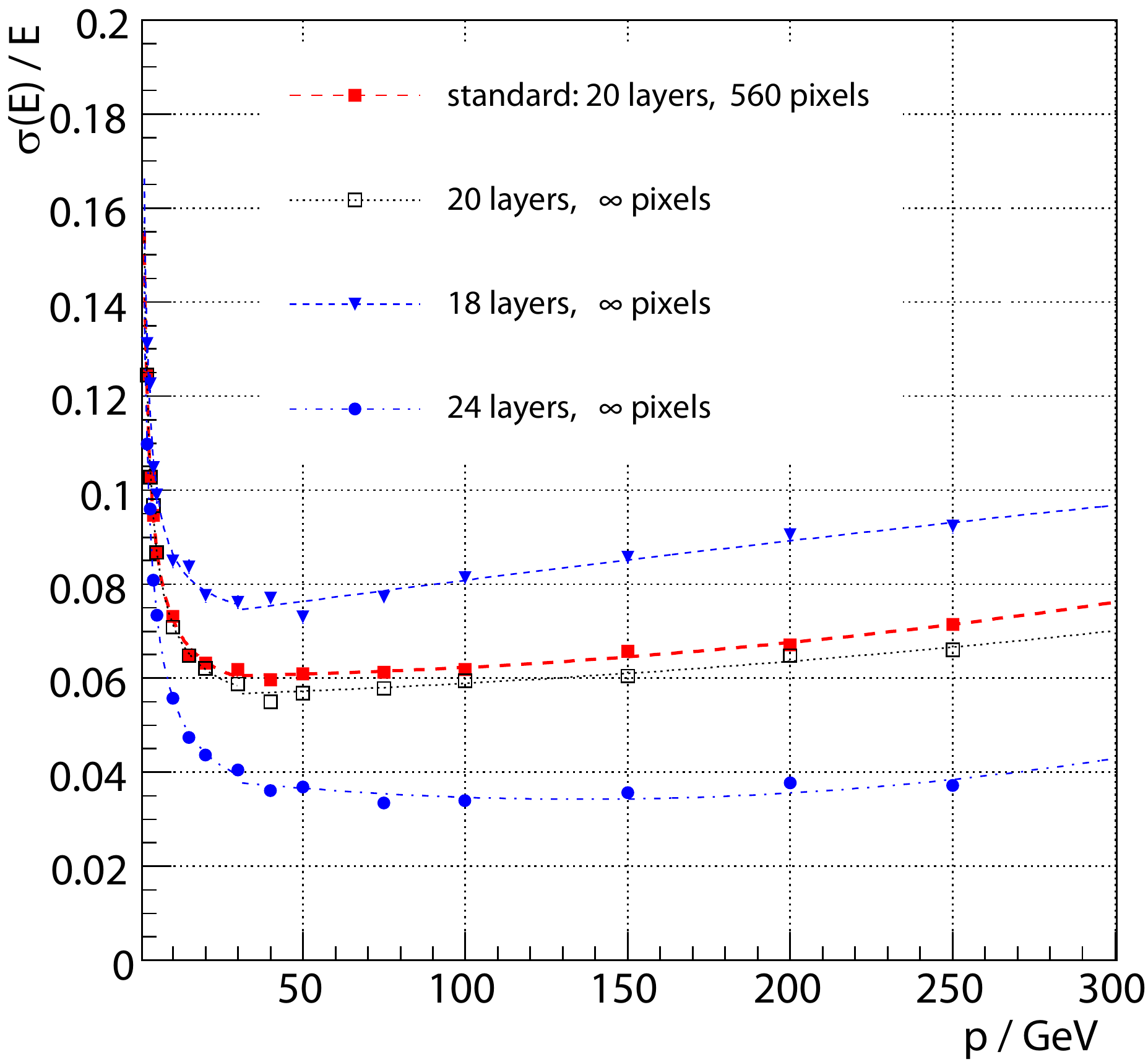}
\end{center}
\caption{ECAL energy resolution as a function of incident momentum,
for the standard configuration with 20~layers and SiPMs with
560~pixels, for the ideal case of a single SiPM with unlimited number
of pixels, and for varying number of layers. The parameterisations are
described in the text.}
\label{fig:ecal_energyres}
\end{figure}
Figure~\ref{fig:ecal_energyres} shows the projected energy resolution
obtained in this way. The energy resolution for the standard
configuration with 20~layers and SiPMs with 560~pixels was fitted to
the parameterisation of (\ref{eq:ecalres}) up to an energy of
$30\,\mathrm{GeV}$ to obtain $a=15.5\,\%\,\cdot\sqrt{\mathrm{GeV}}$ and
$c=5.3\,\%$. Above this energy, two effects worsen the energy
resolution.\\
First of all, the limited depth of the calorimeter cannot
completely be corrected for. This is demonstrated by the additional
curves in the figure showing the resolution for calorimeters with
different numbers of layers. A significant effect is seen already for
a variation of $10\,\%$ of the depth. However, each ECAL layer will
contribute around $30\,\mathrm{kg}$ to the weight budget.\\
The second effect is the limited dynamic range of the SiPMs. As the
figure shows, the energy resolution remains somewhat lower for the ideal case of
an unlimited number of pixels. For this case, the partially reflective
foil is not needed and was replaced by an ideal mirror in the
simulation. This effect is small compared to the first one as the
limited dynamic range is compensated by the ECAL design featuring the
partially reflective foil.\\
To be more quantitative, both the overall thickness of the ECAL and
the number of SiPM pixels used for its readout have been varied around
\begin{figure}[htb]
\begin{center}
\includegraphics[width=0.9\textwidth,angle=0]{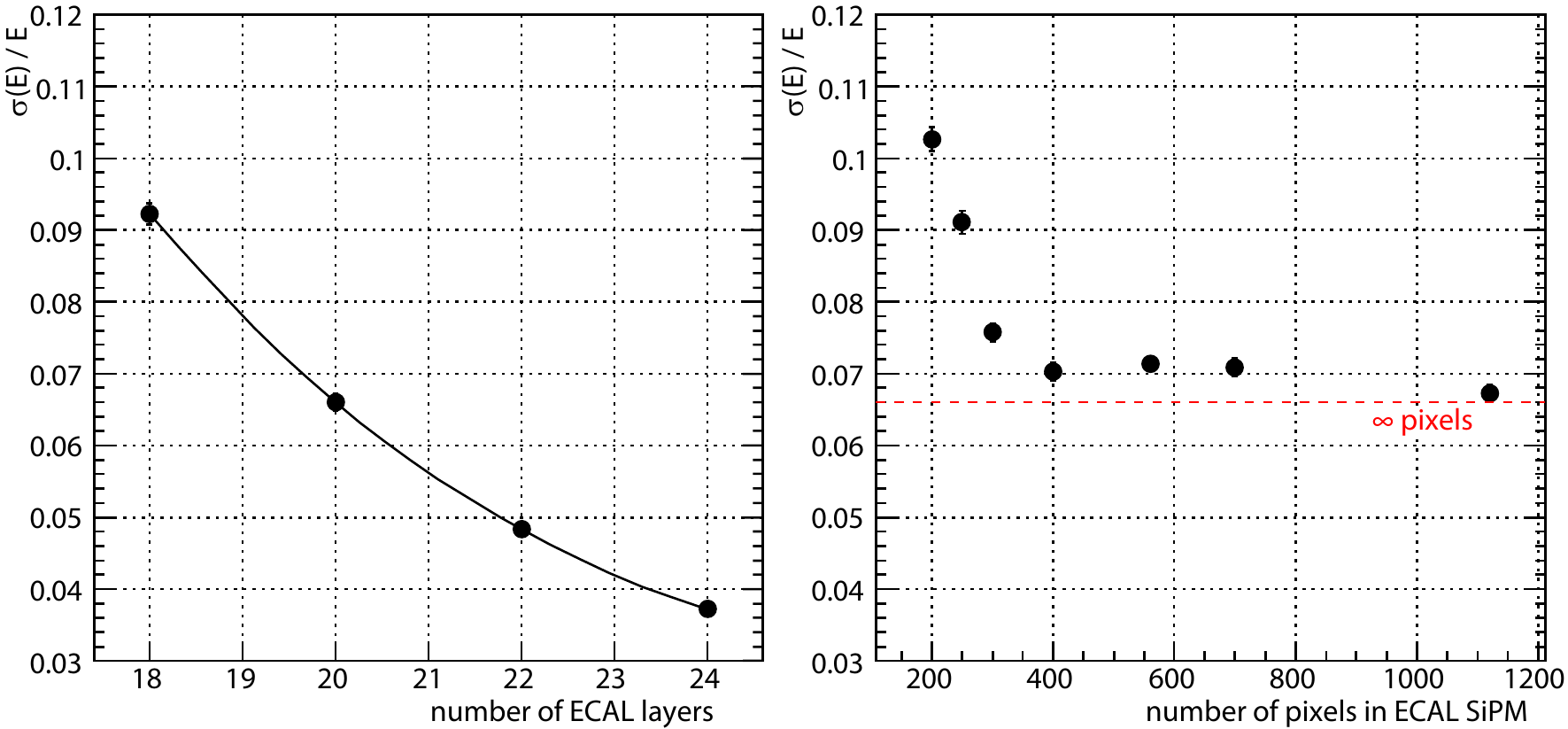}
\end{center}
\caption{ECAL energy resolution at $250\,\mathrm{GeV}$, as a function
of the number of ECAL layers, i.e.~the total thickness, for a fixed
number of 560~pixels ({\it left}),
and for varying numbers of SiPM pixels, at a fixed number of 20~layers
({\it right}). The solid line is a spline interpolation, used to guide
the eye only.}
\label{fig:ecal_energyres_study}
\end{figure}
the values of the standard design to study the corresponding change in
energy resolution
(fig.~\ref{fig:ecal_energyres_study}). While there is a drastic
dependence of the energy resolution on the ECAL thickness, the
resolution cannot be significantly improved by increasing the number
of SiPM pixels above a certain threshold, located around~400 for the
present design.\\
For the reasons described above, a parabolic fit had to
be used to describe the energy resolution
above $30\,\mathrm{GeV}$. For the standard configuration, it reaches a value of $0.07$ at
$250\,\mathrm{GeV}$. Comparing this to the tracker resolution in
figure~\ref{fig:momres} shows that the overall performance of the
detector, especially the $E/p$ matching used for proton
discrimination, will be limited by the tracker because of its considerably
worse resolution.

\subsection{ECAL rejection and efficiency analysis}
\label{sec:ecalrejection}
\begin{figure}[htb]
\begin{center}
\includegraphics[width=\textwidth,angle=0]{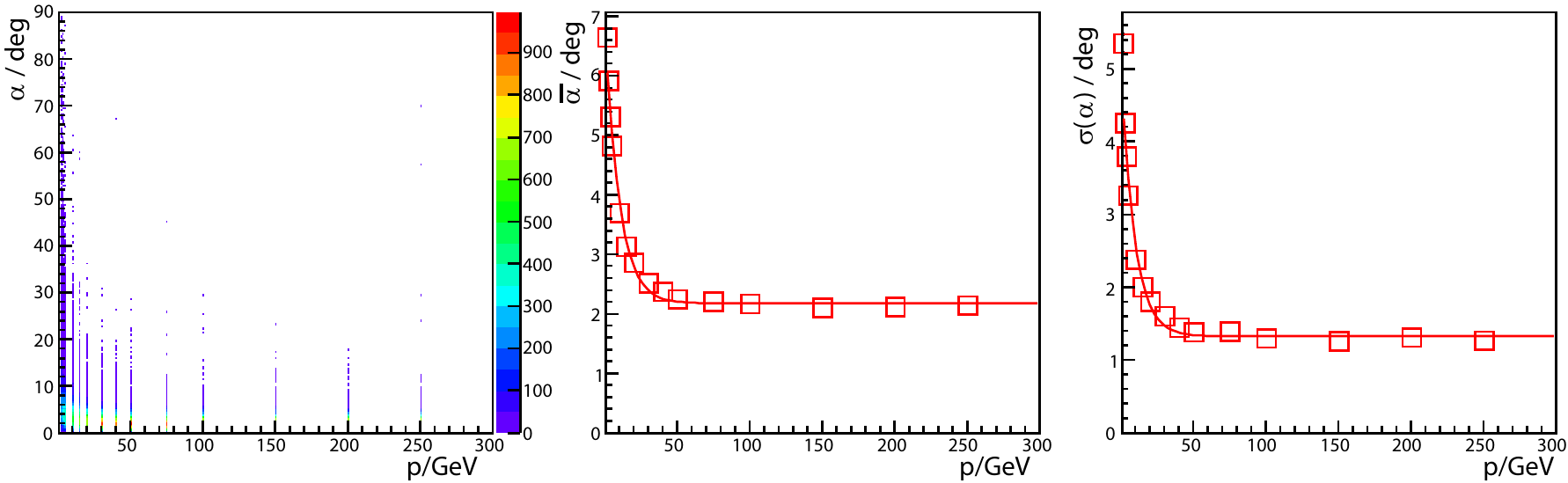}
\end{center}
\caption{Example of procedure for automatic cut determination:
Distribution of reconstructed angle $\alpha$ between ECAL shower and
track found in tracker and TRD for positrons, as a function of
momentum ({\it left}). For each bin in $p$, the projection along the
$\alpha$-axis is calculated, a
Gaussian fit is performed, and the mean values ({\it middle}) and
standard deviations $\sigma$ ({\it right}) are extracted. Suitable
fits to the mean and $\sigma$ values are performed. The
momentum-dependent cut can then be calculated according to the results
of these fits.}
\label{fig:trackanglefits}
\end{figure}
\begin{figure}
\begin{center}
\begin{tabular}{cc}
\includegraphics[width=0.4\textwidth,angle=0]{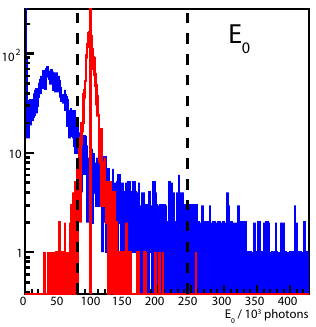}&
\includegraphics[width=0.4\textwidth,angle=0]{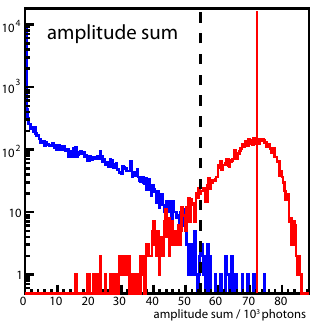}\\
\includegraphics[width=0.4\textwidth,angle=0]{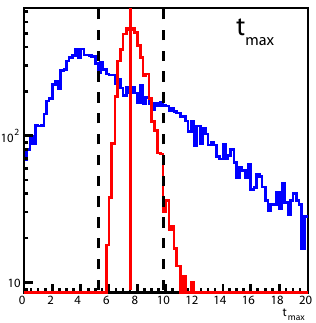}&
\includegraphics[width=0.4\textwidth,angle=0]{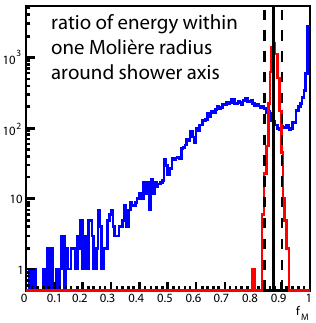}\\
\includegraphics[width=0.4\textwidth,angle=0]{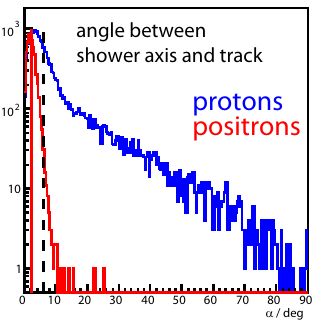}&\\
\\
\end{tabular}
\end{center}
\caption{ECAL cuts: Distributions of the shower variables used for
proton discrimination: shower energy $E_0$ as determined from the
shower fit, sum over layer amplitudes, shower maximum
$t_\mathrm{max}$, ratio $f_M$ of energy within one Moli\`ere radius
around the shower axis, and angle $\alpha$ between shower axis and
reconstructed track. Distributions are shown for positrons (red) and
protons (blue) for a
momentum of $100\,\mathrm{GeV}$. The mean values for positrons
(continuous lines) as well as the cuts (dashed lines) automatically
determined by the analysis procedure are shown, too.}
\label{fig:ecalcutprojections}
\end{figure}
\begin{figure}
\begin{center}
\includegraphics[width=\textwidth,angle=0]{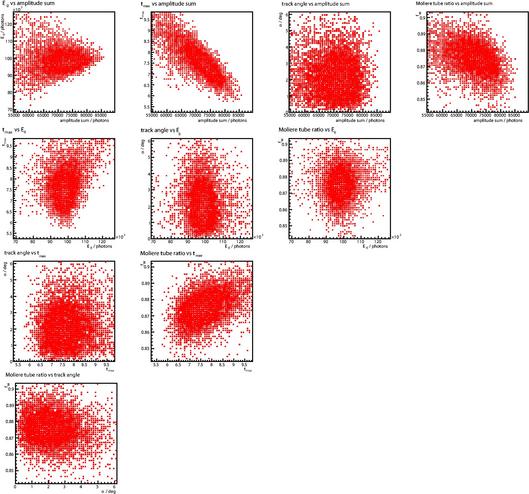}
\end{center}
\caption{Correlations between the five shower variables used for
proton discrimination. Distributions for $100\,\mathrm{GeV}$
positrons are shown.}
\label{fig:showercorr}
\end{figure}
\begin{figure}
\begin{center}
\includegraphics[width=\textwidth,angle=0]{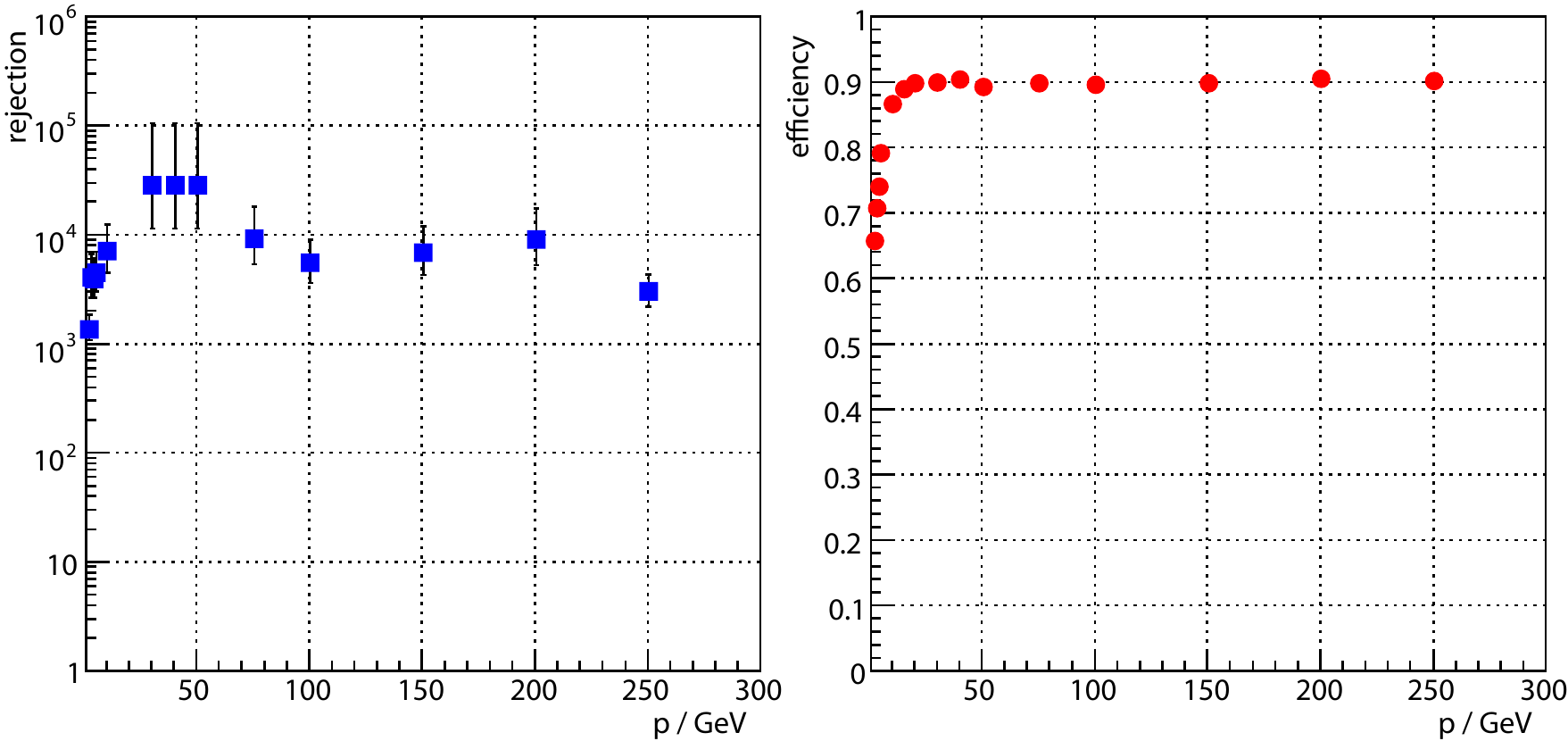}
\end{center}
\caption{Projected ECAL rejection against protons ({\it left}) and
corresponding positron efficiency ({\it right}), as a function of
momentum of the incident particle. The generated momenta provided by
the simulation have been
used for all cuts.}
\label{fig:ecalrejection_pgen}
\end{figure}
\begin{figure}
\begin{center}
\includegraphics[width=\textwidth,angle=0]{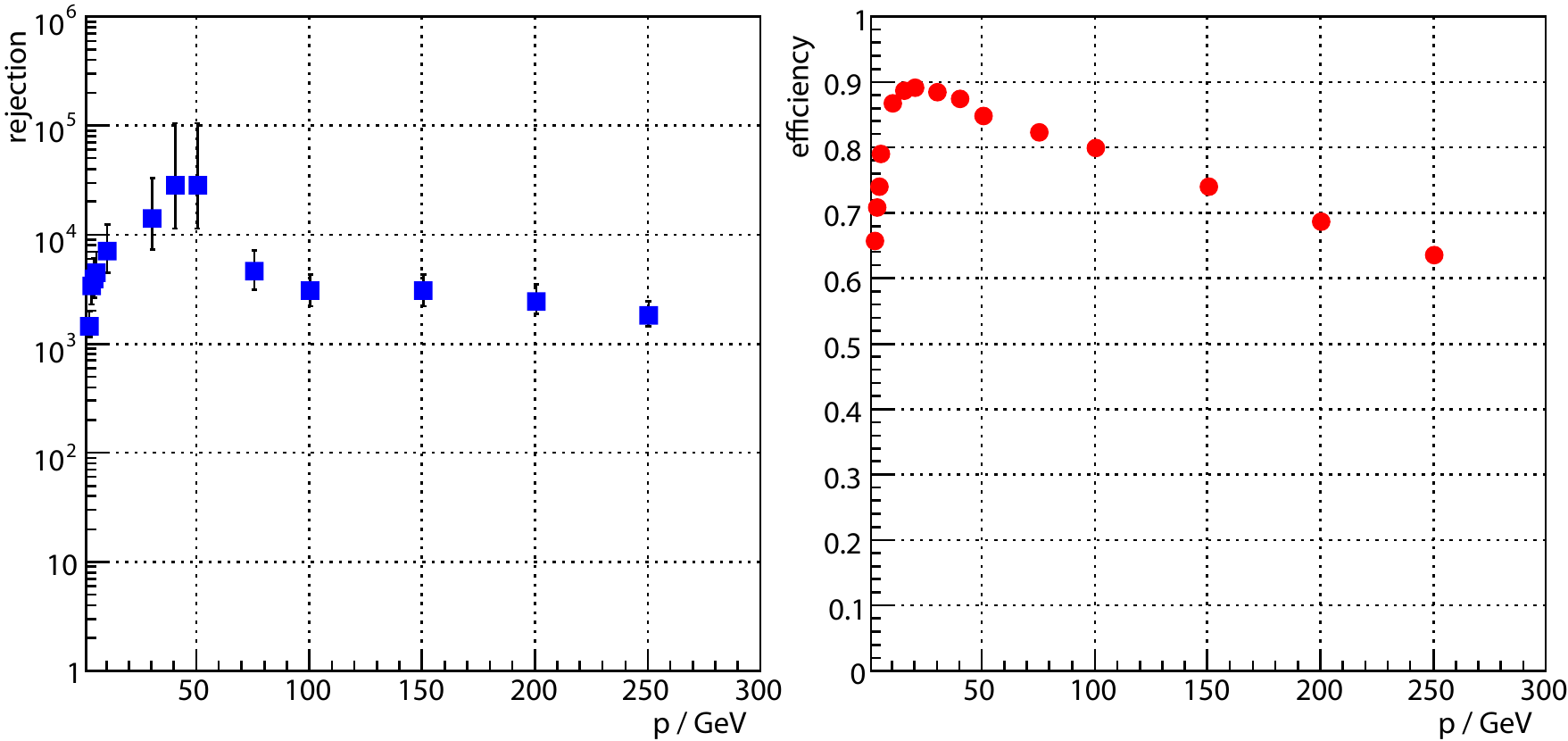}
\end{center}
\caption{Projected ECAL rejection against protons ({\it left}) and
corresponding positron efficiency ({\it right}), as a function of
momentum of the incident particle. The momenta reconstructed by the
tracker and the TRD have been used for all cuts.}
\label{fig:ecalrejection_prec}
\end{figure}
The most important task for the electromagnetic calorimeter is the
reliable suppression of the proton background. In order to study the
projected ECAL rejection, a cut-based shower shape analysis was
performed. Momentum-dependent cuts were applied to five variables on an
event-by-event basis:
\begin{itemize}
\item the shower energy $E_0$ as determined from the longitudinal
shower shape as described in section~\ref{sec:reconstruction},
\item the sum over the amplitudes in each layer,
\item the location $t_\mathrm{max}$ of the shower maximum,
\item the ratio $f_M$ of energy within one Moli\`ere radius around the
shower axis, and
\item the angle $\alpha$ between the shower axis and the reconstructed
track in the tracker
\end{itemize}
A procedure for the automatic determination of the momentum-dependent
cut was developed and the idea is shown for the example of the track
angle $\alpha$ (fig.~\ref{fig:trackanglefits}). The values obtained for the positron
sample are first histogrammed versus the generated momentum $p$. Then,
Gaussian fits are applied to the distributions in each momentum bin to
obtain the mean $\bar{\alpha}$ and standard deviation $\sigma(\alpha)$
as a function of momentum. Suitable parameterisations are then fitted to
both curves, and the momentum-dependent cuts are finally defined at
$\bar{\alpha}(p)\pm{}n_{\alpha,\pm}\,\cdot\,\sigma(p)$. The cut
parameters $n_{i,\pm}$ are chosen appropriately for the various
cuts. As an example, figure~\ref{fig:ecalcutprojections} shows the distributions of the
five shower parameters for positrons and protons of
$100\,\mathrm{GeV}$ momentum, together with the cuts determined for this
momentum. The $E_0$ distribution peaks at lower values for protons
than for positrons. $n_{E_0,-}=2$ is chosen while the small tails
towards large $E_0$-values allow an upper cut of $n_{E_0,+}=15$,
needed to allow for errors in the momentum reconstruction. In the case
of the amplitude sum, only a lower cut of $n_{\mathrm{sum},-}=3$ is
used. The distributions of $t_\mathrm{max}$, $f_M$, and $\alpha$ show
clean peaks in the positron case while the corresponding proton
distributions are smeared out across the parameter ranges. Cuts at
$n_\pm=3$ are used for these variables.\\
For an assessment of the ECAL performance alone, the cuts are calculated
on an event-by-event basis using $p=p_\mathrm{MC}$, while for a
realistic assessment of the overall detector performance,
$p=p_\mathrm{rec}$ has to be used. In this case, the limited momentum
resolution of the tracker worsens the overall performance. An event surviving all five cuts is
classified as a positron or electron, respectively. Correlation plots
of the five shower variables are shown in figure~\ref{fig:showercorr} for the
case of $100\,\mathrm{GeV}$ positrons. Except for the correlation of
the amplitude sum and $t_\mathrm{max}$ which reaches a value of
$-0.8$, the cut variables are largely uncorrelated.\\
The proton rejection and corresponding positron efficiency are plotted
in figures~\ref{fig:ecalrejection_pgen}
and~\ref{fig:ecalrejection_prec} as a function of momentum and using the
generated and reconstructed momenta, respectively. In both cases, the
proton rejection reaches a maximum of roughly $3\cdot10^4$ at around
$50\,\mathrm{GeV}$. It drops to $3\cdot10^3$ and $2\cdot10^3$
at high energies for generated and reconstructed momenta,
respectively. The
electron efficiency is at $90\,\%$ using the generated momenta, but it
becomes considerably worse using the reconstructed momenta, dropping
to $65\,\%$ at $250\,\mathrm{GeV}$. This is mainly due to the
deteriorating momentum resolution of the tracker at high energies
which causes many positron events to lie outside the corridors for the
cuts on $E_0$ and the amplitude sum.
\begin{figure}[htb]
\begin{center}
\includegraphics[width=0.5\textwidth,angle=0]{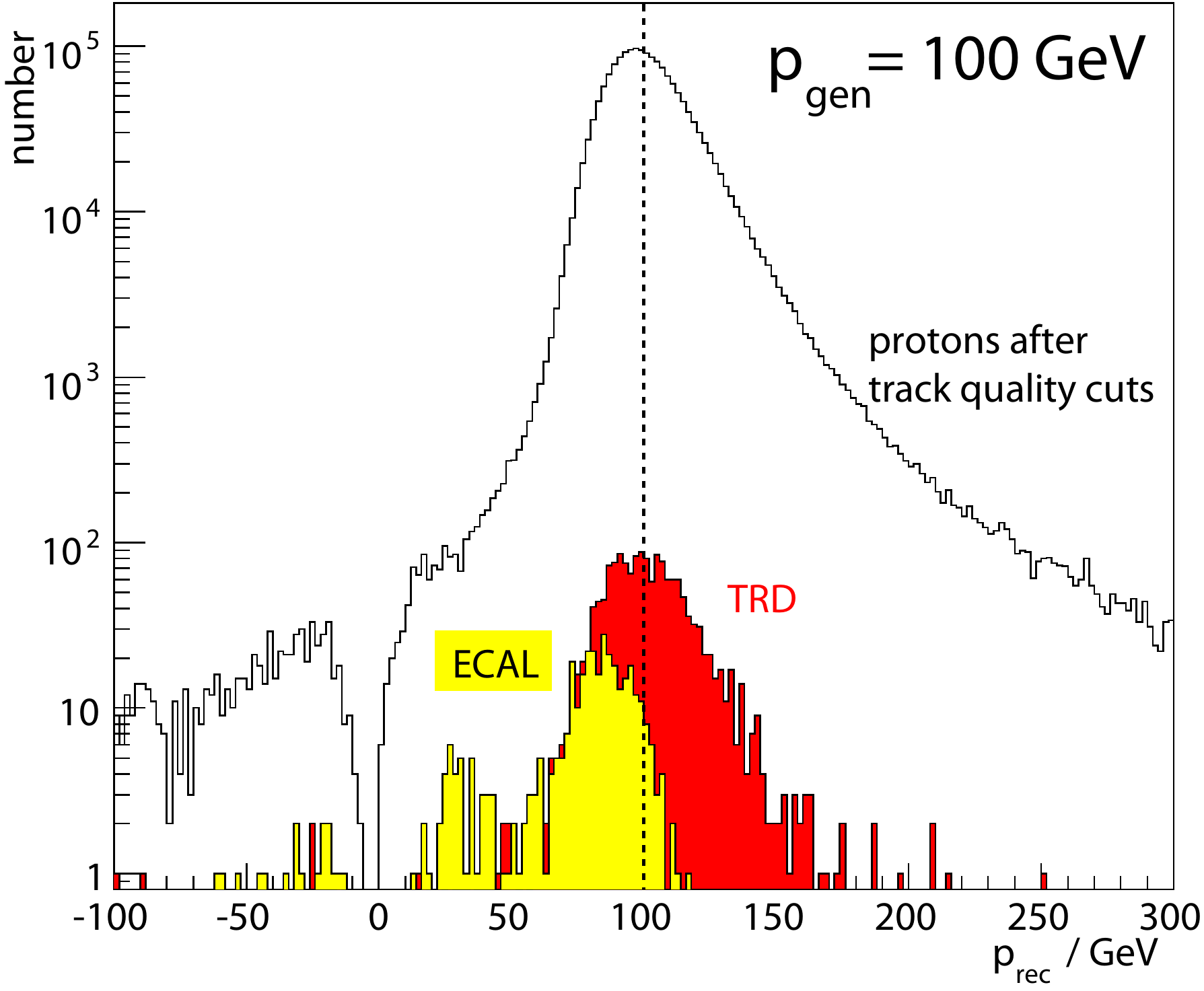}
\end{center}
\caption{Distributions of reconstructed momenta for
$100\,\mathrm{GeV}$ proton events misidentified as positrons or
electrons by the ECAL and the TRD, respectively. The hollow curve
gives the distribution of all tracks passing the simple quality cuts
outlined in section~\ref{sec:momres}.}
\label{fig:ecal_recmom}
\end{figure}
\\
A beneficial effect of using a calorimeter for proton suppression is
illustrated in the distribution
of reconstructed momenta for $100\,\mathrm{GeV}$ proton events which were
misidentified as positrons or electrons by the ECAL (fig.~\ref{fig:ecal_recmom}). Because the ECAL
sets an upper boundary on the energy of a particle, the reconstructed
momenta of particles passing the ECAL cuts will be lower than the
actual momenta. This means that they will contribute to energy bins
with a much higher relative positron flux
(fig.~\ref{fig:protons}). In contrast to this, the corresponding
distribution for protons misidentified by the TRD
(sec.~\ref{sec:trdrej}) simply is determined by the momentum
resolution of the tracker.\\
Distributions like the one in
fig.~\ref{fig:ecal_recmom} give a better picture of the behaviour of the
complete detector than rejections alone. Unfortunately, the amounts of
Monte Carlo statistics needed to obtain them turn out to be prohibitive.

\subsection{TRD rejection}
\label{sec:trdrej}
\begin{figure}[htb]
\begin{center}
\begin{tabular}{cc}
\includegraphics[width=0.5\textwidth,angle=0]{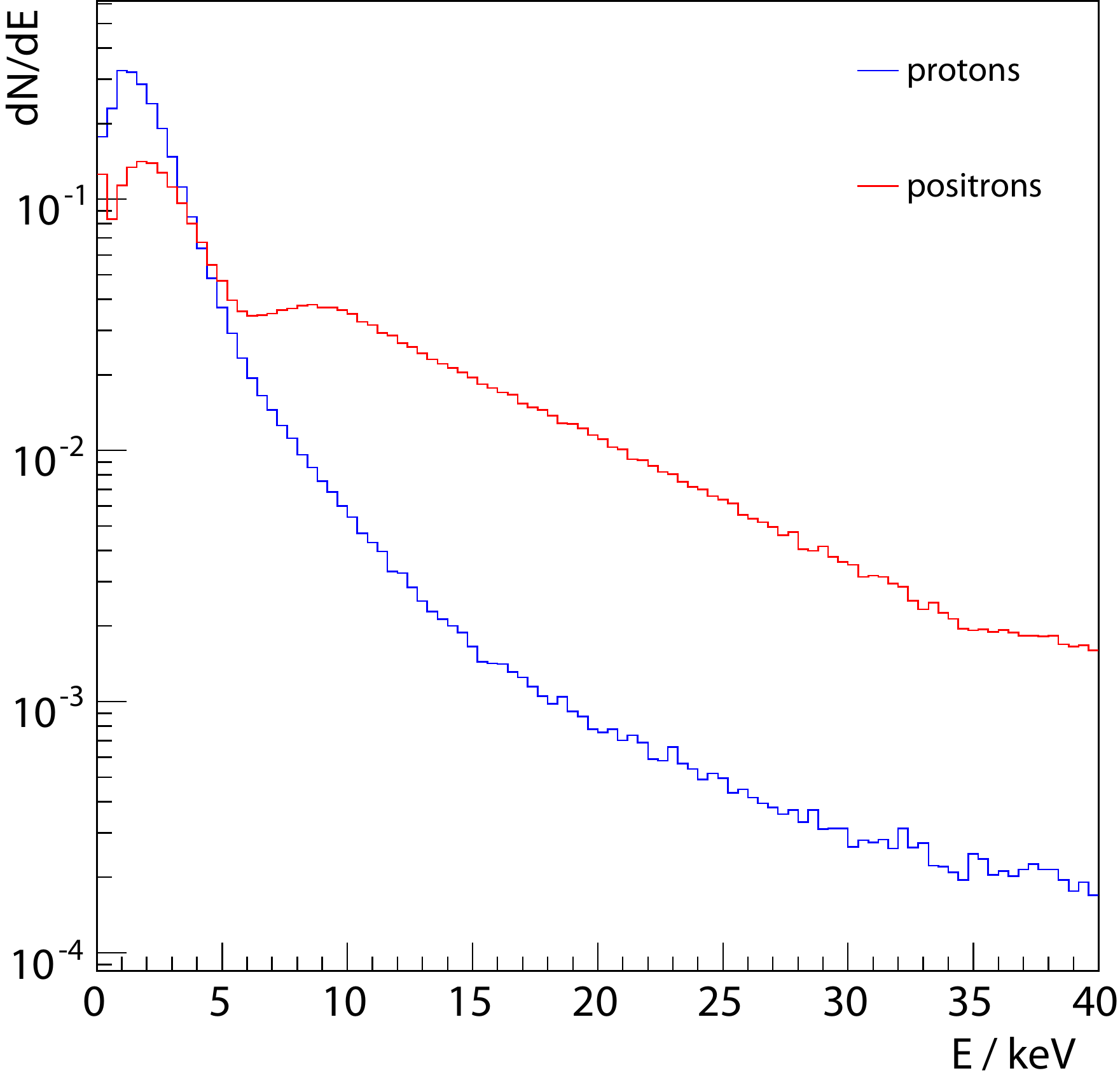}&
\includegraphics[width=0.5\textwidth,angle=0]{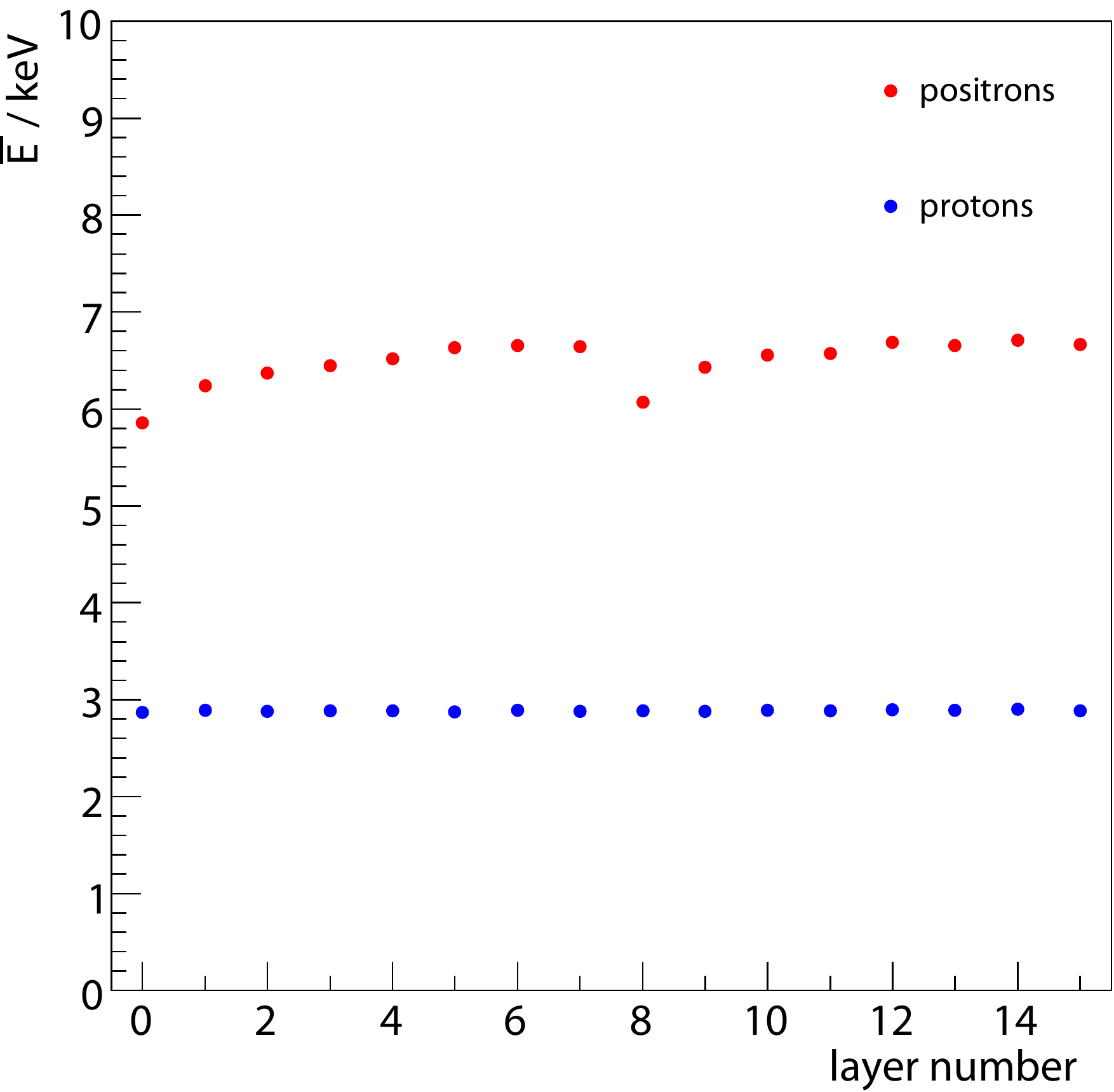}\\
\end{tabular}
\end{center}
\caption{{\it Left:} Distribution of simulated energy depositions in the TRD
tubes, for positrons and protons of
$20\,\mathrm{GeV}$ momentum, respectively. Data from all layers are included in this plot.
{\it Right:} Mean energy depositions in the TRD tubes, as a function
of layer number, for positrons and protons of
$20\,\mathrm{GeV}$ momentum, respectively. The central tracker layers
are located between layers 7~and~8, reducing the mean value for
positrons by absorbing the TR x-ray photons.}
\label{fig:pebs_trd_spectra}
\end{figure}
\begin{figure}[htb]
\begin{center}
\includegraphics[width=0.65\textwidth,angle=0]{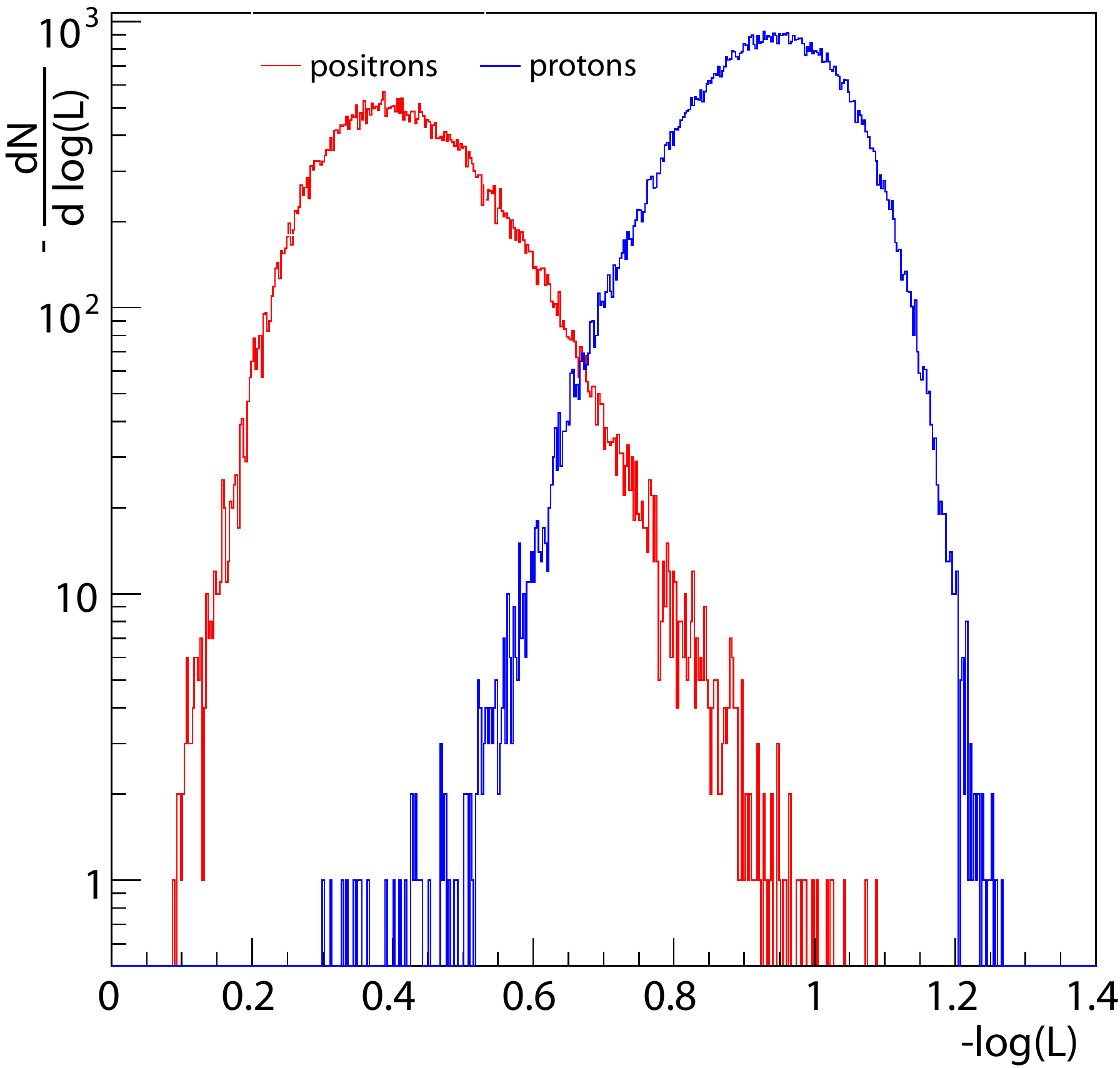}
\end{center}
\caption{Distributions of log-likelihood $-\log(L)$ used for proton
discrimination in the TRD, for positrons and protons of
$20\,\mathrm{GeV}$ momentum.}
\label{fig:pebs_trd_likelihood}
\end{figure}
The determination of the projected TRD performance was performed
analogously to section~\ref{sec:trdtestbeam}. The accumulated distributions
of energy depositions in the TRD tubes, for positrons and protons of
$20\,\mathrm{GeV}$ momentum, respectively, are shown in
figure~\ref{fig:pebs_trd_spectra}. The ionisation peaks for both
species as well as the characteristic shoulder caused by the
transition radiation of positrons are visible. The figure also
contains a plot of the mean energy depositions for each layer. Here,
the two-part structure of the TRD is evident in the positron curve,
where the mean energy deposition reaches its saturation value after a
couple of layers. Between layers 7~and~8, the central tracker layers
absorb the x-ray transition radiation photons generated in the first
half of the TRD. The proton and electron
log-likelihoods $-\log(L)$, defined analogous to section~\ref{sec:trdtestbeam}, are
then calculated on an event-by-event basis according to the incident
energy from the tube energy distributions, like the ones shown in
fig.~\ref{fig:pebs_trd_spectra}. Finally, the proton rejection and
corresponding positron efficiency are determined by setting a cut
in the $-\log(L)$-distributions (fig.~\ref{fig:pebs_trd_likelihood}). As an example, the result for
$20\,\mathrm{GeV}$ protons and positrons is shown in
\begin{figure}[htb]
\begin{center}
\begin{tabular}{cc}
\includegraphics[width=0.5\textwidth,angle=0]{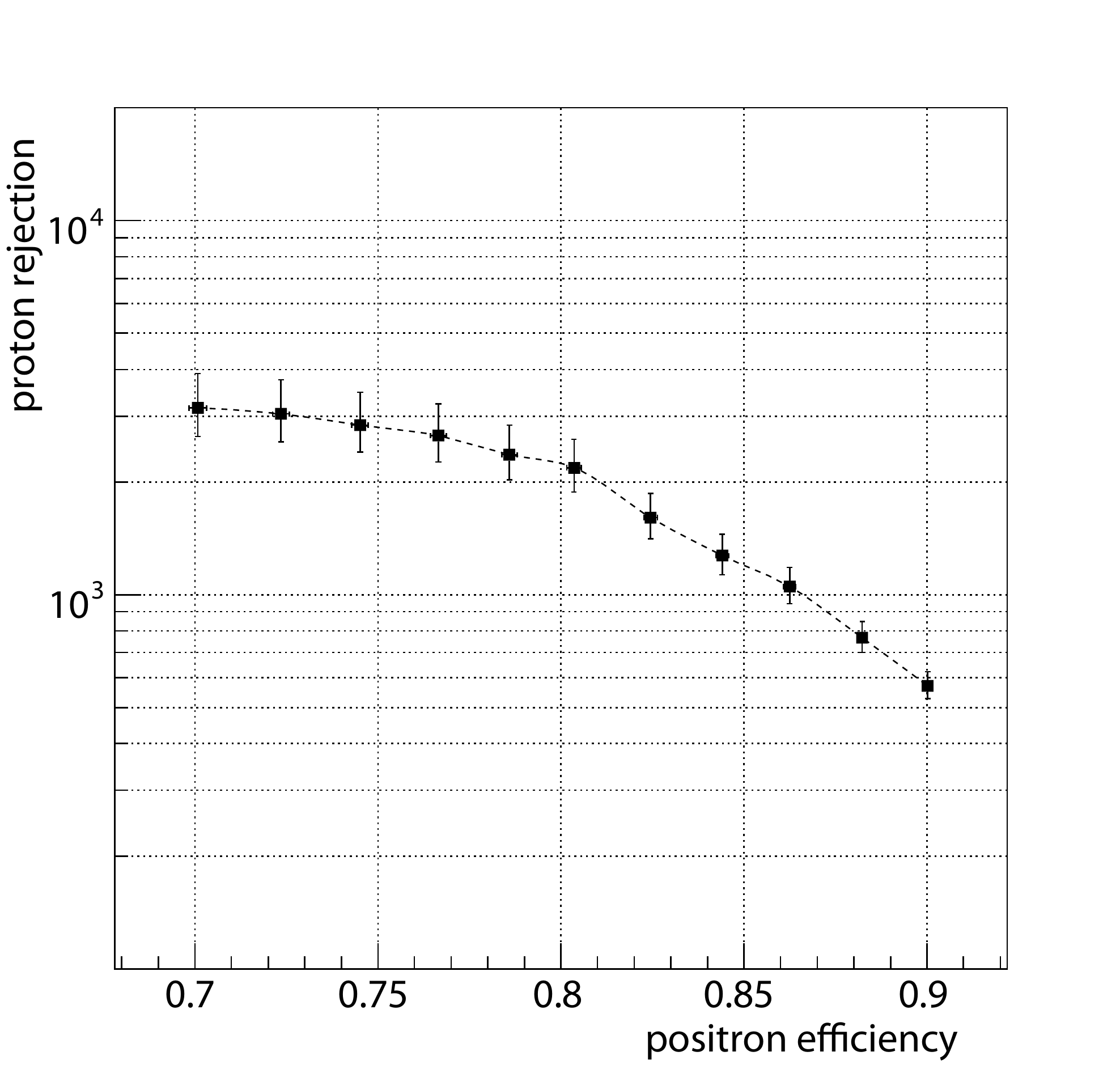}&
\includegraphics[width=0.47\textwidth,angle=0]{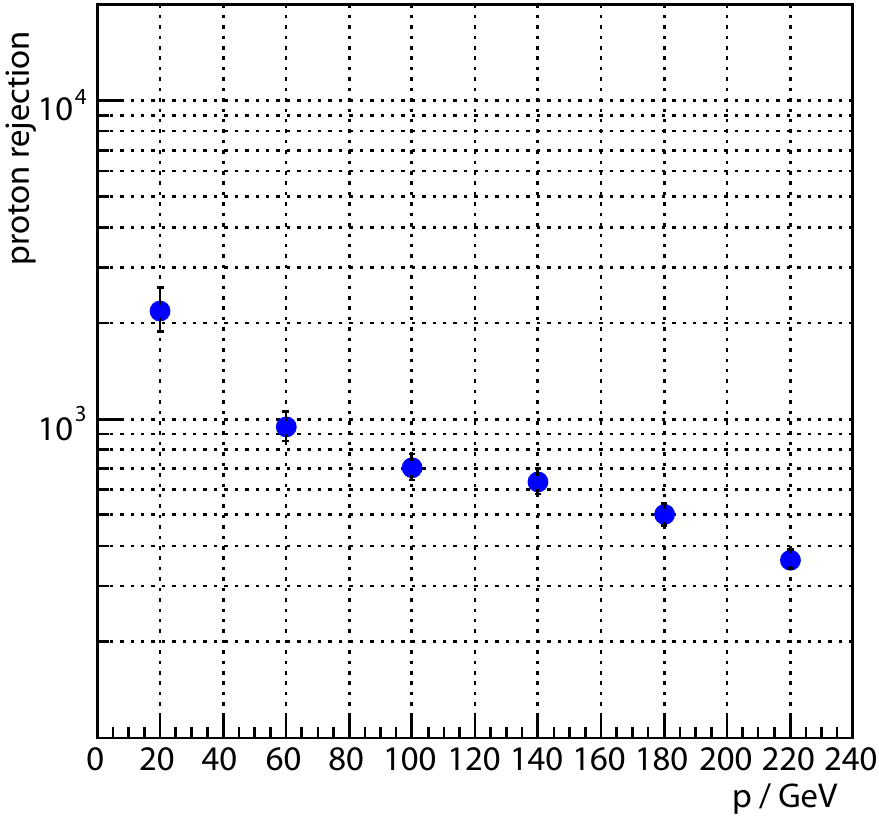}\\
\end{tabular}
\end{center}
\caption{{\it Left:} Rejection achieved by the TRD against $20\,\mathrm{GeV}$ protons, as
a function of the corresponding positron efficiency, according to the simulation.
{\it Right:} Rejection achieved by the TRD against protons, as a
function of incident momentum, at $80\,\%$ positron efficiency.}
\label{fig:pebs_trd_rej}
\end{figure}
figure~\ref{fig:pebs_trd_rej}. At a positron efficiency of $80\,\%$,
the rejection begins to deteriorate and therefore, the proton rejection
is shown for this efficiency as a function of momentum in the right-hand
side of figure~\ref{fig:pebs_trd_rej}. At $100\,\mathrm{GeV}$, the
projected proton rejection is at the level of~700 for $80\,\%$
positron efficiency.

\subsection{Acceptance calculation}
\label{sec:acceptance}
\begin{figure}[htb]
\begin{center}
\begin{tabular}{cc}
\includegraphics[width=0.5\textwidth,angle=0]{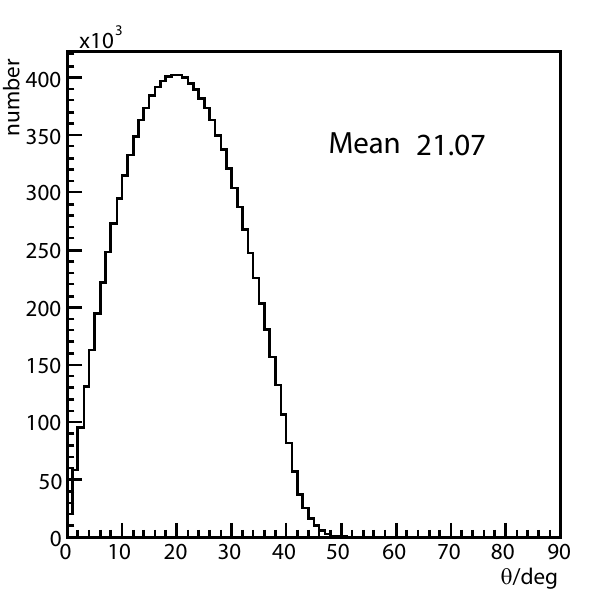}&
\includegraphics[width=0.5\textwidth,angle=0]{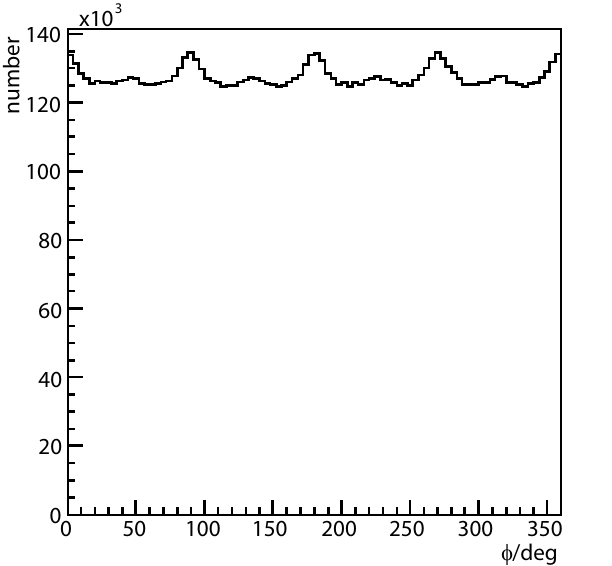}\\
\end{tabular}
\end{center}
\caption{Distribution of zenith angles ({\it left}) and azimuth angles
({\it right}) for straight lines
within the geometrical acceptance of the PEBS detector.}
\label{fig:acceptance_angles}
\end{figure}
The geometric acceptance $A$ determines the gathering power of a cosmic-ray
detector and hence, the size of the statistical errors attributed to a
flux measurement performed by the detector. It is defined as the factor of proportionality between the
counting rate $C$ of a detector and an isotropic flux $\Phi$:
\begin{equation}
\label{eq:accdef}
C=A\Phi
\end{equation}
For an idealised detector, consisting of a given number of detection
planes, the geometric acceptance is given by~\cite{ref:acceptance}
\begin{equation}
\label{eq:accformula}
A=\int_\Omega\mathrm{d}\omega{}F(\omega)\int_S\mathrm{d}\mathbf{\sigma}\cdot\mathbf{\hat{r}}
\end{equation}
where $\mathrm{d}\mathbf{\sigma}$ is the element of surface area of
the last telescope sensor to be penetrated, $S$ is the total area of
the last telescope sensor,
$\mathrm{d}\omega=\mathrm{d}\phi\,\mathrm{d}(\cos\theta)$ is an element
of solid angle, $\Omega$ is the domain of $\omega$, limited by the
other telescope sensors, and $\mathbf{\hat{r}}$ is a unit vector in
the direction of $\omega$. $F(\omega)$ is the angular dependence of
the incident flux, with $F(\omega)=1$ for isotropic incidence.\\
For an ideal telescope consisting of a single planar detector that is
hit by particles from one of its two sides, the
geometrical acceptance is easily calculated from (\ref{eq:accformula})
to be
\begin{equation}
\label{eq:accsingle}
A=\pi{}a
\end{equation}
where $a$ is the surface area of the detector.\\
For a detector such as PEBS, with many detector planes, an analytical
calculation of (\ref{eq:accformula}) becomes
impossible. An elegant approach in this case is a Monte Carlo
integration which proceeds as follows: In the first step, a random
point is chosen at the opening aperture of the detector. A random
direction is chosen from the appropriate angular distribution: The
directions of incidence on a given surface element on the planar aperture will be weighted by
$F(\omega)$ and a factor $\cos\theta$ from
$\mathrm{d}\mathbf{\sigma}\cdot\mathbf{\hat{r}}$. The weighted solid
angle then becomes
\[
\cos\theta{}F(\omega)\,\mathrm{d}\cos\theta\,\mathrm{d}\phi=\frac{1}{2}F(\omega)\,\mathrm{d}\cos^2\theta\,\mathrm{d}\phi
\]
For isotropic incidence, one therefore chooses $\cos^2\theta$ and
$\phi$ randomly from uniform
distributions with the appropriate bounds. In the second step, a
trajectory with the starting point and direction obtained in the first
step is followed through the detector to see if it intersects all
sensitive planes. These two steps are then repeated many times and the
acceptance can finally be calculated as
\begin{equation}
\label{eq:accmc}
A=\frac{\mathrm{number}\;\mathrm{of}\;\mathrm{trajectories}\;\mathrm{hitting}\;\mathrm{all}\;\mathrm{planes}}
{\mathrm{number}\;\mathrm{of}\;\mathrm{trajectories}\;\mathrm{started}}\,\cdot\,A_\mathrm{ap}
\end{equation}
where $A_\mathrm{ap}$ is the acceptance of the aperture, calculated
from (\ref{eq:accsingle}).\\
In order to obtain the acceptance for the PEBS detector, the Monte
Carlo approach was followed, using the dimensions of fig.~\ref{fig:pebs2d}.
\begin{figure}
\begin{center}
\includegraphics[width=\textwidth,angle=0]{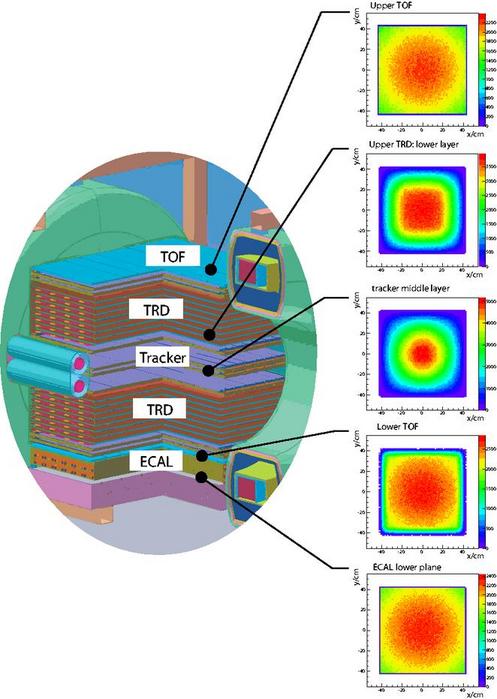}
\end{center}
\caption{Acceptance calculation: Distributions of intersection points
of trajectories found to be within the detector acceptance
on the individual detector planes.}
\label{fig:acceptance_planes}
\end{figure}
Figure~\ref{fig:acceptance_planes} shows the intersection points of
those straight line trajectories that are found to traverse the entire
detector on some of the idealised detector
planes, ranging from the upper TOF layer, which constitutes the
aperture of PEBS, to the lowest ECAL
layer. Figure~\ref{fig:acceptance_angles} shows the zenith and azimuth
angle distributions for the same trajectories. The mean zenith angle
is found to be $21^\circ$. The maxima in the azimuth angle
distribution are spaced by $90^\circ$ and are due to the square shape
of the detector.\\
\begin{table}[htb]
\begin{center}
\begin{tabular}{lrrrr}
&$\langle{}n_H\rangle$&channel number&amplitude&volume per event for
subdetector\\ \hline
Tracker&64&16 bit&16 bit&2048 bit\\
ECAL&85&16 bit&16 bit&2720 bit\\
TRD&22&12 bit&16 bit&616 bit\\
TOF&8&8 bit&16 bit&32 bit\\ \hline
total&&&&5416 bit\\
&&&&$\approx700\,\mathrm{byte}$\\ \hline
\end{tabular}
\end{center}
\caption{Calculation of the expected event size, based on
$20\,\mathrm{GeV}$ protons, with a mean noise level of $0.06$ pixels
in the tracker. The column labeled $\langle{}n_H\rangle$ gives the
mean number of hits to be stored, extracted from the Monte Carlo
simulation.}
\label{tab:data}
\end{table}
The acceptance of PEBS found from a Monte Carlo integration using
100~million trajectories was found to be
$3850\,\mathrm{cm}^2\mathrm{sr}$. The trigger rate that the readout
electronics and data acquisition chain needs to handle can then be
estimated from the proton flux. Taking the BESS 1997 data, taken at
solar minimum, allowing for $10\,\%$ Helium admixture, extrapolating
the data to zero energy and neglecting
all other particle species, an expected trigger rate of roughly
$1.5\,\mathrm{kHz}$ is calculated. Low-energetic particles, whose
fluxes are poorly known and highly variable, and that might saturate the
trigger, are expected to be efficiently absorbed in the remaining
atmosphere at flight altitude. For example, a residual grammage of
$3.7\,\mathrm{g}/\mathrm{cm}^2$ corresponds to $30\,\mathrm{mm}$ of
polycarbonate.\\
\par
Given these numbers, the expected data volume can be calculated
(tab.~\ref{tab:data}). For a flight duration of 40~days, roughly five
billion events will have to be recorded, requiring a disk capacity of
roughly $4\,\mathrm{TB}$.

\chapter{PEBS tracker prototype testbeam campaign}
\label{chapter:pebs_testbeam}
The design of a novel detector technology, such as the scintillating
fibre tracker for PEBS, requires constant experimental checks. In order
to study the response of the key tracker components -- scintillating
fibres read out by silicon photomultipliers -- to minimally ionising particles
with high statistics,
several prototype modules were subjected to a $10\,\mathrm{GeV}$
proton beam at the T9 beamline at the European Organisation for
Nuclear Research, CERN, near Geneva, over the years 2006-2008. Key
questions to be answered by these tests were the photo-electron yield
and the corresponding spatial resolution of the prototypes.\\
The testbeam
results can be compared to the predictions of dedicated Monte Carlo
simulations to see if all relevant effects affecting the behaviour of
the detector have been modelled appropriately. In turn, these
simulations can be incorporated into the full simulation of
PEBS.

\section{First testbeam 2006}
\label{sec:testbeam2006}
For a first proof of principle, two bundles of square fibres of
$300\,\mu\mathrm{m}$ width were used during the first testbeam which
took place in October 2006.

\subsection{Setup description}
\begin{figure}
\begin{center}
\begin{tabular}{cc}
\includegraphics[width=0.4\textwidth,angle=0]{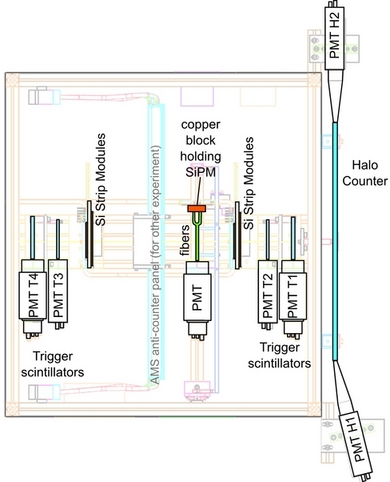}&
\includegraphics[width=0.6\textwidth,angle=0]{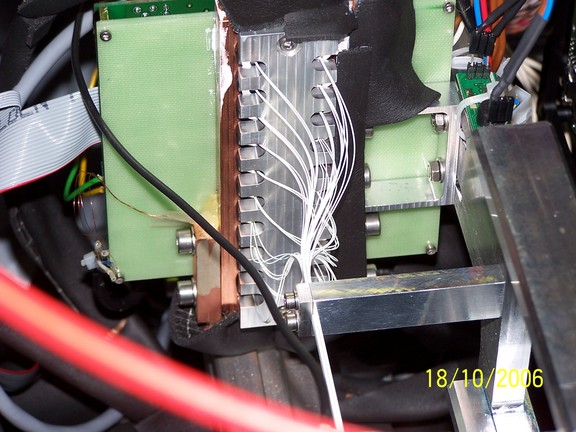}\\
\end{tabular}
\end{center}
\caption{{\it Left:} Schematic drawing of the testbeam
setup. Reprinted, with permission, from~\cite{ref:gregorio}. The beam
would traverse the setup from right to left.
{\it Right:} The two $3\times{}10$ fibre bundles up close. Three
consecutive fibres each were read out by an SiPM sitting in the copper
block visible behind the aluminium frame.}
\label{fig:testbeam_photo_fibrebunch}
\end{figure}

\begin{figure}[htb]
\begin{center}
\includegraphics[width=0.6\textwidth,angle=0]{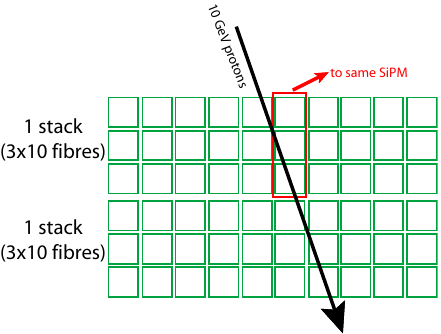}
\end{center}
\caption{Readout scheme used in the 2006~testbeam.}
\label{fig:readoutscheme}
\end{figure}

\begin{figure}[htb]
\begin{center}
\includegraphics[width=0.9\textwidth,angle=0]{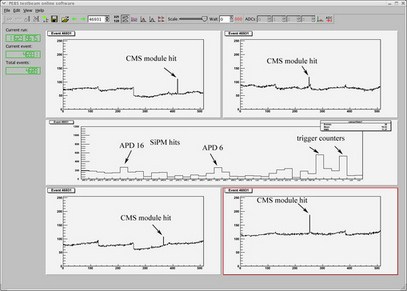}
\end{center}
\caption{Graphical user interface of the online software used for data
acquisition during the testbeam. A golden event is shown here: all
four CMS modules have exactly one cluster and there are hits in two SiPMs
reading out two consecutive fibre slots (numbers 6~and~16 in this
case). In addition, the energy deposition in the trigger counters is visible.}
\label{fig:testbeam_eventdisplay}
\end{figure}

A photograph of the testbeam setup is
\begin{figure}
\begin{center}
\includegraphics[width=0.8\textwidth,angle=0]{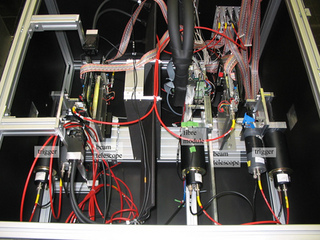}
\end{center}
\caption{A photograph of the setup used in the 2006~testbeam. The fibre bundles are
sitting in front of the central PMT, right of the AMS-02 anticounter
panels. Further out, the beam telescope modules and then the trigger
counters are visible.}
\label{fig:testbeam_photo_setup}
\end{figure}
shown in figure~\ref{fig:testbeam_photo_setup} and
figure~\ref{fig:testbeam_photo_fibrebunch} contains
a schematic drawing of the setup and a close-up view of the fibre
bundles. The fibre bundles were constructed from square,
$300\,\mu\mathrm{m}\,\times\,300\,\mu\mathrm{m}$ multi-clad fibres of
type Bicron BCF-20~\cite{ref:bicronfibres}. Each fibre was covered by a thin layer of white
extra-mural absorber (EMA) coating which is used to suppress optical
crosstalk between adjacent fibres. Each of the two fibre bundles was
10~fibres wide and 3~fibres high and was stabilised by glue. 
Each stack of three consecutive
fibres was read out by an individual SiPM, located inside the copper
block visible in fig.~\ref{fig:testbeam_photo_fibrebunch}. This readout scheme
is illustrated in figure~\ref{fig:readoutscheme}. The first
fibre bundle was read out by Photonique SiPMs of type SSPM-050701GR
while Photonique SSPM-0606EXP SiPMs were used for the second one. The
copper block containing the SiPMs was cooled to approximately
$-15^\circ\mathrm{C}$ using a combination of a Peltier element and
liquid cooling agent flowing through the copper block. The far side of
the fibre bundles, not read out by SiPMs, was connected to a
conventional photomultiplier tube (PMT) which was replaced
during the testbeam by a highly reflective aluminium foil in order to
study the enhancement of the photo-electron yield.\\
Several auxiliary elements completed the testbeam setup. In addition
to the fibre bundles, located at the centre of the setup, five
scintillator panels were used for triggering purposes, located at the
outermost ends of the setup. Two scintillator panels (T1 and T4 in
fig.~\ref{fig:testbeam_photo_fibrebunch}) had dimensions
$100\,\mathrm{mm}\,\times\,50\,\mathrm{mm}$, and two others (T2 and
T3) had dimensions $100\,\mathrm{mm}\,\times\,5\,\mathrm{mm}$, adapted
to the width of the fibre bundles. A veto counter, consisting of
scintillator material with a hole of
$60\,\mathrm{mm}\,\times\,60\,\mathrm{mm}$, was used to select clean
single-track events. All trigger scintillators were connected to
ordinary PMTs.\\
For the determination of the spatial resolution, a reference
measurement is mandatory, preferably with a detector whose resolution
is small compared to the expected result. For that purpose, four
silicon strip detectors identical to those used in the tracker endcap
of the CMS experiment at the Large Hadron Collider at CERN were
employed~\cite{ref:cms,ref:cmsmodules}. The modules were of type W4 and their shape was
trapezoidal because they were
intended for use in an $r\phi$-detector. The active area of each module was
$115.16\,\mathrm{mm}$ long and its width changed from
$58.06\,\mathrm{mm}$ to $71.272\,\mathrm{mm}$ over the length of
the module. With 512~readout strips, the readout pitch $p$ varied from
$113\,\mu\mathrm{m}$ to $139\,\mu\mathrm{m}$. Two modules each were used to determine the trajectory
coordinates parallel and perpendicular to the fibre direction,
respectively. The spatial resolution achieved by one silicon strip
sensor is therefore around $p/\sqrt{12}\approx{}36\,\mu\mathrm{m}$,
but this value becomes considerably smaller once charge sharing
between adjacent strips is taken into account.\\
In addition to the fibre modules for PEBS, two anticounter
scintillator panels for the AMS-02 experiment were tested at the same time~\cite{ref:philip}.\\
The trigger logic was implemented using NIM electronics. The output
signals of T1~and~T4 were discriminated and used as input to an
AND-coincidence unit. The same was done for T2~and~T3. A final
AND-coincidence was formed from these two AND-coincidences and
provided the main trigger. This final coincidence could be blocked by
the veto counter whose two PMTs were connected to a discriminator and
hence to an OR-coincidence. The output signal of this OR-coincidence
was connected to the common inhibit of the final AND-coincidence. A
dead time of $1\,\mu\mathrm{s}$ was started after an AND-coincidence
of T1~and~T4 using a re-triggerable gate generator to ensure a clean event. In addition, a dead time of
$10\,\mathrm{ms}$ was started in parallel to the main trigger signal
to make sure that event processing could be safely completed before
the next trigger.\\
The readout chain was started by the main trigger signal. The SiPM
signals were amplified and inverted and then digitised by LeCroy
2249A CAMAC ADCs whose gates were opened for $100\,\mathrm{ns}$ by the main trigger signal. The
bias voltages for the SiPMs were adjusted separately.\\
The four APV25 chips~\cite{ref:cmsapv} located on each CMS silicon sensor
modules sample their signal each $25\,\mathrm{ns}$ and store it in 192-cell circular pipelines.
An ARC system~\cite{ref:cmsarc}, normally used to test the CMS silicon
modules, was employed to control the APV25 chip and connected to the
same readout PC as the CAMAC controller used to steer the ADCs
digitising the SiPMs. Upon reception of the
main trigger signal, the ARC system would stop the signal sampling in
the APV25 chips and transmit the values in the pipeline slots
corresponding to a predefined offset in time. The data from the ARC
system and the CAMAC ADCs were then assembled into a combined event on
the readout PC and written to disk. The timing of the trigger and
readout system was adjusted such that the SiPM signals arrived during
the gate in the CAMAC ADCs while the offset in the ARC system was
matched to the delay between the passage of the beam particle and the
arrival of the main trigger signal. Coherent events were assured by
linking the entire readout software into one executable.
Many more details on the trigger logic and readout system are found
in~\cite{ref:gregorio}. The overall readout rate was limited by the ARC
system to about $60\,\mathrm{Hz}$. Taking the allocated beam and its
bunch structure into account, an effective trigger rate of roughly
$2\,\mathrm{Hz}$ was achieved over the course of the testbeam.\\
Figure~\ref{fig:testbeam_eventdisplay} shows
a screen shot of the dedicated online software developed for the
testbeam that offered a
graphical user interface to remotely control the data taking, monitor the
incoming raw data, and read the temperature and humidity sensors located
inside the testbeam setup.

\subsection{Analysis procedure}
\begin{figure}[htb]
\begin{center}
\includegraphics[width=0.9\textwidth,angle=0]{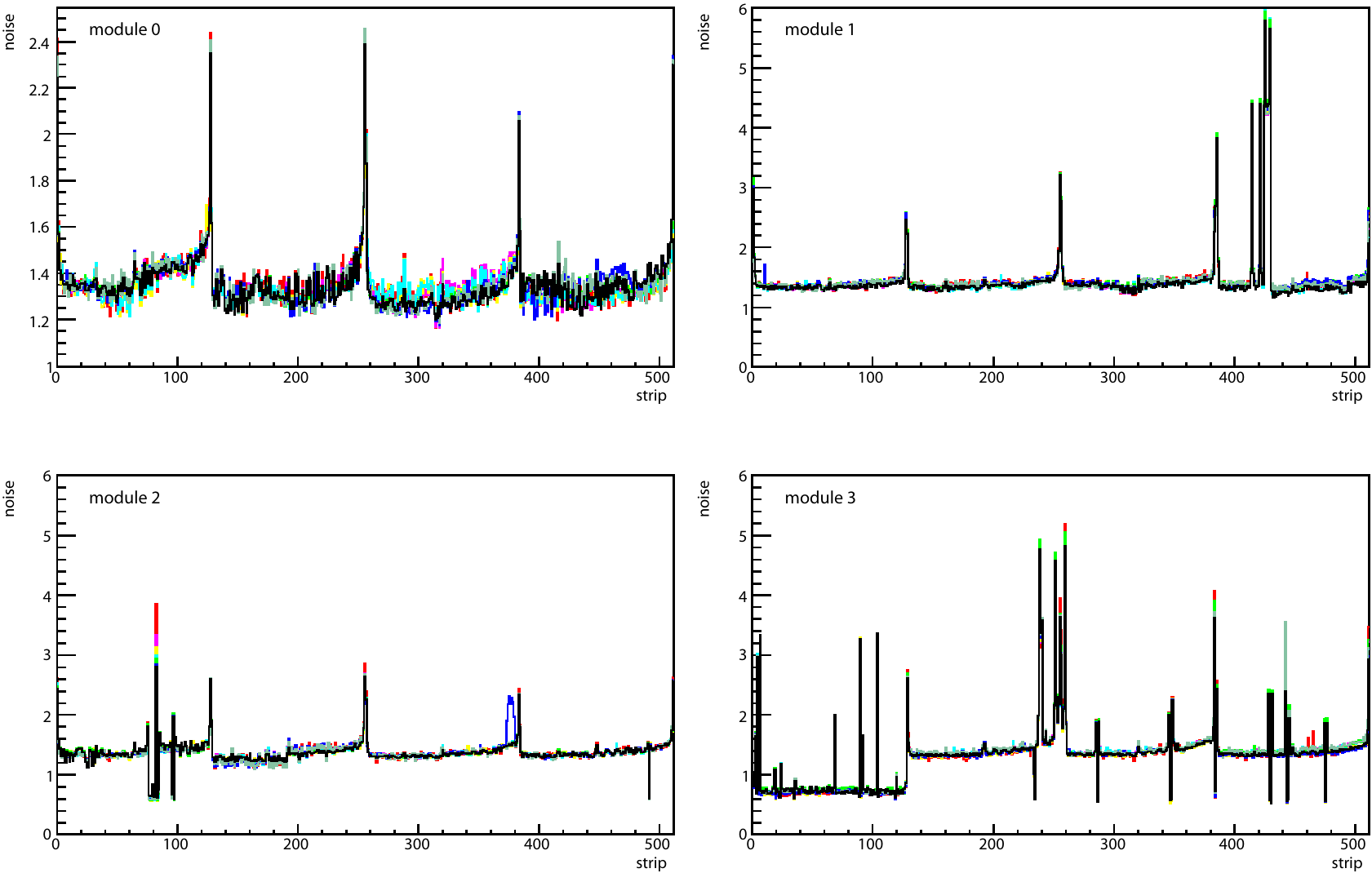}
\end{center}
\caption{Noise values (in ADC counts) for the four CMS beam telescope
modules, as a function of the strip number. Noise obtained during
different pedestal runs over the course of the testbeam is shown in
different colours. The scale for module~0 was magnified to illustrate
the changes over time. High noise values at strips 0, 128, 256, 384,
and~512 mark the boundaries between two APV25 readout chips. Strips with
anomalously high or low noise indicate noisy or dead channels. The
first APV25 on module~3 was inoperative.}
\label{fig:testbeam_cmsnoise}
\end{figure}
\begin{figure}[htb]
\begin{center}
\includegraphics[width=0.8\textwidth,angle=0]{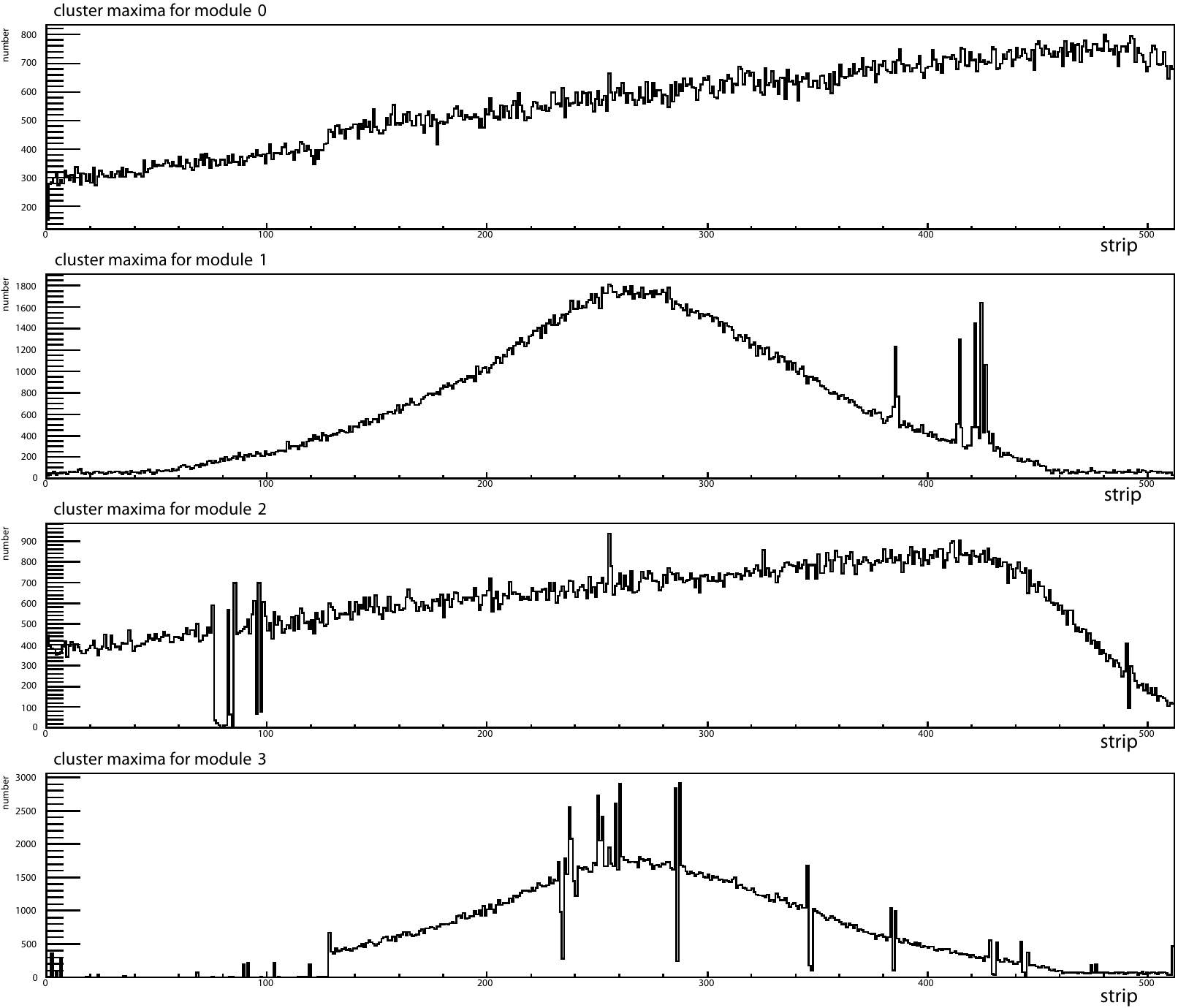}
\end{center}
\caption{Reconstructed cluster positions on the beam telescope
modules. The beam profile determines the shape of the
distributions. In additions, some noisy and some dead strips are
visible, especially on modules 1~and~2, and the first APV25 readout
chip for module~3 is seen to be not operational here. The strips of modules 1~and~3
were oriented parallel to the fibres, while those of modules 0~and~2 were
oriented perpendicular to the fibres.}
\label{fig:testbeam_clustermax}
\end{figure}
The first step in the analysis chain is the identification of clean
single-track events in the beam telescope. The information available
for each event consists of the raw ADC counts for all 2048~channels in
the four beam telescope modules. First, the pedestal and noise values
for each channel are calculated using dedicated pedestal runs that
were taken with random triggers every few hours throughout the
testbeam. Here, the pedestal and noise are defined as the empiric mean
and standard deviation of the ADC counts distribution in each
channel. For a given event, the significance $s_i$ of a channel $i$
belonging to APV25 $j$ is then
calculated as
\begin{equation}
\label{eq:significance}
s_i=\frac{S_i-p_i-c_j}{\sigma_i}
\end{equation}
from the signal $S_i$, pedestal $p_i$ and noise $\sigma_i$ of channel
$i$, all given in ADC counts. $c_j$ is the common-mode noise of the
128~channels belonging to the
given APV25, defined as the median of the pedestal-subtracted
signals. The common-mode noise is a uniform shift in the pedestals of
all channels of a chip that varies from event to event, usually caused
by an external source of noise.
As an example, figure~\ref{fig:testbeam_cmsnoise} shows the noise
values found for the four modules in several pedestal runs scattered
over the duration of the testbeam. The variation in time is small and
a good strip has a noise level of around $1.4$~ADC counts. There are,
however, a number of noisy or dead strips, and one of the modules had
one dead APV25 chip.\\
Next, a cluster finding algorithm is
applied to each module. A~seed strip having at least $s_i\geq{}4$ is
searched and neighbouring clusters with $s_{i\pm{}k}\geq{}2$ are
added to the cluster. A~single-track event is then defined as having
exactly one such cluster on each module in the beam
telescope. Figure~\ref{fig:testbeam_clustermax} depicts the
reconstructed cluster positions for each of the four beam telescope
modules. The strips of modules 1~and~3
were oriented parallel to the fibres, while those of modules 0~and~2 were
oriented perpendicular to the fibres. The distributions are governed
by the vertical and horizontal beam profiles, respectively, as well as
by the noisy and dead strips mentioned above. Here, the cluster
position was calculated as the weighted mean of the participating
strip numbers, using the amplitudes corrected for pedestal and
common-mode noise as weights.\\
The next task is to calculate the intersection point of the particle
track with the fibre module. The trapezoidal shape of the beam
telescope modules complicates this matter. A track fit is performed,
with the goal of finding the trajectory $\vec{x}(z)$ which is taken to
be a straight line. $z$ is the coordinate perpendicular to the planes
of the beam telescope modules. For a given $z$, let $u$ and $v$ be the coordinates of a
point measured in the frame of a beam telescope module. The
coordinates $x$ and $y$ of this point as measured in the frame of the testbeam
setup box can be obtained by using the rotation matrix
\[
R=\left(
\begin{array}{cc}
\cos\alpha&-\sin\alpha\\
\sin\alpha&\cos\alpha\\
\end{array}
\right)
\]
where $\alpha$ is the angle of rotation for the strip at the centre of
a given cluster which can be calculated from the strip number,
the orientation of the module inside the setup, and the module geometry.
The covariance matrix of a point measured on the module is taken to be
\[
V=\left(
\begin{array}{cc}
\sigma_u^2&0\\
0&\sigma_v^2\\
\end{array}
\right)
\]
$\sigma_u$ and $\sigma_v$ are determined by the strip length $l$ and
pitch $d$ at the centre: $\sigma_u=l/\sqrt{12}$ and
$\sigma_v=d/\sqrt{12}$. The covariance matrix $U$ in the $x,y$-frame needed for the
calculation of the $\chi^2$ of the track fit can then be obtained
using standard error propagation~\cite{ref:cowan} as
\begin{equation}
\label{eq:errprop}
U=A\,VA^T
\end{equation}
where $A$ is the matrix of derivatives
\[
A=\left(
\begin{array}{cc}
\frac{\partial{}x}{\partial{}u}&\frac{\partial{}x}{\partial{}v}\\
\frac{\partial{}y}{\partial{}u}&\frac{\partial{}y}{\partial{}v}\\
\end{array}
\right)
\]
Introducing an index $m\in[1,4]$ to label the four clusters on the
track, and letting $\Delta{}x_m$ and $\Delta{}y_m$ be the track
residuals on the $m$-th module, the $\chi^2$ that has to be minimised
during the track fit is given by
\begin{equation}
\label{eq:testbeam_chi2}
\begin{split}
\chi^2&=\sum\limits_{m=1}^4{}\Delta\vec{x}_m^T\,U_m^{-1}\,\Delta\vec{x}_m\\
&=\sum\limits_{m=1}^4{}
\frac{
\Delta{}x_m^2c_m^2\sigma_v^2+
\Delta{}x_m^2s_m^2\sigma_u^2+
\Delta{}y_m^2c_m^2\sigma_u^2+
\Delta{}y_m^2s_m^2\sigma_v^2+
2\Delta{}x_m\Delta{}y_m{}c_ms_m(\sigma_y^2-\sigma_x^2)
}{\sigma_u^2\sigma_v^2}
\end{split}
\end{equation}
where $\Delta\vec{x}_m\equiv(\Delta{}x_m,\Delta{}y_m)$, and
$c_m\equiv\cos\alpha_m$ and $s_m\equiv\sin\alpha_m$.\\
\par
In order to obtain the mean photo-electron yield and spatial
resolution of the fibre modules,
\begin{figure}[htb]
\begin{center}
\begin{tabular}{cc}
\includegraphics[width=0.5\textwidth,angle=0]{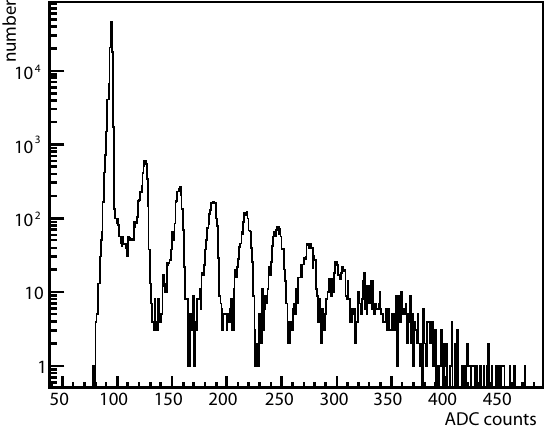}&
\includegraphics[width=0.5\textwidth,angle=0]{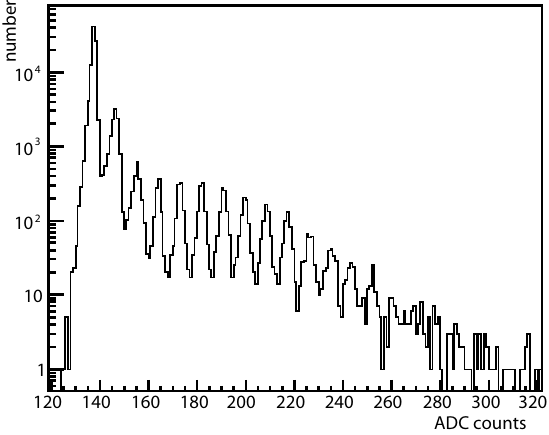}\\
\end{tabular}
\end{center}
\caption{Two examples (fibre channels 1~and~12) of raw ADC spectra observed with SiPMs of
type SSPM-050701GR ({\it left}) and SSPM-0606EXP ({\it right}), for runs
with perpendicular incidence and reflective foil at one end of the
fibre bunches, as used for the pedestal and gain calibration.}
\label{fig:testbeam_rawadcspectra}
\end{figure}
the individual SiPMs have to be calibrated to yield the pedestal $p$,
i.e.~the location of the 0-photo-electron-peak, and the gain $g$, i.e.~the
mean distance between two photo-electron peaks, measured in ADC
counts. The signal $s$ (in photo-electrons) is then calculated from
the amplitude $a$ (in ADC counts) simply by
\begin{equation}
\label{eq:sipmcalibration}
s=\frac{a-p}{g}
\end{equation}
Two examples of the raw ADC spectra observed for runs
with perpendicular incidence and reflective foil at one end of the
fibre bunches are shown in
figure~\ref{fig:testbeam_rawadcspectra}. The individual photo-electron-peaks
are easily identified, and a semi-automatic peak finding procedure was
used for the determination of their locations. Pedestal and gain were
then calculated individually for each SiPM.
\par
For the analysis of the prototype performance, the beam telescope can
now be used to interpolate the intersection point $(x,y)$ of the
particle trajectory with the fibre module.  With the number of
photo-electrons determined for all SiPM channels in a given event,
all fibres whose amplitude exceeded a cut of 2~photo-electrons were
identified and
\begin{figure}[htb]
\begin{center}
\includegraphics[width=1.0\textwidth,angle=0]{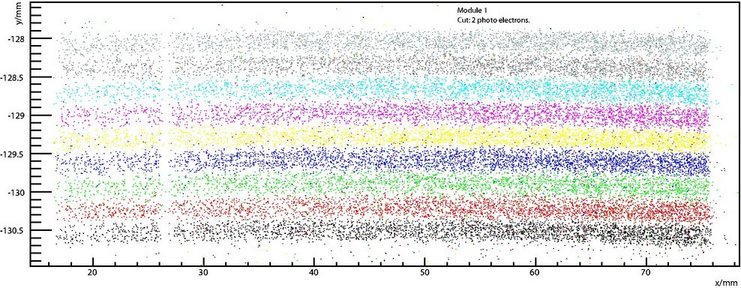}
\end{center}
\caption{Coordinates interpolated from beam telescope tracks for hits
in the fibres of fibre module~1. A cut of 2~photo-electrons was
employed in this case. Each colour corresponds to a single fibre. The
figure may be interpreted as a picture of the fibre bunch, taken by
the beam telescope. The gap at $x\sim{}26\,\mathrm{mm}$ is due to dead
strips in module~2 of the beam telescope, as shown in figure~\ref{fig:testbeam_clustermax}.}
\label{fig:testbeam_fibreProjections_module1}
\end{figure}
the intersection point was histogrammed for those fibres. The results
for one of the fibre bunches are shown in
figure~\ref{fig:testbeam_fibreProjections_module1}. Each fibre is
shown in a different colour and the figure can be interpreted as
showing a picture of the fibre module taken by the beam telescope.
\begin{figure}[htb]
\begin{center}
\begin{tabular}{cc}
\includegraphics[width=0.55\textwidth,angle=0]{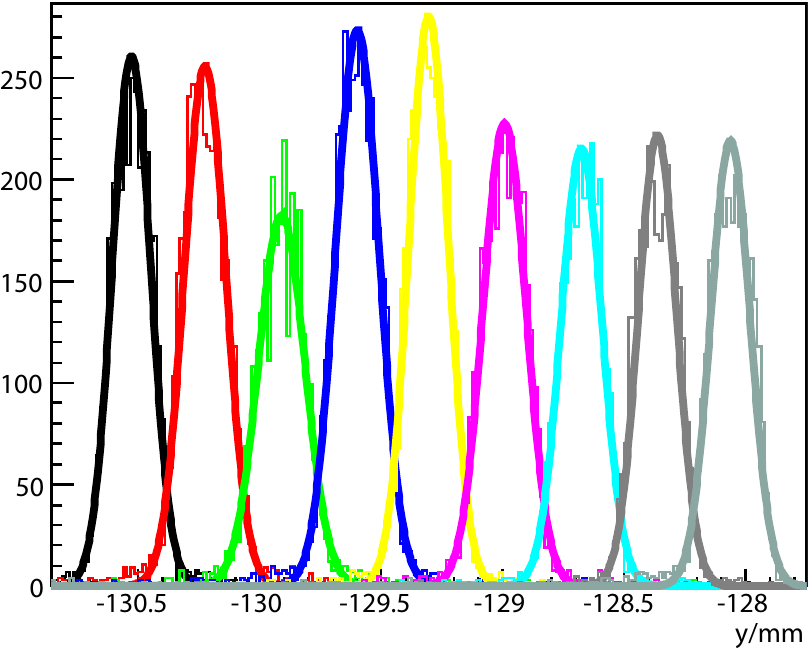}&
\includegraphics[width=0.4\textwidth,angle=0]{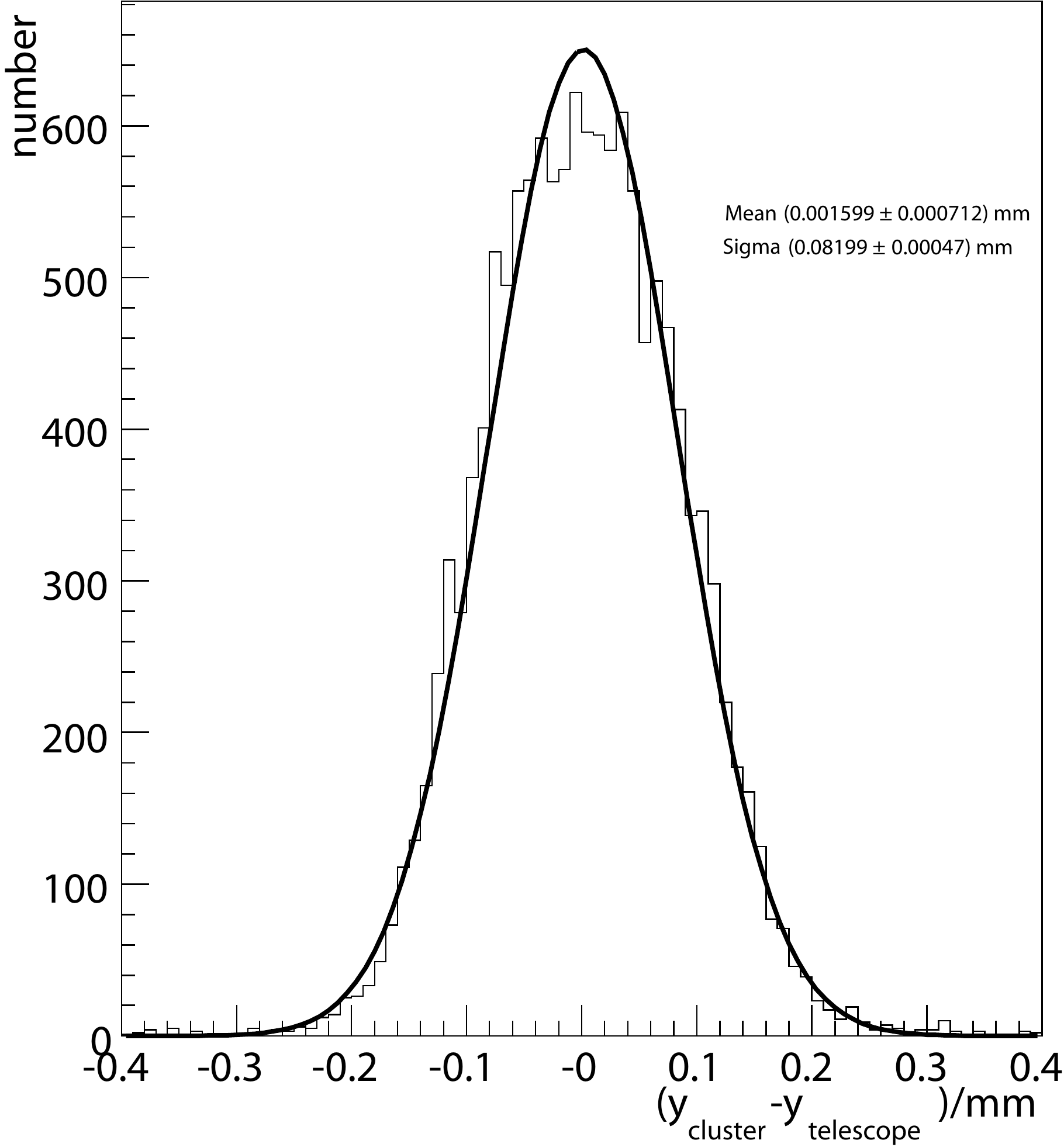}\\
\end{tabular}
\end{center}
\caption{{\it Left:} Determination of the fibre positions from projections
onto the $y$-axis. {\it Right:} Determination of the spatial
resolution from the
distribution of track residuals. Distributions for fibre module~1 are shown
as an example.}
\label{fig:testbeam_trackresiduals}
\end{figure}
The determination of the spatial resolution requires the knowledge of the
fibre positions with respect to the beam telescope. These are measured
by projecting the histograms shown in the figure to the $y$-axis and
fitting a Gaussian to each of them. The mean of the Gaussian is then
taken as the position of the respective fibre, as illustrated in
figure~\ref{fig:testbeam_trackresiduals}.
\begin{figure}[htb]
\begin{center}
\begin{tabular}{cc}
\includegraphics[width=0.52\textwidth,angle=0]{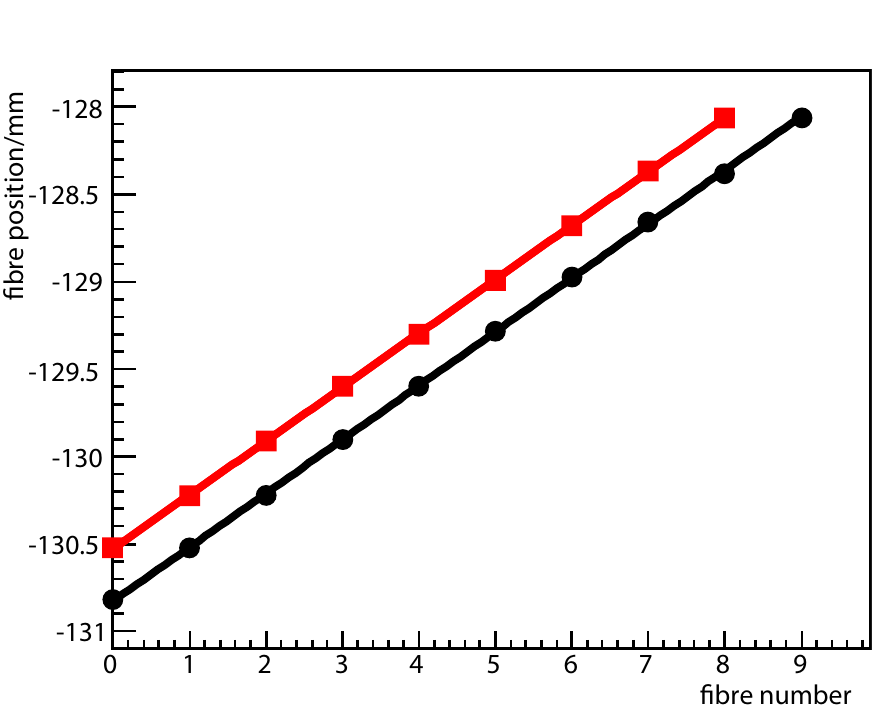}&
\includegraphics[width=0.48\textwidth,angle=0]{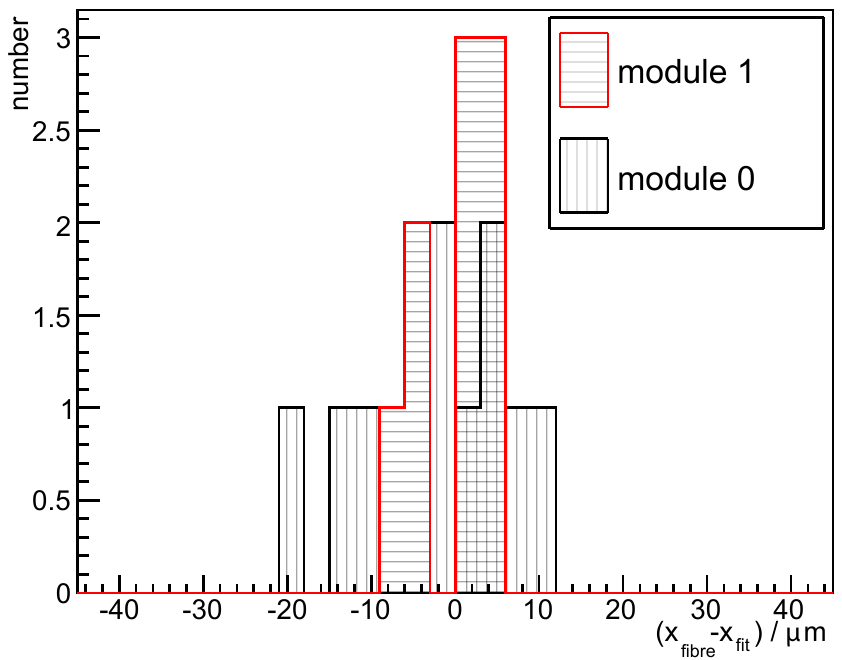}\\
\end{tabular}
\end{center}
\caption{Determination of mean fibre pitch from a
straight-line fit to the positions obtained as shown in
figure~\ref{fig:testbeam_trackresiduals} ({\it left}) and residual
plots for the fits ({\it right}). The mean fibre pitch is
$308.4\,\mu\mathrm{m}$ for both modules, and the root mean square values for the
deviation of the fibre positions from the ideal ones are
$9.2\,\mu\mathrm{m}$ for module~0 and $4.1\,\mu\mathrm{m}$ for module~1.}
\label{fig:testbeam_fibrepos}
\end{figure}
Figure~\ref{fig:testbeam_fibrepos} contains a plot of the fibre
positions so obtained as a function of the fibre number. The slope
determined from a straight-line fit is then the mean distance between
two fibres. The value of $308.4\,\mu\mathrm{m}$ found for both modules
is in excellent agreement with the nominal fibre thickness of
$300\,\mu\mathrm{m}$ considering that a small amount of glue may be
present between the fibres. The figure also shows histograms of the
residuals of the fibre positions with respect to a straight line, and
root mean square values for the
deviation of the fibre positions from the ideal ones are found to be
$9.2\,\mu\mathrm{m}$ and $4.1\,\mu\mathrm{m}$ for the two modules.\\
The determination of the spatial resolution proceeds as follows. Using
an algorithm analogous to the one used for cluster finding on the beam
telescope modules, clusters of neighbouring fibre channels are
identified. Here, a cut of 3~photo-electrons is used for the seed
channel, and 2~photo-electrons for the neighbouring channels. The $y$
coordinate is then calculated as the weighted mean of all $N$ channels
belonging to the cluster,
\begin{equation}
\label{eq:clusterwmean}
y=\frac{\sum\limits_{i=1}^Ns_iy_i}{\sum\limits_{i=1}^Ns_i}
\end{equation}
where the signals $s_i$ are given in photo-electrons and the fibre
positions were determined as in
figure~\ref{fig:testbeam_trackresiduals}. The difference of
the track intersection point as measured by the
beam telescope and the cluster position according to
(\ref{eq:clusterwmean}), $y_\mathrm{cluster}-y_\mathrm{telescope}$, was
then histogrammed and the standard deviation of a Gaussian
distribution fitted to the histogram is called the spatial
resolution. This is illustrated in
figure~\ref{fig:testbeam_trackresiduals}, too. For
perpendicular incidence, spatial resolutions of $82\,\mu\mathrm{m}$
and $88\,\mu\mathrm{m}$ are found for the two modules, read out by
SiPMs of types SSPM-050701GR and SSPM-0606EXP with lower and higher
noise rate, respectively. This has to be compared to the expectation of
$b/\sqrt{12}\approx{}87\,\mu\mathrm{m}$ where $b=300\,\mu\mathrm{m}$
is the fibre width. However, the region of the fibre that will
actually produce scintillation light is reduced somewhat by the
extension of the protective cladding of around $12\,\mu\mathrm{m}$ to
both sides.\\
Using the information provided by the beam telescope,
\begin{figure}[htb]
\begin{center}
\begin{tabular}{cc}
\includegraphics[width=0.5\textwidth,angle=0]{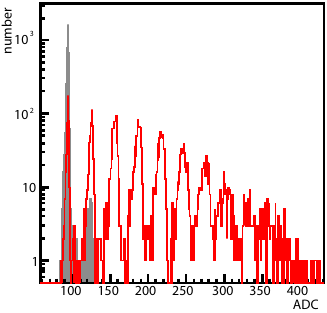}&
\includegraphics[width=0.5\textwidth,angle=0]{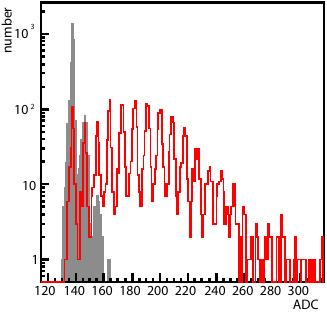}\\
\end{tabular}
\end{center}
\caption{Two examples (fibre channels 1~and~12) of raw ADC spectra observed with SiPMs of
type SSPM-050701GR ({\it left}) and SSPM-0606EXP ({\it right}), for runs
with perpendicular incidence and reflective foil at one end of the
fibre bunches. Events for which the beam telescope indicated a hit in
the corresponding fibres are histogrammed in red, while the grey
histograms show the dark count. The histograms have been scaled to
equal integral. The individual photo-peaks are clearly visible.}
\label{fig:testbeam_adcspectra}
\end{figure}
the signal spectra and noise spectra can be extracted from the raw ADC
spectra. Two examples, for the same channels as in fig.~\ref{fig:testbeam_rawadcspectra}, are shown in
figure~\ref{fig:testbeam_adcspectra}. An event is counted for the
signal spectrum if the beam telescope indicated a hit within
$100\,\mu\mathrm{m}$ of the fibre centre and for the noise spectrum if
the hit was at least $500\,\mu\mathrm{m}$ away from the fibre
centre. For a correct determination of the photo-electron yield, the
effect of optical pixel crosstalk must be taken into
account. Occasionally, photons created during an avalanche in one
of the SiPM pixels will trigger a discharge in one of the neighbouring
pixels, with a probability $p_\mathrm{crosstalk}$. In the analysis
of~\cite{ref:gregorio}, $1.94\,\pm\,0.27$ and $3.12\,\pm\,0.19$ photo-electrons were found for
the SiPMs of type 050701GR without and with reflective foil,
respectively. For the SSPM-0606EXP, the numbers are $3.59\,\pm\,0.24$
and $5.43\,\pm\,0.31$. $p_\mathrm{crosstalk}$ is found to be on the
5-$10\,\%$ level.\\
\par
The mounting containing the fibre modules could be rotated in steps of
$\alpha_\mathrm{slot}=10^\circ$ around an axis parallel to the fibres to study the
dependency of the spatial resolution on the angle of incidence.
\begin{figure}[htb]
\begin{center}
\includegraphics[width=0.5\textwidth,angle=0]{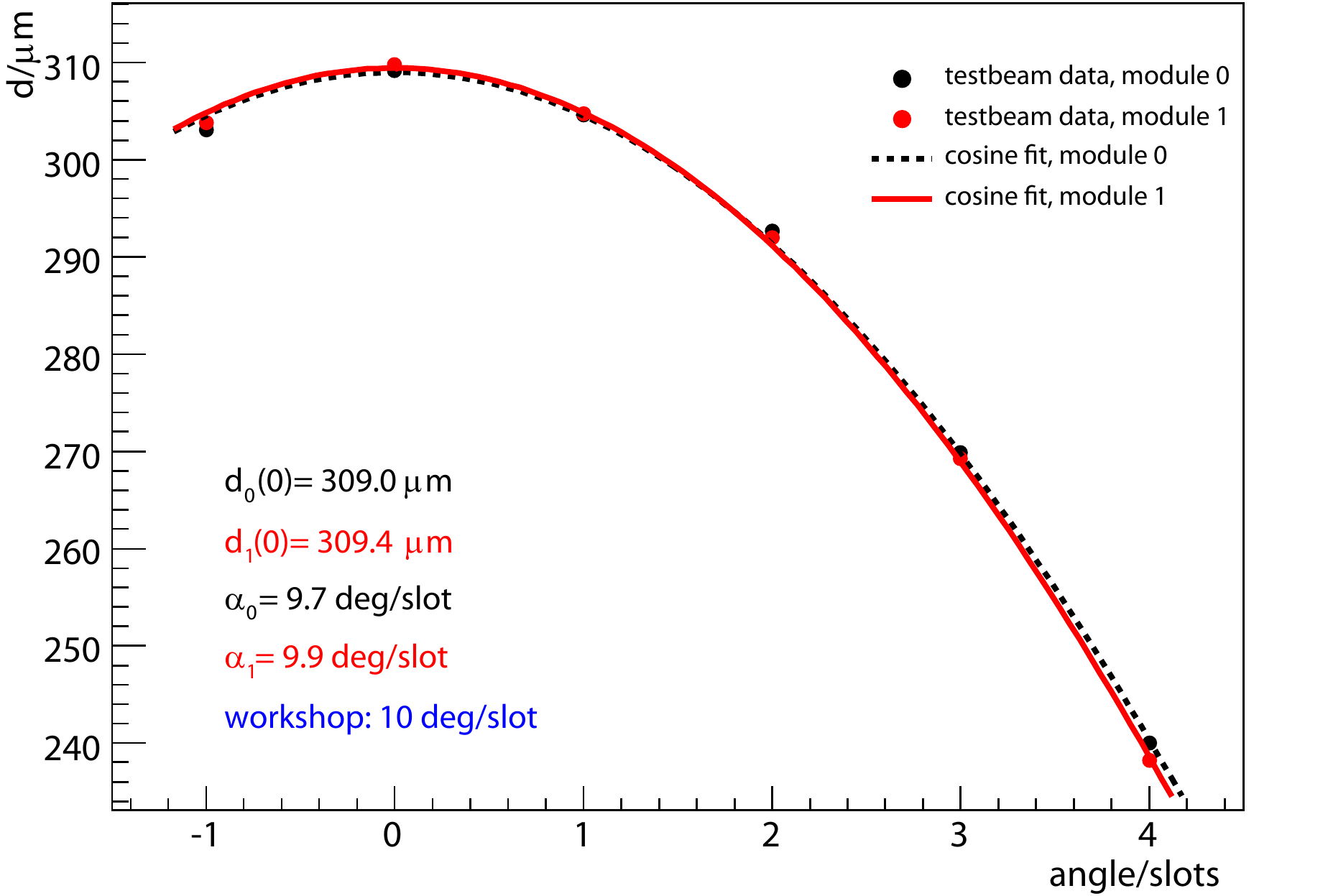}
\end{center}
\caption{Mean distance of adjacent fibre slots, projected into the
beam telescope plane, as a function of rotation angle. The angle is
given as the number of slots moved forward on the rotator arm, with a
nominal value of $10^\circ$/slot.
Allowing for
roughly $10\,\mu\mathrm{m}$ of glue between the fibres, the nominal
values both of the fibre thickness and the angular distance between
two slots are recovered.}
\label{fig:testbeam_distancevsangle}
\end{figure}
Data at nominal values of $-10^\circ$, $0^\circ$, $10^\circ$,
$20^\circ$, $30^\circ$, and $40^\circ$ were taken, but without the
reflective foil. The analysis
procedure outline so far was repeated for each orientation. As a
cross-check, the fibre distance was obtained like in
fig.~\ref{fig:testbeam_fibrepos} and is shown in figure~\ref{fig:testbeam_distancevsangle}
as a function of the nominal angle of rotation. As the measured
distance $d$ is the one projected into the plane of the beam
telescope, it can be expected to follow
\begin{equation}
\label{eq:cosine}
d=d_0\cos(n\alpha_\mathrm{slot})
\end{equation}
where $n$ is the number of slots on the fibre mounting that the
modules were rotated by. Indeed, a fit of a cosine function gives
excellent agreement with the nominal values, as shown in the figure.\\
For the determination of the spatial resolution as a function of the
angle of incidence, treated in the next section, an additional effect
has to be taken into account. As a particle traversing the fibre
\begin{figure}[htb]
\begin{center}
\includegraphics[width=0.5\textwidth,angle=0]{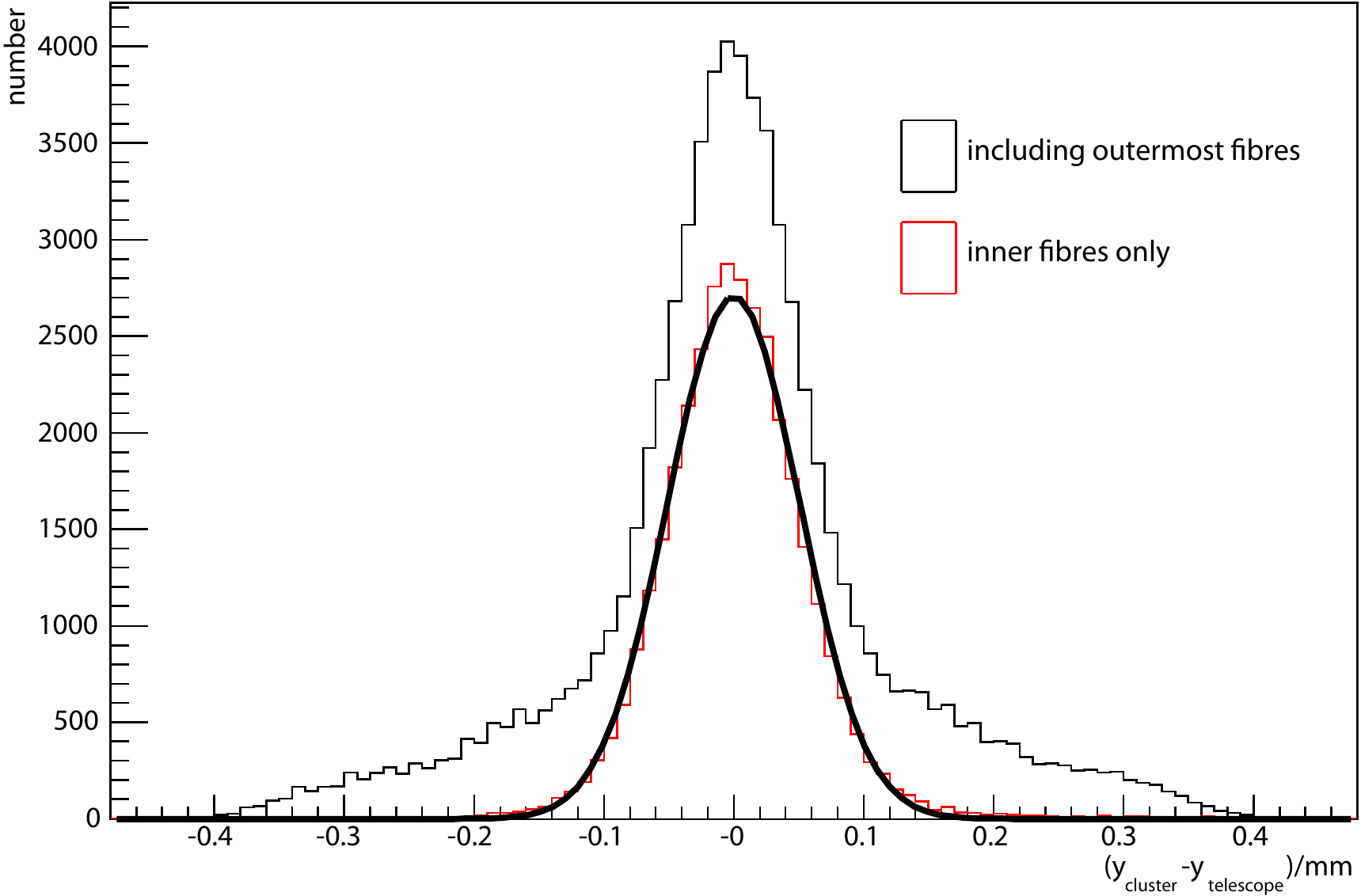}
\end{center}
\caption{The effect of ignoring the outer two fibre slots on each side
on the calculated spatial resolution is shown here. For the inner
fibres, the distribution of the track residuals becomes
Gaussian. Monte Carlo
data at $40^\circ$ rotation of the fibre bunch were used here.}
\label{fig:testbeam_outermostfibres}
\end{figure}
modules with a certain angle will deposit energy across more than one
SiPM channel, the information for tracks grazing the outermost fibres
will be incomplete and the intersection points determined according to
(\ref{eq:clusterwmean}) will be of limited accuracy. This is
illustrated in figure~\ref{fig:testbeam_outermostfibres} showing that
the intersection points of trajectories can only be measured reliably
at high angles if the outermost two fibres are ignored. Naturally,
this can only happen at the expense of tracking efficiency.

\subsection{Prototype performance and comparison to Monte Carlo study}
\begin{figure}[htb]
\begin{center}
\includegraphics[width=0.8\textwidth,angle=0]{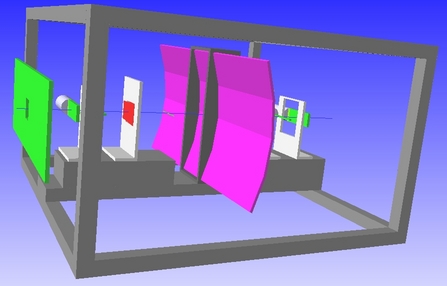}
\end{center}
\caption{Visualisation of the detector setup used in the Geant4
simulation of the testbeam. The veto and trigger scintillators are
shown in green, CMS beam telescope models are painted in red, and the
AMS-ACC panels are pink. The fibre bunch is barely visible on this
scale and is located in front of the ACC panels. Structural elements
and PMTs are shown in grey.}
\label{fig:testbeam_sim_vis}
\end{figure}
For the generalisation of the testbeam results, a Geant4 Monte Carlo
simulation of the testbeam setup was created
(fig.~\ref{fig:testbeam_sim_vis}). All components in or close to the
beam were included, namely the veto and trigger scintillators with
PMTs, beam telescope modules and ACC panels with mounting frames, and
the scintillating fibres, as well as the aluminium carrier frame. The
simulation and digitisation of the scintillating fibres and the SiPMs
used for their readout closely follows the prescriptions used for the
full PEBS simulation as outlined in section~\ref{sec:sipms}.
\begin{figure}[htb]
\begin{center}
\begin{tabular}{cc}
\includegraphics[width=0.5\textwidth,angle=0]{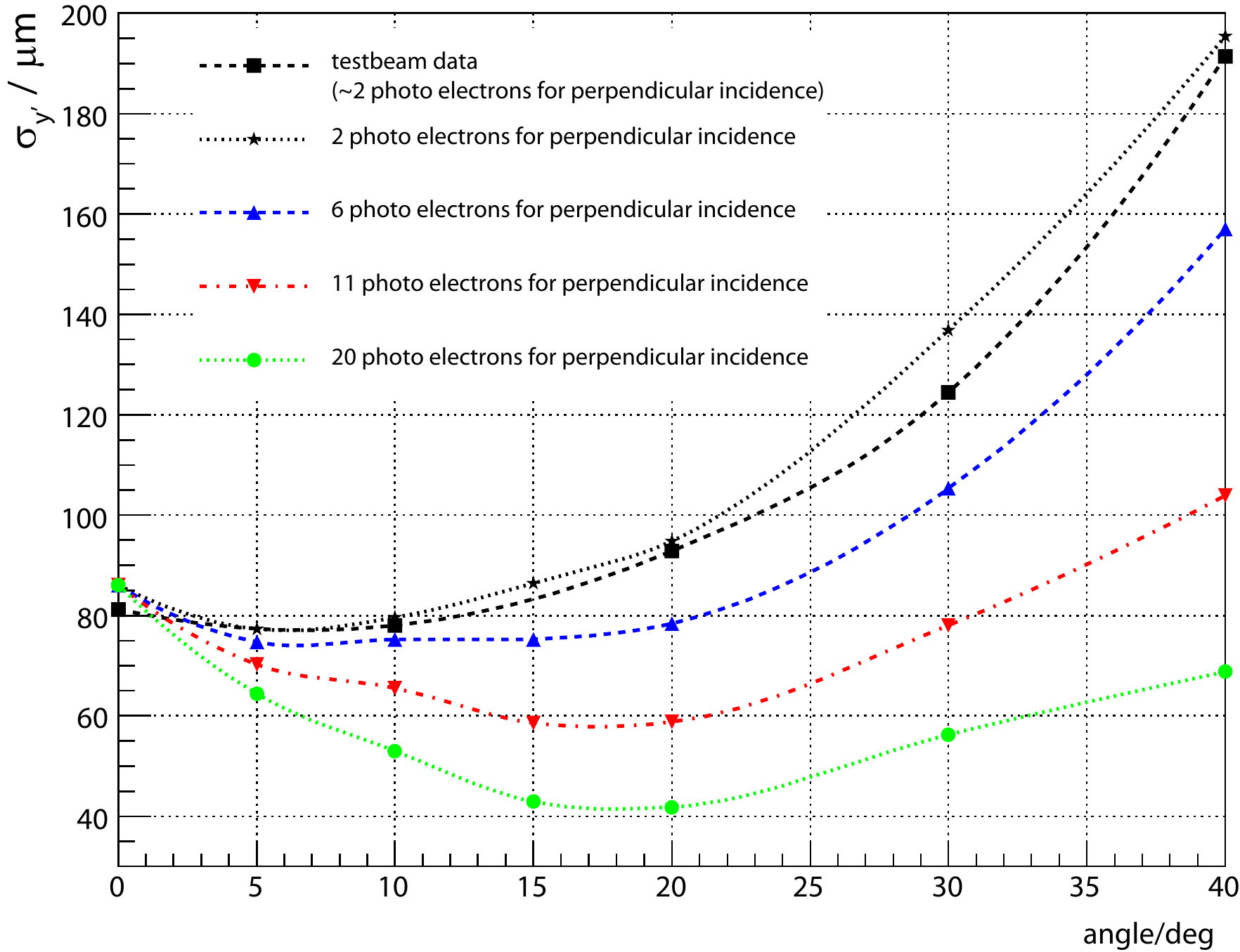}&
\includegraphics[width=0.5\textwidth,angle=0]{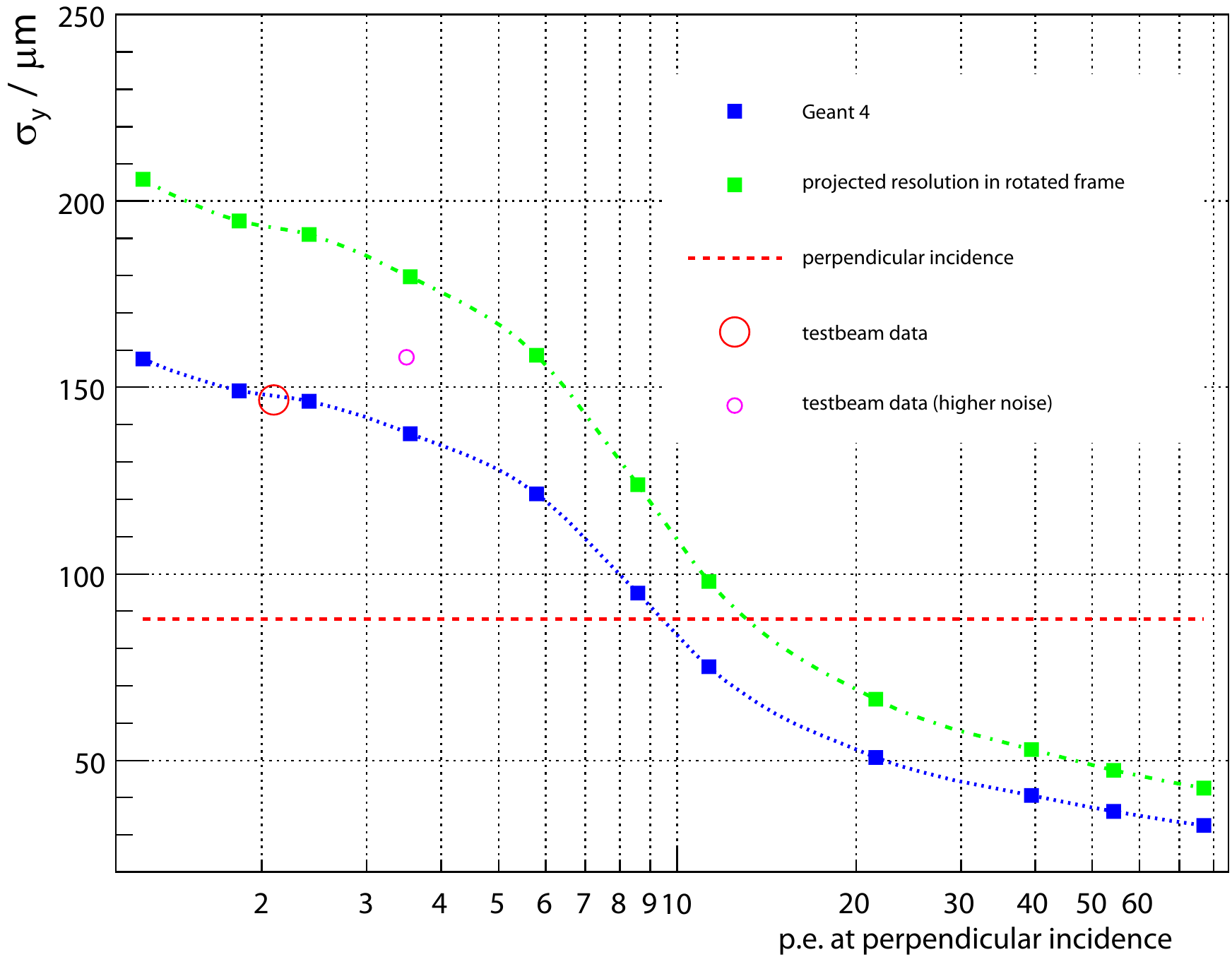}\\
\end{tabular}
\end{center}
\caption{{\it Left:} Spatial resolution as a function of angle of
incidence. Simulation results for various overall light yields, given
by the number of mean photo-electrons at perpendicular incidence, are
shown, together with testbeam results.
{\it Right:} Spatial resolution of the fibre module as obtained from the
Geant4 simulation. The spatial resolution is shown for an angle of
incidence of $40^\circ$ as a function of the photo-electron yield for
perpendicular incidence. The blue curve shows the resolution obtained
in the plane parallel to the beam telescope modules, while the green
curve shows the resolution in the plane parallel to the fibre module,
as calculated from (\ref{eq:projcorr}). The
value at perpendicular incidence is indicated by the horizontal
line. For comparison, the values found in the testbeam are shown, for
the fibre bunch read out by the SSPM-050701GR (big circle) and the
SSPM-0606EXP (small circle). In the latter case, the deviation from
the Monte Carlo curve is explained by the higher noise of this SiPM type.}
\label{fig:testbeam_spatialresolutionmc}
\end{figure}
For an energy deposition in one of the CMS modules, two hits were
generated in the channels adjacent to the intersection point, with
their amplitudes calculated from a simple linear interpolation based
on the position of the energy deposition.
The results were then stored in the same format as the actual testbeam
data so that the same analysis as described in the previous section
can be applied to the simulated data.\\
\par
From a Gaussian fit to the distribution of track residuals
$y_\mathrm{track}-y_\mathrm{cluster}$, the spatial resolution can now
be calculated for both data and simulation
(fig.~\ref{fig:testbeam_spatialresolutionmc}). This has been done for
all angles of incidence at which data were taken during the
testbeam. It must be noted however, that the spatial resolution
determined from the track residuals is measured in the coordinate
frame fixed by the beam telescope. The spatial resolution
$\sigma_{y^\prime}$ measured in the coordinate frame fixed to the
fibre bunches can be calculated by correcting for the projection
effect as follows:
\begin{equation}
\label{eq:projcorr}
\sigma_{y^\prime}=\frac{\sigma_y}{\cos\alpha}
\end{equation}
where $\alpha$ is the angle of rotation of the fibre bunches as
above.\\
Overall, there is good agreement between the simulation and the
testbeam data over the entire range of the angle of incidence. This
indicates a good modelling of the detector physics. The
simulation can therefore be used to extrapolate the spatial resolution for
higher photo-electron yields. The results indicate that there is no
dependence of the spatial resolution on the photo-electron yield for
small angles of incidence. This is expected because a particle will
only deposit energy in a single fibre in this case, leading to a
resolution of $b/\sqrt{12}$. For higher angles however, the cluster
position is calculated as a weighted mean of the fibre positions which
promises an improvement in the spatial resolution as the
discretisation imposed by the fibre width is somewhat removed. On the
other hand, the individual weights are subject to statistical fluctuations that
tend to deteriorate the spatial resolution. The fluctuations become
smaller for higher light yields. Also, the fact that the energy
deposition is distributed across several fibres at higher angles of
incidence means that each individual fibre will receive less
photo-electrons. The interplay of these effects produces a steep
decline in
the spatial resolution for low photo-electron yields at high angles of
incidence. For higher photo-electron yields, the dependence on the
angle of incidence becomes much less severe, and a minimum is reached at
intermediate angles. For very high photo-electron yields, the
distribution of the signal across fibres becomes very effective in
improving the spatial resolution beyond the value of $b/\sqrt{12}$.

\section{Second testbeam 2008}
\label{sec:testbeam2008}
After the first testbeam had been completed, the technology for production of ribbons of
round fibres of $250\,\mu\mathrm{m}$ diameter was developed allowing
the construction of a tracker module prototype according to the design
outlined in section~\ref{sec:tracker_design}.
The second testbeam took place at CERN in June~2008 and lasted two
weeks. Right in time for the testbeam, arrays of SiPMs made by
Hamamatsu became available and were used to create one tracker
prototype module. The response of this module consisting of scintillating fibres of
$250\,\mu\mathrm{m}$ diameter and a Hamamatsu SiPM array to $10\,\mathrm{GeV}$
protons at perpendicular incidence will be studied in the
remainder of this section in order to justify the assumptions made in
section~\ref{sec:sipms}. In addition, a readout board
for the SiPMs based on the VA32 amplifier/shaper chip was
developed. The CMS beam telescope modules were replaced by two silicon
ladders built for the
AMS-02 tracker. This meant that the same DAQ system could now be used
for both the beam telescope and the SiPMs, allowing a readout rate that
was an order of magnitude higher than in the first testbeam.

\subsection{Setup description}
\begin{figure}[htb]
\begin{center}
\includegraphics[width=\textwidth,angle=0]{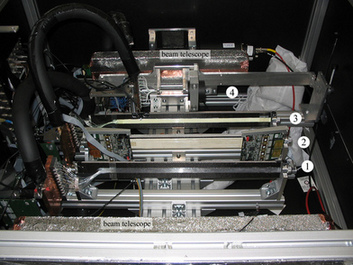}
\end{center}
\caption{A photograph of the 2008~testbeam setup after installation of
the module containing the Hamamatsu SiPM arrays. Four fibre modules
are visible at the centre of the picture: from front to rear, round Kuraray
fibres with individual readout (1), the module read out by Hamamatsu SiPM
arrays, with round Kuraray fibres (front) and round Bicron fibres
(rear) ((2), fig.~\ref{fig:tb08_detailphotos}), round Bicron
fibres with individual readout (3), and square Bicron fibres with individual readout (4). The beam telescope ladders are located
inside the boxes wrapped with aluminium foil. The PMT reading out the
rear trigger counter is visible, too.}
\label{fig:tb08_photo}
\end{figure}
\begin{figure}[htb]
\begin{center}
\begin{tabular}{cc}
\includegraphics[width=0.7\textwidth,angle=0]{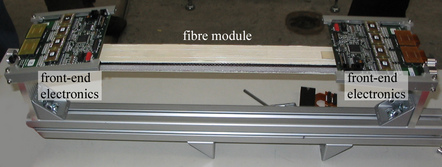}&
\includegraphics[width=0.3\textwidth,angle=0]{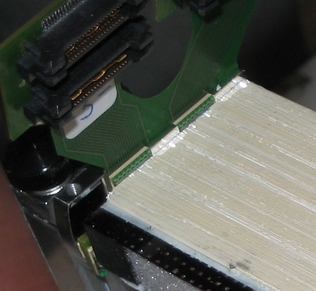}\\
\end{tabular}
\end{center}
\caption{{\it Left:} The fibre module with readout by Hamamatsu SiPM
arrays before installation. The front-end electronics boards used in
the testbeam are
visible at the sides.
{\it Right:} Close-up view of one end of the module before connection
to the front-end electronics board. The SiPM arrays themselves are
sitting in front of the fibres, but their connector pads as well as
part of the reflective foils are visible.}
\label{fig:tb08_detailphotos}
\end{figure}
The testbeam setup (fig.~\ref{fig:tb08_photo}) is based on the one used for the first testbeam. 
Two trigger scintillators, a veto scintillator, two silicon ladders
serving as a beam telescope, and four fibre module prototypes were
mounted to an aluminium frame. The setup was hermetically closed on
all sides to remove any ambient light and equipped with desiccant bags
to reduce humidity to a minimum. The
fibre modules could be rotated around their longitudinal axis to study
their behaviour with respect to different angles of incidence. Again a
$10\,\mathrm{GeV}$ proton beam provided by the T9~beamline at CERN was
used.\\
\par
Five prototype modules were tested:
\begin{itemize}
\item A module made of square Bicron BCF-20 fibres of
$300\,\mu\mathrm{m}$ width with the same
geometry as in the first testbeam for comparison, but this time
read out by individual Hamamatsu S10362-11-100C SiPMs.
\item A module made out of five layers of round Bicron BCF-20 fibres of
$250\,\mu\mathrm{m}$ diameter, the inner 16~fibre columns of which were
read out by individual Hamamatsu S10362-11-100C SiPMs.
\item A module made of five layers of round Kuraray SCSF-81M fibres~\cite{ref:kuraray} of
$250\,\mu\mathrm{m}$ diameter, the inner 16~fibre columns of which were
read out by individual Hamamatsu S10362-11-100C SiPMs.
\item A module with a fibre ribbon consisting of $5\times128$ fibres
on each side, one ribbon made of Bicron fibres, the other one made of
Kuraray fibres, with readout by \mbox{IRST-itc} SiPM arrays located on a
hybrid as sketched in fig.~\ref{fig:trackermodule}. This module was replaced
halfway through the testbeam by 
\item a module with a fibre ribbon consisting of $5\times128$ fibres
on each side, one ribbon made of Bicron fibres, the other one made of
Kuraray fibres, with readout by IRST-itc arrays on one side and
Hamamatsu MPPC~5883 arrays (fig.~\ref{fig:tb08_arrays}) on the other side
(fig.~\ref{fig:tb08_detailphotos}).
Only one Hamamatsu array was operative due to difficulties in
the hybrid production. This one was located in front of the Kuraray fibres.
The data taken by this array will be studied in the following.
\end{itemize}
While the individual SiPMs
were located inside copper blocks that were cooled to around
$-12^\circ\mathrm{C}$, the SiPM arrays were operated at a
temperature that varied between $16^\circ\mathrm{C}$ and
$18^\circ\mathrm{C}$ along with the day-night cycle.
All SiPMs were connected to VA\_32/75 chips. The VA~\cite{ref:ideas} is a
high dynamic range charge-sensitive preamplifier-shaper
circuit, with simultaneous sample and hold. It has a multiplexed
analog readout and is available in versions with different numbers of
input channels and gains. It was used for the readout of both the beam telescope and
the SiPMs here.\\
\par
The AMS-02 ladders~\cite{ref:amsmodules} used as a beam telescope are
made up of double-sided silicon micro-strip detectors. Each ladder is
composed of several silicon sensors of dimensions
$72.045\,\times\,41.360\,\mathrm{mm}^2$ and $300\,\mu\mathrm{m}$
thickness. The sensors have metallisations on both sides, parallel to
the length of the ladder on the front (S-)side and perpendicular on
the rear (K-)side, with readout pitches of $110\,\mu\mathrm{m}$ and
$208\,\mu\mathrm{m}$, respectively. The readout strips of the
individual sensors are daisy-chained. In total, a ladder provides
1024~readout channels, 640 for the S-side and 384 for the K-side. As
the readout strips on the K-side are transverse, a single readout
channel there corresponds to a series of spatial positions due to the
daisy-chaining. The
spatial resolution of an AMS-02 ladder has been demonstrated to be
$10\,\mu\mathrm{m}$ on the S-side and $30\,\mu\mathrm{m}$ on the
K-side~\cite{ref:cristinziani}. The signals induced on the readout
strips are preamplified and shaped by VA64 chips located on the
front-end electronics boards adjacent to the ladders.\\
\par
The DAQ system used in the testbeam is based on components used for
the AMS-02 experiment~\cite{ref:amsdaq}. Two TDR2 boards gather the
signals stored on reception of a trigger signal by the VAs of both the
ladders and the SiPMs. A TBS
board provides the bias voltage for the
silicon ladders while a TPSFE board creates the low voltages needed by
the system. A JINF board serves as event builder, assembling the
information provided by the TDR2s to form a coherent event. This event
is then stored until the JINF is queried by the readout software. The
readout software runs on an ordinary PC connected to the JINF from its
parallel port operated in EPP mode.
All boards are powered and communicate with each other via a common
backplane.\\
The TDR2s can be operated in either raw or compressed mode. For the
SiPM readout, the raw mode is used meaning that the raw data from all
1024 channels, only a fraction of which are actually connected to an
SiPM, are transferred, at the expense of readout rate. Contrary to this,
the silicon ladders are read out in compressed mode. Using
pedestal and noise values calculated for each channel from data taken
in a dedicated calibration run, performed before each physics run,
clusters are identified by a digital signal processor (DSP) located on
the TDR2. A correction for common-mode noise is performed
automatically, too. Only the clusters are then included in the event, resulting
in a significant reduction in event size and corresponding increase in
readout rate. Overall, a readout rate of roughly $200\,\mathrm{Hz}$
was achieved in the testbeam.\\
\par
The two trigger scintillators had dimensions of
$14\times4\times1\,\mathrm{cm}^3$ and were read out by
ordinary PMTs. The veto scintillator was the same as used in the first
testbeam. The trigger scheme was again based on NIM electronics and
had to be optimised for speed as the hold
signal for the VA\_32/75 chips reading out the SiPMs had to arrive
around $65\,\mathrm{ns}$ after the particle crossing as the SiPM
signal pulse reaches its maximum at that time. This was achieved by
using only a single coincidence of the two PMTs reading out the
trigger scintillators. The two PMTs at each side of the veto panel
and the busy signal of the JINF were used as a veto on the
coincidence. A post-event dead-time of $150\,\mu\mathrm{s}$ was started
along with the hold signal, and a pre-event dead-time of
$5\,\mu\mathrm{s}$ was started by a signal from any of the trigger
PMTs.

\subsection{Data analysis}
The analysis presented here deals with the data taken with one
Hamamatsu SiPM array in front of a ribbon of Kuraray
fibres, with $10\,\mathrm{GeV}$ protons at perpendicular
incidence. The fibre ribbon consisted of five layers of fibres of
$250\,\mu\mathrm{m}$ diameter, at a nominal pitch of
$275\,\mu\mathrm{m}$. The spacing of the readout channels on the SiPM
array was $250\,\mu\mathrm{m}$. In total, $672\,003$~events were taken
in this configuration over a time interval of $51.5$~hours.\\
The coordinate system employed here is defined such that the $z$-axis
is parallel to the beam while the $x$-axis is parallel to the S-side strips
of the beam telescope ladders and therefore roughly parallel to the
fibres. Therefore, the fibres are used to measure the $y$-coordinate.\\
The analysis procedure begins with the analysis of the beam telescope
\begin{figure}[htb]
\begin{center}
\begin{tabular}{cc}
\includegraphics[width=0.5\textwidth,angle=0]{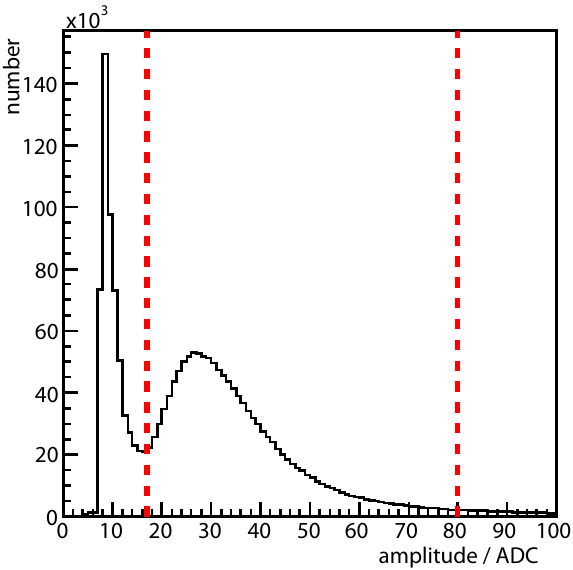}&
\includegraphics[width=0.5\textwidth,angle=0]{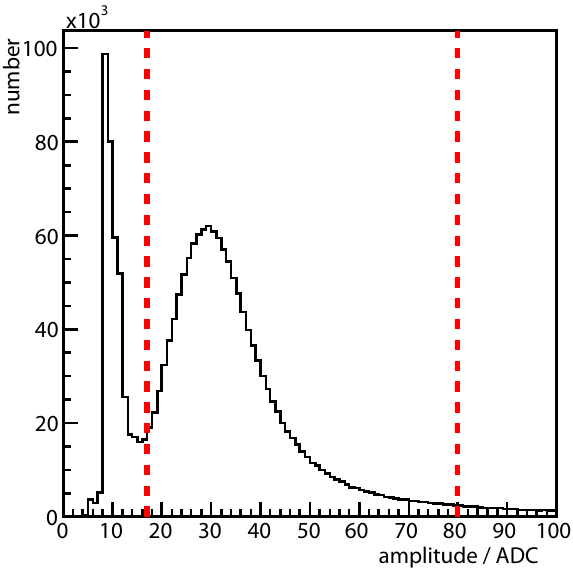}\\
\end{tabular}
\end{center}
\caption{Distributions of cluster amplitudes on the first ({\it left})
and second ({\it right}) beam telescope ladders. Only clusters with
amplitudes between the cut values indicated by the dashed lines are
considered in the analysis.}
\label{fig:tb08_bt_ladderAmp}
\end{figure}
data with the aim of identifying clean single-track events which will
then be used to unambiguously study the behaviour of the fibre module.
\begin{figure}[htb]
\begin{center}
\begin{tabular}{cc}
\includegraphics[width=0.5\textwidth,angle=0]{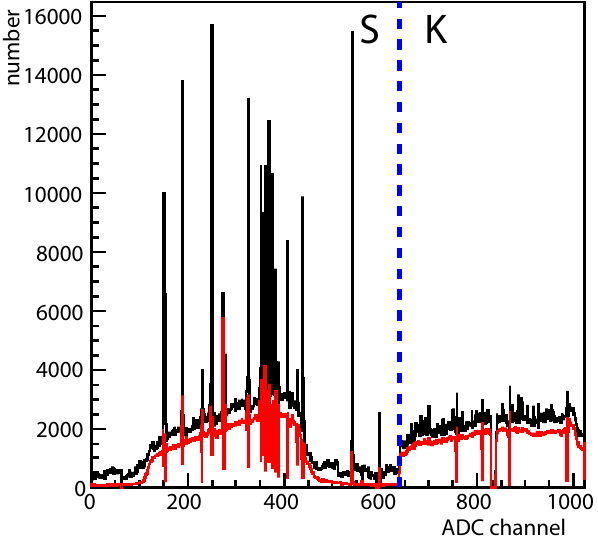}&
\includegraphics[width=0.5\textwidth,angle=0]{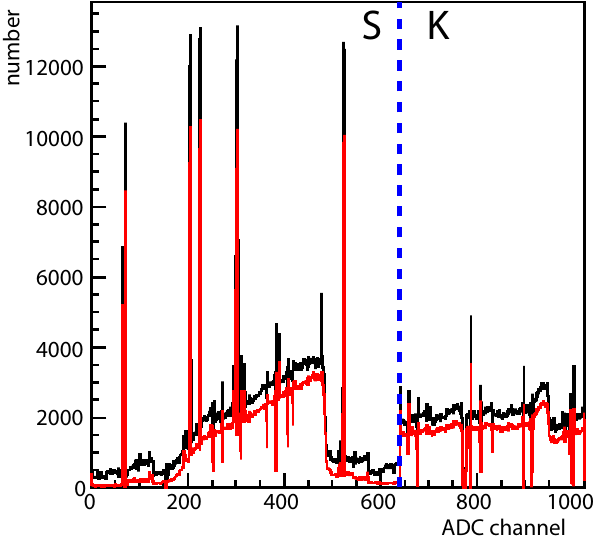}\\
\end{tabular}
\end{center}
\caption{Distributions of cluster positions for the first ({\it left})
and second ({\it right}) beam telescope ladders. The dashed lines
separate the front (S-) and back (K-) sides of the ladders with 640 and 384
readout channels, respectively. Distributions before
cuts are shown in black, distributions after the cut on the cluster
amplitude (fig.~\ref{fig:tb08_bt_ladderAmp}) are shown in red.}
\label{fig:tb08_bt_ladderHit}
\end{figure}
The amplitude distributions (fig.~\ref{fig:tb08_bt_ladderAmp}) for the
two ladders show two distinct components, a contribution from noise
clusters at low amplitudes and a Landau-shaped contribution from
signal clusters. Quality cuts at 17~and 80~ADC counts are applied to
remove noise clusters and clusters with unusually high energies, and
only clusters inside this interval are considered in the following.\\
The effect of this cut is visible in the distributions of the hit
positions for the two ladders
(fig.~\ref{fig:tb08_bt_ladderHit}). Here, the locations of 
clusters are shown, separately for all clusters and for clusters
allowed by the amplitude cuts. Readout channels 0~to~639 correspond to
the front (S-)side while channels 640~to~1023 belong to the rear
(K-)side. For the S-side, the area where one expects to find signal
clusters is limited by the fact that the width of the trigger counters
is smaller than the width of a beam telescope ladder. Indeed, a number
of noisy strips are apparent on both ladders as well as a substantial
number of noise hits outside the acceptance of the beam
telescope. Those can already be removed by a fair amount using the
amplitude cut.\\
In all this, the position $c$ of a cluster of $n$ strips is calculated as a simple
weighted mean
\begin{equation}
\label{eq:tb08_btclusterwmean}
c=\frac{\sum_{i=1}^{n-1}c_iA_i}{\sum_{i=1}^{n-1}A_i}
\end{equation}
At this point, $c$ is still given as a channel number. The $A_i$ are
\begin{figure}[htb]
\begin{center}
\begin{tabular}{cc}
\includegraphics[width=0.5\textwidth,angle=0]{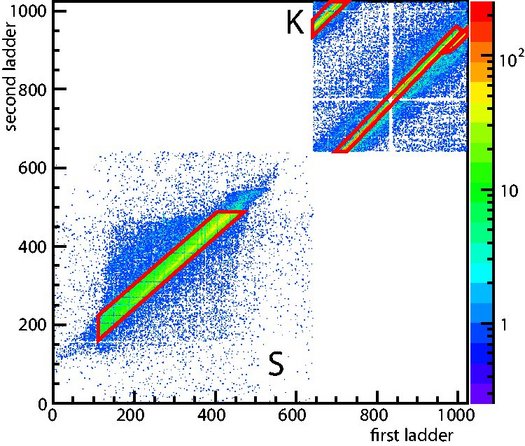}&
\includegraphics[width=0.5\textwidth,angle=0]{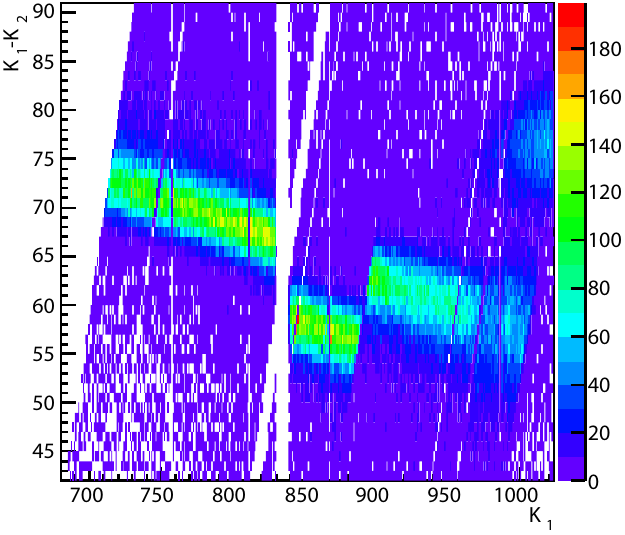}\\
\end{tabular}
\end{center}
\caption{{\it Left:} Positions of clusters on the two beam telescope
ladders. Only events with exactly one cluster remaining after the cuts
on cluster amplitude (fig.~\ref{fig:tb08_bt_ladderAmp}) on each side of each ladder
are included. The collimation of the beam allows the rejection of
spurious hits and only the events within the red cut areas are used
for the analysis of the performance of the fibre modules. The peculiar
structure for the K side (strip number $\geq{}640$) is due to the
daisy-chaining of the readout strips.
{\it Right:} Difference $K_1-K_2$ in back-side cluster
positions, measured in strip numbers, vs.~$K_1$. The offset caused by
the gap between individual silicon wafers is visible at
$K_1=850\sim900$ and the ambiguity in cluster position due to the
daisy-chaining is seen at $K\gtrsim{}1000$.}
\label{fig:tb08_bt_corr_strip_diff}
\end{figure}
the strip amplitudes as provided by the TDR2, i.e.~pedestal subtracted
and corrected for common-mode noise, and the $c_i$ are the
corresponding channel numbers. The TDR2 adds the two neighbouring
strips to an identified cluster for safety reasons. But they do not contain information
about the location of the track. The sum is therefore taken excluding the
outermost two strips.\\
The next requirement imposed on an event is that it have exactly one
signal cluster on each side of each of the two ladders.
In order to further remove spurious tracks from the data sample, the
fact that the proton beam was collimated and of limited angular
aperture can be taken into account
(fig.~\ref{fig:tb08_bt_corr_strip_diff} {\it left}). For particles
passing the setup under a given angle whose variation is limited, a
correlation between the cluster positions $p_1$ and $p_2$ on the first and second
ladders is expected. This is exactly what is seen in the beam
telescope data, and a set of cuts in the $(p_1,p_2)$-area are
defined. Events falling within the cut on the S-side and one of the
cuts of the K-side at the same time are considered clean single-track
events in the following.\\
On the K-side, the effect of the daisy-chaining
of readout channels is apparent and the $x$ coordinate of a passing
track a priori cannot be reconstructed unambiguously. Three distinct
branches in the $(p_1,p_2)$-area can be identified
(fig.~\ref{fig:tb08_bt_corr_strip_diff} {\it left}). However, the
assumption of a smooth beam profile and the limited length of the
trigger counters employed can be used to remove the
ambiguities. Plotting the K-side cluster position $K_1$ on the first
ladder against $K_1-K_2$ where $K_2$ is the K-side cluster position on
the second ladder removes the first ambiguity and reveals a new
problem (fig.~\ref{fig:tb08_bt_corr_strip_diff} {\it right}). The small
gap between the individual silicon sensors in the
ladder causes a shift, exactly located at channel $832=640+384/2$, in
the beam profile that has to be corrected for. An empirical shift of
4~readout channels is introduced for the appropriate clusters. At the
same time, the tracks located in the upper right corner in the figure
are shifted to the leftmost end.\\
Using these shifts and corrections for the K-side, and knowing the
readout pitch of $110\,\mu\mathrm{m}$ and
$208\,\mu\mathrm{m}$
\begin{figure}[htb]
\begin{center}
\begin{tabular}{cc}
\includegraphics[width=0.5\textwidth,angle=0]{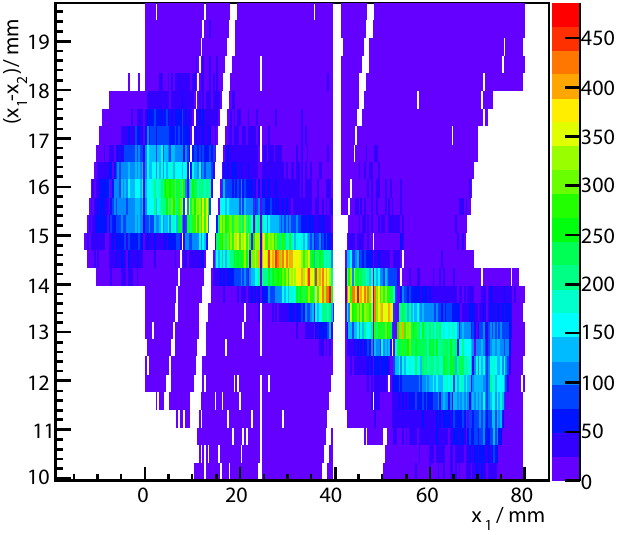}&
\includegraphics[width=0.5\textwidth,angle=0]{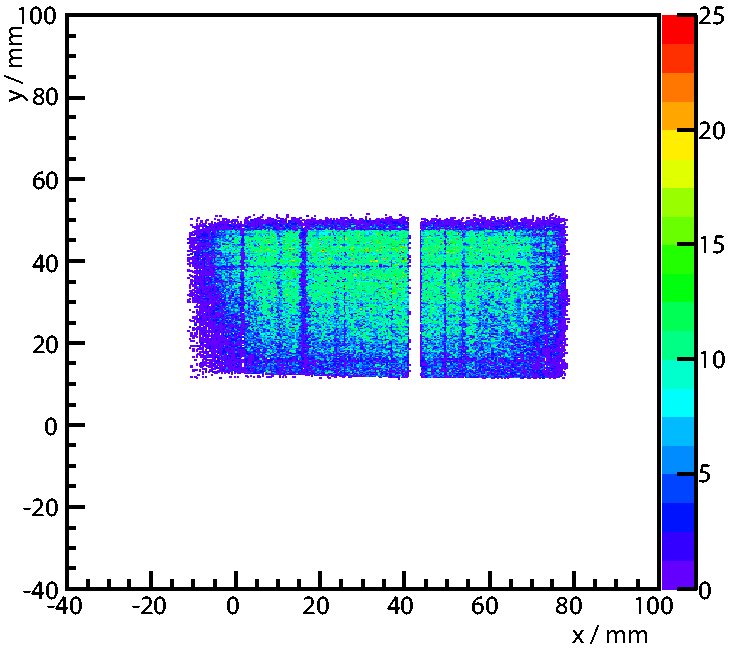}\\
\end{tabular}
\end{center}
\caption{{\it Left:} Correlation of the $x$ coordinates measured by
the two beam telescope ladders after corrections for ambiguity and offset.
{\it Right:} $x$ and $y$ coordinates interpolated at the $z$-position
of the first trigger counter from tracks passing all quality selection
cuts. The exposure is governed by the beam profile and some gaps
are due to noisy or dead strips on the beam telescope.}
\label{fig:tb08_bt_final}
\end{figure}
for the S- and K-sides, respectively,
the $y$ and $x$ coordinates can be calculated for each cluster. A
track object is then created for those events with exactly one signal
cluster on each side of each of the ladders so that the track position
at a given depth $z$ in the setup can be interpolated
unambiguously. The beam profile so reconstructed (fig.~\ref{fig:tb08_bt_final}) turns
out to be smooth and regular. As an example, the figure also shows a
picture of one of the trigger counters taken by the beam
telescope. The exposure is determined by the beam profile and by the
location of a number of dead or noisy strips on either of the beam
telescope ladders.
\begin{figure}[htb]
\begin{center}
\includegraphics[width=\textwidth,angle=0]{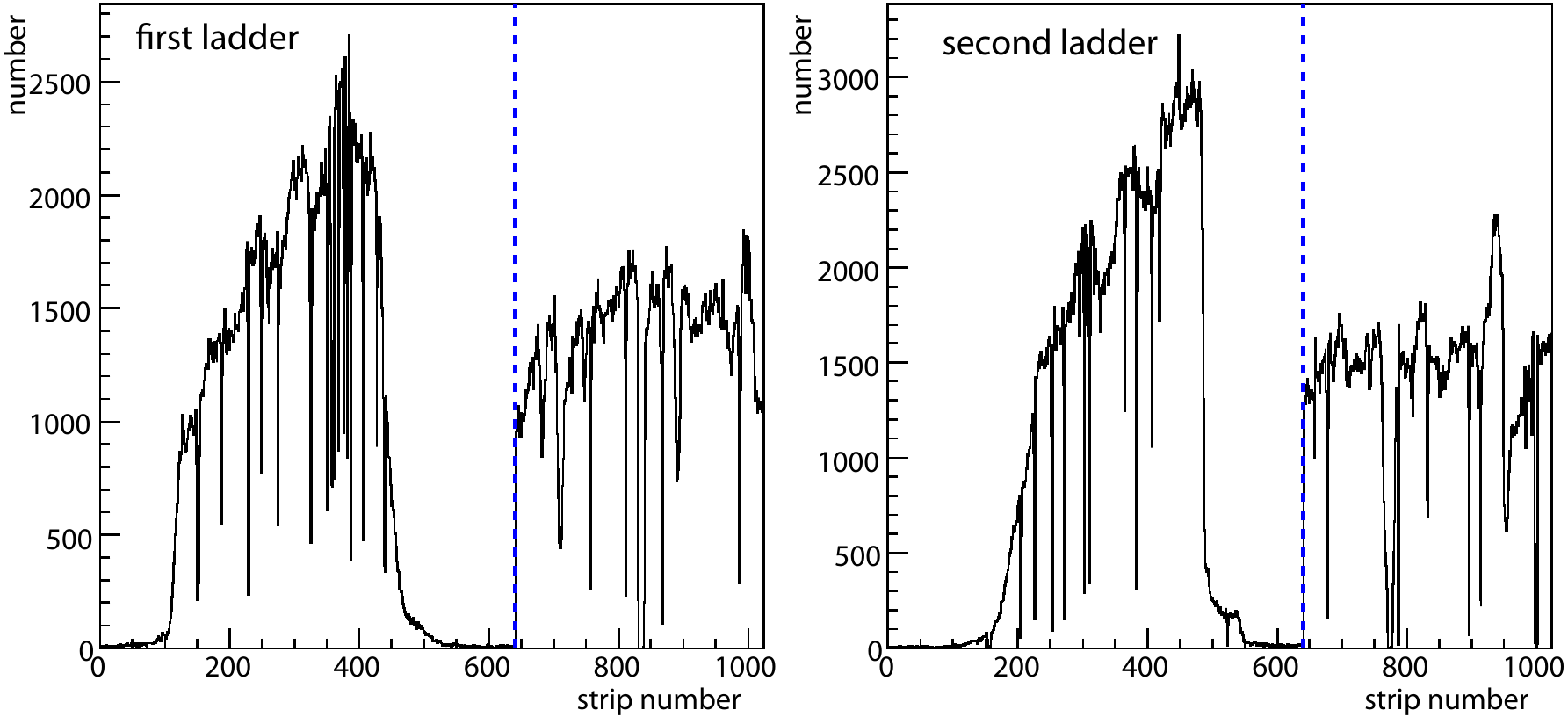}
\end{center}
\caption{Occupancy plots for the two beam telescope ladders after all
quality selection cuts. The boundary between the S~(front) and
K~(back) sides is marked by a dashed line. Apart from the gaps caused
by noisy or dead strips, the distribution of events is uniform on the
K~side and is shaped by the trigger acceptance and beam profile for
the S~side. The spikes and background seen in the initial
distributions (fig.~\ref{fig:tb08_bt_ladderHit}) are gone.}
\label{fig:tb08_bt_occ}
\end{figure}
\\
As a further cross-check, the beam telescope occupancy for clean
single-track events identified as above is examined
(fig.~\ref{fig:tb08_bt_occ}). The distributions are now regular except
for some gaps caused by dead or noisy strips that no longer contribute
to the event sample. In addition, the width of the S-side distribution
is governed by the dimension of the trigger scintillators.\\
\par
Having established the beam telescope as a reference for the fibre
modules and restricting ourselves to clean single-track events in the
following, one can now turn to the analysis of the SiPM data.
\begin{sidewaysfigure}
\begin{center}
\includegraphics[width=\textheight,angle=0]{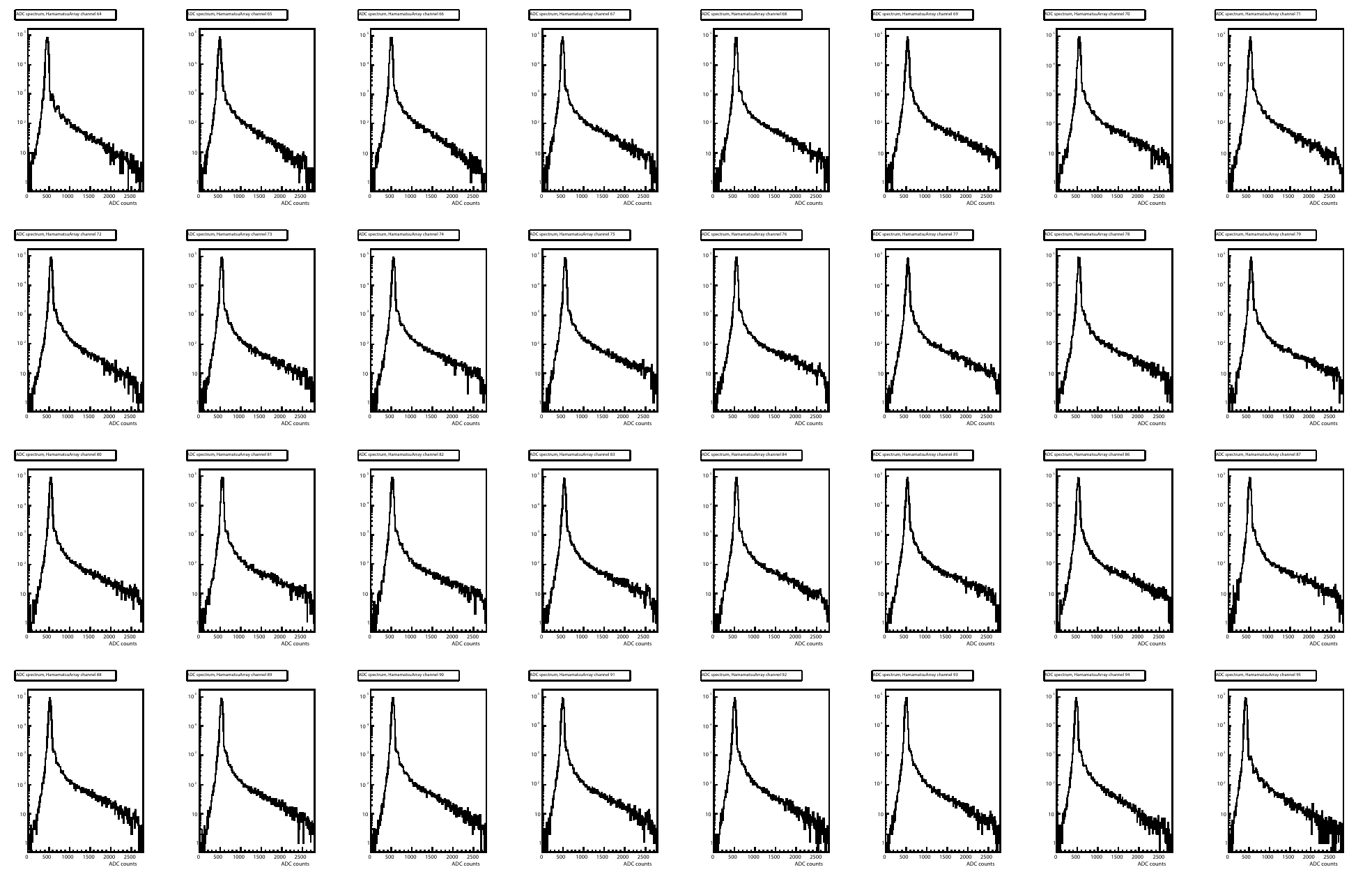}
\end{center}
\caption{Raw ADC spectra of all 32~SiPMs contained in the Hamamatsu
array considered here. All channels were
alive. Figure~\ref{fig:tb08_rawadcmag} contains a magnified version of
two example spectra.}
\label{fig:tb08_rawadc}
\end{sidewaysfigure}
\begin{figure}[htb]
\begin{center}
\includegraphics[width=0.75\textwidth,angle=0]{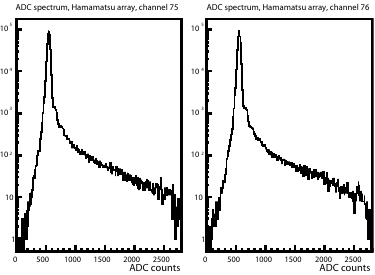}
\end{center}
\caption{Magnified version of two spectra from fig.~\ref{fig:tb08_rawadc}.}
\label{fig:tb08_rawadcmag}
\end{figure}
It starts from the raw ADC data of the SiPM array.
A plot of the raw ADC spectra for all 32~channels contained in the
SiPM array considered (fig.~\ref{fig:tb08_rawadc}) shows that all of
them were functional during the data taking. As the width of an
individual channel is very small compared to the width of the beam
profile and the trigger scintillators, the raw spectra are by far dominated by the pedestal peaks.
The first step in the analysis of the fibre data is the determination of
the positions of the fibres in the coordinate frame defined by the beam telescope.
To this end, a simple amplitude cut well above the pedestal and
occasional noise hits, taken here at 1000~ADC counts, is applied and
the interpolated track position for the $z$ of the fibre module is
\begin{figure}[htb]
\begin{center}
\begin{tabular}{cc}
\includegraphics[width=0.5\textwidth,angle=0]{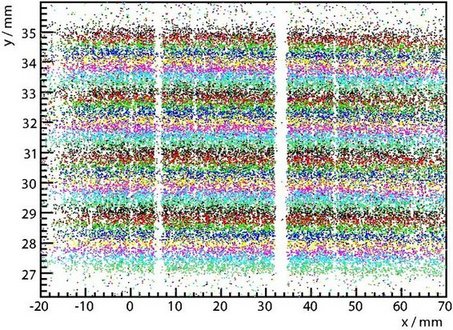}&
\includegraphics[width=0.5\textwidth,angle=0]{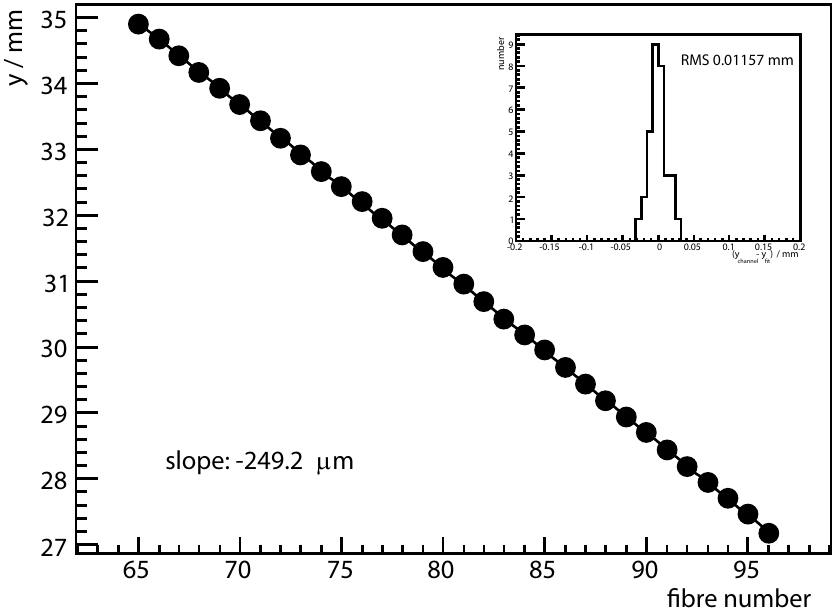}\\
\end{tabular}
\end{center}
\caption{{\it Left:} Intersection points of tracks interpolated to the
$z$~position of the fibre module, for fibre hits exceeding an
amplitude of 1000~ADC counts. Fibres are marked by different colours.
{\it Right:} Measured $y$ positions from Gaussian fits to the
projections of the above, as a function of the fibre number. A line
fit gives the measured pitch of the readout channels, in excellent
agreement with the nominal value of $250\,\mu\mathrm{m}$. The inset
contains the distribution of the fit residuals.}
\label{fig:tb08_xy}
\end{figure}
plotted using separate colours for the readout channels (fig.~\ref{fig:tb08_xy} {\it left}). The individual channels
are not as well separated as in
fig.~\ref{fig:testbeam_fibreProjections_module1} for the simple reason
that, due to the positioning of the round fibres in the tightest
arrangement, a track crossing the fibre module will usually generate
scintillation light that is then accumulated on at least two adjacent
SiPM array channels. For a first estimate
of the fibre positions, the scatter plot for each fibre is projected
to the $y$-axis and the mean obtained from a Gaussian fit is taken as
the fibre position. As a cross-check, this position is plotted as a
function of the channel number (fig.~\ref{fig:tb08_xy} {\it right}).
The slope of a straight-line fit to the resulting curve
matches the nominal pitch of the readout channels very well.
\begin{figure}[htbp]
\begin{center}
\includegraphics[width=\textwidth,angle=0]{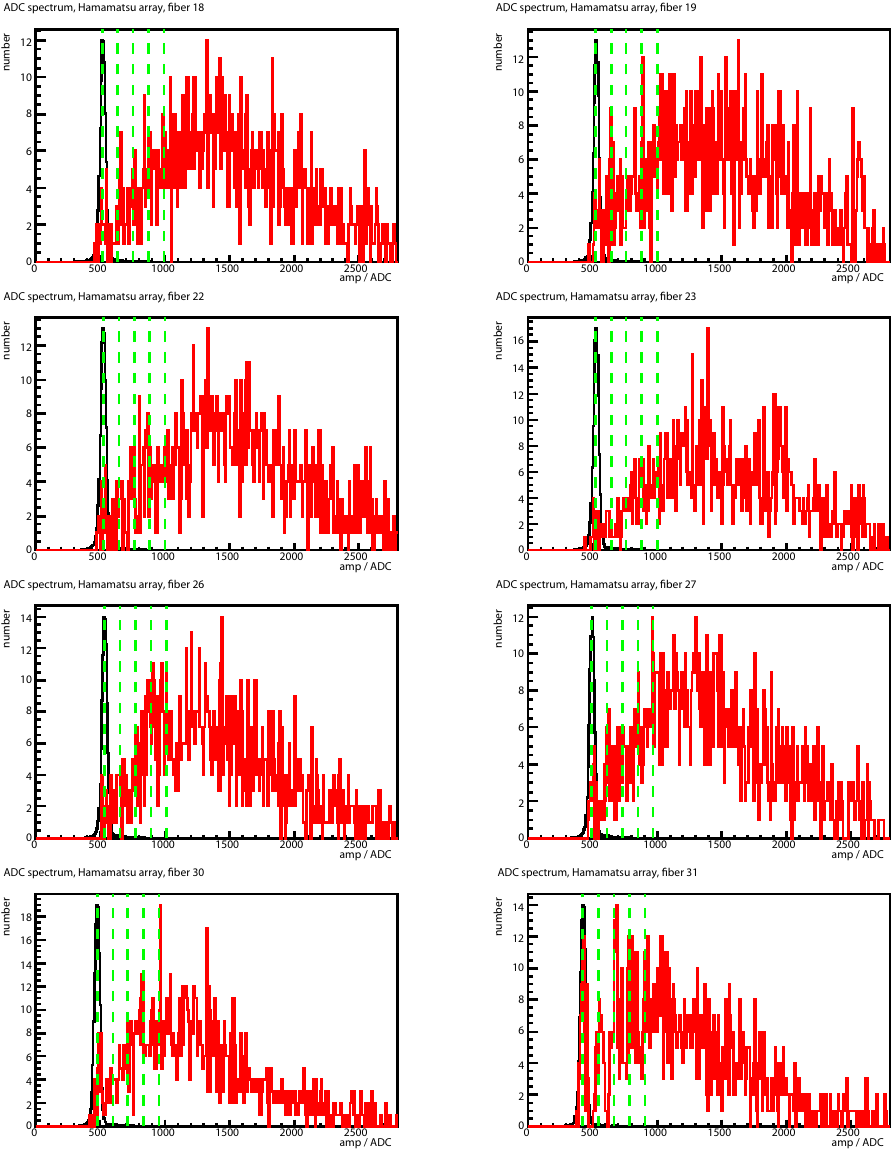}
\end{center}
\caption{Dark and signal spectra for eight of the 32 SiPM
channels. The signal spectra (red) are filled for hits within
$100\,\mu\mathrm{m}$ of the mean $y$ position (fig.~\ref{fig:tb08_xy})
of a given fibre, the dark spectra (black) are obtained from events in
which the beam does not pass through the area subtended by the SiPM array. The dashed
lines indicate the expected peak positions for a gain of $120$~ADC
counts. The dark spectra have been scaled to have the same maximum as
the signal spectra.}
\label{fig:tb08_signals}
\end{figure}
\\
Using the fibre positions so obtained, the signal and background
spectra for all the SiPM channels can be
found~(fig.~\ref{fig:tb08_signals}). To this end, events in which the
track passed within $\Delta{}y\leq{}100\,\mu\mathrm{m}$ of the fibre
position were used to fill the signal spectrum of the respective
fibre, while the background spectra were filled from events with a
track passing well outside the area subtended by the SiPM array under
study. Two observations are in order. First, the noise level of the
Hamamatsu SiPMs turns out to be very low, with anything besides the
pedestal peak barely visible on a linear scale. Second, the photo-electron-peak
structure is not as clearly discernible as expected from
figs.~\ref{fig:tracker_amps} and~\ref{fig:testbeam_rawadcspectra}. This
can be attributed to the sensitivity of the gain of the SiPMs to small
variations in the bias voltage over the course of the
testbeam. However, a calibration of the SiPMs is still possible. The
pedestal position for a given SiPM channel is easily found from the
location of the peak in the dark spectrum. The gain can be determined
from the positions of the first few peaks in the signal spectra with
an accuracy of roughly $10\,\%$. It turns out to be nearly uniform
across the SiPM array and is taken to be 120~ADC counts in the
following. The number of photo-electrons for a given channel is then
calculated according to~(\ref{eq:sipmcalibration}).\\
\par
For a determination of the spatial resolution of the module, a
possible tilt angle between the longitudinal axis of the fibres and
the S-side strips of the beam telescope ladders has to be corrected
for. For this purpose, the projections of the fibre scatter plots in
\begin{figure}[htb]
\begin{center}
\includegraphics[width=0.75\textwidth,angle=0]{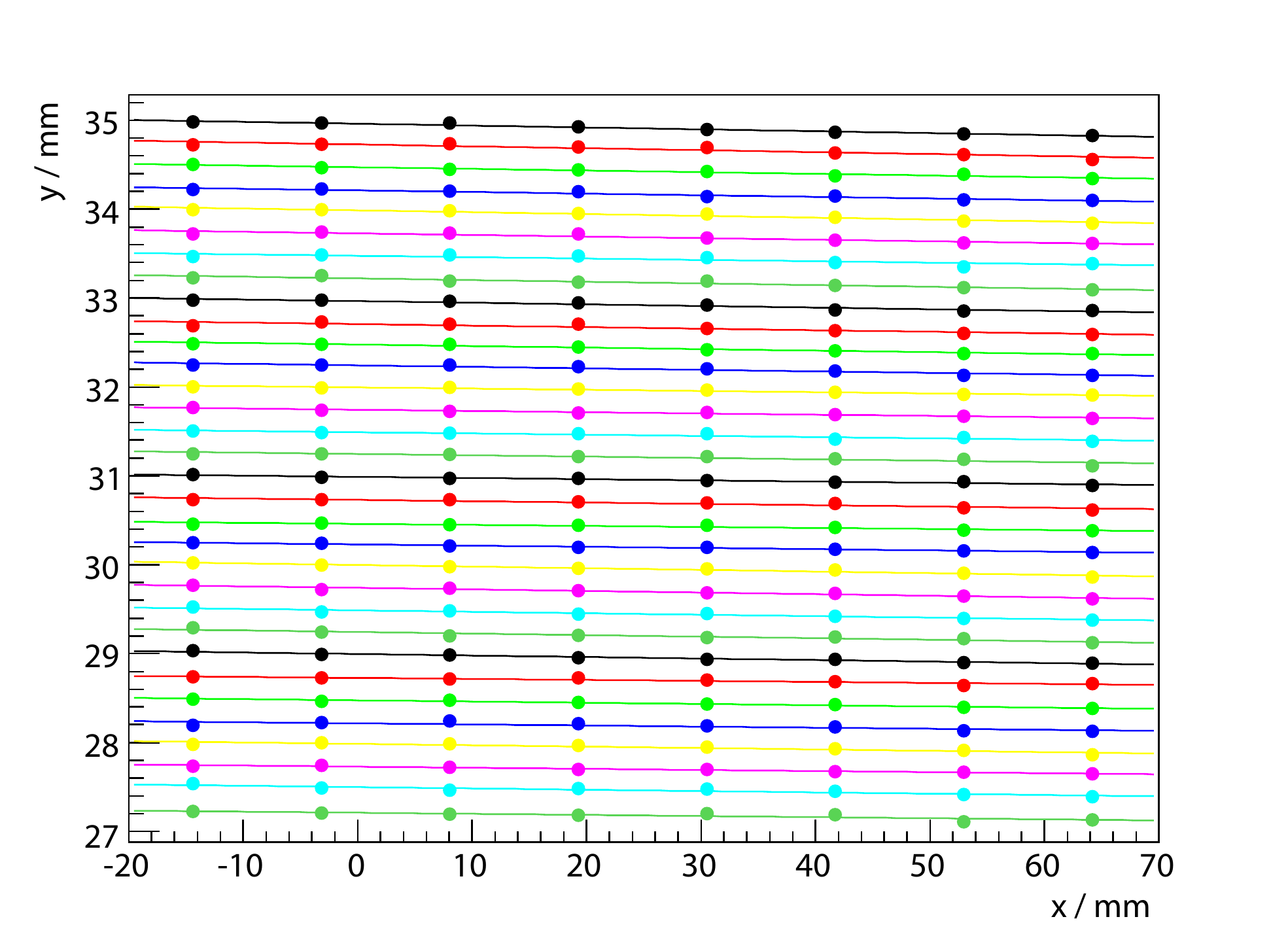}
\end{center}
\caption{Fibre positions along $x$, the coordinate along the fibres,
taken from Gaussian fits to slices of the $(x,y)$-distributions
(fig.~\ref{fig:tb08_xy} {\it left}). A straight line is fitted to each fibre.}
\label{fig:tb08_fc}
\end{figure}
\begin{figure}[htb]
\begin{center}
\begin{tabular}{cc}
\includegraphics[width=0.5\textwidth,angle=0]{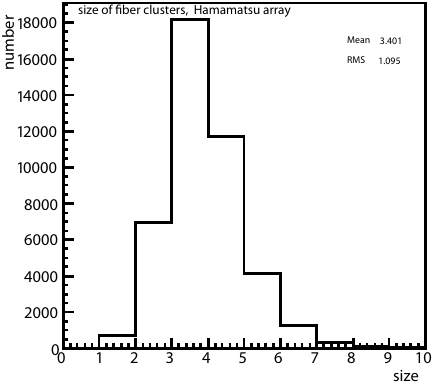}&
\includegraphics[width=0.5\textwidth,angle=0]{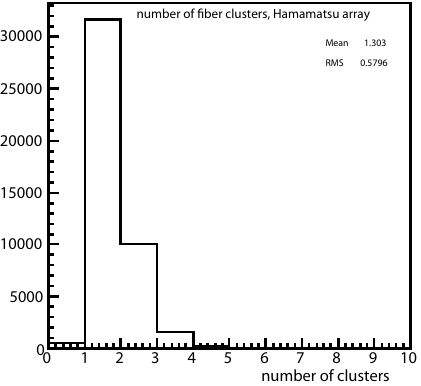}\\
\end{tabular}
\end{center}
\caption{{\it Left:} Distribution of the size of fibre clusters, for
tracks inside the fiducial area, and taking only the cluster with the
highest amplitude into account.
{\it Right:} Distribution of the number of fibre clusters found for
the Hamamatsu array considered here, for tracks inside the fiducial volume.}
\label{fig:tb08_clusterstats}
\end{figure}
fig.~\ref{fig:tb08_xy} are calculated for slices along $x$
(fig.~\ref{fig:tb08_fc}).
The mean values found in Gaussian fits to these projections give the
mean fibre location $y$ in a given interval in $x$. The fibres are
then parameterised as straight lines and a fit is done for each of
them to determine the fibre parameterisation $y_i(x)$ for fibre $i$. A typical tilt
angle of $1.5\,\mathrm{mrad}$ was found in this way. Knowing the
locations of the fibres, a fiducial area can be defined excluding the
outermost two fibres covered by the SiPM array, to make sure that
clusters on the edge of the active area are not included in the
distributions. Only tracks
crossing the module within the fiducial area are considered.\\
In the next step in the analysis, fibre clusters are
identified. Starting from a channel with at least
$1.5$~photo-electrons, neighbouring channels are added to the cluster
as long as
their amplitude exceeds $0.5$~photo-electrons. The position of the
fibre cluster is then calculated as
\begin{equation}
\label{eq:clusterwmean08}
y_\mathrm{cl}(x)=\frac{\sum\limits_{i}s_iy_i(x)}{\sum\limits_{i}s_i}
\end{equation}
where the sum is taken only over the channel with the highest
amplitude and its immediate neighbours at most. For the determination of the
spatial resolution, only the cluster with the highest total amplitude
is considered. The distribution of the cluster size
(fig.~\ref{fig:tb08_clusterstats} {\it left}) shows that clusters are
typically rather broad. This might be due to the rather large gap, on the order of $250\,\mu\mathrm{m}$,
between the fibre end and the silicon photomultiplier because of a
protective layer of epoxy coating on the
SiPM~\cite{ref:gregoriophd}. As a consequence, light exiting a given
fibre might reach a neighbouring SiPM channel. This effect
is the reason for the prescription of taking the sum in
(\ref{eq:clusterwmean08}). The low threshold used in the cluster
finding means that there are occasional noise clusters present, too
(fig.~\ref{fig:tb08_clusterstats} {\it right}), but it ensures an
overall tracking efficiency of $99\,\%$.\\
From the interpolated track position and the measured cluster
position, the track residual $y_\mathrm{cl}-y_\mathrm{tr}$ can be
calculated.
\begin{figure}[tb]
\begin{center}
\includegraphics[width=0.85\textwidth,angle=0]{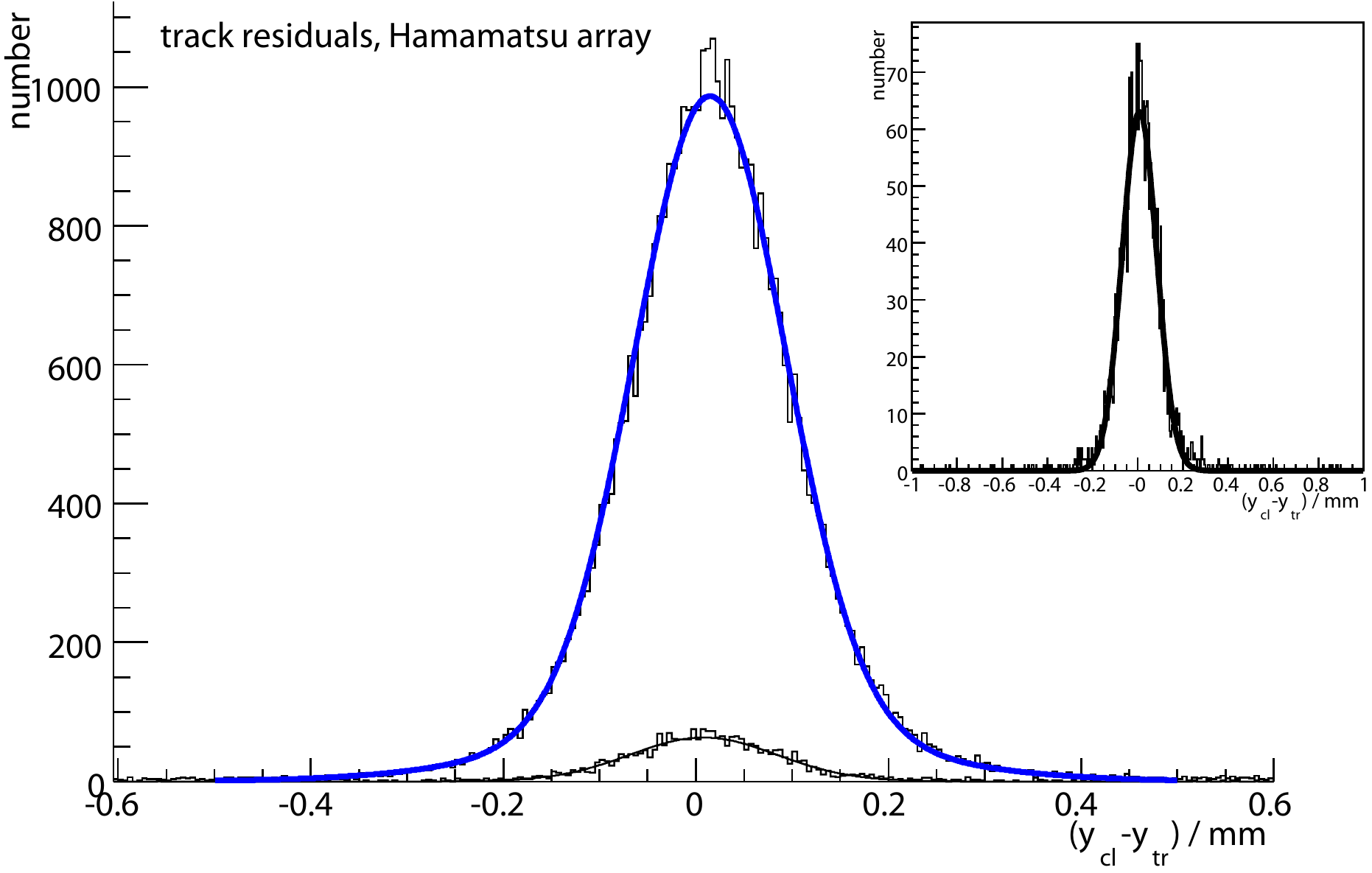}
\end{center}
\caption{Residuals of fibre cluster positions, using only the
innermost three channels for the calculation of the fibre cluster position, with respect to the
interpolated track position. For the full dataset and a fiducial
area excluding the outermost two fibre channels on each side of the
array, a spatial resolution of $89\,\mu\mathrm{m}$ is obtained from a
fit of a sum of two Gaussians (blue curve). The corresponding tracking efficiency is $99\,\%$. For a dataset comprising only
the first twelve hours of data-taking and restricted to a fiducial area
given by $x\in[10,30]\,\mathrm{mm}$, a resolution of $76\,\mu\mathrm{m}$ is
obtained with almost no non-Gaussian tails (black curve and inlet).}
\label{fig:tb08_trackres}
\end{figure}
The resulting distribution (fig.~\ref{fig:tb08_trackres}) is well
described by a sum of two Gaussians. Each Gaussian is determined by
its amplitude $A_i$, mean $\mu_i$ and standard deviation
$\sigma_i$. For a robust estimation of the parameters, $\mu_1$ and
$\sigma_1$ are first calculated from a fit taking only an inner window
of width $d$ around the mean of the track residuals histogram into
account. In a second step, $\sigma_1$ and $\mu_1=\mu_2$ are kept
fixed and only $\sigma_2$ and the amplitudes $A_{1,2}$ are varied. The
function
\begin{equation}
\label{eq:twogaussians}
N(y)=A_1\exp\left(-\frac{1}{2}\left(\frac{y-\mu_1}{\sigma_1}\right)^2\right)+
A_2\exp\left(-\frac{1}{2}\left(\frac{y-\mu_2}{\sigma_2}\right)^2\right)
\quad\mathrm{with}\quad{}y=y_\mathrm{cl}-y_\mathrm{tr}
\end{equation}
is fitted to the entire histogram. The average spatial resolution is
then defined as
\begin{equation}
\label{eq:spatialreswmean}
\sigma=\sqrt{\frac{A_1\sigma_1^2+A_2\sigma_2^2}{A_1+A_2}}
\end{equation}
$d$ is chosen such that the
$\chi^2$ of this final fit is minimized, however the result for
the spatial resolution is rather insensitive to the exact value of $d$.\\
A fit according to this procedure yields widths of
$78\,\mu\mathrm{m}$ and $184\,\mu\mathrm{m}$ with relative
amplitudes of $93\,\%$ and $7\,\%$, respectively.
Taking the weighted mean according to (\ref{eq:spatialreswmean}) results in
an overall spatial resolution of $89\,\mu\mathrm{m}$. Plotting the
track residuals as a function of the time~$t$ and the coordinate $x$
\begin{figure}[htb]
\begin{center}
\includegraphics[width=\textwidth,angle=0]{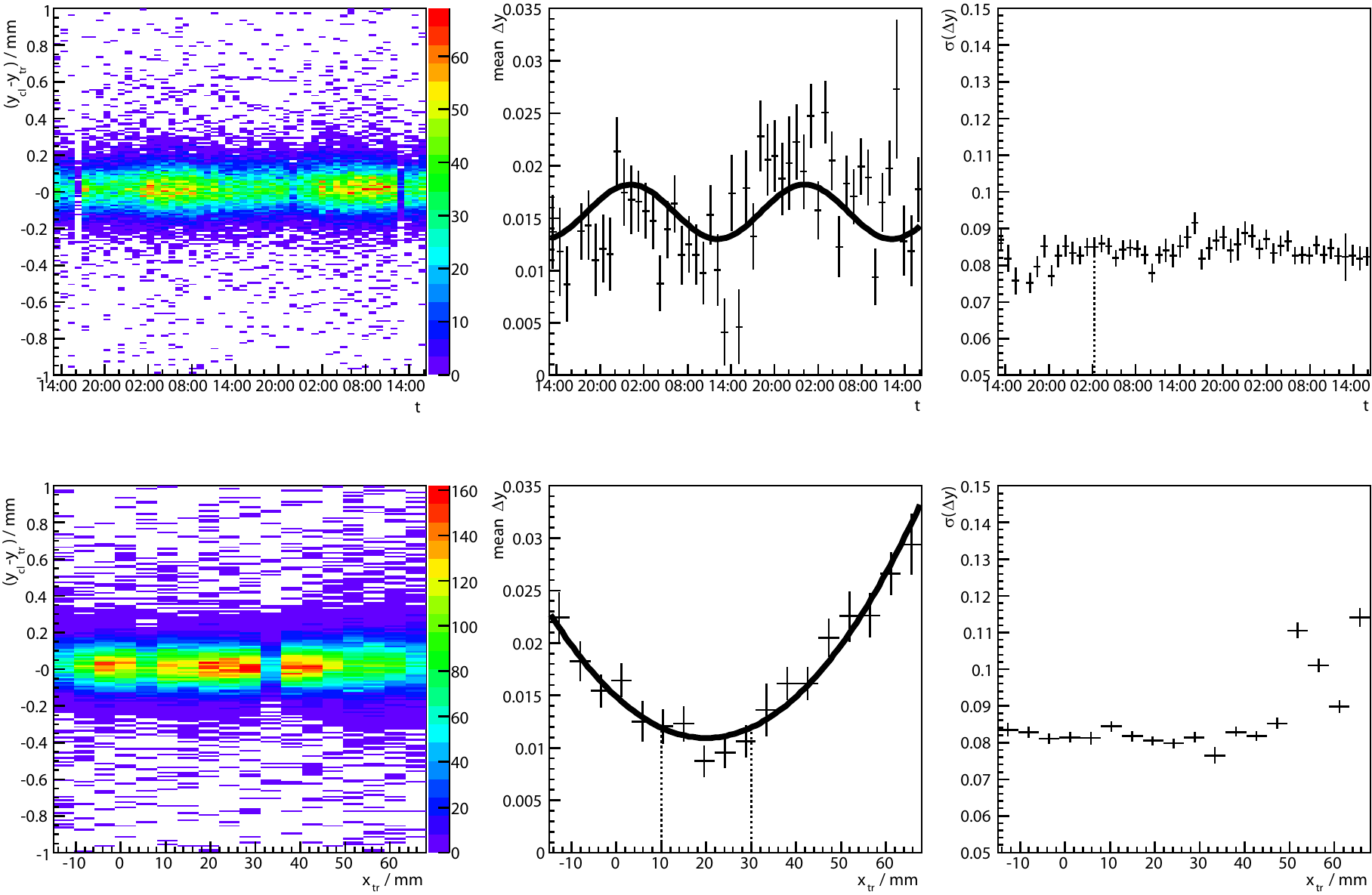}
\end{center}
\caption{Track position residuals as a function of time ({\it top})
and position along the fibre ({\it bottom}). Both mean ({\it middle}) and standard
deviation ({\it right}) obtained from Gaussian fits to slices of the respective
distributions are plotted. The variations of the mean values are
fitted with eq.~(\ref{eq:timevar}) and a parabola, respectively. The
dashed lines bound the intervals for which the black curve in
fig.~\ref{fig:tb08_trackres} is obtained.}
\label{fig:tb08_trackressyst}
\end{figure}
along the fibres (fig.~\ref{fig:tb08_trackressyst}) reveals
systematic effects. The mean track residual moves with time inside a
range of roughly $20\,\mu\mathrm{m}$ and follows a periodic trend with
a period of 1~day, as described by
\begin{equation}
\label{eq:timevar}
\Delta{}y(t)=\Delta{}y_0+A\,t\,\sin\left(\frac{2\pi}{1\,\mathrm{d}}\,\cdot\,t+\phi_0\right)
\end{equation}
which indicates a temperature effect. A dependence on $x$ is seen with
the same order of magnitude and the shift of the mean track residual
is well described by a parabola. In fact, restricting the fiducial area to a
small range in $x$ and the first twelve hours of data taking, a
spatial resolution of $76\,\mu\mathrm{m}$ and a distribution with very
small non-Gaussian tails are obtained (black curve and inlet in
fig.~\ref{fig:tb08_trackres}).\\
This value reflects the result found for the inner Gaussian when
looking at the complete event sample. Apparently, this intrinsic
resolution is deteriorated by effects causing non-Gaussian
tails. While the exact nature of these effects is still under
investigation~\cite{ref:gregoriophd}, the results of this study point
to two possible sources: First, as the behaviour of the silicon
photomultipliers depends sensitively on the bias voltage, variations
in this value, possibly correlated with temperature fluctuations, over the course of the measurement might cause the SiPMs
to work in a non-ideal way. The bias voltage should therefore be
monitored carefully during future testbeam measurements. A second effect
may be a slight
departure of the fibre geometry from the straight line shape assumed in
this analysis.\\
To obtain the spatial resolution of the fibre module alone, the
limited resolution of the beam telescope needs to be corrected
for. For the testbeam setup, a Geant4-based Monte Carlo simulation
assuming a resolution of $10\,\mu\mathrm{m}$~\cite{ref:cristinziani}
for the two silicon ladders shows that the estimated resolution with
which the track can be reconstructed at the position of the fibre
module is on the order of $30\,\mu\mathrm{m}$ for $10\,\mathrm{GeV}$
protons with perpendicular
incidence~\cite{ref:gregoriophd}. Quadratically subtracting this value
yields intrinsic and total resolutions of $70\,\mu\mathrm{m}$ and
$84\,\mu\mathrm{m}$, respectively.\\
The simulation presented in section~\ref{sec:design_study_pebs} gives a spatial
resolution of $43\,\mu\mathrm{m}$ for perpendicularly incident muons
under ideal circumstances. 
The improvement to be gained from the ongoing
study~\cite{ref:gregoriophd,ref:romanphd} of the detector design,
construction and operation is therefore substantial.\\
\par
Before the mean number of photo-electrons can be calculated, it is
necessary to determine the probability $\epsilon$ for inter-pixel
crosstalk which turns out to be larger than in the 2006~testbeam. The
basic idea is to look for the broadening of the dark
spectra of the SiPM channels compared to the expectation for
Poissonian noise. A toy Monte Carlo study was created
in~\cite{ref:gregoriophd} to reproduce the measured dark spectra. It
implements the pulse shape of an SiPM as sampled by a
VA chip at random times, with the number of samples drawn from a
Poisson distribution whose mean is determined by the noise rate. For
each fired pixel, an additional pixel is fired with the probability
$\epsilon$, taking the limited number of pixels into account. The
output spectrum is calculated using a given gain. It turns out
\begin{figure}[htb]
\begin{center}
\includegraphics[width=0.9\textwidth,angle=0]{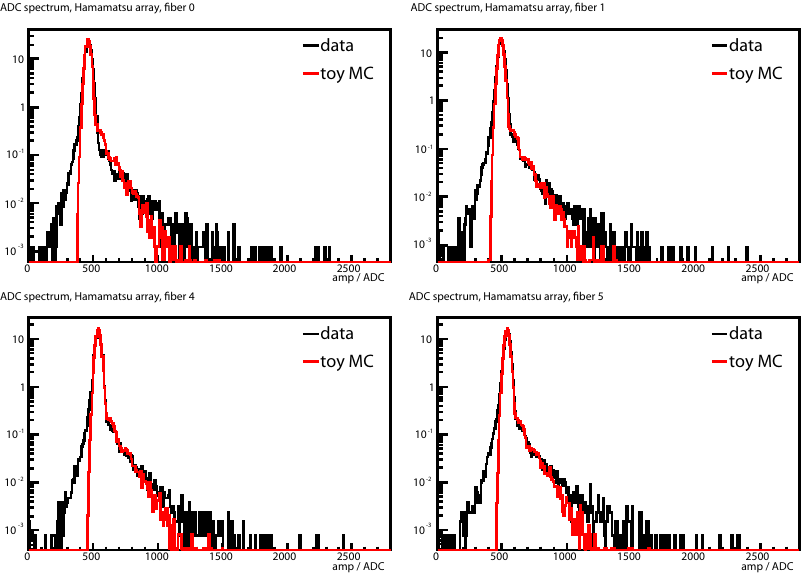}
\end{center}
\caption{Dark spectra for four of the 32~SiPM array channels, compared to
predictions of the toy Monte Carlo study described in the text, for a
crosstalk probability of $0.4$, a noise rate of $160\,\mathrm{kHz}$,
and a gain of~120.}
\label{fig:tb08_toymc}
\end{figure}
\begin{figure}[htb]
\begin{center}
\includegraphics[width=0.5\textwidth,angle=0]{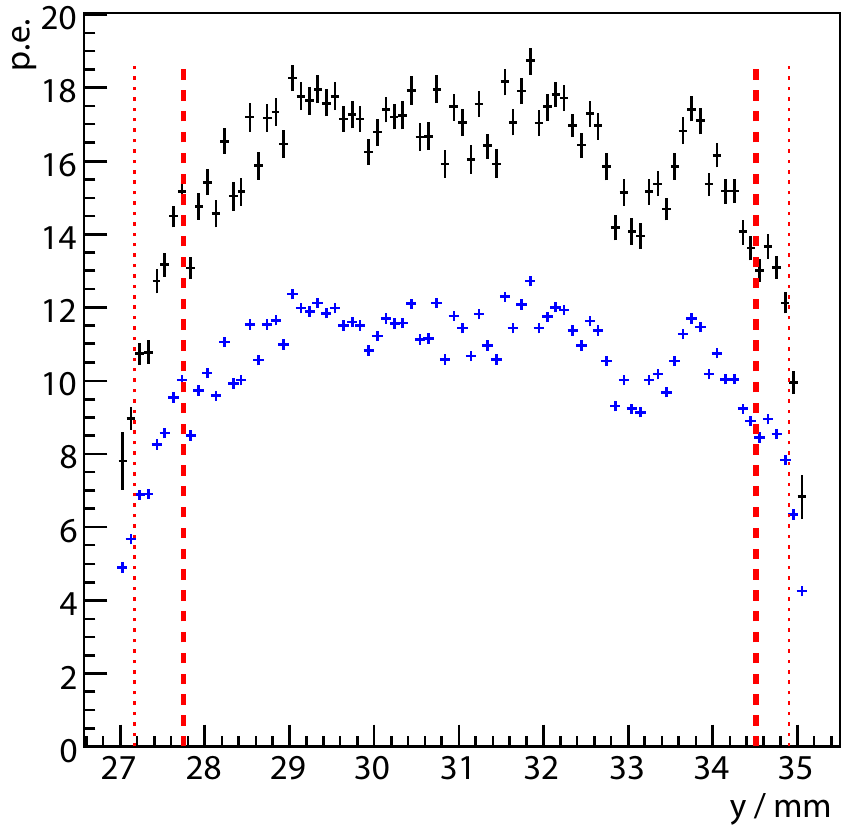}
\end{center}
\caption{Mean cluster amplitudes as a function of the interpolated track
position across the fibre module (black). After correction for inter-pixel
crosstalk and the limited number of pixels, the blue curve is
obtained, shown here without error bars for clarity. The positions of
the outermost fibres are indicated by the thin dashed lines, while the
thick dashed lines mark the approximate boundaries of the fiducial
area. A mean corrected number of $10.7$~photo-electrons is found
within the outer boundary.}
\label{fig:tb08_clusteramp}
\end{figure}
(fig.~\ref{fig:tb08_toymc}) that reasonable agreement between the
measured dark spectra and the toy Monte Carlo simulation is achieved
for a crosstalk probability of $\epsilon=0.4$, a noise rate of $160\,\mathrm{kHz}$,
and a gain of~120~ADC counts.\\
As shown in section~\ref{sec:momres}, the mean number of
photo-electrons per cluster is an important input value to the full
detector
simulation. As a function of the position across the module, the
average cluster amplitudes (in fired pixels) are calculated
(fig.~\ref{fig:tb08_clusteramp}). 
Using (\ref{eq:urnmodel}) to correct for the limited number
$N_\mathrm{pix}=80$ of pixels
and the result that crosstalk increases the number of fired pixels on
average by a factor of $1/(1-\epsilon)$~\cite{ref:vladikecal}, the
corrected number $N_\mathrm{pe}$ of photo-electrons is calculated from
the number $N_H$ of fired pixels as
\begin{equation}
\label{eq:pixelcorr}
N_\mathrm{pe}=\frac{\log\left(1-\frac{N_H}{N_\mathrm{pix}}\right)}{\log\left(1-\frac{1}{N_\mathrm{pix}}\right)}\,\cdot\,(1-\epsilon)
\end{equation}
Within the range subtended by the Hamamatsu SiPM array, a mean of
$10.7$~photo-electrons is found. This number was used in the simulation
results presented in chapter~\ref{chapter:pebs_design_study}.

\chapter{Constraining supersymmetry with cosmic-ray data}
\label{chapter:susyscan}
The indirect search for dark matter is the most important task for the
PEBS and AMS-02 detectors. In this chapter, the improvement to
be expected from PEBS or AMS-02 data over the currently available data will be evaluated. In order
to quantify this improvement, a certain model for dark matter has to
be chosen. The signal fluxes of positrons, antiprotons and others, as
well as the dark matter relic density and other observables, can
then be calculated in the framework of the model. Of the many candidates for dark matter, the supersymmetric
neutralino is by far the most
popular~\cite{ref:turner,ref:kamionkowski,ref:darksusyprop,ref:baltz_positron_excess,ref:kanewangwang,ref:hoopertaylor,ref:hoopersilk}.
Therefore, this
case was chosen as an example for this study, and in addition, it was restricted to the
mSUGRA model whose limited number of free parameters makes it a
convenient playground for the study of SUSY phenomenology. Assuming
that mSUGRA is realised in nature and that at least some part of the
dark matter is made up of neutralinos, the study proceeds as follows:
First, some more details about the model are outlined and
the connection to cosmic-ray and other observables is described.
A large fraction of the mSUGRA parameter space has been systematically scanned. After compiling
a good part of the presently available data with implications for the
models included in these scans, the situation once PEBS or AMS-02 data are
available will be studied, focusing on two benchmark parameter
points.\\
Very recently, the positron fraction data of the PAMELA
experiment have become available. In the final
section~\ref{sec:msugra_pamela} of this chapter, it will be discussed
how the conclusions drawn from the study presented here may be altered
in light of these new data.

\section{Model description}
\subsection{mSUGRA and observables}
\begin{figure}[htb]
\begin{center}
\includegraphics[width=0.5\textwidth,angle=0]{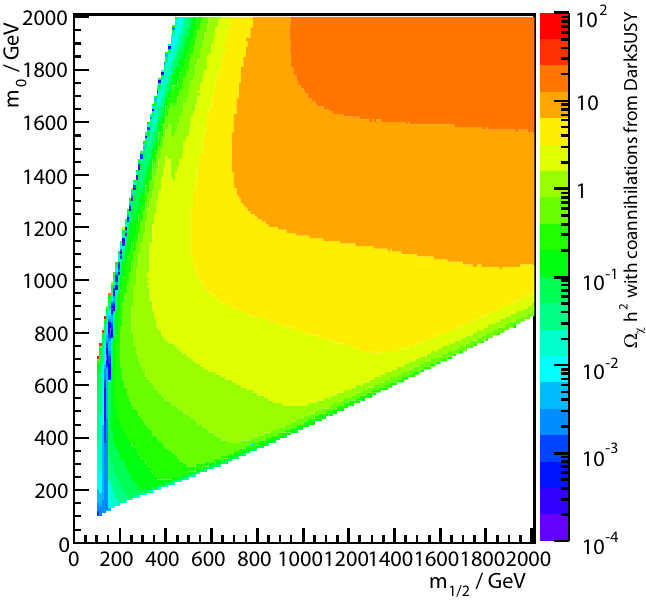}
\end{center}
\caption{Neutralino relic density $\Omega_{\chi}h^2$ in the
$m_{1/2}$-$m_0$-plane, for $\tan\beta=40$ and
$m_t=170.9\,\mathrm{GeV}$. The parameter space is bounded by the
regions where no electroweak symmetry breaking occurs (at large values
of $m_0$) or where the LSP would be a charged particle (at large values
of $m_{1/2}$).}
\label{fig:omega}
\end{figure}
\begin{figure}[htb]
\begin{center}
\begin{tabular}{cc}
\includegraphics[width=0.5\textwidth,angle=0]{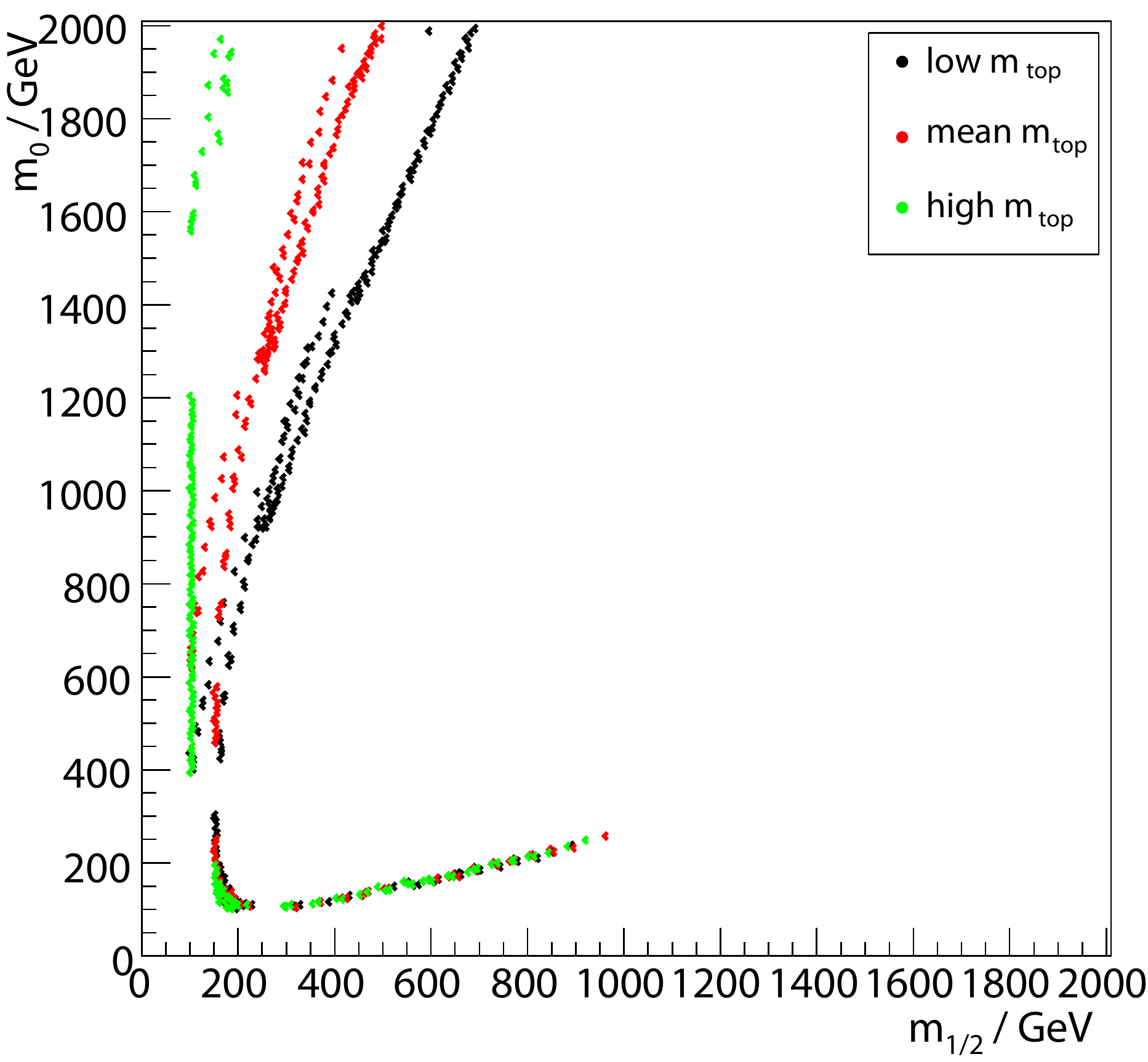}&
\includegraphics[width=0.5\textwidth,angle=0]{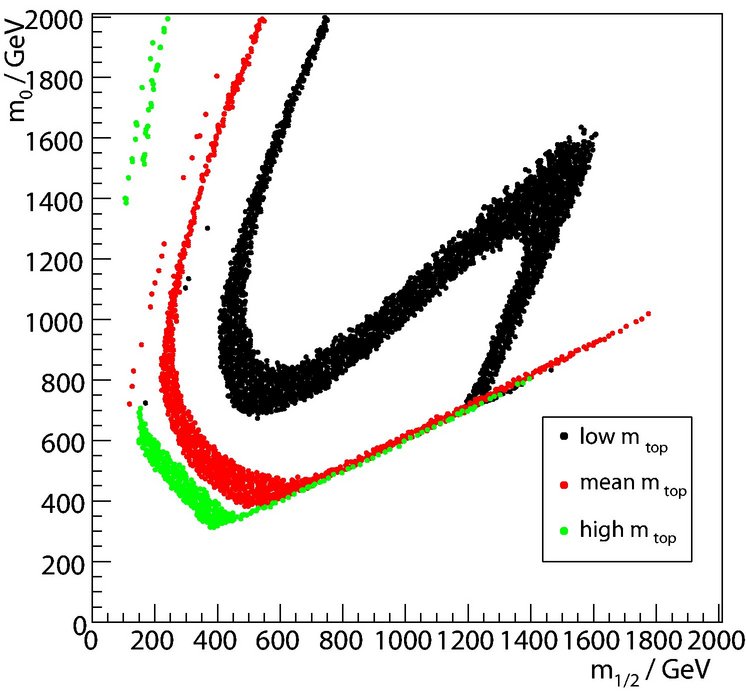}\\
\end{tabular}
\end{center}
\caption{mSUGRA parameter points fulfilling the relic density
constraint to within $3\sigma$, for $\tan\beta=20$ ({\it left}) and
$\tan\beta=50$ ({\it right}), depending on the mass of the top
quark. Points for top quark masses of $167.18\,\mathrm{GeV}$,
$170.9\,\mathrm{GeV}$ and $174.62\,\mathrm{GeV}$
are shown.}
\label{fig:mtop_relic_density}
\end{figure}
As stated in section~\ref{sec:dmmsugra}, a single point in the mSUGRA parameter space is characterised by the soft
SUSY breaking scalar and fermionic mass parameters $m_0$ and $m_{1/2}$
at the GUT scale, the ratio of the Higgs vacuum expectation values,
$\tan\beta$, the trilinear scalar coupling $A_0$ at the GUT scale, and
the sign of the Higgs mass parameter $\mu$. The ISAJET~7.75
software package~\cite{ref:isajet} is used to solve the renormalisation group
equations and calculate the mass spectrum of supersymmetric particles and
their couplings at the electroweak scale. Various observables,
described in detail below, are then calculated with the help of the
DarkSUSY~4.1~\cite{ref:ds}, micrOMEGAs~2.0.6~\cite{ref:mm}, and
ISATOOLS~\cite{ref:isajet} codes. In particular, the DarkSUSY software predicts
electron, antiproton and $\gamma$-ray fluxes from neutralino
annihilations in the Galactic halo.\\
\par
Several observables can be used to exclude regions of the mSUGRA parameter
space. The most important ones are the neutralino relic abundance, the
anomalous magnetic moment of the muon, and the $b\rightarrow{}s\gamma$
branching ratio.\\
\par
The data on temperature anisotropies in
the cosmic microwave background gathered by the WMAP
experiment~\cite{ref:wmap}, in combination with other cosmological
observations such as the spatial distribution of galaxies, give a
density of the cold, non-baryonic matter of~\cite{ref:pdg}
\begin{equation}
\label{eq:cosmconstr}
\Omega_\mathrm{nbm}h^2=0.106\,\pm\,0.008
\end{equation}
where $h$ is the Hubble constant in units of
$100\,\mathrm{km}\,\mathrm{s}^{-1}\,\mathrm{Mpc}^{-1}$ and $\Omega$
denotes a density in units of the critical density.\\
The calculation of the present-day relic density of neutralinos is an
involved task. The procedure followed by DarkSUSY is outlined
in~\cite{ref:relicdensitycalc} and starts with the Boltzmann equation
\begin{equation}
\label{eq:boltzmann_ds}
\frac{\mathrm{d}n}{\mathrm{d}t}+3Hn=-\langle\sigma_\mathrm{eff}v\rangle(n^2-n^2_\mathrm{eq})
\end{equation}
for the summed number density $n=\sum_{i=1}^N{}n_i$ of the neutralino
($n_1$) and the $N-1$
supersymmetric particles that will eventually decay to the neutralino
if R-parity is conserved. The second term on the left-hand side comes
from the dilution due to the expansion of the Universe, and
\begin{equation}
\label{eq:reldens_sigmaeff}
\langle\sigma_\mathrm{eff}v\rangle= \sum\limits_{ij}\langle\sigma_{ij}v_{ij}\rangle\frac{n^\mathrm{eq}_i}{n^\mathrm{eq}}\,\frac{n^\mathrm{eq}_j}{n^\mathrm{eq}}
\end{equation}
and brackets denote thermal averaging and
$v_{ij}=\sqrt{(p_i\cdot{}p_j)^2-m_i^2m_j^2}/E_iE_j$. $n^\mathrm{eq}$
denotes number densities at thermal equilibrium. The annihilation
cross sections entering (\ref{eq:reldens_sigmaeff}) not only include
$\chi^0_i\chi^0_j$-annihilations ($i,j=1..4$) to all possible final
states, but also coannihilations between the neutralinos, charginos
and sfermions.\\
Figure~\ref{fig:omega} shows the predicted relic neutralino density
for a scan of the $m_{1/2}$-$m_0$-plane, for $\tan\beta=40$ and
$m_t=170.9\,\mathrm{GeV}$. It varies over many orders of magnitude and
the cosmological constraint (\ref{eq:cosmconstr}) allows only a
tiny portion of that space. It turns out however that the allowed
region changes with $\tan\beta$. Also, the relic density predicted for
a certain mSUGRA parameter point
depends strongly on the mass $m_t$ of the top quark. This is
illustrated in figure~\ref{fig:mtop_relic_density} where parameter
points in the $m_{1/2}$-$m_0$-plane yielding a value of the relic
density within $3\sigma$ of the value quoted above are plotted for two
values of $\tan\beta$.
For high values of $\tan\beta$, the appearance of the rapid
annihilation funnel, where the
$\chi\chi\rightarrow{}A\rightarrow{}f\bar{f}$ cross section is large,
depends on the value adopted for $m_t$. The latest result from
the Tevatron Electroweak Working Group is used~\cite{ref:mtop08}:
\begin{equation}
m_t=(172.6\,\pm\,0.8\,\pm\,1.1)\,\mathrm{GeV}
\end{equation}
Adding the statistical and systematic errors in quadrature yields
$\sigma_{m_t}=1.36\,\mathrm{GeV}$.
$m_t$ is varied in our scans, but in order to keep the number of
parameters limited, $A_0=0$ is set in the following.\\
\par
\begin{figure}[htb]
\begin{center}
\begin{tabular}{cc}
\includegraphics[width=0.52\textwidth,angle=0]{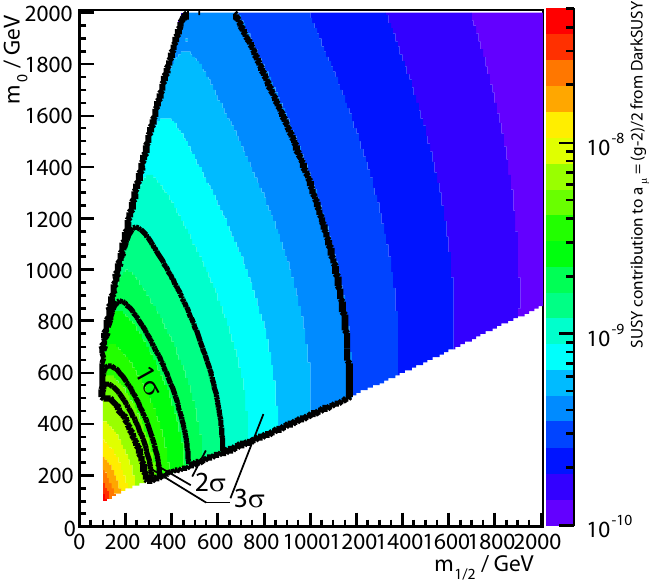}&
\includegraphics[width=0.45\textwidth,angle=0]{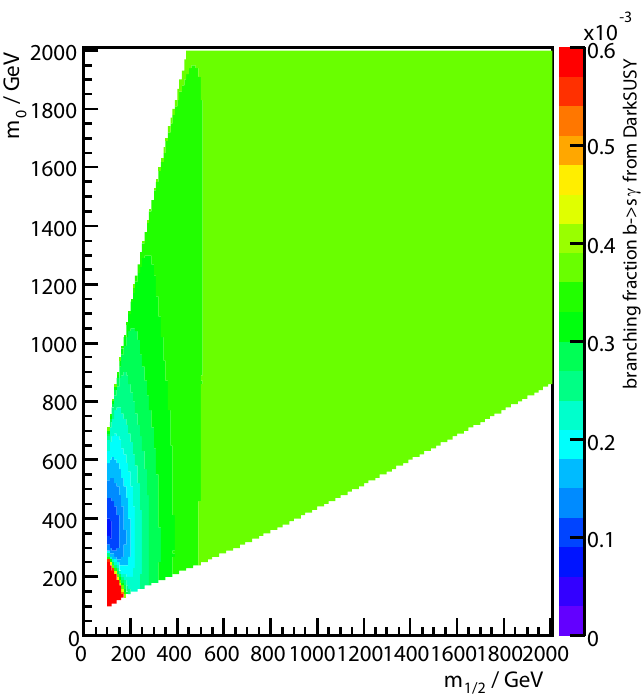}\\
\end{tabular}
\end{center}
\caption{{\it Left:} SUSY contribution to $a_\mu$ in the
$m_{1/2}$-$m_0$-plane, for $\tan\beta=40$ and
$m_t=170.9\,\mathrm{GeV}$. The preferred region is indicated at the $1\sigma$,
$2\sigma$ and $3\sigma$-levels. {\it Right:} Branching fraction for $b\rightarrow{}s\gamma$ decays in the
$m_{1/2}$-$m_0$-plane, for $\tan\beta=40$ and
$m_t=170.9\,\mathrm{GeV}$,
as calculated by DarkSUSY.}
\label{fig:muon_bsg}
\end{figure}

The magnetic moment of the muon is related to its intrinsic
spin by the gyromagnetic ratio $g_\mu$:
\begin{equation}
\label{eq:gmu}
\vec{m}=g_\mu\,\frac{e}{2m_\mu}\vec{S}
\end{equation}
While the Dirac equation predicts $g=2$ for a structureless
spin-$\frac{1}{2}$ particle, quantum loop effects lead to a small
deviation, parameterised by the anomalous magnetic moment
$a_\mu\equiv{}(g_\mu-2)/2$~\cite{ref:pdg}. $a_\mu$ has been measured
to enormous precision by the E821 experiment at the Brookhaven
National Lab that finds for the charge average~\cite{ref:e821}
\begin{equation}
\label{eq:amuexp}
a_\mu^\mathrm{exp}=(11659208.0\,\pm\,5.4\,\pm\,3.3)\cdot{}10^{-10}
\end{equation}
\begin{figure}[htb]
\begin{center}
\begin{tabular}{cccc}
\includegraphics[width=0.2\textwidth,angle=0]{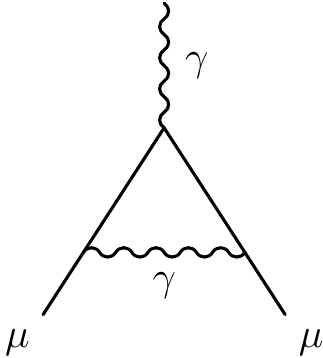}&
\includegraphics[width=0.2\textwidth,angle=0]{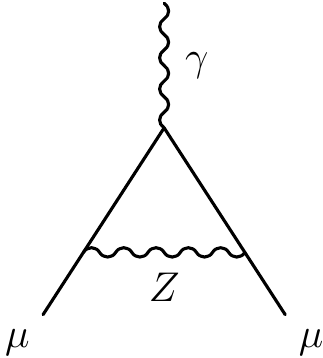}&
\includegraphics[width=0.2\textwidth,angle=0]{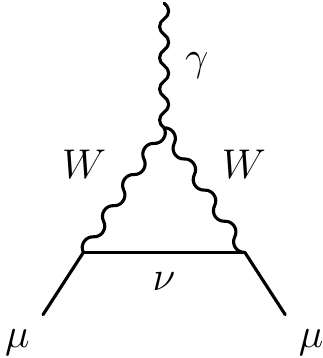}&
\includegraphics[width=0.2\textwidth,angle=0]{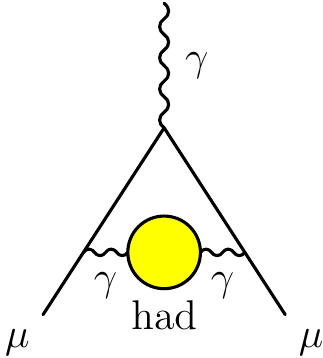}\\
\end{tabular}
\end{center}
\caption{Representative diagrams contributing to $a_\mu^\mathrm{SM}$:
first-order QED ({\it left}), lowest-order weak ({\it middle}) and
lowest-order hadronic ({\it right}). Figure taken from~\cite{ref:pdg}.}
\label{fig:amu_diagr}
\end{figure}
The standard model prediction for $a_\mu$ can be calculated from
diagrams like those shown in figure~\ref{fig:amu_diagr} and is the sum
$a_\mu^\mathrm{SM}=a_\mu^\mathrm{QED}+a_\mu^\mathrm{EW}+a_\mu^\mathrm{had}$
of the contributions from QED, loop diagrams involving $W^\pm$, $Z$
and Higgs particles, and hadronic loops. The latter is determined from
$\sigma(e^+e^-\rightarrow\mathrm{hadrons})$ data and dominates the
theoretical error. Overall, the standard model prediction is~\cite{ref:pdg}
\begin{equation}
\label{eq:amusm}
a_\mu^\mathrm{SM}=(11659178.8\,\pm\,5.8)\cdot{}10^{-10}
\end{equation}
This means that there is a $3.4\sigma$ deviation of
\begin{equation}
\label{eq:amudiff}
\Delta{}a_\mu=a_\mu^\mathrm{exp}-a_\mu^\mathrm{SM}=(29.2\,\pm\,8.56)\cdot{}10^{-10}
\end{equation}
of the measured and theoretical values. The theoretical value
gets amended in a supersymmetric theory leading to regions
in the mSUGRA parameter space where the discrepancy is cancelled and that
can therefore be considered to be preferred by the experimental
results (fig.~\ref{fig:muon_bsg}). The sign of the supersymmetric
contribution
to $a_\mu$ is identical to the sign of $\mu$, so that only the case
$\mu>0$ is considered in the following.\\
\par
The branching ratio $BR$ of the penguin decay
$b\rightarrow{}s\gamma$ is sensitive to supersymmetric
contributions caused by additional loop diagrams containing the
superpartners. From~\cite{ref:pdg}, one calculates for the mean
experimental value:
\begin{equation}
BR(b\rightarrow{}s\gamma)=(3.63\,\pm\,0.50)\cdot{}10^{-4}
\end{equation}
The results given by mircOMEGAs for this observable turn out to be
negative for some regions in the parameter space. Therefore,
only the calculation of DarkSUSY was used.\\
\par
\begin{figure}[htb]
\begin{center}
\includegraphics[width=1.0\textwidth,angle=0]{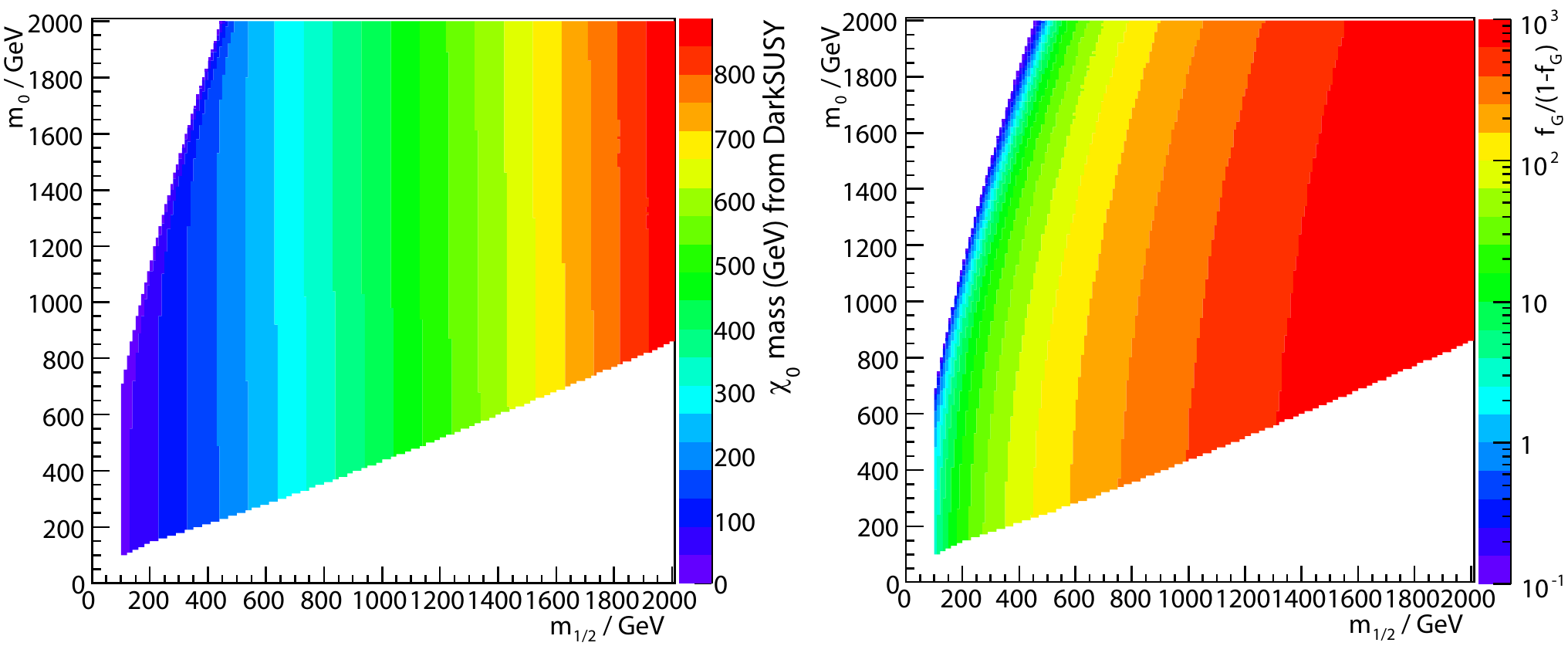}
\end{center}
\caption{Neutralino mass ({\it left}) and content $f_G/(1-f_G)$ as defined by (\ref{eq:neutrfractions}) ({\it right}) in the
$m_{1/2}$-$m_0$-plane, for $\tan\beta=40$ and $m_t=170.9\,\mathrm{GeV}$.}
\label{fig:neutralino}
\end{figure}

Direct searches by the experiments at the LEP collider set lower
bounds on sparticle masses, thereby already providing some constraints
on the mSUGRA parameter space. For example, the data indicate a neutralino
mass $m_{\chi^0}>46\,\mathrm{GeV}$ and a chargino mass
$m_{\chi^\pm}>94\,\mathrm{GeV}$ at the $95\,\%$ confidence
level~\cite{ref:pdg}. Limits for other sparticle species also exist but
do not play an important role in the context of this study.
Figure~\ref{fig:neutralino}
illustrates the fact that $m_\chi$ is roughly proportional to
$m_{1/2}$. The figure also shows the ratio of the gaugino and higgsino fractions of the
neutralino for a given scan in the $m_{1/2}$-$m_0$-plane, which
determines the preferred annihilation channels.\\
\par
A number of experiments are searching for WIMPs directly, looking for
elastic collisions of WIMPs with nuclei in terrestrial
targets~\cite{ref:baudis}. The differential rate for WIMP elastic
scattering off nuclei can be expressed as
\begin{equation}
  \label{eq:wimprate}
  \frac{\mathrm{d}R}{\mathrm{d}E_R}=N_T\frac{\rho_0}{m_W}\int^{v_\mathrm{max}}_{v_\mathrm{min}}\mathrm{d}\mathbf{v}f(\mathbf{v})\frac{\mathrm{d}\sigma}{\mathrm{d}E_R}
\end{equation}
where $N_T$ denotes the number of target nuclei, $m_W$ is the WIMP
mass, $f(\mathbf{v})$ is the WIMP velocity distribution in the Earth
frame and $\mathrm{d}\sigma/\mathrm{d}E_R$ is the WIMP-nucleus
differential cross section.
$\rho_0=0.3\,\frac{\mathrm{GeV}}{\mathrm{cm}^3}$ is the local density. $v_\mathrm{min}$ is determined by the WIMP
and nuclear masses and the energy threshold of the detector, and
$v_\mathrm{max}$ is the escape velocity of the WIMP.\\
Two separate components contribute to the differential WIMP-nucleus
cross section, an effective scalar coupling between the WIMP and the
nucleus and an effective coupling between the spin of the WIMP and the
total spin of the nucleus. Therefore,
\begin{equation}
  \label{eq:wimpcs}
  \frac{\mathrm{d}\sigma}{\mathrm{d}E_R}\propto\sigma^0_{SI}F^2_{SI}(E_R)+\sigma^0_{SD}F^2_{SD}(E_R)
\end{equation}
where $\sigma^0_{SI,SD}$ are the spin-independent and spin-dependent WIMP-nucleus cross sections in the
limit of zero momentum transfer and $F^2_{SI,SD}(E_R)$ denote the
nuclear form factors.\\
As a WIMP in the $\mathrm{GeV}$-$\mathrm{TeV}$ mass range will deposit a recoil energy
of around $50\,\mathrm{keV}$ and the predicted event rates for
neutralinos can be as low as $10^{-6}$ per kg and day, a WIMP detector
must have low energy threshold, low background and high target
mass. In the detector, the recoil energy is transformed into a
measurable signal, such as charge, light or phonons, and powerful
background discrimination can be achieved by observing two signals
simultaneously. Classes of WIMP detectors currently deployed include
cryogenic detectors at $\mathrm{mK}$ temperatures, such as
CDMS~\cite{ref:cdmslimit}, liquid noble element detectors,
e.g.~XENON10~\cite{ref:xenon10limit}, and superheated liquid detectors,
such as COUPP~\cite{ref:coupplimit}. While a convincing WIMP signal has
not been observed yet, the limits imposed by some of these detectors
are already quite stringent (fig.~\ref{fig:directdetection}).

\subsection{Cosmic rays}
As has been shown, neutralino annihilation in the Galactic halo leads to the production
of positrons (and electrons), antiprotons and $\gamma$-rays in the
$\mathrm{GeV}$-range. This assumed primary source of cosmic rays has
to be distinguished from the so-called secondary component. Cosmic-ray
particles, mostly protons, react with the interstellar matter creating
stable secondary particles and $\gamma$-rays via $\pi^0$-production in
the process. $\gamma$-rays are also produced by bremsstrahlung
processes, synchrotron radiation
of electrons in the Galactic magnetic field and inverse
Compton-scattering of electrons on background photons from starlight
and the cosmic microwave background. Because of their small mass, electrons quickly lose
energy due to synchrotron and bremsstrahlung radiation.\\
For the evaluation of the secondary cosmic-ray component,
the Galprop package as described in
section~\ref{sec:galprop} is used. The conventional model was chosen to
calculate the background positron, antiproton and $\gamma$-ray
fluxes. At the same
time as investigating a possible primary signal, the PEBS or AMS-02 positron
data will be used to refine the propagation model itself. It will
therefore be assumed that the secondary component is known with good
precision and the study will be restricted to this one model.\\
The calculation of the expected primary flux of positrons from
neutralino annihilation for a given model is done by DarkSUSY and
proceeds as follows: In a first step, the local annihilation rate is
calculated as $1/2(\rho_\chi(\vec{x})/m_\chi)^2\sigma_\mathrm{ann}v$.
For this purpose, the
standard modified isothermal profile for the halo density $\rho_\chi$ is used, given by
\begin{equation}
\rho_\chi(r)=\rho_0\,\cdot\,\frac{1+\left(\frac{r_0}{a_h}\right)^2}{1+\left(\frac{r}{a_h}\right)^2}
\end{equation}
with the local density
$\rho_0=0.3\,\frac{\mathrm{GeV}}{\mathrm{cm}^3}$, the galactocentric
distance of the sun $r_0=8.5\,\mathrm{kpc}$ and the length scale
$a_h=3.5\,\mathrm{kpc}$. Other halo profiles are frequently used in
the literature and they can differ greatly in their behaviour near the
galactic centre. Nevertheless, as one is primarily interested in the
positrons, which quickly lose energy and therefore originate in the
solar neighbourhood, the influence of the choice of halo model on the expected
fluxes can be expected to be small.\\
In a second step, the production rate of positrons is estimated by
folding together the branching ratio into a given two-body final state
with the Monte Carlo simulation of the hadronisation and/or decay of
that state as implemented in DarkSUSY. The relevant final states for positrons are
$\ell^+\ell^-$, $q\bar{q}$,
$W^+W^-$, $Z^0Z^0$, $W^\pm{}H^\mp$, $ZH_1^0$, $ZH_2^0$, $H_1^0H_3^0$ and
$H_2^0H_3^0$ at tree-level, and $Z\gamma$ and $gg$ at one-loop
level. For the hadronisations and decays, the results from a
Pythia~6.154~\cite{ref:pythia} simulation are tabulated and
interpolated.\\
The third step is the propagation of the signal flux through the
interstellar medium to obtain the local interstellar flux. The
default implementation is used, an analytical solution of a diffusion equation
with energy losses and cylindrical symmetry as outlined
in~\cite{ref:darksusyprop}, that is implemented in DarkSUSY.
In a more sophisticated approach, an additional interface from
DarkSUSY to Galprop will be necessary to treat both background and
signal fluxes by the same propagation model.\\
Lastly, the local interstellar signal fluxes obtained from Galprop and
DarkSUSY ($\Phi_i^\mathrm{sig}(e^+)=\Phi_i^\mathrm{sig}(e^-)$) are
subjected to solar and geomagnetic modulation. Solar modulation is
modelled by the force-field approximation introduced in
section~\ref{sec:solarmod} and equation~(\ref{eq:geomod}) is used for
cutting off the fluxes due to the geomagnetic effect.
From a fit to the AMS-01 electron spectrum (fig.~\ref{fig:electrons}), the solar modulation
parameter $\phi$ as well as the cutoff rigidity
$R_c$ and steepness $\gamma_c$ are
extracted for the background flux.
Using the same $\phi$ for electrons and positrons, a
different set of cutoff parameters is then fitted for the positrons to the AMS-01
positron flux.
The goal of this procedure is to provide a good description of the
positron fraction data at lower energies, where the SUSY signal does
not contribute significantly, so that the $\chi^2$ used below
to compare different models to the positron data is governed by the
signal region at higher energies.

\section{Constraints on mSUGRA parameter space from currently
available data}
\label{sec:constraints}
In this section, it will be investigated which regions of the mSUGRA parameter
space are compatible with measurements of the various observables
discussed in the previous section. Especially interesting is the question how
much information the presently available positron data can contribute
to this problem.
The results of this section will guide us in our choice of example
scenarios for the discussion of the physics potential of PEBS and AMS-02 in the
following sections. The new positron fraction data from PAMELA will be
left aside for the moment. They will briefly be considered in
section~\ref{sec:msugra_pamela}.\\
\par
In order to evaluate the constraining power of the positron fraction
data, the (mSUGRA,$m_t$)-parameter space was scanned as described below
and the $\chi^2$ was calculated, defined as follows:
\begin{eqnarray}
\label{eq:chi2posfrac}
\chi^2_{e^+/(e^++e^-)}=\sum\limits_i\left(\frac{\Delta{}f_i}{\sigma_i}\right)^2\quad\mathrm{with}\quad\Delta{}f_i\equiv{}f_i^\mathrm{model}-f_i^\mathrm{data}\quad\mathrm{and}\\
f_i^\mathrm{data/model}\equiv\frac{\Phi_i^\mathrm{data/model}(e^+)}{\Phi_i^\mathrm{data/model}(e^+)+\Phi_i^\mathrm{data/model}(e^-)}
\end{eqnarray}
where $\Phi$ denotes fluxes and $i$ denotes energy bins. The
positron fraction from data $f_i^\mathrm{data}(e^+)$ is the weighted mean of
most of the existing data, as presented in figure~\ref{fig:positronfraction}.
The model fluxes are calculated as the sum of the modulated background flux and
the boosted modulated signal flux:
\begin{equation}
\label{eq:boostedflux}
\Phi_i^\mathrm{model}(e^\pm)=\Phi_i^\mathrm{bg,mod}(e^\pm)+f_b\cdot\Phi_i^\mathrm{sig,mod}(e^\pm)
\end{equation}
The boost factor $f_b$ in equation (\ref{eq:boostedflux}) is used as
the only free parameter in a minimisation of $\chi^2_{e^+/(e^++e^-)}$
for a given point in the mSUGRA parameter space.
The asymmetric errors $\sigma^+$ and $\sigma^-$ of the cosmic-ray data
are treated according to the recommendation of~\cite{ref:pdg}:
\begin{equation}
\sigma_i=
\left\{
\begin{array}{ll}
\sigma^+_i &\quad\Delta{}f_i>\sigma_i^+\\
\frac{\sigma_i^+-\sigma_i^-}{\sigma_i^++\sigma_i^-}\,\Delta{}f_i+\frac{2\sigma_i^+\sigma_i^-}{\sigma_i^++\sigma_i^-}
&\quad\-\sigma_i^-<\Delta{}f_i<\sigma_i^+\\
\sigma^-_i &\quad\Delta{}f_i<-\sigma_i^-\\
\end{array}
\right.
\end{equation}
For the positron fraction and all results quoted here, the
following grid\footnote{The range adopted for $m_t$ was motivated by
the confidence interval current at the time the analysis was
performed~\cite{ref:mtop07}.} was scanned:
\begin{itemize}
\item $m_0$ and $m_{1/2}$ from $100\,\mathrm{GeV}$ to
$2000\,\mathrm{GeV}$ in steps of $10\,\mathrm{GeV}$,
\item $\tan\beta$ from 10 to 60 in steps of 10, and
\item $m_t$ from $167.18\,\mathrm{GeV}$ to $174.62\,\mathrm{GeV}$ in
steps of $0.465\,\mathrm{GeV}$.
\end{itemize}
In addition, a small part of the coannihilation region was scanned
using a finer sampling of $\tan\beta$ in order to obtain smoother
$\chi^2$-contours. Roughly four million parameter sets are included in
the scans.
\begin{figure}[htb]
\begin{center}
\includegraphics[width=1.0\textwidth,angle=0]{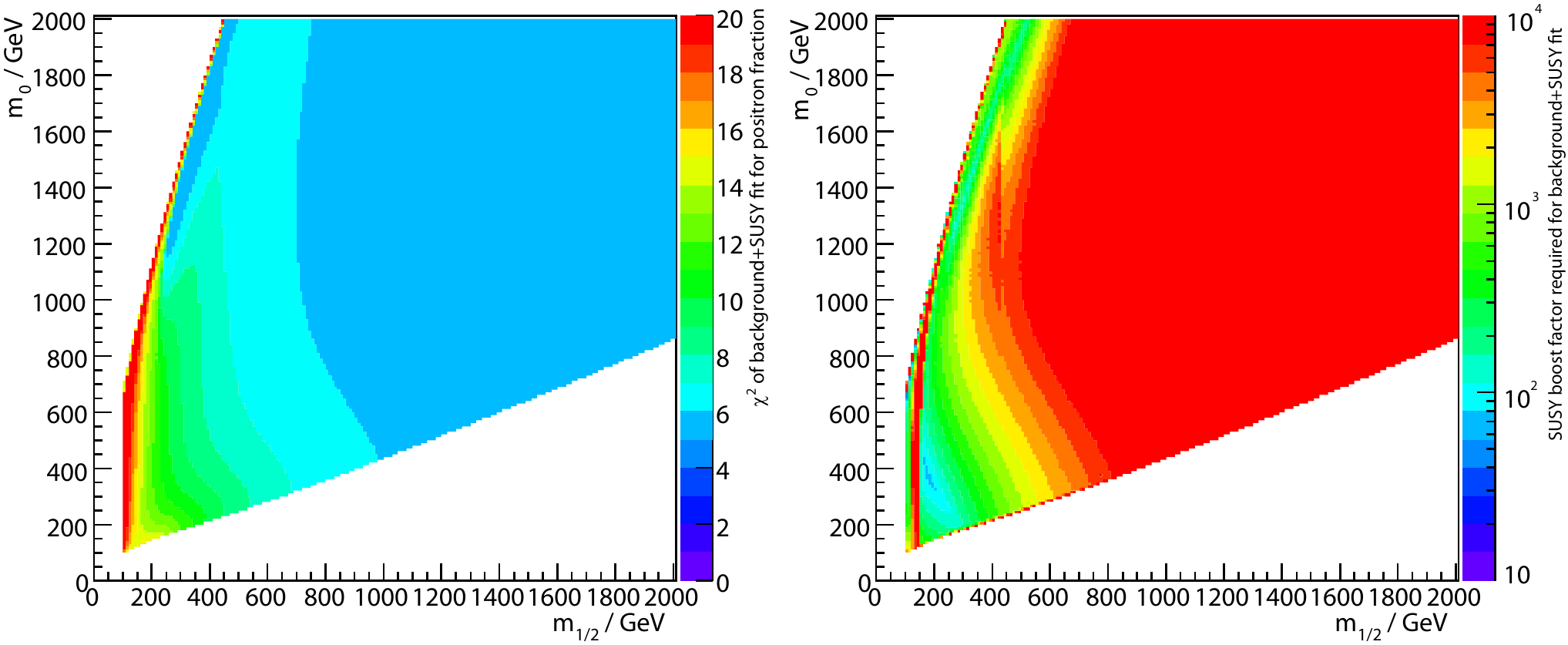}
\end{center}
\caption{$\chi^2_{e^+/(e^++e^-)}$ and corresponding boost factor for the positron
fraction, for $\tan\beta=40$ and $m_t=170.9\,\mathrm{GeV}$.}
\label{fig:scanposfrac}
\end{figure}
\par
\begin{figure}[htb]
\begin{center}
\begin{tabular}{cc}
\includegraphics[width=0.52\textwidth,angle=0]{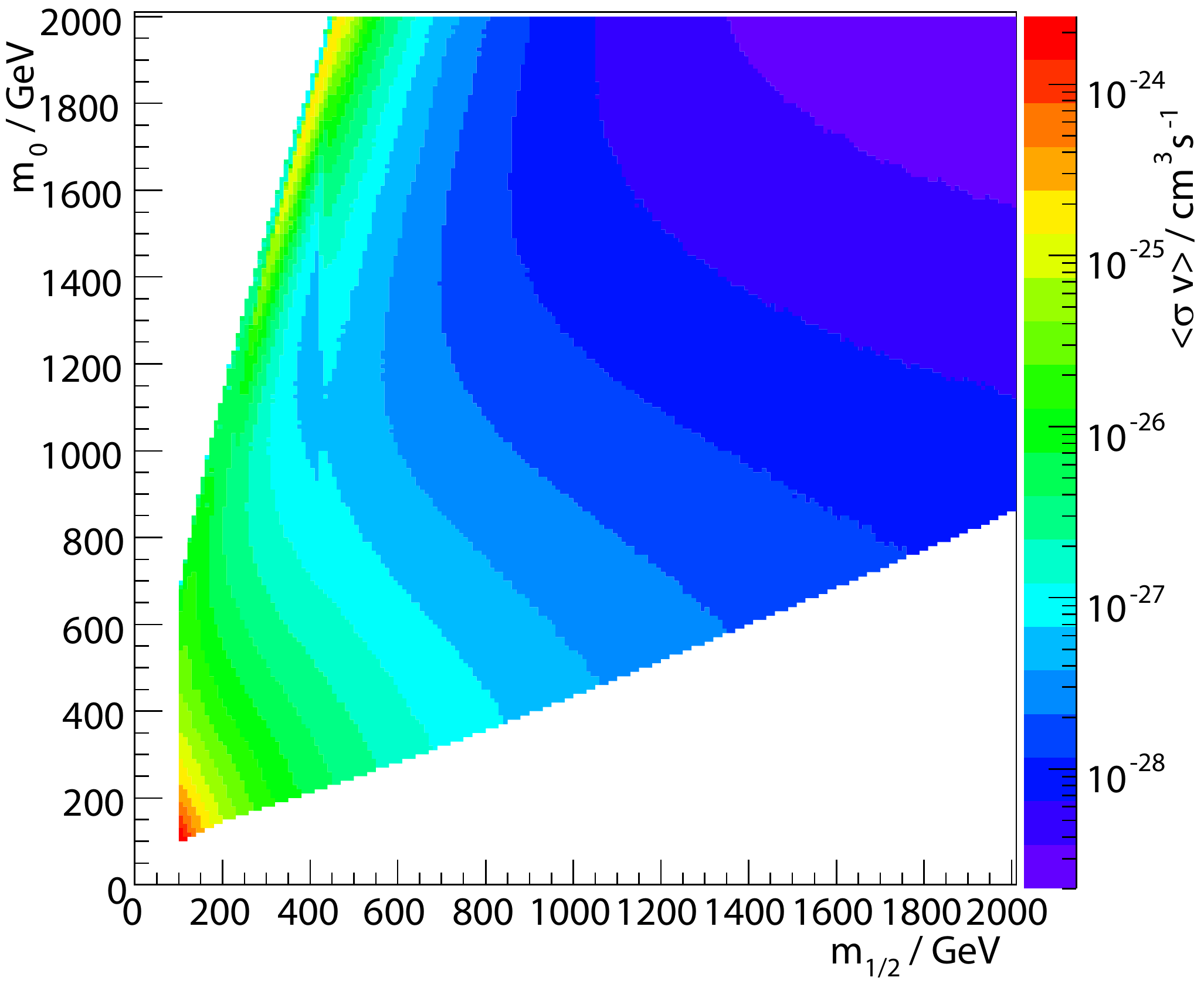}&
\includegraphics[width=0.44\textwidth,angle=0]{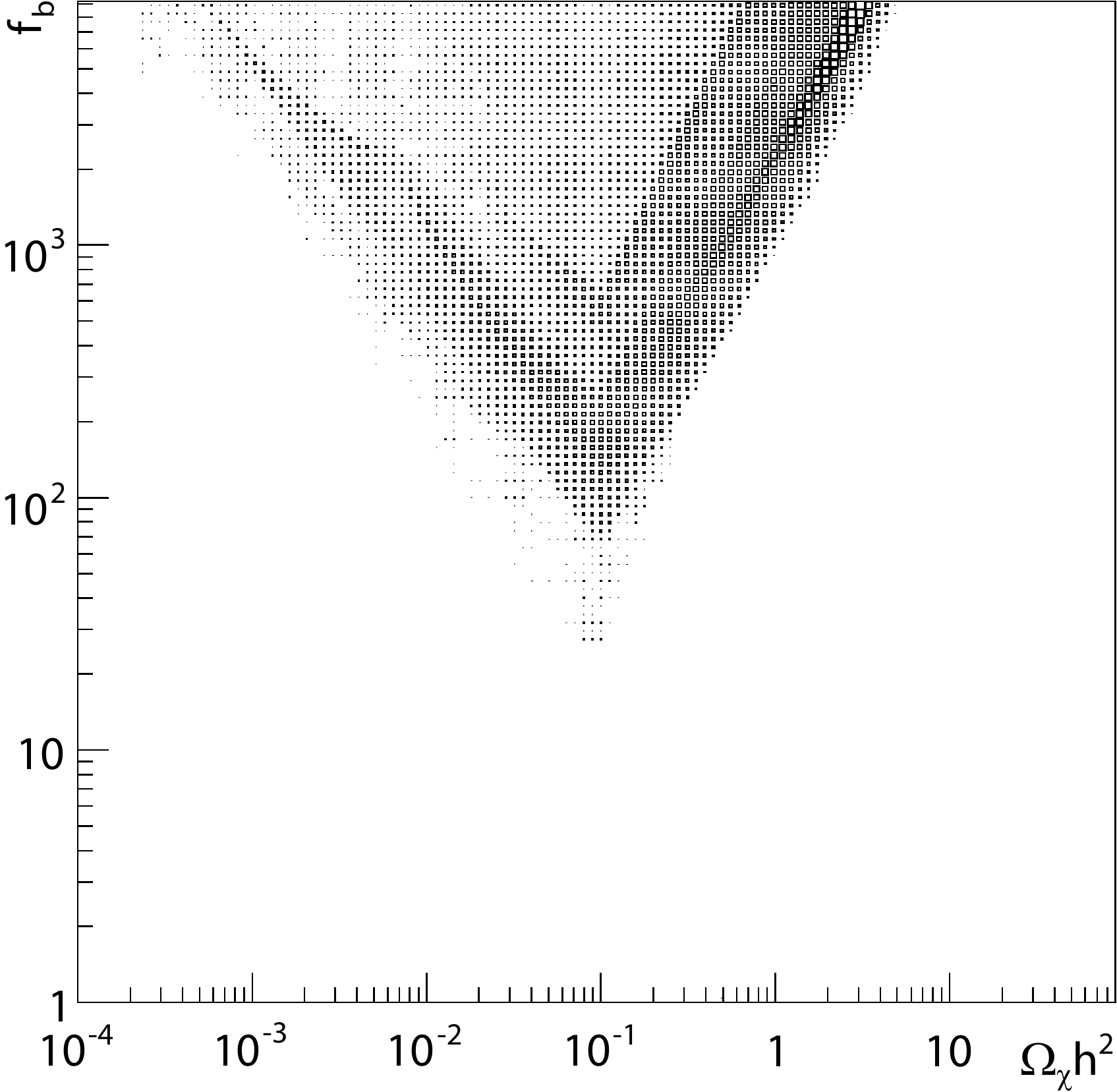}\\
\end{tabular}
\end{center}
\caption{{\it Left:} $\langle\sigma_\mathrm{ann}v\rangle$ calculated
by DarkSUSY, in the $m_{1/2}$-$m_0$-plane, for $\tan\beta=40$ and
$m_t=170.9\,\mathrm{GeV}$. {\it Right:} Best-fit positron fraction
boost factor $f_b$ plotted against
the corresponding neutralino relic density $\Omega_\chi{}h^2$ for
mSUGRA models considered here. The boost factor rises towards low
relic densities due to the rescaling according to
eq.~(\ref{eq:bf_rescaling}).}
\label{fig:bf_omega}
\end{figure}

Figure~\ref{fig:scanposfrac} illustrates the behaviour of
$\chi^2_{e^+/(e^++e^-)}$ for given values of $\tan\beta$ and $m_t$:
Except for very low values of $m_{1/2}$, the $\chi^2$-contour is
essentially flat and therefore lacks predictive power. The figure
also shows the corresponding boost factor obtained in the minimisation
of (\ref{eq:chi2posfrac}). It varies over many orders of magnitude
along with the calculated value of
$\langle\sigma_\mathrm{ann}v\rangle$ (fig.~\ref{fig:bf_omega} {\it
left}) because the overall signal amplitude is fixed by the positron
fraction data.\\
The boost factor $f_b$ included in (\ref{eq:boostedflux}) can be
interpreted as a measure of the clumpiness of the dark matter. Under
the influence of their own gravitational attraction, dark matter
particles will form clumps thus enhancing the local WIMP density over
that found for a smooth distribution. Since the annihilation rate
scales as the square of the number density, the {\it local} boost
factor for a given clump is:
\begin{equation}
\label{eq:bf}
f_b=\frac{\langle\rho^2\rangle_\mathrm{cl}}{\langle\rho\rangle^2_\mathrm{cl}}
\end{equation}
The overall factor that the positron flux is boosted by compared to a
smooth distribution is strongly affected by the propagation process as
the influence of clumps decreases with increasing distance.\\
It was shown recently that it is difficult to accommodate boost
factors differing much from unity in the light of
$\Lambda$CDM  N-body simulation results~\cite{ref:bfs}. For models that 
already seem unlikely, boost factors no larger than~20 can be
obtained. Such a configuration requires either a dark matter clump
very close to the solar system or extremely peaked density profiles
for the clumps. Other scenarios, such as dark
matter density spikes around intermediate mass
black holes~\cite{ref:imbhs} can yield boost factors of up to a few
thousand.\\
Naively, one will expect models with lower neutralino relic density
$\Omega_\chi{}h^2$ to have higher signal fluxes because of the higher
annihilation cross sections, as implied by
equation~(\ref{eq:reldens_crosssection}). In turn, this means lower
boost factors. Figure~\ref{fig:bf_omega} shows the best-fit boost factors to the
positron and electron signal flux obtained in the mSUGRA models
considered in this study, plotted against the corresponding
$\Omega_\chi{}h^2$.
Assuming that the Galactic dark halo density
equals the neutralino relic density, the neutralino density
in models featuring a neutralino relic density that cannot account for
the entire density of non-baryonic matter must be rescaled as
follows~\cite{ref:baltz_positron_excess}:
\begin{equation}
\label{eq:rescaling}
\rho_\chi(r)=\left(\frac{\Omega_\chi{}h^2}{\Omega_\mathrm{nbm}h^2}\right)\rho(r)
\end{equation}
It was chosen to apply this rescaling for all models with
$\Omega_\chi{}h^2\leq{}0.1$. This affects the boost factor as
\begin{equation}
\label{eq:bf_rescaling}
f_b\rightarrow{}f_b\left(\frac{\Omega_\chi{}h^2}{0.1}\right)^{-2}
\end{equation}
As is obvious from the figure, this means that the lowest boost
factors are found in models that have neutralino relic densities small
enough as to just require no rescaling. The figure also implies that
it is difficult to accommodate a scenario that explains the positron
fraction excess as being due to neutralino annihilations without
needing boost factors larger than~100 or so in the mSUGRA model. The
Sommerfeld enhancement effect~\cite{ref:silksommerfeld} that has been
discussed in the literature to boost the annihilation cross section at
non-relativistic speeds is not important in the context of this
study. Typically, it becomes sizeable for dark matter masses above
$\sim1\,\mathrm{TeV}$, but this range is not considered here.\\
\par
\begin{figure}[htb]
\begin{center}
\includegraphics[width=1.0\textwidth,angle=0]{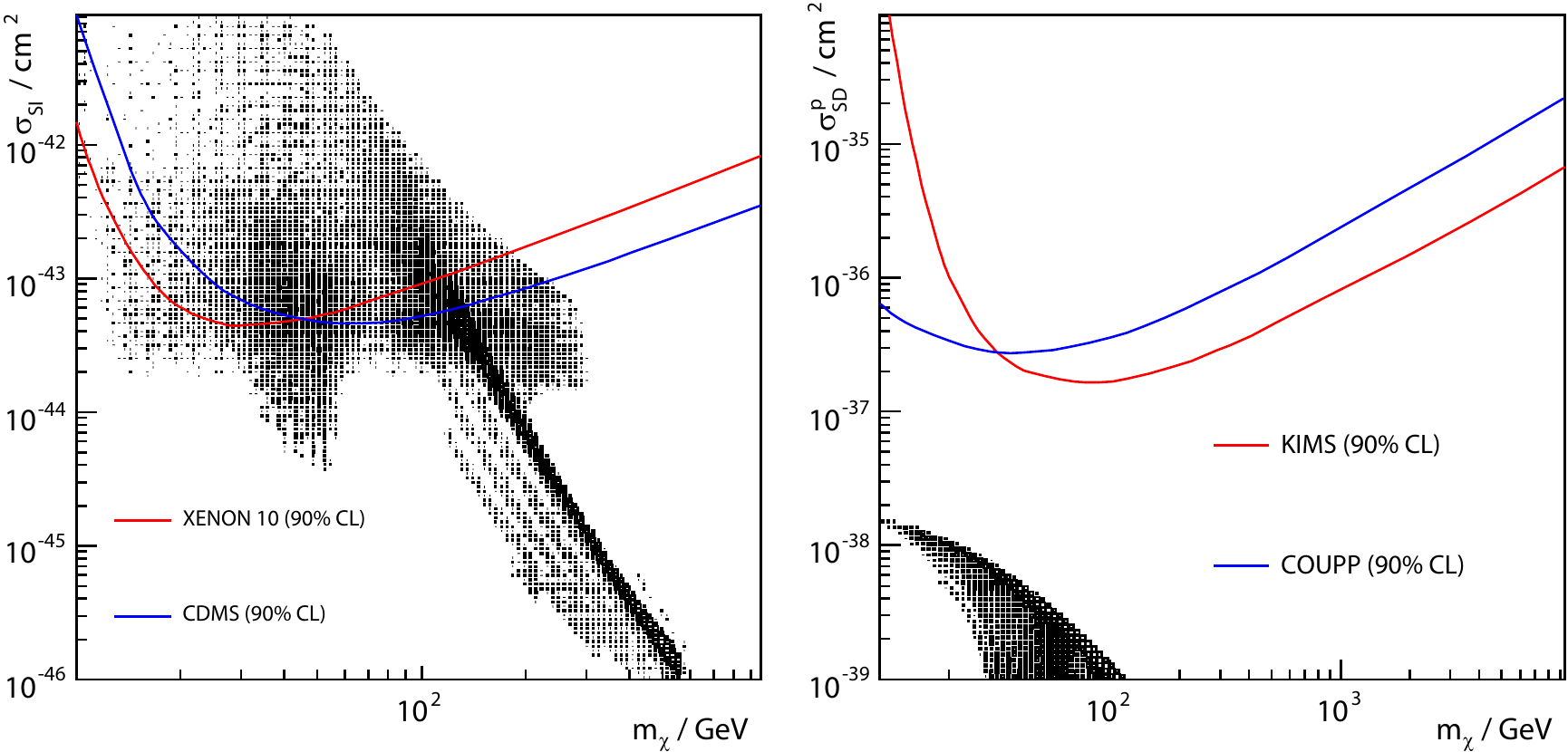}
\end{center}
\caption{mSUGRA prediction for WIMP-nucleon cross sections, for
effective scalar (spin-independent) couplings ({\it left}) and pure
proton spin-dependent couplings ({\it right}). Models allowed by the
constraints on $\Omega_\chi{}h^2$, $\Delta{}a_\mu$ and
$BR(b\rightarrow{}s\gamma)$ at the $3\sigma$-level each are
included. Also shown are the currently best limits, provided by the
XENON10~\cite{ref:xenon10limit}, CDMS~\cite{ref:cdmslimit},
COUPP~\cite{ref:coupplimit} and KIMS~\cite{ref:kimslimit} experiments,
respectively.}
\label{fig:directdetection}
\end{figure}

Figure~\ref{fig:directdetection} shows the latest results of some
direct detection experiments. The limits obtained on the
spin-independent cross section and the cross section for pure proton
spin-dependent couplings are drawn as a function of $m_\chi$. mSUGRA
models allowed by the constraints on $\Omega_\chi{}h^2$,
$\Delta{}a_\mu$ and $BR(b\rightarrow{}s\gamma)$ at the $3\sigma$-level
each are included. 
The measurements by the CDMS and XENON10 experiments already exclude
significant amounts of the mSUGRA parameter space. Especially points
belonging to the focus point region or the bulk annihilation region at
low values of $m_0$ and $m_{1/2}$ are excluded in this way. On the other hand,
measurements of the spin-dependent couplings still need at least an
order of magnitude more sensitivity.\\
\par
\begin{figure}[htb]
\begin{center}
\includegraphics[width=0.5\textwidth,angle=0]{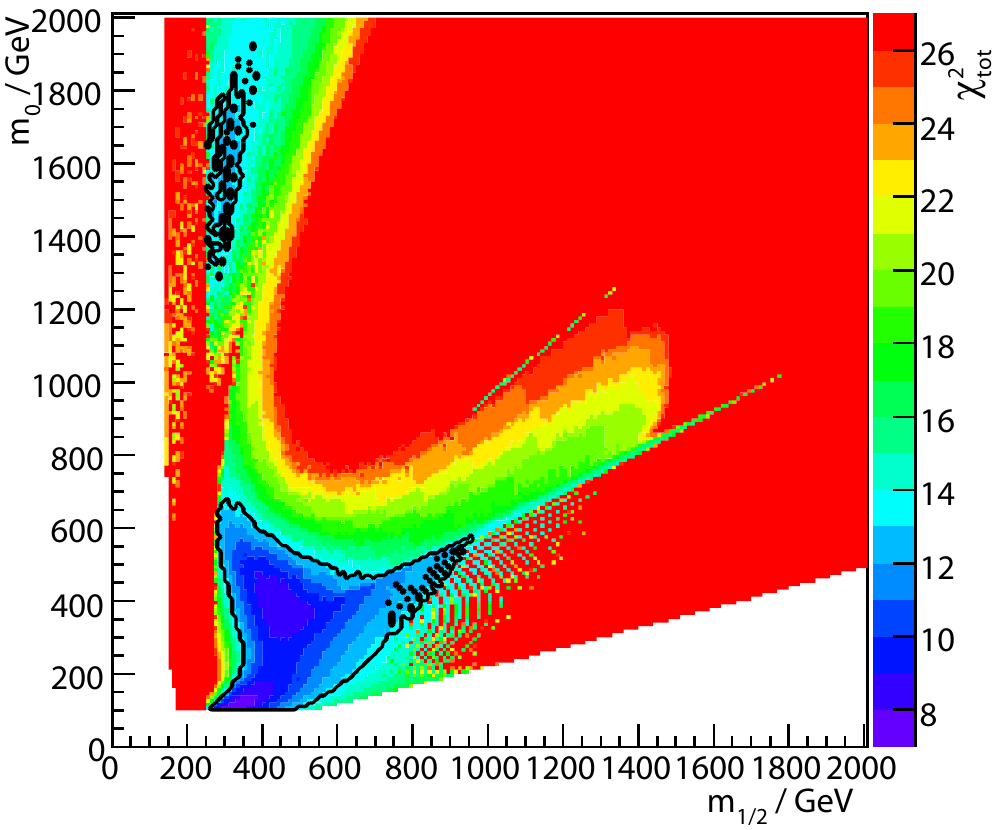}
\end{center}
\caption{$\chi^2_\mathrm{tot}$-contour in the
  $m_{1/2}$-$m_0$-plane. The minimum obtainable for any value of
  ($\tan\beta$,$m_t$) in the scans is plotted in each point. Also
  included is the contour for the parameters within the intervals of
  $90\,\%$ confidence level around the minimum. Only
  parameter points allowed by the direct detection experiments and LEP
  mass limits are included.}
\label{fig:contour_totchi2}
\end{figure}

All the information gathered so far can now be assembled in
a condensed way. For this purpose, a total $\chi^2$ for
all mSUGRA parameter points included in the scans is calculated:
\begin{equation}
  \label{eq:totalchi2}
\begin{array}{l}
\chi^2_\mathrm{tot}=\\
\chi^2_\mathrm{e^+/(e^++e^-)}+
\left(\frac{m_t^\mathrm{scan}-m_t}{\sigma_{m_t}}\right)^2+\\
\Theta(\Omega^\mathrm{scan}_\chi{}h^2-\Omega_\mathrm{nbm}h^2)\cdot\left(\frac{\Omega^\mathrm{scan}_\chi{}h^2-\Omega_\mathrm{nbm}h^2}{\sigma_{\Omega_\mathrm{nbm}h^2}}\right)^2+\\
\left(\frac{\Delta{}a_\mu^\mathrm{scan}-\Delta{}a_\mu}{\sigma_{\Delta{}a_\mu}}\right)^2+
\left(\frac{BR^\mathrm{scan}_{b\rightarrow{}s\gamma}-BR_{b\rightarrow{}s\gamma}}{\sigma_{BR(b\rightarrow{}s\gamma)}}\right)^2
\end{array}
\end{equation}
where the superscript $^\mathrm{scan}$ denotes a value at a given
mSUGRA parameter point and the Heaviside function in front of the
relic density term makes sure that only neutralino relic densities in
excess of the constraint contribute to the $\chi^2$.\\
One proceeds by calculating $\chi^2_\mathrm{tot}$ for all mSUGRA
parameter points in the scans and looking at $\chi^2_\mathrm{tot}$ in the two-dimensional
projections of the (mSUGRA,$m_t$)-parameter space. The most
interesting one, the $m_{1/2}$-$m_0$-plane, is shown in
figure~\ref{fig:contour_totchi2}. For a given point in that plane, the minimum
value of $\chi^2_\mathrm{tot}$ obtainable for any pair of
($\tan\beta$,$m_t$) is plotted, but
only  parameter points allowed by the direct detection experiments and
LEP mass limits are included.
The contour for the parameters within an interval of
$90\,\%$ confidence level around the minimum, where the
number of free parameters considered is three ($m_0$, $m_{1/2}$, and
$\tan\beta$), is drawn, too~\cite{ref:cowan}.
It is stressed that the figure does not provide any evidence for the
existence of neutralino dark matter, but rather may indicate some
hints on where to look assuming that mSUGRA is realised in nature.\\
In this sense, low values of $\chi^2_\mathrm{tot}$ are found for several
regions in the mSUGRA parameter space~\cite{ref:baer}:
The lowest values can be found for the $\tilde{\tau}$-coannihilation
regions at low values of $m_0$, where
$\chi\tilde{\tau}$-coannihilations contributed significantly to the
neutralino annihilation rate in the early Universe. Acceptable values
of $\chi^2_\mathrm{tot}$ are also found in the focus point region at
large $m_0$, near the boundary of the region forbidden by the absence
of electroweak symmetry breaking, and in the $A$-annihilation funnel
occurring for large $\tan\beta$. Unfortunately, only weak conclusions
regarding the neutralino mass can be drawn from these results
(fig.~\ref{fig:neutralino_PP1}).\\
For illustration purposes, two reference points were chosen, both with
acceptable values of $\chi^2_\mathrm{tot}$, 
called PP1 and PP2 in the following, giving a typical example of mSUGRA
phenomenology in the preferred region:
\begin{equation}
  \label{eq:pp1}
\mathrm{PP1:}\quad m_0=1560\,\mathrm{GeV}\quad
m_{1/2}=260\,\mathrm{GeV}\quad \tan\beta=40\quad m_t=172.76\,\mathrm{GeV}
\end{equation}
\begin{equation}
  \label{eq:pp2}
\mathrm{PP2:}\quad m_0=100\,\mathrm{GeV}\quad
m_{1/2}=310\,\mathrm{GeV}\quad \tan\beta=20\quad m_t=172.76\,\mathrm{GeV}
\end{equation}
PP2 is very close to the lowest value of $\chi^2_\mathrm{tot}$ in the scans and lies
in the $\tilde{\tau}$-coannihilation region, while PP1 is in the focus
point region offering better prospects for discovery with PEBS and
AMS-02 as will be seen.
Neutralino annihilation is dominated by the
$\chi\chi\rightarrow{}W^+W^-$ channel in this case. Explaining the
observed excess in the positron fraction requires large boost factors,
as seen above, and best-fit values of 152 and 1492 are obtained for PP1 and
PP2, respectively.
\par
\begin{figure}[htb]
\begin{center}
\includegraphics[width=1.0\textwidth,angle=0]{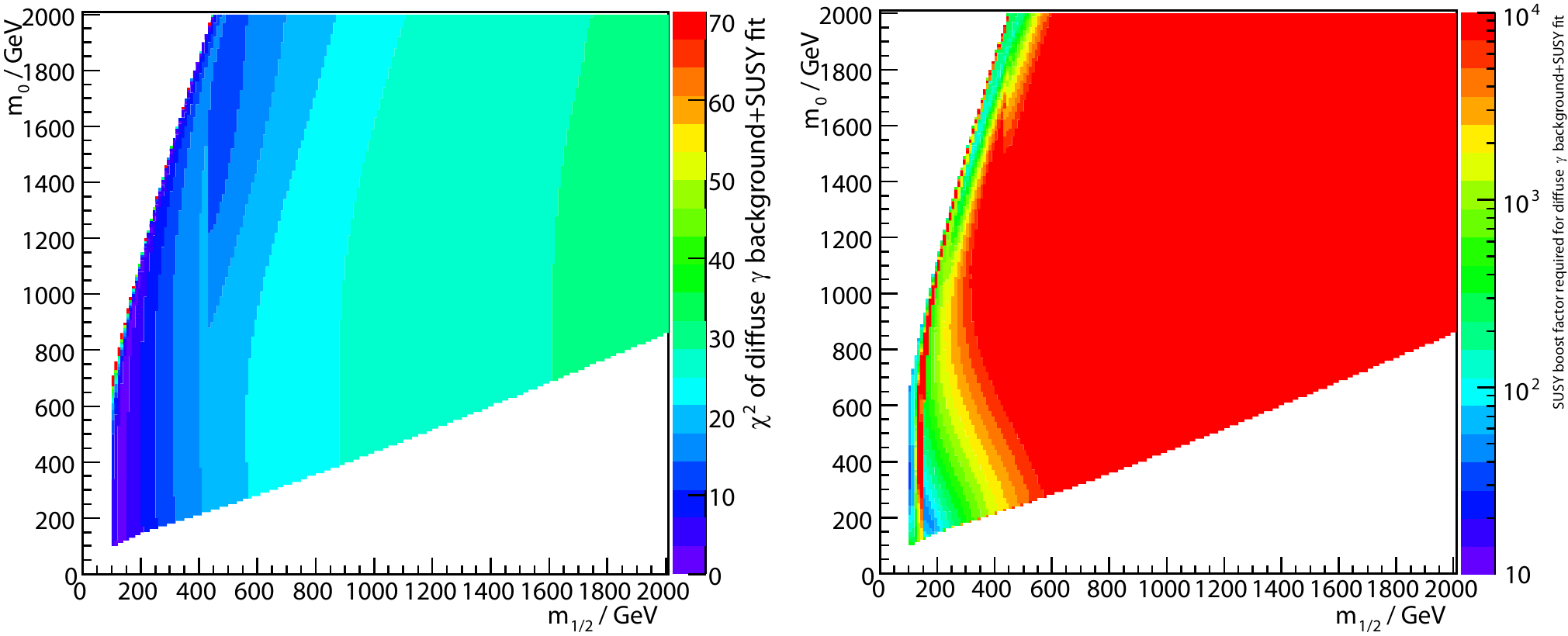}
\end{center}
\caption{$\chi^2_\gamma$ and corresponding boost factor for $\gamma$-ray
data, for $\tan\beta=40$ and $m_t=170.9\,\mathrm{GeV}$.}
\label{fig:scangamma}
\end{figure}
\begin{figure}[htb]
\begin{center}
\begin{tabular}{cc}
\includegraphics[width=0.5\textwidth,angle=0]{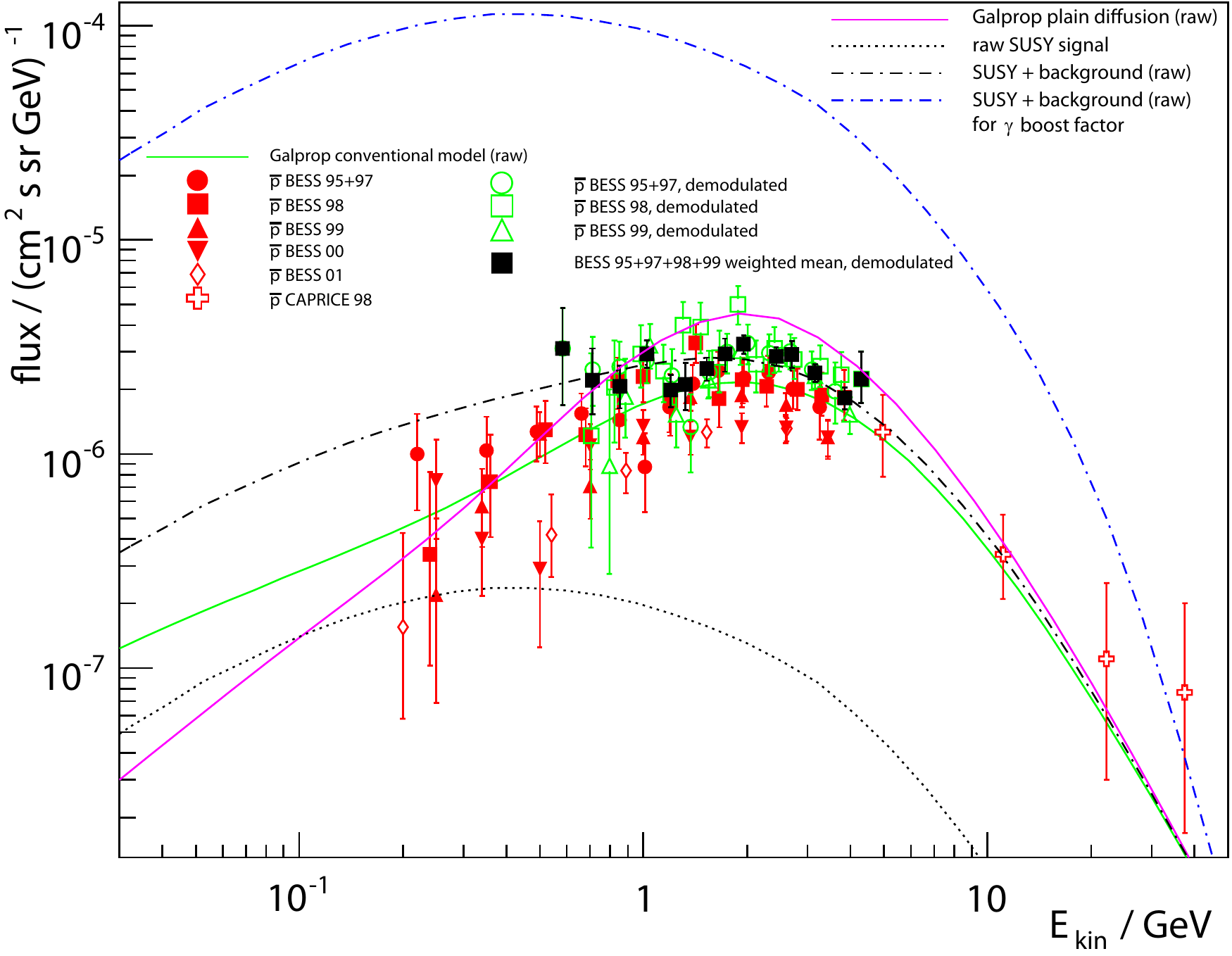}&
\includegraphics[width=0.5\textwidth,angle=0]{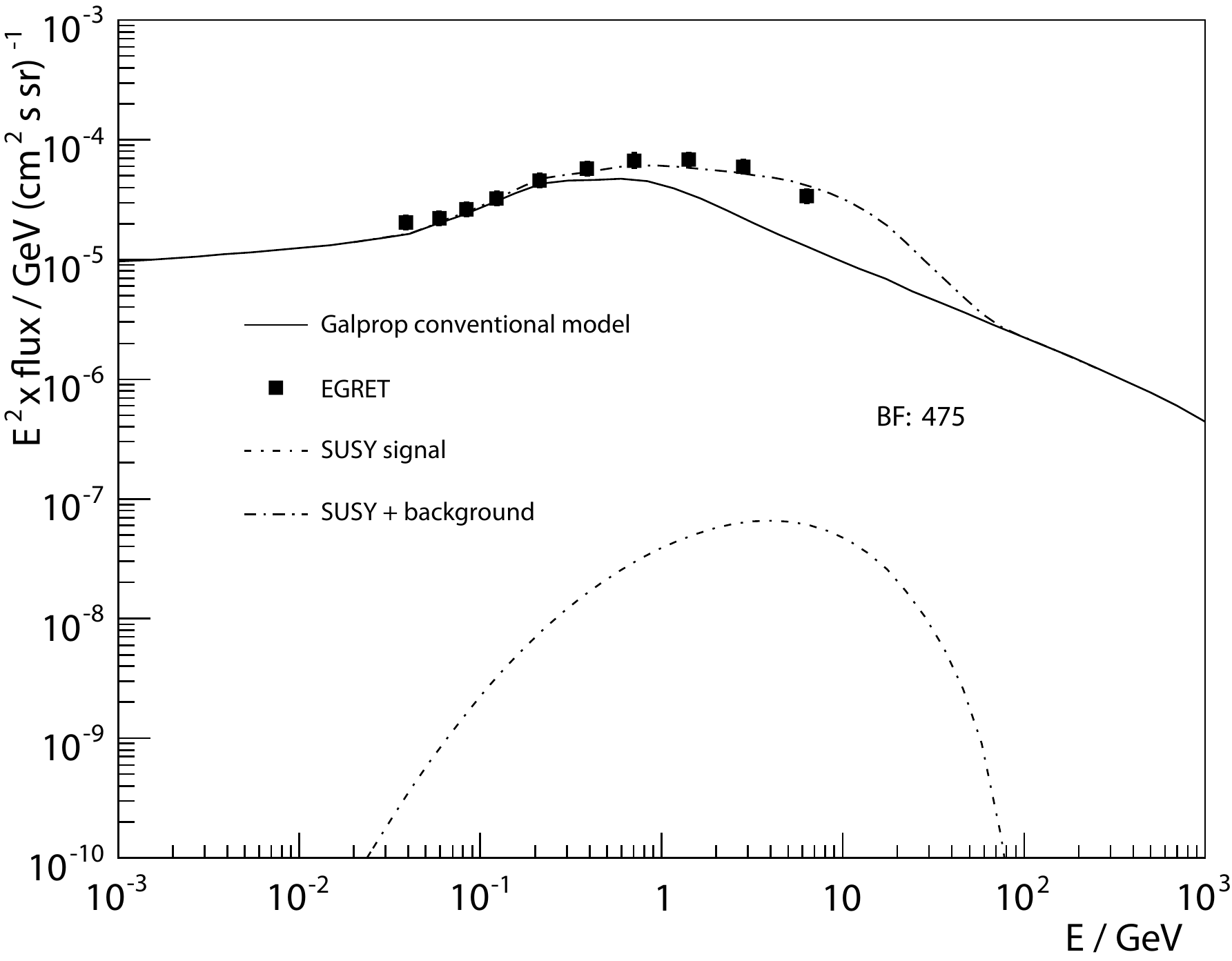}\\
\end{tabular}
\end{center}
\caption{Antiproton data ({\it left}) and data for diffuse $\gamma$-rays from the
  inner Galaxy ($330^\circ<l<30^\circ$, $|b|<5^\circ$) ({\it right}). Antiproton data are from
  BESS~\cite{ref:bess98,ref:bess99,ref:bess9597,ref:bess01}
  and CAPRICE~\cite{ref:capricepbar}. For better comparison, the weighted
  mean of the demodulated BESS data for the years 1995-99 is also
  shown. The raw spectra predicted by Galprop's conventional and plain
diffusion models are shown, together with the signal from neutralino
annihilation. The predicted spectrum obtained with the boost factor
for the best fit to the $\gamma$-ray data is included. $\gamma$-ray
data are from EGRET~\cite{ref:egret}. The best-fit spectra obtained in
the PP1 benchmark model are also shown.}
\label{fig:pbargammaPP1}
\end{figure}

Before turning to the detection prospects for PEBS and AMS-02 for these benchmark
models, a discussion of the situation for other promising dark matter probes,
namely $\gamma$-rays, antiprotons and antideuterons follows.\\
The $\gamma$-ray component in the cosmic rays is unique in that it
always points directly back to its origin. $\pi^0$s created in the
decay chain of WIMP annihilations will constitute an additional source
of $\gamma$-rays. The best data available so far in the GeV-range was
taken by the EGRET detector~\cite{ref:egret} on board the Compton Gamma
Ray Observatory (fig.~\ref{fig:pbargammaPP1}).\\
Again, the $\chi^2$ of a background+model hypothesis was calculated, as
\begin{equation}
  \label{eq:gammachi2}
\chi^2_\gamma=\sum\limits_i\left(\frac{\Phi^\mathrm{model}_i-\Phi^\mathrm{data}_i}{\sigma_i}\right)^2
\end{equation}
where
$\Phi^\mathrm{model}_i=\Phi^\mathrm{bg}_i+f_b^\gamma\cdot\Phi^\mathrm{sig}_i$.
The boost factor for the $\gamma$-rays is allowed to differ from the
one for positrons and electrons. This is due to the fact that the
latter originate from regions of space in our vicinity because
of their large energy losses whereas the former sample the entire
Galaxy. The boost factor is chosen as a free parameter minimising
$\chi^2_\gamma$ for a given mSUGRA parameter
point. Figure~\ref{fig:scangamma} shows the $\chi^2_\gamma$ obtained and the
corresponding boost factors for the same parameters as in
fig.~\ref{fig:scanposfrac}. While the boost factors tend to be even
higher than in the positron case, the $\chi^2$ shows a preference for
areas with relatively low values of $m_{1/2}$.\\
\par
A brief discussion of the diffuse $\gamma$-ray data in the context of
the spectrum of antiprotons follows. The latter has been measured in continuously repeated
flights of the BESS
experiment~\cite{ref:bessexp}. Figure~\ref{fig:pbargammaPP1} shows the
EGRET $\gamma$-ray data from the
inner Galaxy and the best fit of the SUSY signal in the benchmark PP1
model to the data. A rather high boost factor of roughly~500 is
needed. The figure also compares the BESS antiproton data to the
secondary prediction and the expected SUSY signal. For better
comparison, the antiproton data for the years 1995-1999, when solar
activity was low, were demodulated according to the force-field
approximation and a binned weighted mean was calculated for
them using the procedure outlined in~\cite{ref:barlow}.
These can then be compared to the unmodulated background and
signal fluxes. As the figure shows, the conventional Galprop model
slightly underproduces antiprotons, while the plain diffusion model
predicts somewhat higher fluxes than are observed. However, no need
for an additional signal component is apparent. In fact, as
energy losses during propagation are negligible for antiprotons and
they therefore originate from the entire Galaxy, the boost factors for
$\gamma$-rays and antiprotons should be comparable. This assumption
leads to an antiproton flux that is higher than the measured one by
orders of magnitude (fig.~\ref{fig:pbargammaPP1}). This problem has
been pointed out before~\cite{ref:deboercritique} and shows that
interpreting the EGRET excess in terms of SUSY dark matter is
problematic. In fact, simpler explanations for the excess have been put
forward. It has been suggested to be due to
a systematic effect in the sensitivity determination of the EGRET
detector~\cite{ref:egretsystematics} and it was demonstrated that 
an optimised model of cosmic-ray propagation can be found that
fits the EGRET data at all energies but requires an upward
normalisation of the propagated fluxes by a factor of a few over the locally
observed ones~\cite{ref:galpropoptimized}. Therefore, it was chosen not
to include the $\gamma$-ray data in the $\chi^2_\mathrm{tot}$-scan.\\
A principle comment on the approach taken here is in order. If, as in the
case of the antiprotons, the spectral shapes of the background and
signal components are similar, tuning the background model alone to
fit the data will only lead to valid results if the signal component
is small. While this cannot be assumed a priori, the $\gamma$-ray
spectrum, which is dominated by the $\pi^0$ contribution in the inner
region of the Galaxy, clearly suggests that it is justified in this
case as antiprotons are created in the same reactions as the $\pi^0$s.\\
\par
Antideuterons constitute an interesting probe for neutralino dark
matter because they are not plagued by the problems seen for the
antiprotons in fig.~\ref{fig:pbargammaPP1}, namely the high flux and
similar spectral shape of the secondary component with respect to the
annihilation signal~\cite{ref:dbarbg}. 
\begin{figure}[htb]
\begin{center}
\begin{tabular}{cc}
\includegraphics[width=0.5\textwidth,angle=0]{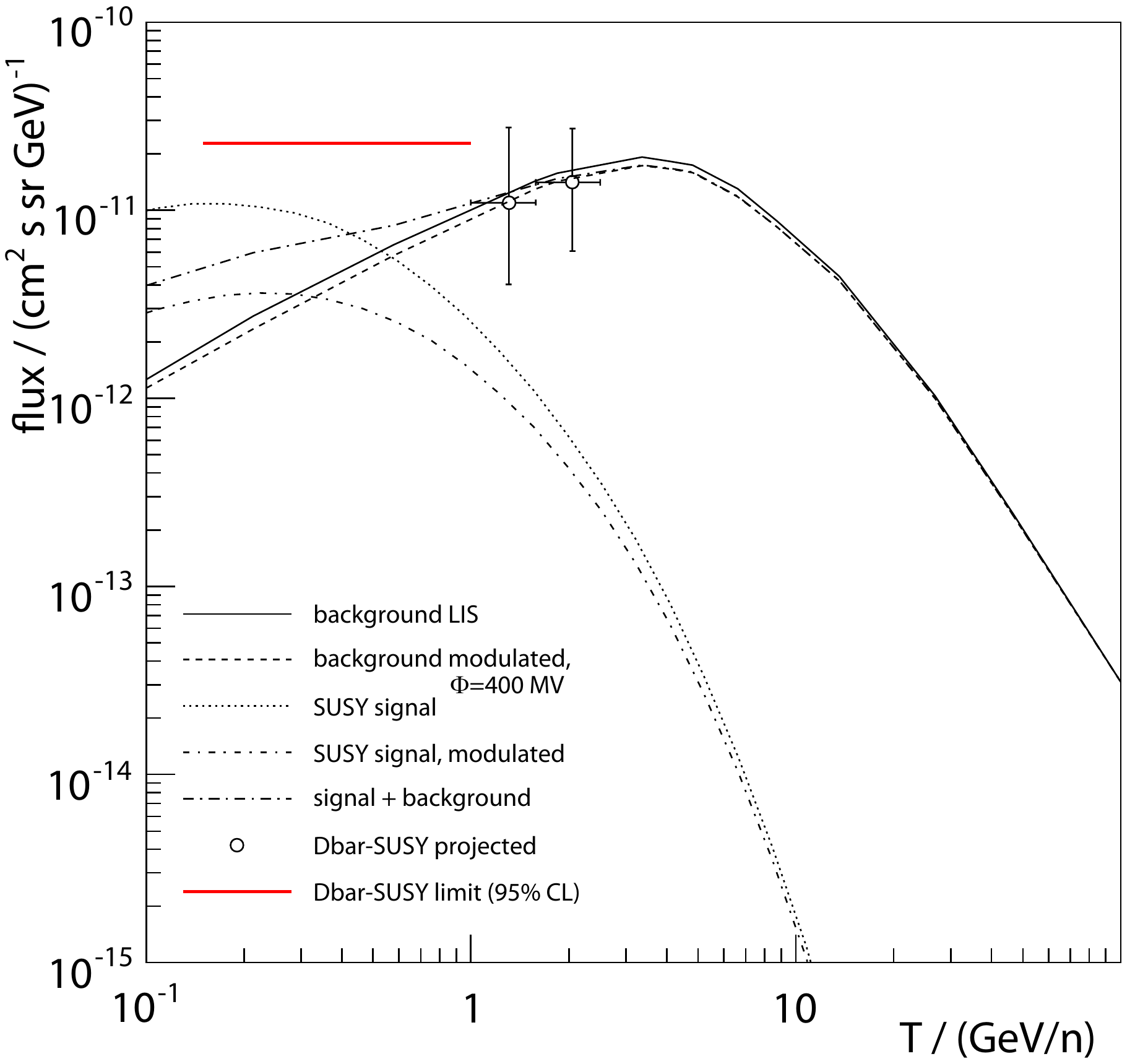}&
\includegraphics[width=0.5\textwidth,angle=0]{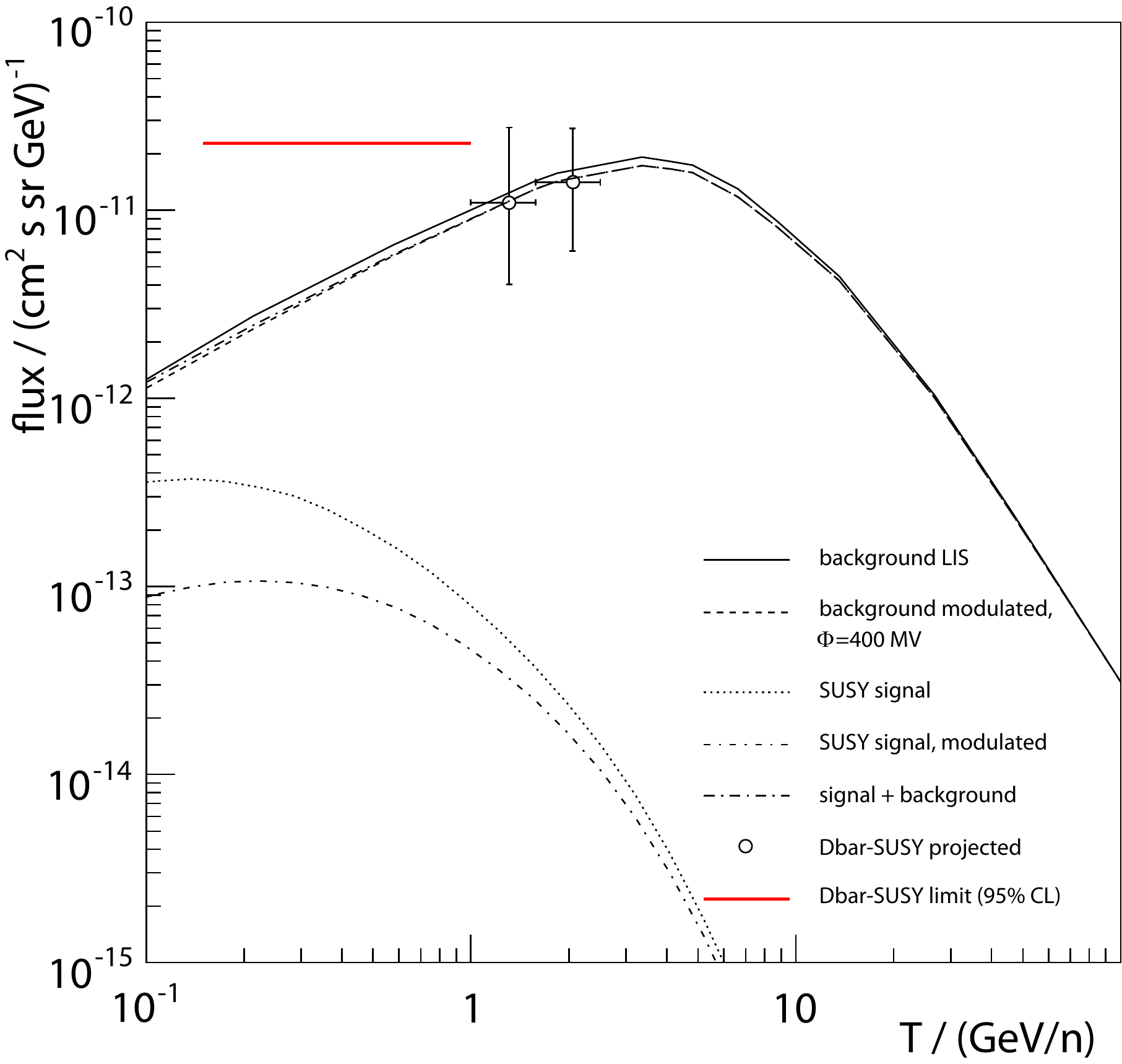}\\
\end{tabular}
\end{center}
\caption{Antideuteron signal from neutralino annihilation for the PP1
({\it left})
and PP2 ({\it right}) benchmark models, together with the background
prediction of~\cite{ref:dbarbg}. Both local interstellar and solar
modulated ($\phi=400\,\mathrm{MV}$) fluxes are shown. The statistical errors that an experiment with a geometric acceptance of
$1.8\,\mathrm{m}^2\mathrm{sr}$ and a measurement time of
$100\,\mathrm{d}$, like the proposed Dbar-SUSY mission~\cite{ref:dbarsusy}, will be able to achieve are shown. At low energies,
the flux drops below the detection threshold, and the expected 95\%~CL
limits are included, too.}
\label{fig:dbar}
\end{figure}
This is because spallation reactions taking place when cosmic-ray
protons interact with the interstellar matter create very few
low-energy particles, and low-energy secondary antideuterons are even
further suppressed. On the other hand, the fusion of an antiproton and
an antineutron will only be successful if their relative velocity is
low. This is the case for neutralino
annihilations. Figure~\ref{fig:dbar} shows the predicted $\bar{D}$
fluxes from $\chi\chi$-annihilation as calculated by DarkSUSY for the
benchmark models PP1 and PP2, together with a prediction for the
secondary background. While the detection of low-energy antideuterons
would constitute a very strong hint at the existence of neutralino
dark matter, the expected flux is prohibitively low.

\section{Projected improvements with PEBS}
\label{sec:projectedpebs}
\begin{figure}[htb]
\begin{center}
\begin{tabular}{cc}
\includegraphics[width=0.52\textwidth,angle=0]{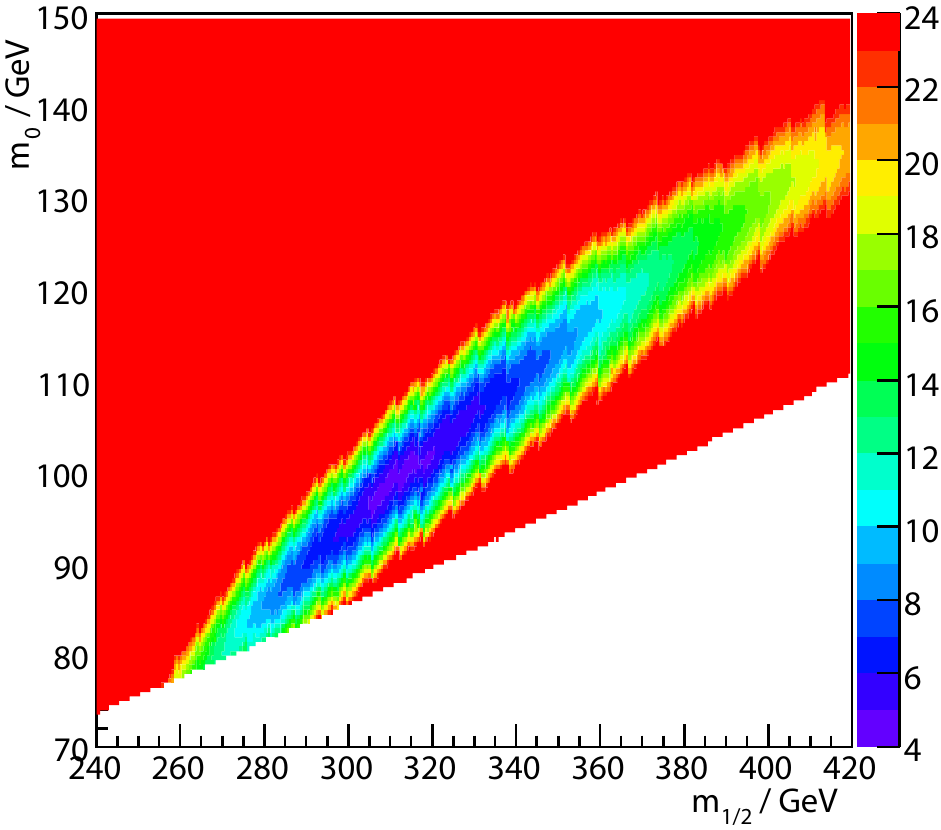}&
\includegraphics[width=0.46\textwidth,angle=0]{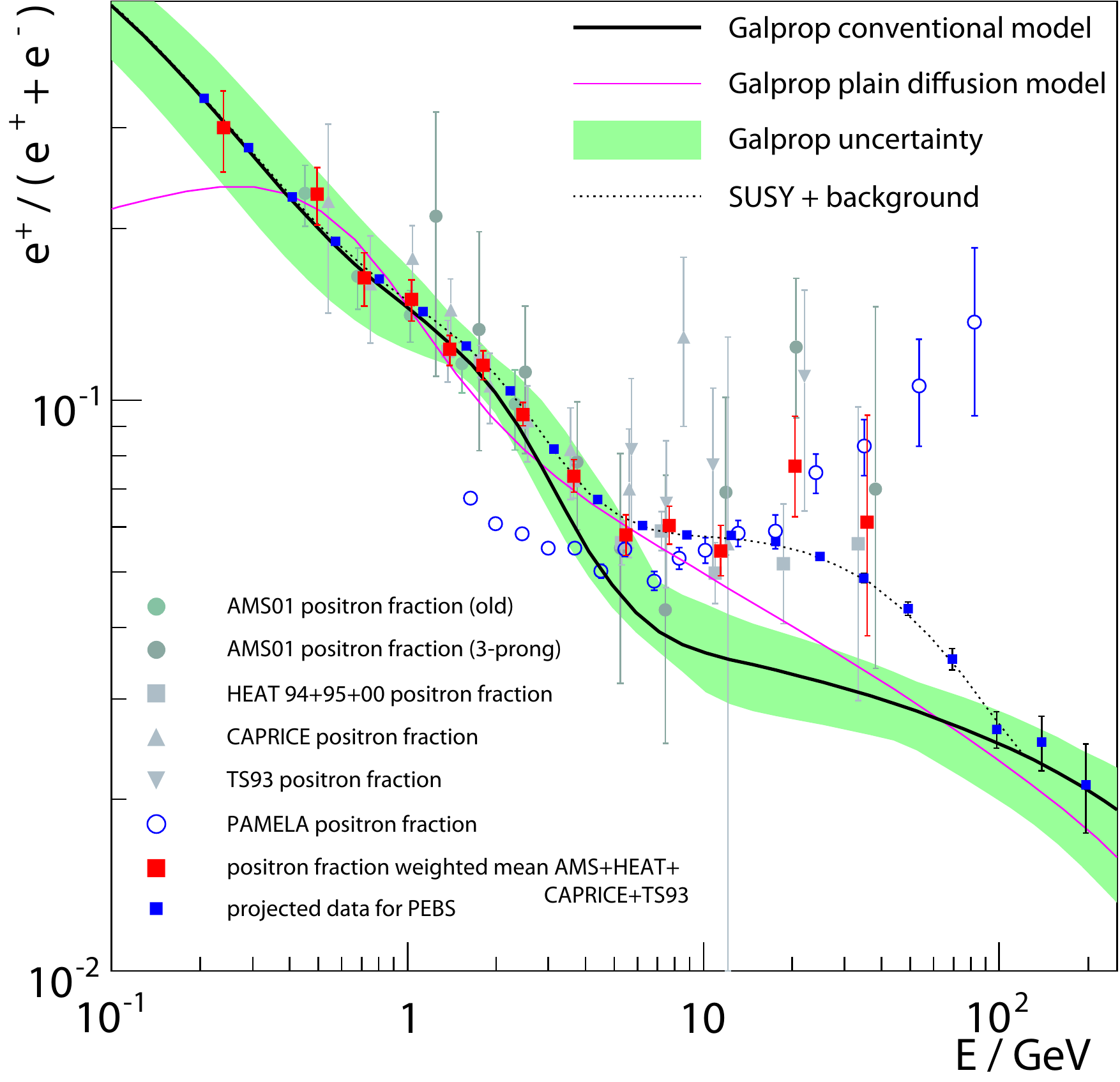}\\
\end{tabular}
\end{center}
\caption{$\chi^2_{e^+/(e^++e^-)}$-contour for a fit to the projected PEBS
  data for the parameter point PP2 in the $m_{1/2}$-$m_0$-plane, for $\tan\beta=20$ and
  $m_t=172.76\,\mathrm{GeV}$ ({\it left}) and projected positron
  fraction for a measurement of 40 days assuming an acceptance of
  $3850\,\mathrm{cm}^2\mathrm{sr}$ ({\it right}). Only the signal
region above $5\,\mathrm{GeV}$ was considered for the calculation of
the $\chi^2_{e^+/(e^++e^-)}$.}
\label{fig:pebs_PP2}
\end{figure}

Having established the most likely regions of the mSUGRA parameter space
and having chosen two representative parameter points,
one can now turn to examining the prospects for dark matter detection with
PEBS.\\
The positron fraction and its statistical errors to be expected in
the PP1 and PP2 models for a detector with a geometric acceptance of
$3850\,\mathrm{cm}^2\mathrm{sr}$ and an exposure of 40~days, as
\begin{figure}[htb]
\begin{center}
\begin{tabular}{cc}
\includegraphics[width=0.52\textwidth,angle=0]{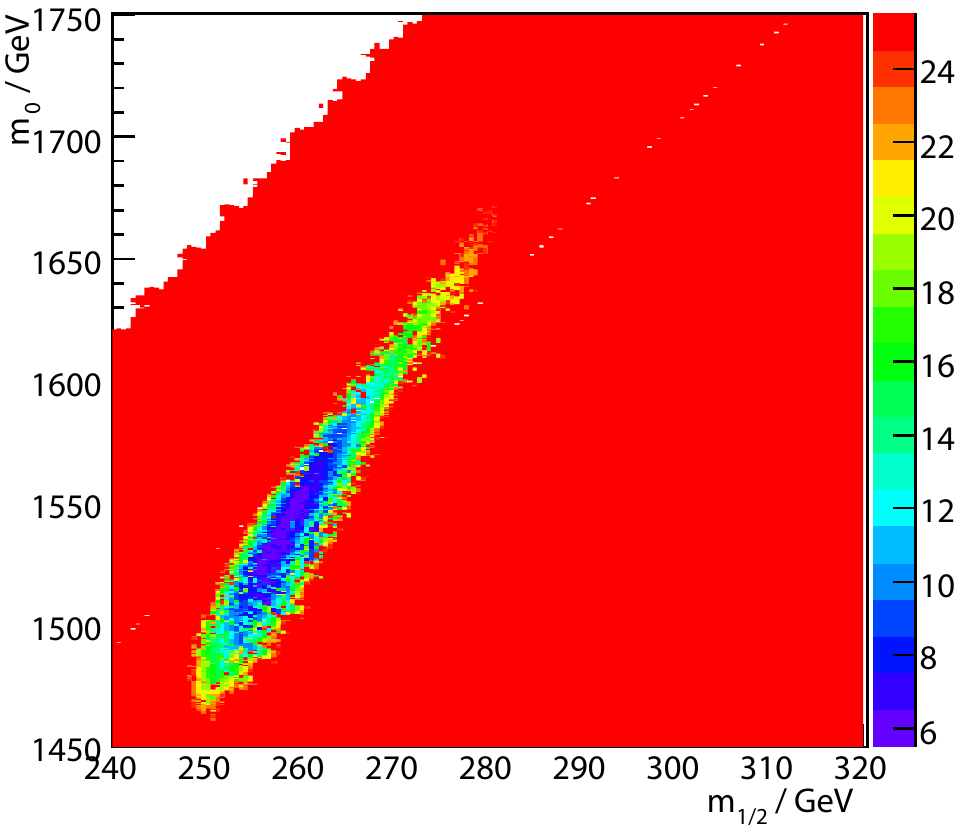}&
\includegraphics[width=0.46\textwidth,angle=0]{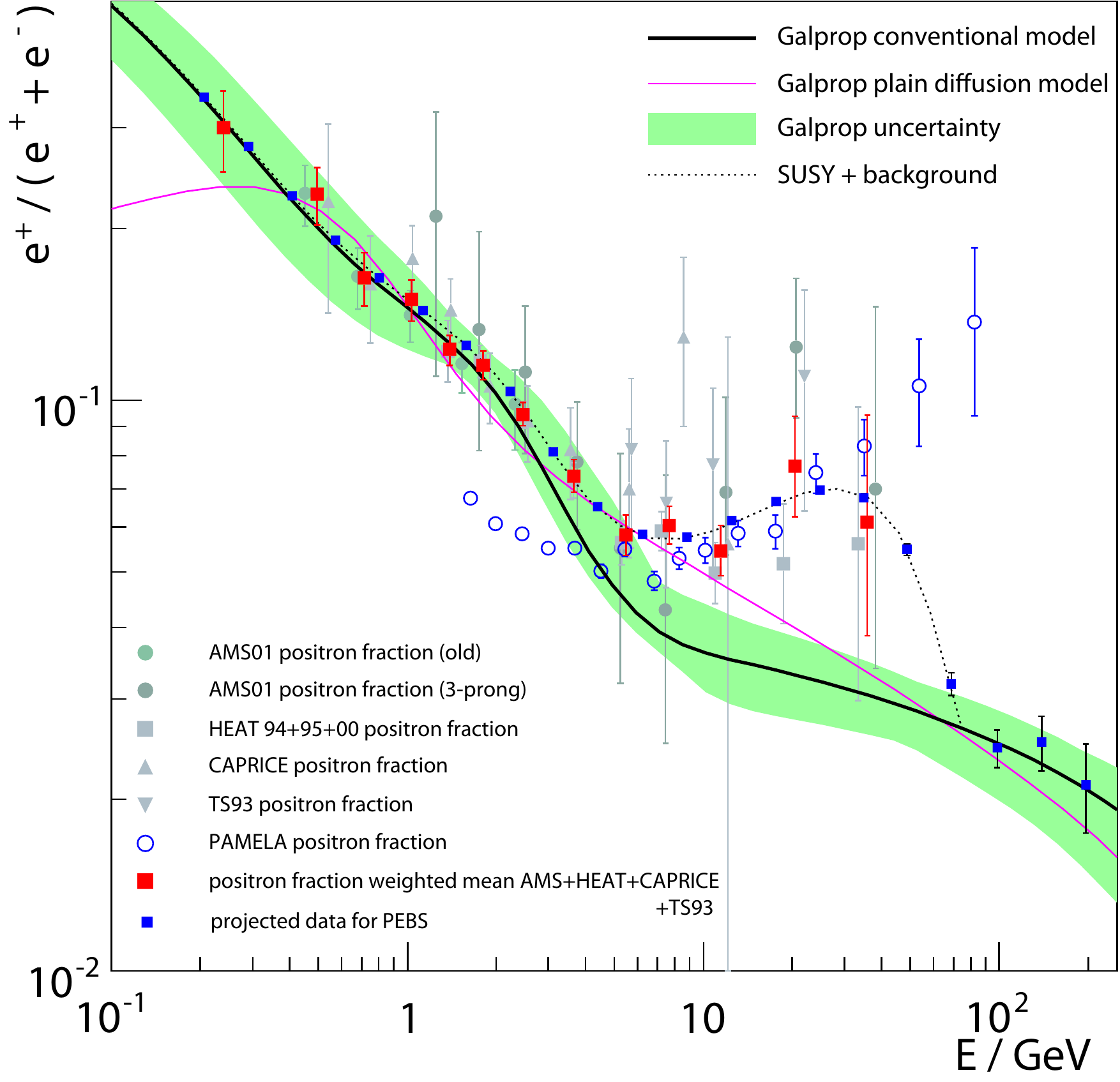}\\
\end{tabular}
\end{center}
\caption{$\chi^2_{e^+/(e^++e^-)}$-contour for a fit to the projected PEBS
  data for the parameter point PP1 in a small part of the $m_{1/2}$-$m_0$-plane, for $\tan\beta=40$ and
  $m_t=172.76\,\mathrm{GeV}$ ({\it left}) and projected positron
  fraction for a measurement of 40 days assuming an acceptance of
  $3850\,\mathrm{cm}^2\mathrm{sr}$ ({\it right}). Only the signal
region above $5\,\mathrm{GeV}$ was considered for the calculation of
the $\chi^2_{e^+/(e^++e^-)}$.}
\label{fig:pebs_PP1}
\end{figure}
projected for PEBS, are depicted in figures~\ref{fig:pebs_PP1}
and~\ref{fig:pebs_PP2}, respectively.
The improvement to be expected over the existing measurements, also
shown in the figures, is twofold. First, the statistical uncertainties
in the region below $100\,\mathrm{GeV}$ will be completely
negligible. Second, and probably even more important, the improved
energy and charge sign resolution, combined with the larger
acceptance, of the next generation of detectors will extend the energy
range of the measurements into the currently unexplored regime above
this level. This will allow to check for the distinctive return to
\begin{figure}[htb]
\begin{center}
\includegraphics[width=0.8\textwidth,angle=0]{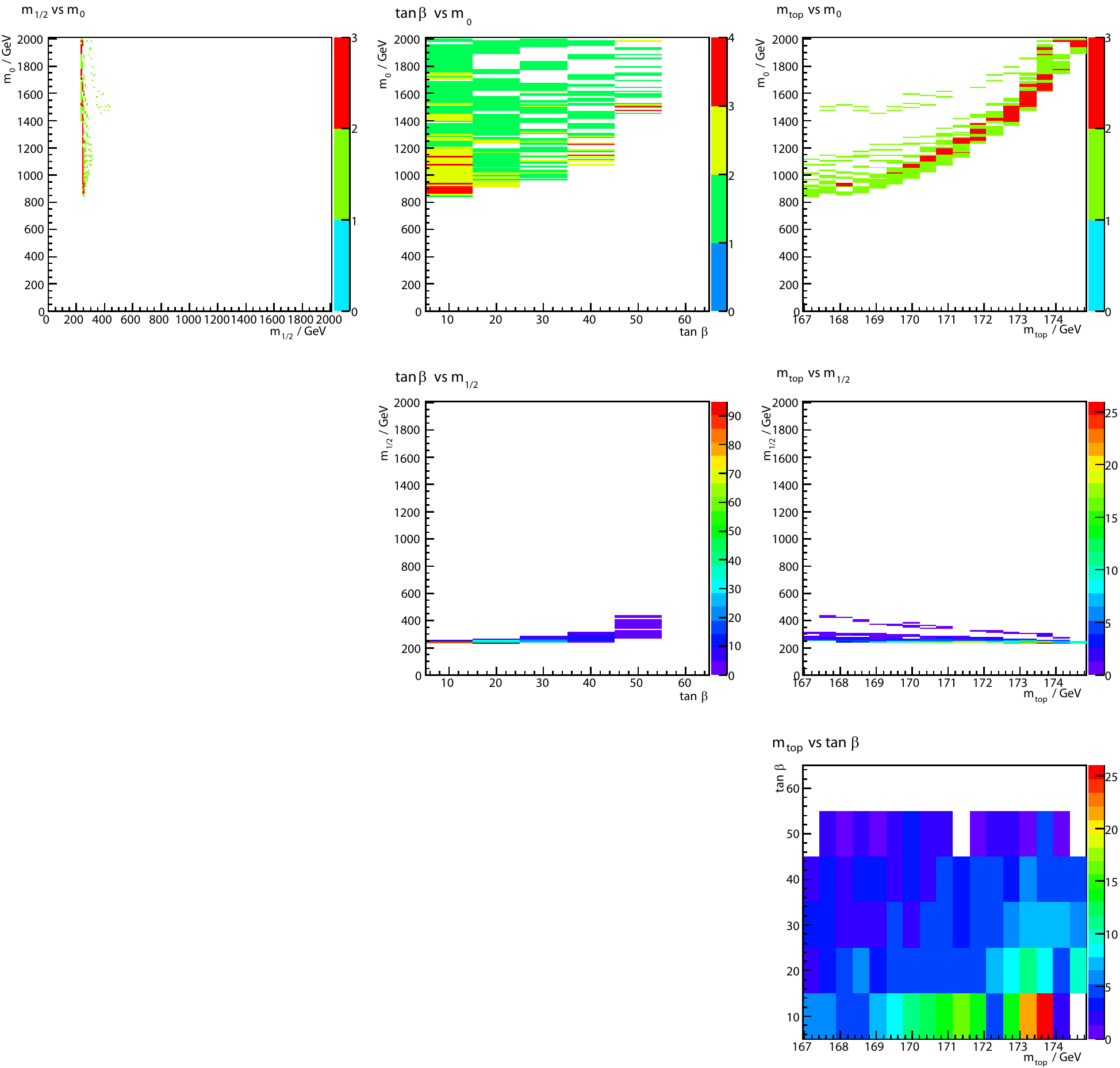}
\end{center}
\caption{Remaining mSUGRA parameter space from projected PEBS positron
fraction alone, for the PP1 parameter point. The $z$-axis here counts number of
models. Only models yielding a $\chi^2_{e^+/(e^++e^-)}$ in the $90\,\%$
confidence level interval around the minimum are included.}
\label{fig:msugra_projections_PP1}
\end{figure}
the background curve which essentially constitutes a smoking gun for
some sort of particle dark matter. In our model, the location of the
edge
- though smeared out during the propagation - is a measure for the
neutralino mass, while its steepness and overall form indicate the preferred
\begin{figure}[htb]
\begin{center}
\includegraphics[width=0.8\textwidth,angle=0]{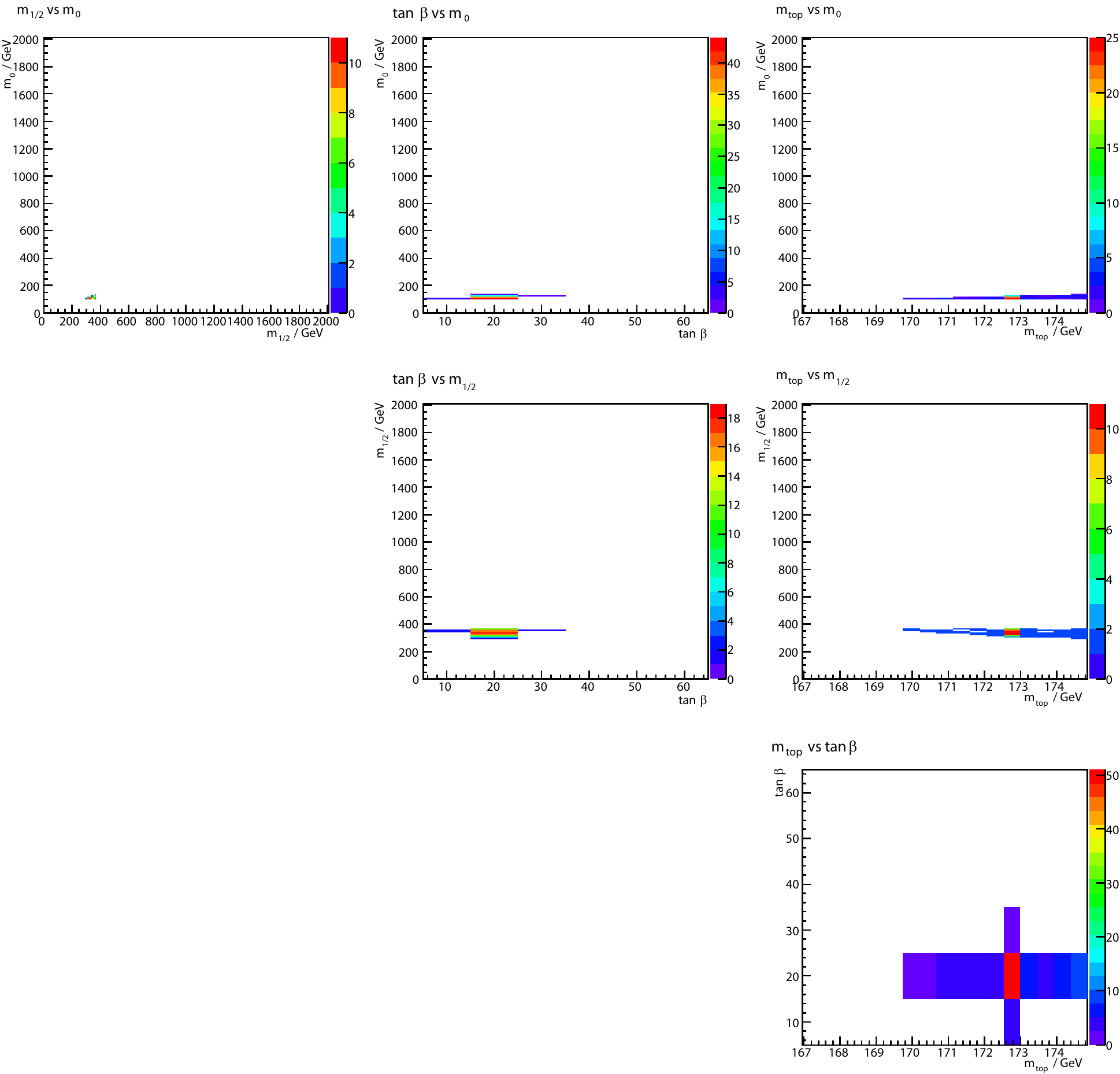}
\end{center}
\caption{Remaining mSUGRA parameter space from projected PEBS positron
fraction alone, for the PP2 parameter point. The $z$-axis here counts number of
models. Only models yielding a $\chi^2_{e^+/(e^++e^-)}$ in the $90\,\%$
confidence level interval around the minimum are included.}
\label{fig:msugra_projections_PP2}
\end{figure}
decay channel. In the case of PP1, the dominant reaction
$\chi\chi\rightarrow{}WW$ leads to the rather prominent slope.\\
Figures~\ref{fig:pebs_PP1} and~\ref{fig:pebs_PP2} also contain the
$\chi^2_{e^+/(e^++e^-)}$-contour for projected PEBS data in a
finer sampling of the GUT mass scales around PP1 and PP2, respectively. For a given
($\tan\beta$,$m_t$) pair, these values can -- in principle -- be very
well constrained using the positron fraction alone. This is in stark
contrast to the present-day situation (fig.~\ref{fig:scanposfrac}).\\
\par
Looking at the entire parameter space included in the scans,
figures~\ref{fig:msugra_projections_PP1}
and~\ref{fig:msugra_projections_PP2} show those points yielding a projected
$\chi^2_{e^+/(e^++e^-)}$ within the $90\,\%$ confidence level interval around
the minimum, for PP1 and PP2, respectively. The problem of displaying the
four-dimensional parameter space has been solved here by showing the
six possible two-dimensional projections, one for each pair of
parameters. While neither $\tan\beta$
nor $m_t$ will be constrained at either point,
the quantity that can be determined with great accuracy from the
upcoming positron detectors is the neutralino mass.
\begin{figure}[htb]
\begin{center}
\includegraphics[width=0.5\textwidth,angle=0]{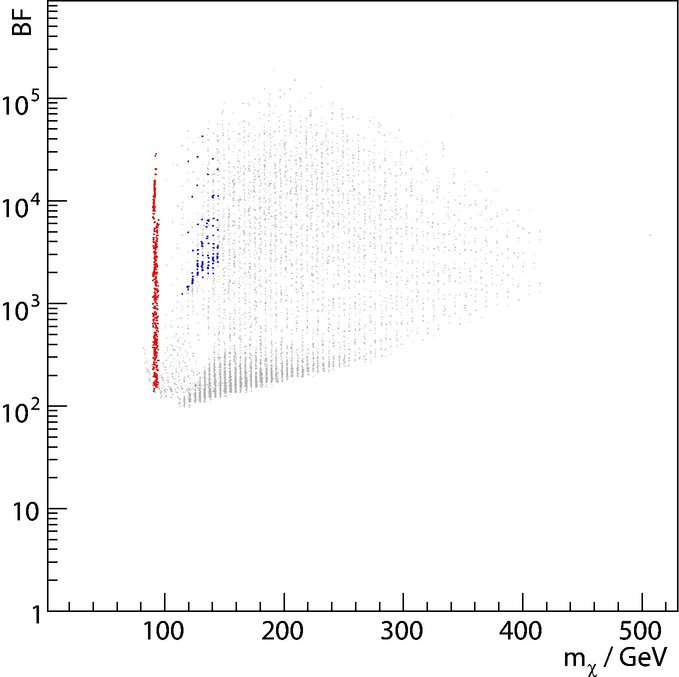}
\end{center}
\caption{Neutralino mass and positron fraction boost factor for points
included in figure~\ref{fig:msugra_projections_PP1} (large red dots) and for points
included in figure~\ref{fig:msugra_projections_PP2} (large blue dots) and for points
corresponding to the $90\,\%$ CL interval in figure~\ref{fig:contour_totchi2}
(small grey dots).}
\label{fig:neutralino_PP1}
\end{figure}
This is illustrated in figure~\ref{fig:neutralino_PP1} showing
$m_\chi$ for all points included in figs.~\ref{fig:contour_totchi2}, \ref{fig:msugra_projections_PP1} and \ref{fig:msugra_projections_PP2}.
The resolution obtainable in principle is on the order of $5\,\mathrm{GeV}$ for PP1 while it is
$25\,\mathrm{GeV}$ for PP2, due to the flatter shape of the signal
flux at this point. These results indicate that the mass resolution
obtainable in practice will be limited both by the energy resolution of
the detector, which was not taken into account here, and the systematic uncertainties in the propagation
model. Interestingly, the corresponding boost factors vary over several orders of magnitude.\\
If supersymmetry is realised in nature, the LHC collider at CERN
\begin{figure}[htb]
\begin{center}
\includegraphics[width=1.0\textwidth,angle=0]{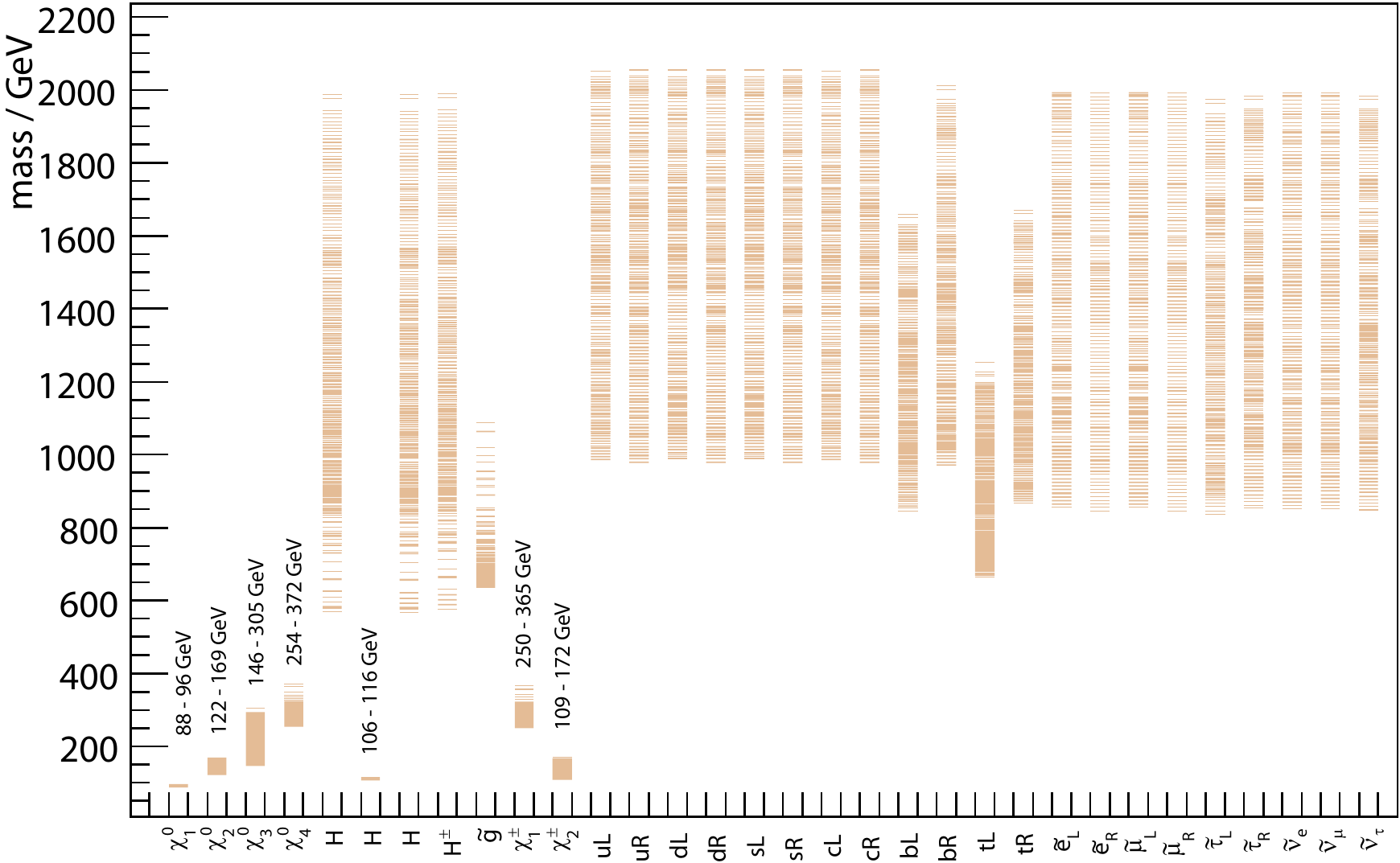}
\end{center}
\caption{Sparticle mass spectrum for points included in fig.~\ref{fig:msugra_projections_PP1}.}
\label{fig:msugra_masses_PP1}
\end{figure}
will have a good chance of creating supersymmetric particles, among
them the neutralino. A coherent picture of the dark matter will
ideally include the establishing of candidate particles by a collider,
proof of its existence in Earth's vicinity by direct detection
experiments and evidence for its presence throughout the Galaxy by
indirect searches. Independent verification of as many properties of
the dark matter as possible by the different approaches will play a
key role in this picture. An example for such an interplay would be the
prediction of sparticle masses expected from cosmic-ray observations
which in turn could be verified by the LHC. For the benchmark
parameter point PP1, the predicted mass spectrum of sparticles is
depicted in figure~\ref{fig:msugra_masses_PP1}. Again, all points yielding a projected
$\chi^2_{e^+/(e^++e^-)}$ within the $90\,\%$ confidence level interval around
the minimum have been included. Unfortunately, while the neutralino
and chargino masses can be predicted to the order of 10 to
$100\,\mathrm{GeV}$, the other sparticle masses still vary over a
range of $1000\,\mathrm{GeV}$.\\
As a cross-check regarding the sensibility towards the choice of the
propagation model, it was tried to fit the projected PEBS data at PP1
(generated with the conventional Galprop model) with any mSUGRA model
on top of the alternative plain-diffusion model. No parameter set
giving a $\chi^2$ below a few thousand was found in this way. One may
thus hope that ambiguities arising from the choice of propagation
model will not play a major role. Instead, it may be possible to draw
conclusions about cosmic-ray propagation (sec.~\ref{sec:galprop}) and the presence of a primary
signal at the same time.\\
\par
Figures~\ref{fig:pebs_PP1} and~\ref{fig:pebs_PP2} also contain the new
positron fraction data measured by the PAMELA experiment. A comparison
to the signal spectrum expected for the mSUGRA model at PP1 shows that
it matches the data quite well at intermediate energies, but fails to meet the data points at
the highest two energies which indicate an even steeper increase with
energy. The PP2 model is clearly disfavoured by the PAMELA data
because the positron fraction it predicts is much too shallow to
follow the rise apparent in the PAMELA data towards high energies.

\section{Comparison to AMS-02}
\label{sec:susyams02}
\begin{figure}[htb]
\begin{center}
\begin{tabular}{cc}
\includegraphics[width=0.52\textwidth,angle=0]{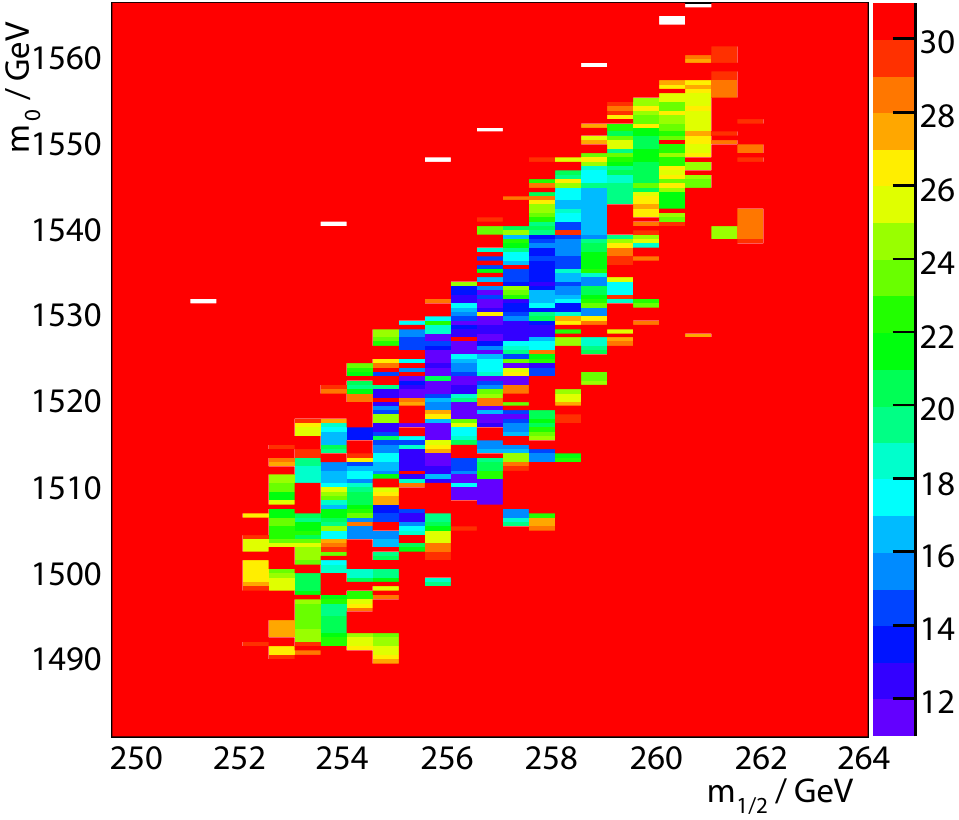}&
\includegraphics[width=0.46\textwidth,angle=0]{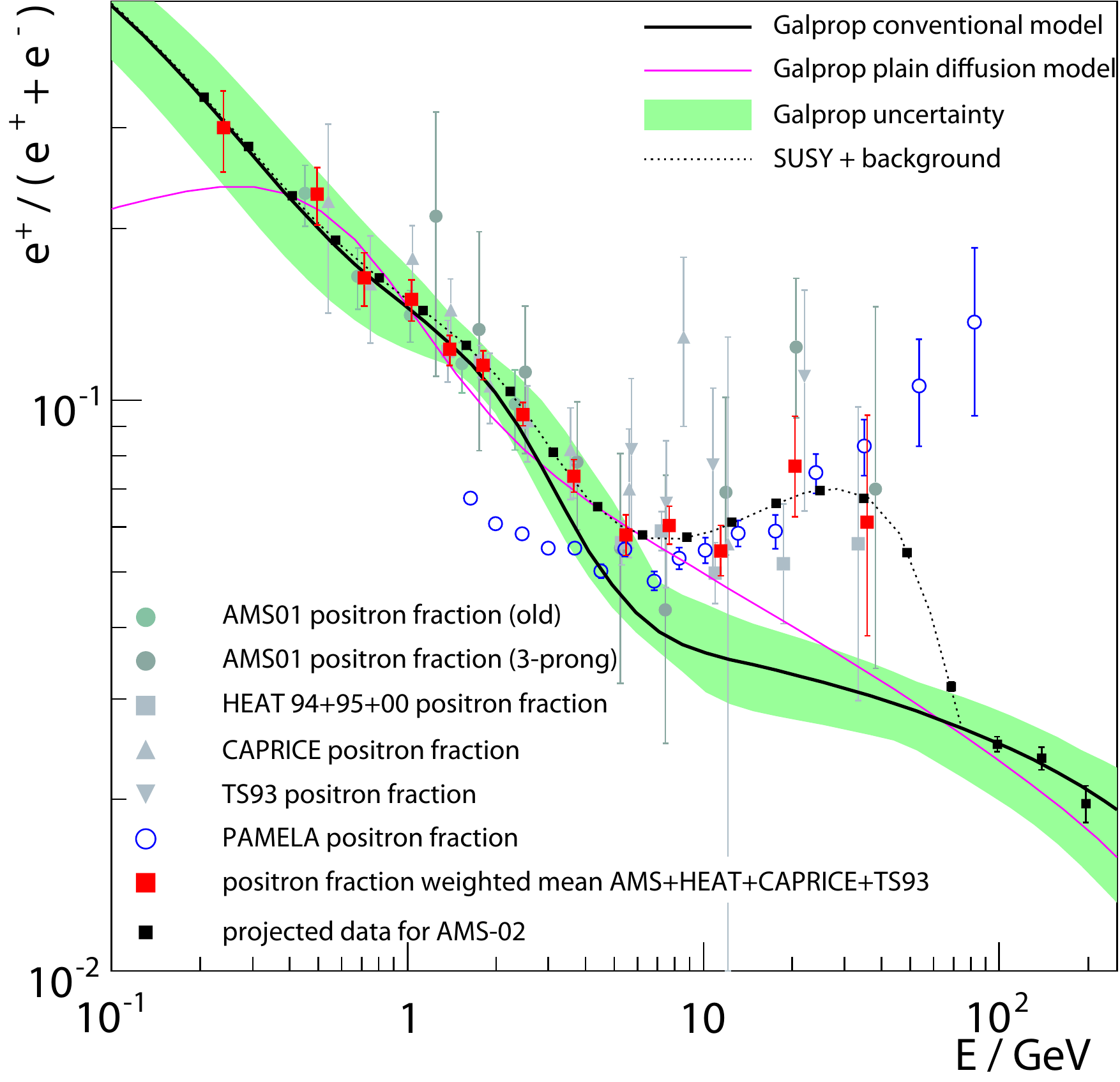}\\
\end{tabular}
\end{center}
\caption{$\chi^2_{e^+/(e^++e^-)}$-contour for a fit to the projected AMS-02
  data for the parameter point PP1 in a small part of the $m_{1/2}$-$m_0$-plane, for $\tan\beta=40$ and
  $m_t=172.76\,\mathrm{GeV}$ ({\it left}) and projected positron
  fraction for a measurement of 3~years assuming an acceptance of
  $875\,\mathrm{cm}^2\mathrm{sr}$ ({\it right}). Only the signal
region above $5\,\mathrm{GeV}$ was considered for the calculation of
the $\chi^2_{e^+/(e^++e^-)}$. Note the reduced scale as compared to fig.~\ref{fig:pebs_PP1}.}
\label{fig:ams_PP1}
\end{figure}
\begin{figure}[htb]
\begin{center}
\begin{tabular}{cc}
\includegraphics[width=0.52\textwidth,angle=0]{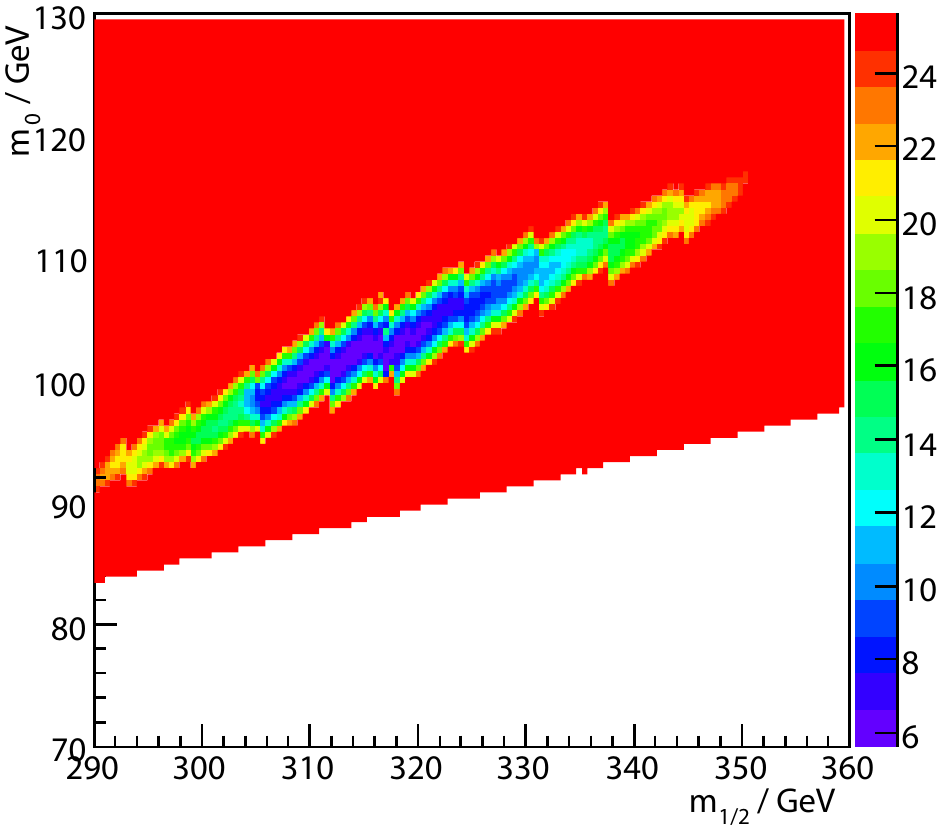}&
\includegraphics[width=0.46\textwidth,angle=0]{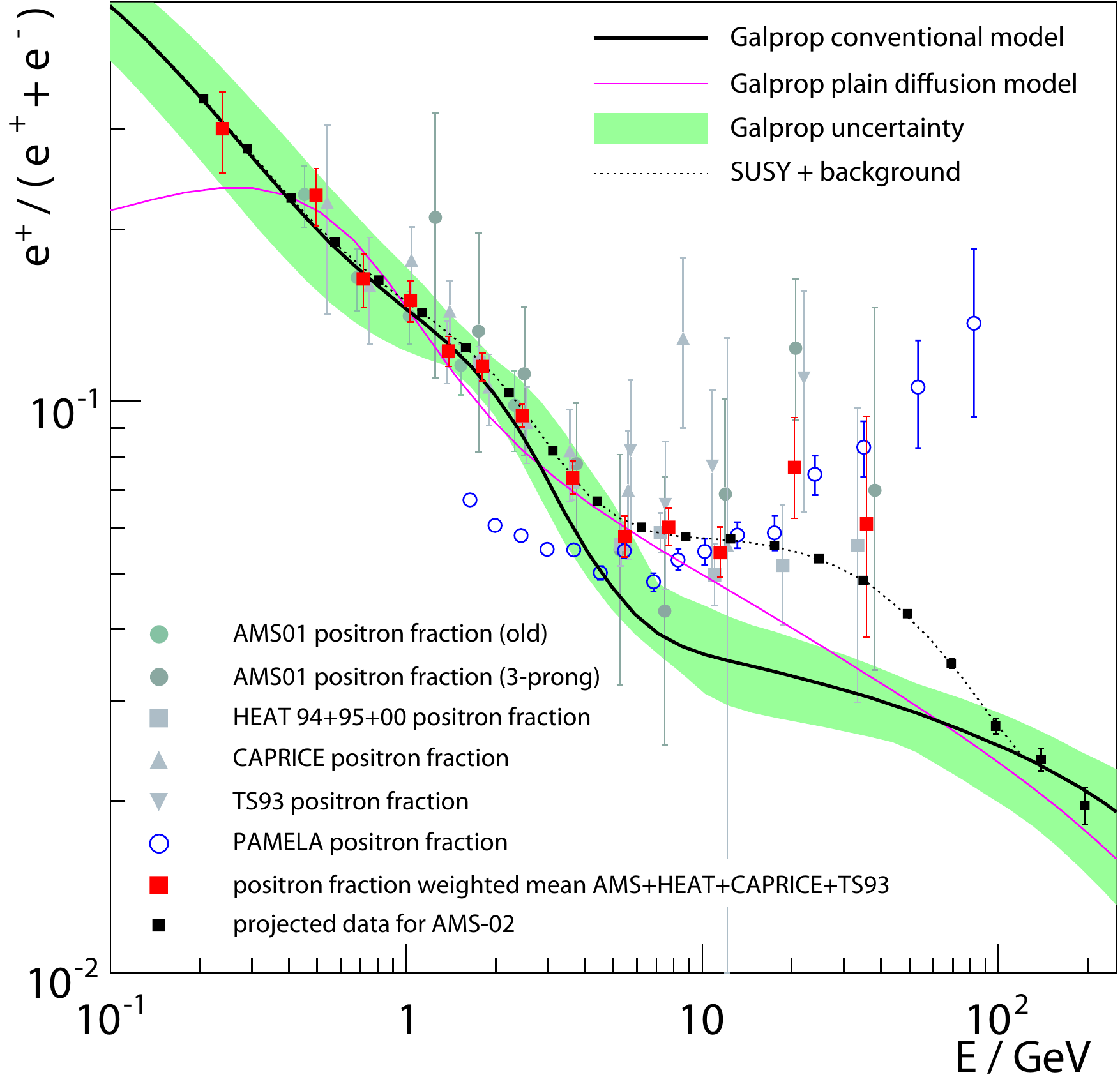}\\
\end{tabular}
\end{center}
\caption{$\chi^2_{e^+/(e^++e^-)}$-contour for a fit to the projected AMS-02
  data for the parameter point PP2 in a small part of the $m_{1/2}$-$m_0$-plane, for $\tan\beta=20$ and
  $m_t=172.76\,\mathrm{GeV}$ ({\it left}) and projected positron
  fraction for a measurement of 3~years assuming an acceptance of
  $875\,\mathrm{cm}^2\mathrm{sr}$ ({\it right}). Only the signal
region above $5\,\mathrm{GeV}$ was considered for the calculation of
the $\chi^2_{e^+/(e^++e^-)}$.}
\label{fig:ams_PP2}
\end{figure}
To put the projected performance of PEBS into perspective, a study of
the performance of AMS-02, with a measurement time of three years and
an acceptance of $875\,\mathrm{cm}^2\,\mathrm{sr}$, was done for the
two benchmark parameter points PP1 and PP2 (figs.~\ref{fig:ams_PP1}
and~\ref{fig:ams_PP2}). AMS-02 will be able to gather more statistical
power than PEBS and the $\chi^2$-contours are correspondingly
smaller. On the other hand, taking the systematic uncertainties,
e.g.~the limited energy resolution of the calorimeters employed and
the limited knowledge of the expected secondary background, into
accout, it becomes clear that the interpretation of the measurements
will be ultimately limited by the systematics at this level of
statistical accuracy.
\begin{figure}[htb]
\begin{center}
\includegraphics[width=0.5\textwidth,angle=0]{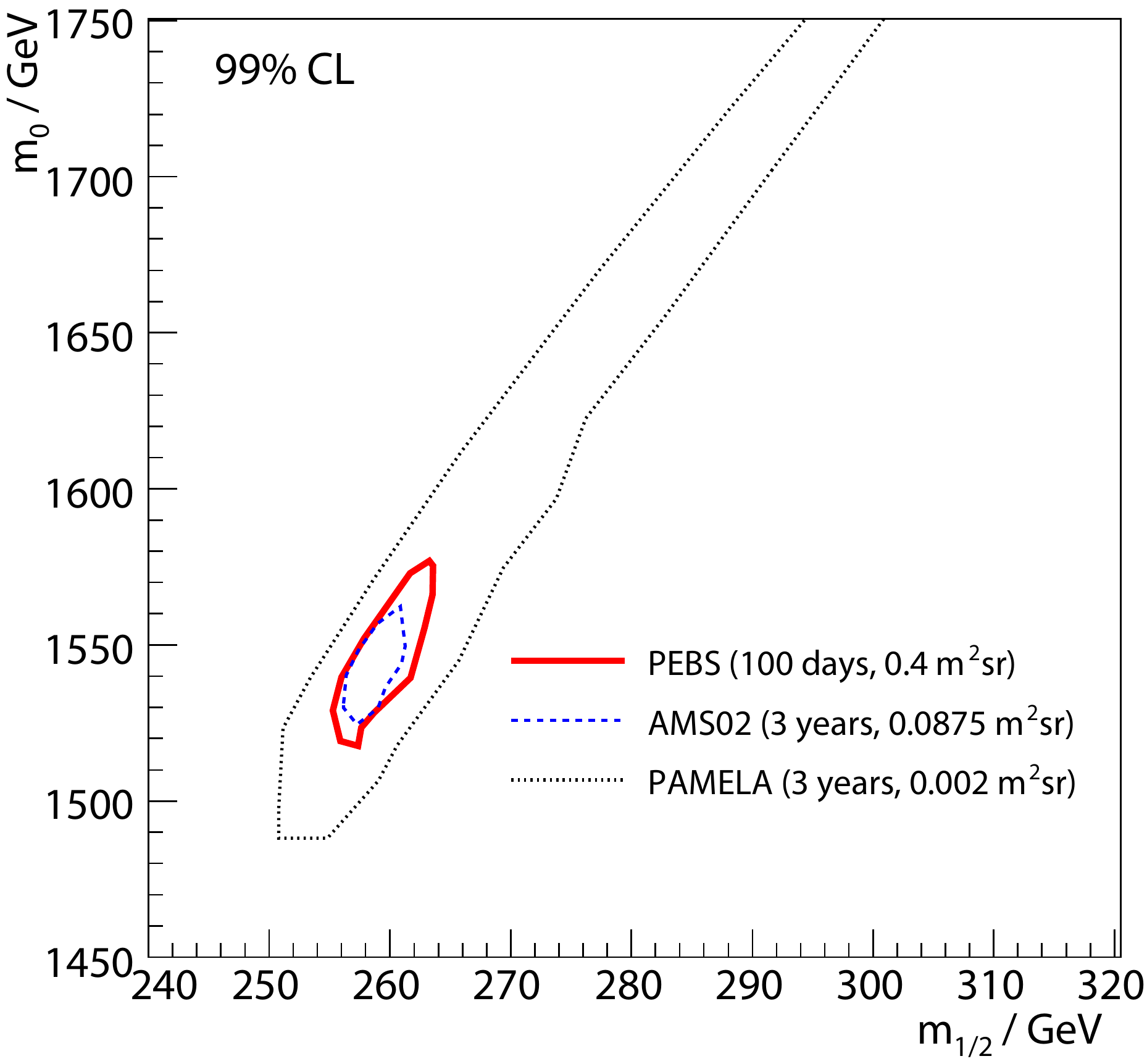}
\end{center}
\caption{$99\,\%$ confidence level areas derived from the $\chi^2_{e^+/(e^++e^-)}$-contours for fits to the projected
  data from PAMELA, PEBS, and AMS-02, for the parameter point PP1 in a small part of the $m_{1/2}$-$m_0$-plane, for $\tan\beta=40$ and
  $m_t=172.76\,\mathrm{GeV}$. The statistics that PEBS could
accumulate in a series of flights for a total measurement time of
100~days was used. The PAMELA contour refers to the projected data
based on acceptance and mission duration, for comparison to the other
projects, not to the actual data.}
\label{fig:conflevel}
\end{figure}
Neglecting these effects, a comparison of the projected constraining
power of PAMELA, PEBS, and AMS-02 (fig.~\ref{fig:conflevel}) for the
PP1 benchmark case shows that
PEBS and AMS-02 perform almost equally well, while the smaller
acceptance of PAMELA leads to significantly larger confidence level
contours in mSUGRA parameter space.

\section{mSUGRA dark matter in the light of PAMELA data}
\label{sec:msugra_pamela}
\begin{figure}[htb]
\begin{center}
\begin{tabular}{cc}
\includegraphics[width=0.51\textwidth,angle=0]{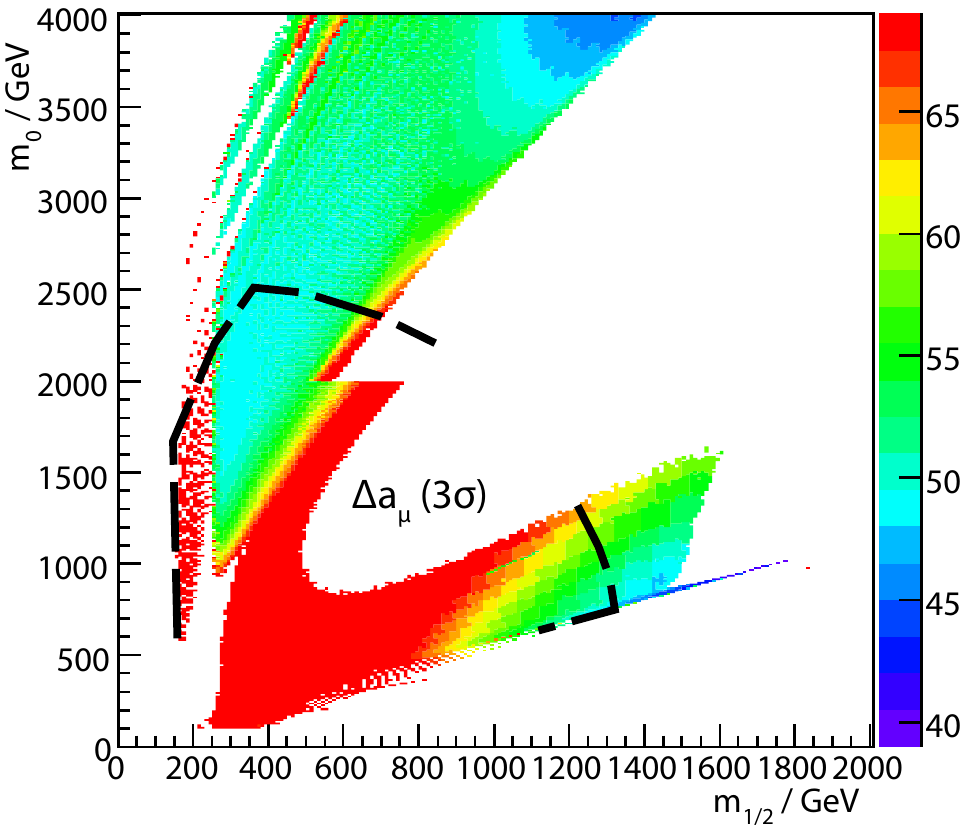}&
\includegraphics[width=0.45\textwidth,angle=0]{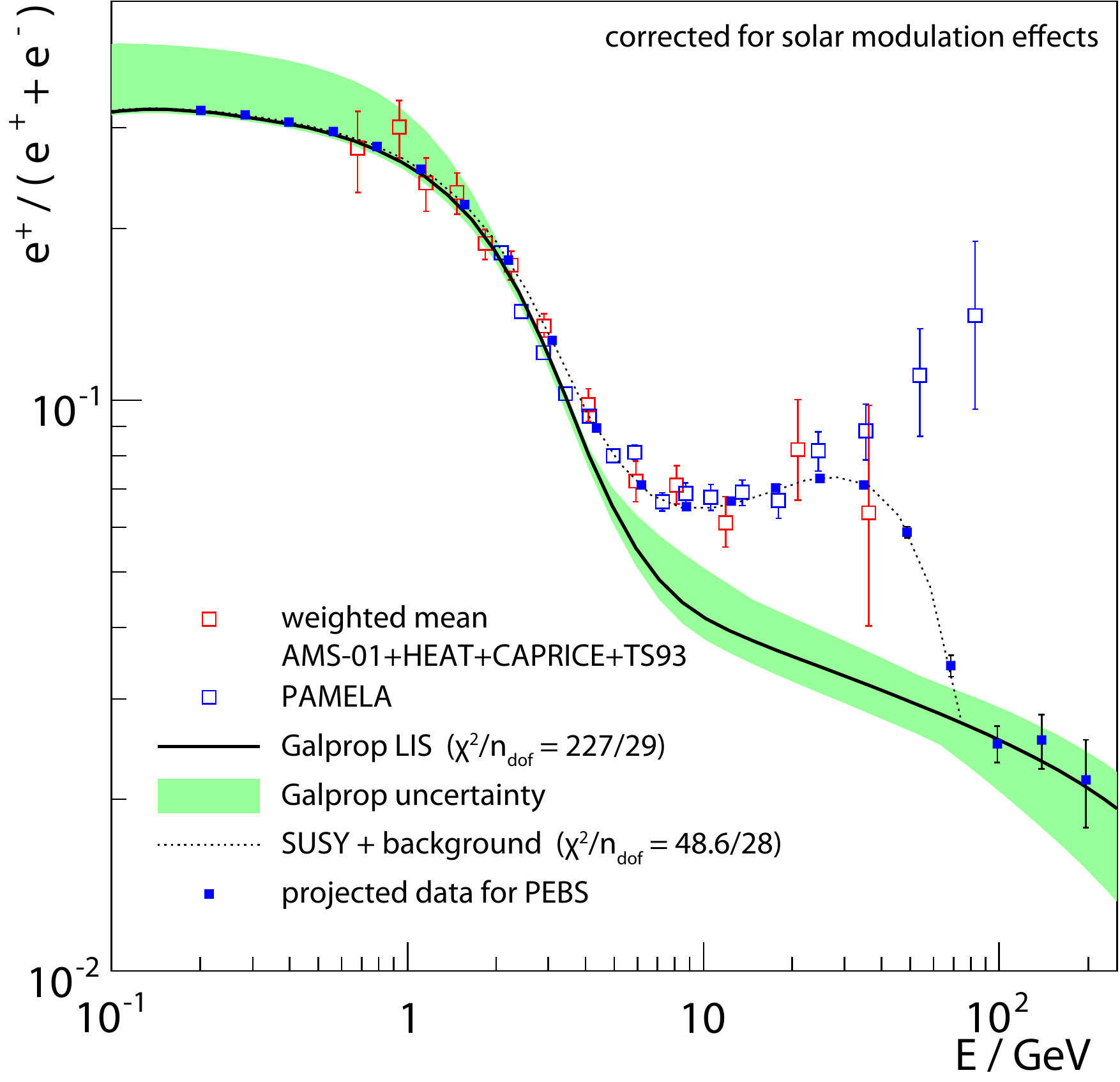}\\
\end{tabular}
\end{center}
\caption{{\it Left:} $\chi^2$ with respect to the
  positron fraction data including PAMELA and the top-quark mass, as defined in the
  text, in the $m_{1/2}$-$m_0$-plane. The correction for solar
  modulation according to section~\ref{sec:pamela_modulation} was
  applied to the data and the local interstellar spectra of the Galprop
  conventional model were used. The remaining statistical uncertainty of
  the background model was taken into account in the calculation of the
  $\chi^2$. Only models fulfilling the
  constraints on relic density and $BR(b\rightarrow{}s\gamma)$ at the
  $3\sigma$-level, as well as existing mass limits and direct
  detection limits, and having a best-fit boost factor of less than
  $10^4$ are included. Models additionally giving a value of
  $\Delta{}a_\mu$ falling within the preferred region at the
  $3\sigma$-level lie within the region bounded from above by the
  dashed line.
{\it Right:} Best-fit positron fraction with respect to the
data of AMS-01, HEAT, CAPRICE, TS93, and PAMELA, for the PP3 mSUGRA
model.}
\label{fig:pamela_dm}
\end{figure}
As found in section~\ref{sec:projectedpebs}, the mSUGRA models
considered so far have difficulties in describing the PAMELA data that
have very recently become available. These results have confirmed the
excess of positrons with respect to models of purely secondary
production at high energies. They have subsequently created a surge of
publication activity. The most popular mechanisms invoked to explain
the data are the annihilation or decay of particle dark matter and an
origin in nearby young pulsars. There, electrons
are accelerated in the quasi-static electric fields of the pulsar's
magnetosphere to ultra-relativistic energies. They will in turn
radiate energetic synchrotron photons which give rise to $e^\pm$-pairs
by pair production in the dense magnetic fields.
Ref.~\cite{ref:profumo_pulsars} contains an extensive list of recent
work on these subjects.\\
To put the PAMELA results into perspective, the scans of mSUGRA
parameter space were extended to cover also the range of higher $m_0$,
specifically $2000\,\mathrm{GeV}\leq{}m_0\leq4000\,\mathrm{GeV}$, but
in the limited ranges of $10\leq\tan\beta\leq50$ and $170.435\,\mathrm{GeV}\leq{}m_t\leq174.62\,\mathrm{GeV}$,
in order to keep the necessary computing time acceptable.\\
The result is a $\chi^2$-scan of the extended mSUGRA parameter space
(\ref{fig:pamela_dm} {\it left}), with the $\chi^2$ defined as
\begin{equation}
  \label{eq:pamela_chi2}
  \chi^2\equiv\chi^2_{\mathrm{positron}\:\mathrm{fraction}}+\left(\frac{m_t^\mathrm{scan}-m_t}{\sigma_{m_t}}\right)^2
\end{equation}
The first term refers to the deviation of a given model from the
positron fraction data of AMS-01, HEAT, CAPRICE, TS93, and PAMELA,
analogous to (\ref{eq:chi2posfrac}). 
The local interstellar spectra obtained by applying the correction for solar
modulation effects described in section~\ref{sec:pamela_modulation}
were used. The statistical uncertainty of the Galprop conventional
model was evaluated as in section~\ref{sec:galprop}, by looking at the
B/C-data and the low-energy positron fraction data. The $\sigma_i^2$
values used in the calculation of the $\chi^2$ were then calculated as
\begin{equation}
\label{eq:sigmai}
\sigma_i^2=\sigma_{i,\mathrm{data}}^2+\sigma_{i,\mathrm{bg}}^2
\end{equation}
taking the statistical uncertainties both of the data and of the
background calculation into account. This is important as the PAMELA
data at intermediate energies now have statistical uncertainties that
are small compared to the background ones. The boost factor was chosen as to give
the best $\chi^2$. The second term in (\ref{eq:pamela_chi2}) punishes
models with a top-quark mass differing from the measured value. Only
models fulfilling the
constraints on relic density and $BR(b\rightarrow{}s\gamma)$ at the
$3\sigma$-level, as well as existing mass limits and direct
detection limits, and having a best-fit boost factor of less than
$10^4$ are considered. The PAMELA data favour either a focus-point
scenario, at large values of $m_0$, or a coannihilation scenario, at
large values of $m_{1/2}$. The remaining parameter space with $\chi^2$-values close to
the minimum is large and extends beyond the range covered in the
scans. However, when it is additionally required that supersymmetry
cancel the discrepancy in the magnetic moment of the muon at the
$3\sigma$-level, two portions
of parameter space remain, located between the values
$1300\,\mathrm{GeV}\leq{}m_0\leq2500\,\mathrm{GeV}$ and
$250\,\mathrm{GeV}\leq{}m_{1/2}\leq700\,\mathrm{GeV}$,
or between the values
$600\,\mathrm{GeV}\leq{}m_0\leq1300\,\mathrm{GeV}$ and
$900\,\mathrm{GeV}\leq{}m_{1/2}\leq1350\,\mathrm{GeV}$.\\
\par
As an example, the minimum
within the former region is found at the parameter point
\begin{equation}
  \label{eq:pp3}
\mathrm{PP3:}\quad m_0=2040\,\mathrm{GeV}\quad
m_{1/2}=390\,\mathrm{GeV}\quad \tan\beta=40\quad m_t=172.295\,\mathrm{GeV}
\end{equation}
The best-fit positron fraction at PP3 (fig.~\ref{fig:pamela_dm} {\it right}) has
a boost factor of~1510. The lightest neutralino has a mass of
$m_\chi=93\,\mathrm{GeV}$ and annihilates predominantly through the
channels $\chi\chi\rightarrow{}WW$ and $\chi\chi\rightarrow{}ZZ$ with
branching ratios of $83\,\%$ and $10\,\%$, respectively. The
quality-of-fit achieved in this model is typical for the mSUGRA models
within the regions quoted above. While a good description of the PAMELA
positron fraction data is possible at intermediate energies, the data
points at the two highest energies show deviations of $2.5\sigma$ and
$2.6\sigma$, respectively. The PAMELA data therefore arguably still
indicate a steeper increase than is possible from
dark matter annihilations in the mSUGRA model. This scenario could
also be tested with the statistics and energy range accessible to both
PEBS and AMS-02.\\
It should be stressed that this scenario requires a drastically lower
boost factor (on the order of a few) for antiprotons than for electrons lest the antiproton
flux from neutralino annihilations exceeds the measured one. This
is a rather universal conclusion concerning any kind of dark matter
which decays or annihilates into standard model gauge bosons or
quarks. As a consequence, the dark matter candidates studied in the
recent literature, e.g.~in \cite{ref:cirelli,ref:cholis,ref:yin} to name
only a few, are constructed such that they preferably or exclusively
yield leptons.

\chapter{Conclusions and outlook}
\label{chapter:conclusions}
The longstanding problem concerning the nature of dark matter might finally be
solved within the next years. The Large Hadron Collider will commence
operations in summer~2009, several direct detection experiments are
underway and are pushing their limits to lower and lower cross
sections, and the PAMELA detector is in orbit and the first
publications on the positron fraction have proven to be intriguing. At the same
time, AMS-02 is nearing completion and offers outstanding potential for
precision spectroscopy of cosmic rays. And the Fermi Gamma-ray Space
Telescope has been launched in June~2008 and might observe the
$\gamma$-ray signals of dark matter annihilations~\cite{ref:fermi}.\\
\par
It is important to remember that the dark matter conundrum cannot be
solved by any of these different approaches alone. If neutralinos are
produced by the accelerator, who will be able to tell whether they
really constitute the dark matter? A signal in the direct detection
experiments is necessary to confirm the presence of dark matter in the
solar neighbourhood. On the one hand, a~signature in the cosmic-ray spectra, and quite
possibly in $\gamma$-ray telescopes and neutrino observatories, would
go a long way towards establishing the nature of dark matter in the
Milky Way and possibly its neighbours. On the other hand, it seems
unlikely that the underlying model and its properties can be
pinpointed from these measurements alone, without the studies possible
at accelerators.\\
\par
The main focus of this thesis has been a design study for a potential
new player in the field, named Positron Electron Balloon Spectrometer
(PEBS). The design process has relied to a large extent on the
flexible and expandable Monte
Carlo simulation of the PEBS detector presented here. It allows the
variation of the design parameters and a prediction of the overall
detector performance. A reconstruction program was created, too, to
study the simulated events as one would do with those from the real detector. It
includes algorithms for track finding and fitting and the
determination of shower parameters in the calorimeter. An analysis suite was written
to study event properties on a statistical basis and extract the key
figures of merit for a given detector design.\\
\par
Intended for a measurement of the cosmic-ray positron fraction
on one or more flights at high altitude using a
long-duration balloon, PEBS has an unprecedentedly high acceptance of
almost $0.4\,\mathrm{m}^2\,\mathrm{sr}$. A first launch could take
place in~2012. Using a superconducting
magnet to create a mean magnetic field of $0.8\,\mathrm{T}$ and a
scintillating fibre tracker with silicon photomultiplier readout, it will
allow reliable charge-sign and momentum measurements up to at least
$100\,\mathrm{GeV}$. The novel silicon photomultiplier has received great
attention in the literature in recent years because of its many
advantages compared to ordinary photomultiplier tubes. The combination
with ultra-thin scintillating fibres of $250\,\mu\mathrm{m}$ diameter
results in the concept of a novel device for tracking of charged
particles, with applications beyond PEBS. Such a tracker could be robust, cover large areas and come
at a moderate price while offering spatial resolution on the order of
$70\,\mu\mathrm{m}$ or better. First prototype modules have been
subjected to proton testbeams at CERN and the proof of principle for the fibre tracker has been established.
The analysis of the data
gathered has provided crucial input for the PEBS simulation.\\
\par
The enormous challenge of reliably identifying positrons in front of
the vast proton background is tackled by a combination of two
independent subdetectors for particle identification. The
electromagnetic calorimeter will consist of layers of tungsten
absorber interleaved with scintillator bars, read out by
silicon photomultipliers. An analysis of the shower shape and a
comparison of the energy measured by the calorimeter to the momentum
obtained from the tracker allows for protons to be rejected at the
level of~1000 at efficiencies around $70\,\%$. A similar factor is
provided by the transition radiation
detector whose design is based on the one used for the AMS-02
detector. Using testbeam data acquired with a prototype for the
AMS-02-TRD, the accuracy of the simulation of transition radiation and
ionisation losses provided by the commonly used Geant4 package has been
studied. Excellent agreement was found between the transition
radiation spectra in data and simulation. Small discrepancies at the
$25\,\%$-level are
present in the tails of the proton energy loss spectra but this makes
the predicted proton rejections uncertain by a factor of two.\\
\par
For the case that dark matter is made up of neutralinos as predicted
by the popular mSUGRA model, the projected performance of PEBS and
AMS-02 has been studied. Looking first at the presently available data
on the positron fraction,
it was shown that the data show an excess in the positron fraction
at energies above roughly $10\,\mathrm{GeV}$ which seems impossible to
explain with common models for cosmic-ray propagation in the
Galaxy. The data could be fit with an mSUGRA model but both the
limited statistics and energy range of the available data do not allow for discrimination of
different models. While the PAMELA data are already starting to make inroads into the mSUGRA parameter
space, this situation might change dramatically with
precision data from AMS-02 or PEBS with their big exposure and energy
range. A scan of mSUGRA models was performed and models allowed by
observational constraints, most notably the neutralino relic
abundance, were identified. From a fit to the data, the most likely
signal normalisation, related to the amount of clumpiness in the
distribution of dark matter and expressed in terms of the boost
factor, was determined. It turns out that the required boost
factors are quite large and that they must differ significantly from those applicable
to antiprotons. Nevertheless, a $\chi^2$-value can then be calculated for a
fit of a given signal hypothesis to the projected data for PEBS and
AMS-02, determined by the acceptance and measurement time, and the range of models within a reasonable distance from the
$\chi^2$-minimum can be identified. The volume of parameter space of
the given model that will remain after such a procedure was shown to be
very small. In fact, the interpretation of the positron fraction measurements
will be ultimately limited by the systematic uncertainties at the level of
statistical accuracy provided by PEBS or AMS-02.\\
\par
This thesis is merely a snapshot of ongoing work. On the
software side, laboratory
and testbeam measurements of prototypes for the various subdetectors
will be used to make the simulation more and more accurate. More
layers of detail have to be added. For example, a
realistic map of the magnetic field has to be included and more
sophisticated track reconstruction routines have to be implemented as a
consequence. The physics programme for PEBS has to be developed
systematically. On the hardware side, the research and development
programme is actively pursued in several laboratories across
Europe. If it continues to be successful, the construction of PEBS will
be a challenging endeavour requiring many skilled individuals. With
luck, a significant scientific result may be achieved that will resonate
with both experts and the general public.

%~~~~~~~~~~~~~~~~~~~~~~~~~~~~~~~~

\pagestyle{empty}
\chapter*{Acknowledgements}
First and foremost, I would like to thank Professor Stefan Schael who
supervised the work for this thesis during the past four years. He let
me play a major role in the development of the PEBS project, for which
he is the driving force.
In countless meetings and discussions, he inspired and motivated me by
his passion for physics. He also gave me the opportunity to
attend several international conferences and schools which proved to
be invaluable and truly memorable experiences.\\
\par
Professor Christopher Wiebusch kindly agreed to serve as the second
referee for this thesis.\\
\par
The smallness of the PEBS group in Aachen is more than compensated for by
the talents of the gifted individuals forming part of it. Dr~Thomas
Kirn was always available for enlightening discussions about detector
physics and beyond. Because of his guidance and experience, the
testbeam measurements at CERN were both successful and
enjoyable. Gregorio Roper has the power to work magic with
any equipment that comes his way. I am especially indebted to him for
the hard work he did in preparing and conducting the testbeam
measurements described here and for the discussions about the analysis
of the data gathered there. Roman Greim put a lot of effort into the
preparation of the 2008~testbeam and has a very patient ear for
discussions about almost any subject.\\
\par
Dr~Jan Olzem is a source of knowledge in so many
areas, from physics to computing and far beyond, and he is never
reluctant to share it copiously. The collaboration with him in the
early stages of the work for this thesis was as efficient as it was
pleasant. Dr~Thorsten Siedenburg is exactly the
right person to ask when you
have run out of ideas. Admirably, he never gives up searching for the
answer to a problem. Philip von~Doetinchem calculated the
expected atmospheric backgrounds for PEBS. Over the years that we have
worked together,
starting our theses at the same institute at almost the same time, he
has become a close and trusted friend.\\
\par
Michael Wlochal and Waclaw Karpinski and their respective teams in the
mechanical and electronics workshops provided most of the hardware for
the testbeam measurements described here. In addition, together with
Arndt Schultz von Dratzig, they turned the wild ideas of the
physicists into an actual design for PEBS.\\
\par
Bastian Beischer and Niko Zimmermann are extremely talented and
resourceful students
who contributed to many different aspects of this work.\\
\par
I am deeply indebted to Dr~Vladik Balagura. His short visit to Aachen
brought a huge boost to the ECAL design, as well as to the ECAL simulation
and analysis. He taught me a great deal about calorimetry and inspired
me with his keen analytical thinking.\\
\par
It is a great pleasure to thank Dr~Giovanni Ambrosi and Dr~Philippe
Azzarello for their kind hospitality during my visit to Perugia. They
repeatedly allowed us to use their equipment and taught me how to
operate the beam telescope used in the 2008~testbeam.\\
\par
Professor Volker Blobel allowed me to use his software for robust
parameter estimation and patiently answered many questions related to it.\\
\par
Dr~Chan Hoon Chung did an earlier analysis of the cosmic-ray positron
fraction data in Aachen and shared his insights in cosmic-ray
propagation and SUSY codes with me.\\
\par
Although not directly involved with PEBS,
Dr~Chan Hoon~Chung, Dr~Alexander Furgeri, Andr\'e Goerres, Dr~Sonia
Natale, Dr~Jan Olzem, Dr~Andrei
Ostapchouk, and Dr~Georg Schwering helped to make
the testbeam campaign a success, by helping with the organisation or by
spending many hours on shifts at CERN.\\
\par
Dr~Thomas Kress, Dr~Michael Bontenackels, and Christoph Kukulies
permanently kept the computing facilities in Aachen in excellent
shape. Without the impressive batch cluster maintained by them, the
results in this thesis would have taken so much longer to obtain.\\
\par
Dr~Georg Schwering is the capable head of administration of the
institute. I do not remember any of my stupid questions that he did
not answer patiently.\\
\par
Thanks to my officemates over the years, Jan, Hendrik, Georgos, Roman
and Katja, the office was always a warm and happy place.\\
\par
Roman, Philip and especially Jan and Thomas read the
manuscript of this thesis carefully and improved it substantially.\\
\par
I owe a lot to the people I studied with, especially Klaus, Max,
Martin, Philipp, Philip, Stefan and Katja. The countless hours we
spent discussing and solving problems shaped my view of physics and,
in fact, the world. Thank you for all the fun.\\
\par
I cannot hope to repay with words what my family, Hans-Peter, Marlies
and Frauke, and what my future family, Astrid, did for me over the years. I could always
count on their unwavering support.
\cleardoublepage
{\selectlanguage{german}
{\large\bf Lebenslauf}\\[0.5cm]
\begin{tabular}{ll}
{\bf Pers\"onliche Daten}&\\[0.5cm]
Name&Henning Gast\\
Geburtsdatum&24.~M\"arz~1979\\
Geburtsort&D\"usseldorf\\[1cm]
{\bf Ausbildung}&\\[0.5cm]
seit 12/2007&wissenschaftlicher Mitarbeiter am I.~Physikalischen
Institut~B\\
&der RWTH Aachen\\
seit 12/2004&Promotion am I.~Physikalischen Institut B der RWTH Aachen,\\
&Stipendiat im Graduiertenkolleg\\
&\glqq{}Elementarteilchenphysik an der
TeV-Skala\grqq\\
12/2004&Diplom in Physik, Thema der Arbeit: \glqq{}Untersuchung der\\
&kosmischen H\"ohenstrahlung mit dem AMS01-Detektor\\
&im Weltraum\grqq\\
09/1999 - 12/2004&Studium der Physik an der RWTH Aachen\\
07/1998 - 04/1999&Wehrdienst, 2./JaboG 33 in B\"uchel\\
1998&Abitur\\
1989 - 1998&Georg-B\"uchner-Gymnasium, Kaarst\\
1985 - 1989&Grundschule Lichtenvoorder Stra"se, Kaarst-B\"uttgen\\
\end{tabular}
}

%~~~~~~~~~~~~~~~~~~~~~~~~~~~~~~~~

\begin{thebibliography}{99999}
\setlength{\itemsep}{-1mm}
%%%
%%% intro
%%%

%%%
%%% cosmology and dark matter
%%%
\bibitem{ref:kolb}
E.W.~Kolb and M.S.~Turner, The Early Universe, Addison-Wesley, 1990
\bibitem{ref:bergstroem_goobar}
L.~Bergstr\"om and A.~Goobar, Cosmology and Particle Astrophysics, Springer, 2006
\bibitem{ref:samtleben}
D.~Samtleben et al., Annu.~Rev.~Nucl.~Part.~Sci. {\bf 57} (2007) 245-283
\bibitem{ref:penzias}
A.A.~Penzias and R.W.~Wilson, ApJ {\bf 142} (1965) 419-421
\bibitem{ref:pdg}
W.-M. Yao et al. (Particle Data Group), J.~Phys.~G {\bf 33}, 1 (2006)
and 2007 partial update for edition 2008
\bibitem{ref:wmap_descr}
C.L.~Bennett et al., ApJ {\bf 583} (2003) 1-23
\bibitem{ref:sn1a}
B.~Leibundgut, Annu.~Rev.~Astron.~Astrophys. {\bf 39} (2001) 67-98
\bibitem{ref:bao}
W.J.~Percival et al., Mon.~Not.~R.~Astron.~Soc. {\bf 381} (2007) 1053-1066
\bibitem{ref:wmap_obs}
G.~Hinshaw et al., {\tt arXiv:0803.0732v2}, accepted for publication in ApJS
\bibitem{ref:wmap_cosm}
E.~Komatsu et al., {\tt arXiv:0803.0547v2}, accepted for publication in ApJS
\bibitem{ref:dmreview}
G.~Bertone et al., Phys.~Rep. {\bf 405} (2005) 279-390
\bibitem{ref:seljak}
U.~Seljak et al.,
JCAP 0610 (2006) 014
\bibitem{ref:kolbkk}
E.W.~Kolb and R.~Slansky, Phys.~Lett.~B {\bf 135} (1984) 378-382
\bibitem{ref:kkdm}
G.~Servant and T.M.P.~Tait, Nucl.~Phys.~B {\bf 650} (2003) 391-419
\bibitem{ref:axionrev}
S.J.~Asztalos et al., Annu.~Rev.~Nucl.~Part.~Sci. {\bf 56} (2006) 293-326
%%%
%%% SUSY
%%%
\bibitem{ref:kane}
G.L.~Kane et al., Phys.~Rev.~D {\bf 49} (1994) 6173
\bibitem{ref:aitchison}
I.J.R.~Aitchison, Supersymmetry and the MSSM: An Elementary Introduction, {\tt arXiv:hep-ph/0505105v1}
\bibitem{ref:tata}
X.~Tata, University of Hawaii report UH-511-872-97, {\tt arXiv:hep-ph/9706307v1}
\bibitem{ref:chamseddine}
A.H.~Chamseddine et al., Phys.~Rev.~Lett. {\bf 49} (1982) 970-974
\bibitem{ref:hall}
L.~Hall et al., Phys.~Rev.~D {\bf 27} (1983) 2359-2378
\bibitem{ref:barbieri}
R.~Barbieri et al., Phys.~Lett.~B {\bf 119} (1982) 343-347
\bibitem{ref:nath}
P.~Nath et al., Nucl.~Phys.~B {\bf 227} (1983) 121-133
\bibitem{ref:edsjophd}
J.~Edsj\"o, Aspects of Neutrino Detection of Neutralino Dark Matter, PhD thesis, Uppsala University 1997, {\tt arXiv:hep-ph/9704384}


%%%
%%% cosmic rays
%%%
\bibitem{ref:ginzburg}
V.L.~Ginzburg and S.I.~Syrovatskii, The Origin of Cosmic Rays, Pergamon Press, 1964
\bibitem{ref:gaisser}
T.K.~Gaisser, Cosmic Rays and Particle Physics, Cambridge University Press, 1990
\bibitem{ref:longair}
M.S.~Longair, High Energy Astrophysics, Cambridge University Press, 1992, reprinted with corrections 2004
\bibitem{ref:stanev}
T.~Stanev, High Energy Cosmic Rays, Springer, 2004
\bibitem{ref:hess}
V.F.~Hess, Physik.~Zeitschr. {\bf 13} (1912) 1084-1091
\bibitem{ref:anderson}
C.D.~Anderson, Phys.~Rev. {\bf 43} (1933) 491
\bibitem{ref:anderson_muon}
S.H.~Neddermeyer and C.D.~Anderson, Phys.~Rev. {\bf 51} (1937) 884
\bibitem{ref:street_muon}
J.C.~Street and E.C.~Stevenson, Phys.~Rev. {\bf 52} (1937) 1003
\bibitem{ref:perkins_pion}
D.H.~Perkins, Nature {\bf 159} (1947) 126
\bibitem{ref:kaon}
G.D.~Rochester and C.C.~Butler, Nature {\bf 160} (1947) 855
\bibitem{ref:pa_agn}
The Pierre Auger collaboration, Science {\bf 318} (2007) 938
\bibitem{ref:hesssnr}
F.~Aharonian et al., Astron.~Astrophys. {\bf 464} (2007) 235-243, {\tt arXiv:astro-ph/0611813v1}
\bibitem{ref:maurinprop}
D.~Maurin et al., ApJ {\bf 555} (2001) 585-596
\bibitem{ref:galpropposelec}
I.V.~Moskalenko and A.W.~Strong, ApJ {\bf 493} (1998) 694-707
\bibitem{ref:galpropnucl}
A.W.~Strong and I.V.~Moskalenko, ApJ {\bf 509} (1998) 212-228
\bibitem{ref:galproppbar}
I.V.~Moskalenko et al., ApJ {\bf 565} (2002) 280-296
\bibitem{ref:galpropmhd}
V.S.~Ptuskin et al., ApJ {\bf 642} (2006) 902-916
\bibitem{ref:galproprev}
A.W.~Strong et al., Annu.~Rev.~Nucl.~Part.~Sci. {\bf 57} (2007) 285-327
\bibitem{ref:solarmod}
L.J.~Gleeson and W.I.~Axford, ApJ {\bf 154} (1968) 1011-1026
\bibitem{ref:cutoff}
T.~Mizuno et al., Nucl.~Sci.~Symp.~Conf.~Rec. (2001) IEEE Vol.~1, 442-446
\bibitem{ref:auger}
J.~Abraham et al., Nucl.~Instr.~Meth.~A {\bf 523} (2004) 50-95
\bibitem{ref:philip}
P.~von~Doetinchem, PhD thesis in preparation, RWTH Aachen
\bibitem{ref:p_ams}
J.~Alcaraz et al., Phys.~Lett.~B {\bf 490} (2000) 27-35
\bibitem{ref:he_ams}
J.~Alcaraz et al., Phys.~Lett.~B {\bf 494} (2000) 193-202
\bibitem{ref:electrons_heat}
M.A.~DuVernois et al., ApJ {\bf 559} (2001) 296-303
\bibitem{ref:ams01old}
J.~Alcaraz et al., Phys.~Lett.~B {\bf 484} (2000) 10-22
\bibitem{ref:bess98}
T.~Maeno et al., Astropart.~Phys. {\bf 16} (2001) 121-128
\bibitem{ref:capricepbar}
M.~Boezio et al., ApJ {\bf 561} (2001) 787
\bibitem{ref:egret}
S.D.~Hunter et al., ApJ {\bf 481} (1997) 205-240
\bibitem{ref:p_bess93}
J.Z.~Wang et al., ApJ {\bf 564} (2002) 244-259
\bibitem{ref:p_bess}
Y.~Shikaze et al., Astropart.~Phys. {\bf 28} (2007) 145-167
\bibitem{ref:p_pamela}
M.~Casolino et al., {\tt arXiv:astro-ph/0810.4980v1}
\bibitem{ref:electrons_nishimura}
J.~Nishimura et al., Adv.~Space Res. {\bf 26} (No 11) (2000) 1827-1830
\bibitem{ref:electrons_bets}
S.~Torii et al., ApJ {\bf 559} (2001) 973-984
\bibitem{ref:ams01new}
M.~Aguilar et al., Phys.~Lett.~B {\bf 646} (2007) 145-154
\bibitem{ref:heat}
J.J.~Beatty et al., Phys.~Rev.~Lett. {\bf 93} (2004) 241102
\bibitem{ref:caprice}
M.~Boezio et al., ApJ {\bf 532} (2000) 653-669
\bibitem{ref:ts93}
R.L.~Golden et al., ApJ {\bf 457} (1996) L103-L106
\bibitem{ref:olzem}
J.~Olzem, Signatures of SUSY Dark Matter at the LHC and in the Spectra of Cosmic Rays, PhD thesis, RWTH Aachen, 2007
\bibitem{ref:pamela_posfrac}
O.~Adriani et al., {\tt arXiv:astro-ph/0810.4995v1}

%%% B/C
\bibitem{ref:bc_ace}
A.J.~Davis et al., Proc.~ACE 2000 Symposium, AIP, 421
\bibitem{ref:bc_cm}
J.H.~Caldwell and P.~Meyer, Proc.~15th ICRC (1977),1,243
\bibitem{ref:bc_cw}
J.H.~Chapell and W.R.~Webber, Proc.~17th ICRC (1981) 2,59
\bibitem{ref:bc_dm}
R.~Dwyer and P.~Meyer, ApJ {\bf 322} (1987) 981-991
\bibitem{ref:bc_imp8}
M.~Garcia-Munoz et al., ApJ {\bf 280} (1984) L13-L17
\bibitem{ref:bc_gw}
M.~Gupta and W.R.~Webber, ApJ {\bf 340} (1989) 1124-1134
\bibitem{ref:bc_heao3}
J.J.~Engelmann et al., Astron. Astrophys. {\bf 233} (1990) 96-111
\bibitem{ref:bc_isee3}
K.E.~Krombel and M.E.~Wiedenbeck, ApJ {\bf 328} (1988) 940-953
\bibitem{ref:bc_juliusson}
E.~J\'uliusson, ApJ {\bf 191} (1974) 331-348
\bibitem{ref:bc_lw}
J.A.~Lezniak and W.R.~Webber, ApJ {\bf 223} (1978) 676-696
\bibitem{ref:bc_maehl}
R.C.~Maehl et al., Astrophysics and Space Science {\bf 47} (1977) 163-184
\bibitem{ref:bc_orth}
C.D.~Orth et al., ApJ {\bf 226} (1978) 1147-1161
\bibitem{ref:bc_simon}
M.~Simon et al., ApJ {\bf 239} (1980) 712-724
\bibitem{ref:bc_ulysses}
M.A.~DuVernois et al., Astron. Astrophys. {\bf 316} (1996) 555-563
\bibitem{ref:bc_voyager}
A.~Lukasiak et al., Proc.~26th ICRC (1999) OG1.1.12
%%%
%%%
%%%
\bibitem{ref:cowan}
G.~Cowan, Statistical Data Analysis, Oxford University Press, 1998
\bibitem{ref:clem}
J.M.~Clem et al., ApJ {\bf 464} (1996) 507-515
\bibitem{ref:babcock}
H.D.~Babcock, ApJ {\bf 130} (1959) 364
\bibitem{ref:sunspotnumber}
Solar Influences Data Analysis Center, {\tt http://sidc.oma.be/sunspot-data/}
\bibitem{ref:snodgrass}
H.B.~Snodgrass et al., Solar Physics {\bf 191} (2000) 1-19
\bibitem{ref:durrant}
C.J.~Durrant and P.R.~Wilson, Solar Physics {\bf 214} (2003) 23-29
\bibitem{ref:pamela_pbarp}
O.~Adriani et al., Phys.~Rev.~Lett. {\bf 102} (2009) 051101
\bibitem{ref:pbarp_besspolar}
K.~Abe et al., Phys.~Lett.~B {\bf 670} (2008) 103-108
\bibitem{ref:bess99}
Y.~Asaoka et al., Phys.~Rev.~Lett. {\bf 88(5)} (2002) 051101


%%%
%%% PEBS intro
%%%
\bibitem{ref:vcihenning}
P.~von Doetinchem, H.~Gast, T.~Kirn, G.~Roper Yearwood, and S.~Schael, Nucl.~Instr.~Meth.~A {\bf 581} (2007) 151-155
\bibitem{ref:icrchenning}
H.~Gast, P.~von Doetinchem, T.~Kirn, G.~Roper Yearwood, and S.~Schael, Proc.~$30^\mathrm{th}$ International Cosmic Ray Conference,
M\'erida,~Mexico, July~2007, Vol.~2, 293-296
\bibitem{ref:vcigregorio}
H.~Gast, T.~Kirn, G.~Roper Yearwood, and S.~Schael, Nucl.~Instr.~Meth.~A {\bf 581} (2007) 423-426
\bibitem{ref:pamelaoverview}
P.~Picozza et al., PAMELA -- A payload for antimatter matter exploration and light-nuclei astrophysics, Astropart.~Phys. {\bf 27} (2007) 296-315
\bibitem{ref:pamelalaunch}
M.~Casolino et al., J.~Adv.~Space~Res. (2007), doi:10.1016/j.asr.2007.07.023
\bibitem{ref:pamelainflight}
P.~Papini et al., Nucl.~Instr.~Meth.~A {\bf 588} (2008) 259-266
\bibitem{ref:pamelaabundances}
V.~Malvezzi, Nucl.~Instr.~Meth.~A {\bf 588} (2008) 250-254
\bibitem{ref:pamelaspatialres}
S.~Straulino et al., Nucl.~Instr.~Meth.~A {\bf 556} (2006) 100-114
\bibitem{ref:pamelaecal}
M.~Boezio et al., Astropart.~Phys. {\bf 26} (2006) 111-118
\bibitem{ref:amsbattiston}
R.~Battiston, Nucl.~Instr.~Meth.~A (2008), doi:10.1016/j.nima.2008.01.044
\bibitem{ref:gregorio}
G.~Roper, Development of a high resolution tracking detector with SiPM readout, graduate thesis, RWTH~Aachen, April~2007
\bibitem{ref:gregoriophd}
G.~Roper, PhD thesis in preparation, RWTH~Aachen
\bibitem{ref:cream}
H.S.~Ahn et al., Nucl.~Instr.~Meth.~A {\bf 579} (2007) 1034-1053
\bibitem{ref:atic}
A.D.~Panov et al., Adv.~Space~Res. {\bf 37} (2006) 1944-1949
\bibitem{ref:bessprogram}
A.~Yamamoto et al., Nucl.~Phys.~B (Proc.~Suppl.) {\bf 166} (2007) 62-67
\bibitem{ref:capriceinstr}
R.L.~Golden et al., Nucl.~Instr.~Meth.~A {\bf 306} (1991) 366-377
\bibitem{ref:heatinstr}
S.W.~Barwick et al., Nucl.~Instr.~Meth.~A {\bf 400} (1997) 34-52
\bibitem{ref:tracer}
M.~Ave et al., ApJ {\bf 678} (2008) 262-273
\bibitem{ref:isomax}
M.~Hof et al., Nucl.~Instr.~Meth.~A {\bf 454} (2000) 180-185

%%%
%%% PEBS
%%%
\bibitem{ref:g4}
S.~Agostinelli et al., Nucl.~Instr.~Meth.~A {\bf 506} (2003) 250-303
\bibitem{ref:gluckstern}
R.L.~Gluckstern, Nucl.~Instr.~Meth. {\bf 24} (1963) 381-389
\bibitem{ref:roman}
R.~Greim, Lineare Silizium-Photomultiplier-Arrays, graduate thesis (in German), RWTH~Aachen, September~2008
\bibitem{ref:sipm1}
B.~Dolgoshein et al., Nucl.~Instr.~Meth. A {\bf 563} (2006), 368-376
\bibitem{ref:sipm2}
M.~McClish et al., Nucl.~Instr.~Meth. A {\bf 567} (2006), 36-40
\bibitem{ref:sipm3}
P.~Buzhan et al., Nucl.~Instr.~Meth. A {\bf 567} (2006), 353-355
\bibitem{ref:sipm_renker}
D.~Renker, Nucl.~Instr.~Meth. A {\bf 567} (2006), 48-56
\bibitem{ref:bicronfibres}
BICRON catalogue, Saint Gobain Ceramics \& Plastics, Inc.,~2005
\bibitem{ref:vladikecal}
V.~Balagura et al., Nucl.~Instr.~Meth. A {\bf 564} (2006) 590-596
\bibitem{ref:vladikprivate}
V.~Balagura, private communication
\bibitem{ref:urnmodel}
N.L.~Johnson and S.~Kotz, Urn models and their application, Wiley~1977
\bibitem{ref:calorimetry_fabjan}
C.W.~Fabjan and F.~Gianotti, Calorimetry for Particle Physics, CERN-EP/2003-075
\bibitem{ref:wigmans}
R.~Wigmans, Calorimetry, Oxford University Press, 2000
\bibitem{ref:berger}
C.~Berger, Elementarteilchenphysik, Springer, 2002
\bibitem{ref:barlowbook}
R.~Barlow, Statistics: A Guide to the Use of Statistical Methods in the Physical Sciences, Wiley VCH 1994
\bibitem{ref:jackson}
J.D.~Jackson, Classical Electrodynamics, 3$^\mathrm{rd}$ edition, John Wiley \& Sons, 1999
\bibitem{ref:dolgoshein_trd}
B.~Dolgoshein, Nucl.~Instr.~Meth. A {\bf 326} (1993) 434-469
\bibitem{ref:egorytchev}
V.~Egorytchev et al., Nucl.~Instr.~Meth. A {\bf 453} (2000) 346-352
\bibitem{ref:struczinski}
C.W.~Fabjan and W.~Struczinkski (sic!), Phys.~Lett.~B {\bf 57} (1975) 483-486
\bibitem{ref:ams02}
P.~v.~Doetinchem et al., Nucl.~Instr.~Meth. A {\bf 558} (2006) 526-535
\bibitem{ref:fopp}
S.~Fopp, Entwicklung und Bau eines auf Proportionalkammern basierenden \"Ubergangsstrahlungsdetektors
f\"ur das AMS-02-Weltraumexperiment, PhD thesis (2004), RWTH Aachen
\bibitem{ref:g4physics}
Geant4 Physics Reference Manual
%{\tt http://geant4.web.cern.ch/geant4/UserDocumentation/UsersGuides/PhysicsReferenceManual/html/index.html}
\bibitem{ref:g4hadr}
J.~Apostolakis et al., CERN-LCGAPP-2007-02
\bibitem{ref:g4binarycascade}
G.~Folger et al., Eur.~Phys.~J.~A {\bf 21} (2004) 407
\bibitem{ref:trdstudy}
B.~Beischer et al., Nucl.~Instr.~Meth. A {\bf 583} (2007) 485-493
\bibitem{ref:tr_spectra}
A.~Andronic et al., Transition radiation spectra of electrons from 1 to 10 GeV/$c$ in regular and irregular radiators,
Nucl.~Instr.~Meth.~A {\bf 558} (2006) 516-525
\bibitem{ref:g4pai}
J.~Apostolakis et al., Nucl.~Instr.~Meth. A {\bf 453} (2000) 597-605
\bibitem{ref:grichine_nim_02}
V.M.~Grichine, Nucl.~Instr.~Meth. A {\bf 484} (2002) 573-586
\bibitem{ref:g4trpackage}
V.M.~Grichine, S.S.~Sadilov, Nucl.~Instr.~Meth. A {\bf 563} (2006) 299-302
\bibitem{ref:grichinetrd}
V.M.~Grichine, Phys.~Lett.~B {\bf 525} (2002) 225-239
\bibitem{ref:orboeck}
J.~Orboeck, The final 20-Layer-Prototype for the AMS Transition Radiation Detector:
Beamtests, Data-Analysis, MC-Studies, PhD thesis (2003), RWTH Aachen
\bibitem{ref:grichine_nima_04}
V.M.~Grichine, S.S.~Sadilov, Nucl.~Instr.~Meth. A {\bf 522} (2004) 122-125
\bibitem{ref:allison_pai}
W.W.M.~Allison, J.H.~Cobb, Ann.~Rev.~Nucl.~Part.~Sci. {\bf 30} (1980) 253-298
\bibitem{ref:beamx7}
L.~Gatignon, The West Experimental Area at the CERN
SPS, CERN~SL-2000-016~EA
\bibitem{ref:hough}
R.O.~Duda and P.E.~Hart, Comm.~ACM {\bf 15} no.~1 (1972) 11-15
\bibitem{ref:blobel_robustfit}
V.~Blobel, Nucl.~Instr.~Meth. A {\bf 566} (2006) 14-17
\bibitem{ref:gsl}
GNU Scientific Library, {\tt http://www.gnu.org/software/gsl/}
\bibitem{ref:acceptance}
J.D.~Sullivan, Nucl.~Instr.~Meth. {\bf 95} (1971) 5-11

%%%
%%% Testbeam
%%%
\bibitem{ref:cms}
S.~Chatrchyan et al. (CMS collaboration), JINST {\bf 3} (2008) S08004
\bibitem{ref:cmsmodules}
L.~Borrello et al., CMS note 2003/020
\bibitem{ref:cmsapv}
M.~Friedl et al., Nucl.~Instr.~Meth.~A {\bf 488} (2002) 175-183
\bibitem{ref:cmsarc}
%M.~Axer et al., CMS internal note 2001/046
M.~Axer et al., Nucl.~Instr.~Meth. A {\bf 518} (2004) 321-323
\bibitem{ref:kuraray}
Kuraray Co Ltd, Scintillation Materials
\bibitem{ref:ideas}
Gamma Medica - Ideas (Norway) AS, {\tt http://www.ideas.no}
\bibitem{ref:amsmodules}
P.~Azzarello, Tests And Production Of The AMS-02 Silicon Tracker Detectors, PhD thesis, Universit\'e de Gen\`eve, 2004
\bibitem{ref:cristinziani}
M.~Cristinziani, Search for Heavy Antimatter and Energetic Photons in Cosmic Rays with the AMS-01 Detector in Space, PhD thesis, Universit\'e de Gen\`eve, 2002
\bibitem{ref:amsdaq}
A.~Kounine et al., AMS-2 DAQ software organisation, xDR and JINx nodes, European Organization for Nuclear Research draft, unpublished AMS document
\bibitem{ref:romanphd}
R.~Greim, PhD thesis in preparation, RWTH~Aachen

%%%
%%% Susyscan
%%%
\bibitem{ref:turner}
M.S.~Turner and F.~Wilczek, Phys.~Rev.~D {\bf 42} (1990) 1001-1007
\bibitem{ref:kamionkowski}
M.~Kamionkowski and M.S.~Turner, Phys.~Rev.~D {\bf 43} (1991) 1774-1780
\bibitem{ref:darksusyprop}
E.A.~Baltz and J.~Edsj\"o, Phys.~Rev.~D {\bf 59} (1998) 023511
\bibitem{ref:baltz_positron_excess}
E.A.~Baltz et al., Phys.~Rev.~D {\bf 65} (2002) 063511
\bibitem{ref:bfs}
J.~Lavalle et al., A\&A {\bf 479} (2008) 427-452
\bibitem{ref:kanewangwang}
G.L.~Kane et al., Phys.~Lett.~B {\bf 536} (2002) 263-269
\bibitem{ref:hoopertaylor}
D.~Hooper et al., Phys.~Rev.~D {\bf 69} (2004) 103509
\bibitem{ref:hoopersilk}
D.~Hooper and J.~Silk, Phys.~Rev.~D {\bf 71} (2005) 083503
\bibitem{ref:isajet}
H.~Baer et al., {\tt arXiv:hep-ph/0312045}, latest version available
at {\tt http://www.hep.fsu.edu/$\sim$isajet}
\bibitem{ref:ds}
P.~Gondolo et al., JCAP 0407 (2004) 008
\bibitem{ref:mm}
G.~B\'elanger et al., Comput.~Phys.~Commun. {\bf 176} (2007) 367\\
G.~B\'elanger et al., Comput.~Phys.~Commun. {\bf 174} (2006) 577\\
G.~B\'elanger et al., Comput.~Phys.~Commun. {\bf 149} (2002) 103
\bibitem{ref:wmap}
D.N.~Spergel et al., ApJ Suppl.~Ser. {\bf 170} (2007) 377-408
\bibitem{ref:relicdensitycalc}
J.~Edsj\"o and P.~Gondolo, Phys.~Rev.~D {\bf 56} (1997) 1879
\bibitem{ref:mtop08}
Tevatron Electroweak Working Group, {\tt arXiv:0803.1683v1}
\bibitem{ref:e821}
G.W.~Bennett et al., Phys.~Rev.~D {\bf 73} (2006) 072003
%\bibitem{ref:misiak}
%M.~Misiak et al., Phys.~Rev.~Lett. {\bf 98} (2007) 022002
\bibitem{ref:baudis}
L.~Baudis, submitted to Proc.~SUSY07, {\tt arXiv:0711.3788v1}
% direct detection limits
\bibitem{ref:cdmslimit}
Z.~Ahmed et al., {\tt arXiv:0802.3530} (UCLA 2008 DM conference)
\bibitem{ref:xenon10limit}
J.~Angle et al., Phys.~Rev.~Lett. {\bf 100} (2008) 021303
\bibitem{ref:coupplimit}
E.~Behnke et al., Science {\bf 319} (2008) 933
\bibitem{ref:kimslimit}
H.S.~Lee et al., Phys.~Rev.~Lett. {\bf 99} (2007) 091301
%%
\bibitem{ref:pythia}
T.~Sj\"ostrand, Comp.~Phys.~Comm. {\bf 82} (1994) 74-89
\bibitem{ref:mtop07}
Tevatron Electroweak Working Group, {\tt arXiv:hep-ex/0703034v1}
%\bibitem{ref:dokuchaev}
%V.~Berezinsky et al., Phys.~Rev.~D {\bf 77} (2008) 083519
\bibitem{ref:imbhs}
P.~Brun et al., Phys.~Rev.~D {\bf 76} (2007) 083506
\bibitem{ref:silksommerfeld}
M.~Lattanzi and J.~Silk, {\tt astro-ph:0812.0360v1}
\bibitem{ref:baer}
H.~Baer et al., JCAP08 (2004) 005
\bibitem{ref:bessexp}
Y.~Ajima et al., Nucl.~Instr.~Meth.~A {\bf 443} (2000) 71-100
\bibitem{ref:barlow}
R.~Barlow, MAN/HEP/04/02, {\tt arXiv:physics/0406120}
\bibitem{ref:dbarsusy}
R.~Battiston, Proposal for an ASI - DLR research program on Astroparticle Physics from LDB at the North Pole, talk given at 1st workshop
on science and technology through long duration balloons, Rome, Italy, June 2008
\bibitem{ref:deboercritique}
L.~Bergstr\"{o}m et al., JCAP05 (2006) 006
\bibitem{ref:egretsystematics}
F.W.~Stecker et al., Astropart.~Phys. {\bf 29} (2008) 25-29
\bibitem{ref:galpropoptimized}
A.W.~Strong et al., ApJ {\bf 613} (2004) 962-976
\bibitem{ref:dbarbg}
%F.~Donato et al., Phys.~Rev.~D {\bf 62} (2000) 043003
F.~Donato et al., Torino U. preprint DFTT 06/2007, {\tt arXiv:0803.2640v1}
\bibitem{ref:bess9597}
S.~Orito et al., Phys.~Rev.~Lett. {\bf84(6)} (2000) 1078
\bibitem{ref:bess01}
K.~Yamato et al., Phys.~Lett.~B {\bf 632} (2006) 475-479
\bibitem{ref:profumo_pulsars}
S.~Profumo, {\tt arXiv:astro-ph/0812.4457v1}
\bibitem{ref:fermi}
E.A.~Baltz et al., JCAP07 (2008) 013
\bibitem{ref:cirelli}
M.~Cirelli et al., {\tt arXiv:hep-ph/0809.2409v3}, to appear in Nucl.~Phys.~B
\bibitem{ref:cholis}
I.~Cholis et al., {\tt arXiv:astro-ph/0811.3641v1}
\bibitem{ref:yin}
P.~Yin et al., {\tt arXiv:hep-ph/0811.0176v2}

\end{thebibliography}
\end{document}